\documentclass[a4paper,12pt]{thesis_eng}             
\pdfoutput=1

\usepackage{natbib}         
\bibpunct{(}{)}{;}{a}{,}{,} 

 \usepackage[spanish,czech,english]{babel} 
\usepackage[utf8]{inputenc} 

\def\firstname{Edgar S.}
\def\familyname{Carlin Ram\'irez}
\def\FileAuthor{\firstname \familyname}
\def\FileTitle{\firstname \familyname's Ph.D. Thesis}
\def\FileSubject{Ph.D. Thesis}
\def\FileKeyWords{\firstname \familyname, Ph.D., Thesis}

\usepackage{ifpdf}              
\usepackage{url}                
\usepackage[toc,page]{appendix} 
\usepackage{overrulehere}       
\ifpdf     
  \usepackage[pdftex,bookmarks=false, breaklinks, baseurl=http://, pdfpagemode=UseNone, pdfstartview=XYZ, pdfstartpage=1,
   colorlinks= true, 
   linkcolor=blue,  
   citecolor=blue,  
   urlcolor=blue]   
   {hyperref}       
  \hypersetup{ 
    pdfauthor   = \FileAuthor, 
    pdftitle    = \FileTitle,  
    pdfsubject  = \FileSubject,
    pdfkeywords = \FileKeyWords}
\else
  \usepackage[dvips]{hyperref}
\fi
\usepackage{verbatim}           
\usepackage[Lenny]{fncychap}    

\usepackage{type1cm}

\usepackage{lettrine}

\usepackage{latexsym}   
\usepackage{amssymb}    
\usepackage{amsfonts}   
\usepackage{amstext}    
\usepackage{amsmath}    
\usepackage{mathtools}
\usepackage{braket}
\usepackage{bm}

\usepackage{graphicx}
\usepackage{color}
\usepackage{empheq} 
\usepackage{MnSymbol} 
\usepackage[labelfont=scriptsize,aboveskip=0.1pt]{caption}
\usepackage[labelfont=scriptsize,aboveskip=0.1pt]{subcaption}
\usepackage{floatfig}
\usepackage{multirow}                
\usepackage{multicol}                
\usepackage{deluxetable}             

\usepackage{footnote}                
\makesavenoteenv{tabular}
\usepackage{overrulehere}            
\usepackage{enumitem}

\def\j13{\textit{Swift}\,J1357.2-0933}   
\def\j18{XTE\,J1859+226}                 
\def\j11{XTE\,J1118+480}                 
\def\a0 {A0620-003}                      
\def\v4 {V404\,Cyg}                      
\def\nv93{Nova Vel\,93}                  
\def\no77{Nova Oph\,77}                  
\def\nm91{Nova Mus\,91}                  

\def\deg{\,\mbox{$^{\circ}$}}            
\def\arcsec{\hbox{$^{\prime\prime}$}}  

\def\2mdiag{(J-H)-(H-Ks)}                

\definecolor{orange}{RGB}{255,127,0}
\definecolor{gold}{RGB}{255,215,0}
\definecolor{goldenrod}{RGB}{218,165,32}

\newcommand{\clearemptydoublepage}{\newpage{\pagestyle{empty}\cleardoublepage}}

\def\nar{\aaref@jnl{NewAR}}               

\newcommand{\sixj}[6] {\begin{Bmatrix}
#1&#2&#3\\
#4&#5&#6 \end{Bmatrix}}

\renewcommand{\sixj}[6]
{ \left\{\begin{array}{ccc}
#1&#2&#3\\
#4&#5&#6
\end{array}\right\} }

\begin{document}
\initfloatingfigs
\frenchspacing   
\frontmatter

\def\titleth{\bf {\LARGE Generation and Transfer 
of Polarized Radiation \\
in Hydrodynamical Models 
of the \\
Solar Chromosphere.}}
\def\authorth{\textbf{Edgar S. Carlin Ram\'irez}}
\def\submitdateth{\ifcase\the\month\or
  enero\or febrero\or marzo\or abril\or mayo\or junio\or
  julio\or agosto\or septiembre\or octubre\or noviembre\or diciembre\fi
  \space de\space \number\the\year}

\def\supervisorB{Prof. Andr\'es Asensio Ramos}
\def\supervisorA{Dr. Javier Trujillo Bueno}
\def\supervisors{ Dr. Andr\'es Asensio Ramos \& Prof. Javier Trujillo Bueno  }

\title{\titleth}
\author{\authorth} 

\date{}                  
\titulobonito            

\vspace*{\fill}   
{\small
\noindent
Examination date: December 2013. \\  
Thesis supervisors:\\
\supervisors \\
\\
\copyright\, \authorth, \space \number\the\year
\\
ISBN: xx-xxx-xxxx-x\\
Dep\'osito legal: TF-1207/2013\\
\\
\\
Some of the material included in this document has been already
published in 
\textit{The Astrophysical Journal}. \\
Parte del material incluido en este documento ya ha sido publicado 
en  
\textit{The Astrophysical Journal}.\\
}

\clearemptydoublepage
\hyphenation{averaged}


 \section*{\centering Resumen}
\addcontentsline{toc}{chapter}{Resumen}

\selectlanguage{spanish}  

El principal objetivo de esta tesis ha sido investigar el efecto que los gradientes de velocidad vertical tienen en las se\~nales de polarizaci\'on por scattering formadas en la cromosfera solar. Seguimos un enfoque te\'orico basado en la s\'intesis espectral de se\~nales de polarizaci\'on en modelos din\'amicos del Sol en calma. Los movimientos macrosc\'opicos nunca hab\'ian sido considerados en el tratamiento de se\~nales de polarizaci\'on producidas por procesos de scattering y efecto Hanle. Esto es especialmente importante en la cromosfera solar, dado su fuerte dinamismo y su reducida intensidad de campo magn\'etico. El estudio se centra en el an\'alisis
de las l\'ineas del triplete infrarrojo del Ca {\sc ii} (en $8498$, $8542$ y $8662$ {\AA}). La metodolog\'ia de s\'intesis de perfiles de Stokes permite confrontar los modelos cromosf\'ericos con observaciones.

Resolvimos el problema NLTE del transporte y generaci\'on de radiaci\'on polarizada en sistemas at\'omicos multinivel usando modelos de atm\'osfera solar con creciente nivel de realismo: atm\'osferas Milne-Eddington, atm\'osfera compuesta por \'atomos de dos niveles, modelos semiemp\'iricos con velocidades ad-hoc, series temporales de modelos hidrodin\'amicos y una captura instant\'anea de una simulaci\'on MHD tridimensional. Para ello incluimos la acci\'on de los campos de velocidad sobre la polarizaci\'on at\'omica. Primero estudiamos el impacto de los gradientes de velocidad en la anisotrop\'ia del campo de radiaci\'on, la cual controla la polarizaci\'on por scattering; luego mostramos los efectos de modulaci\'on en amplitud, desplazamiento espectral y asimetrizaci\'on que los gradientes de velocidad vertical y las ondas de choque cromosf\'ericas producen sobre los perfiles de de polarizaci\'on lineal a campo cero; finalmente, estudiamos la polarizaci\'on emergente en geometr\'ia de forward scattering incluyendo el efecto Hanle producido por el campo magn\'etico.

 As\'i sintetizamos la primera tomograf\'ia de un modelo de la cromosfera de Sol en calma que combina mapas de polarizaci\'on por scattering y efecto Hanle junto con mapas de polarizaci\'on circular producidos por efecto Zeeman. Nos centramos en el uso de estos mapas para el diagn\'ostico de la topolog\'ia espacial del campo magn\'etico, del estado termodin\'amico de la atm\'osfera y de la estratificaci\'on de velocidades. Estudiamos tambi\'en la relevancia de la termodin\'amica y la din\'amica en el c\'alculo de la orientaci\'on del campo magn\'etico en presencia de las ambig\"uedades de $90\deg$ y $180\deg$. Adem\'as, simulamos observaciones degradando los mapas de polarizaci\'on resultantes tal y como har\'ia el futuro telescopio espacial Solar-C con observaciones reales, reconstruy\'endolas posteriormente mediante varios m\'etodos (por ejemplo, PCA). Encontramos que Solar-C y EST deber\'ian ser capaces de medir comportamientos similares a los simulados en esta tesis para el triplete IR del Ca {\sc ii}. 

Dedicamos un cap\'itulo a las herramientas y procedimientos t\'ecnicos desarrollados: c\'odigos de transporte radiativo, m\'etodo de adaptaci\'on de redes num\'ericas para mejorar la convergencia de los c\'odigos de transporte, c\'odigo de an\'alisis de componentes principales, herramienta de c\'alculo de funciones respuesta en l\'ineas cromosf\'ericas y t\'ecnicas de visualizaci\'on y an\'alisis tridimensional.

 \clearemptydoublepage    

 \section*{\centering Summary}
\addcontentsline{toc}{chapter}{Summary}

\selectlanguage{english}

The main goal of this thesis has been to investigate the effect that the macroscopic vertical velocity fields have on the scattering polarization signals formed in the solar chromosphere. We followed a theoretical approach based on the spectral synthesis of scattering polarization signals in dynamic models of the quiet Sun. Until now, the impact of macroscopic motions had never been considered in the treatment of the polarization signals produced by scattering processes and the Hanle effect. This is especially important in the solar chromosphere, given its strong dynamism and reduced magnetic field intensity. This investigation focuses in the analysis of the Ca {\sc ii} IR triplet lines (at $8498$, $8542$ and $8662$ {\AA}). The methodology of spectral synthesis allows to confront chromospheric models with real observations.

We solved the multilevel, non-LTE radiative problem of the generation and transfer of polarized radiation in increasingly realistic atmosphere models: Milne-Eddington atmospheres, atmospheres composed by two-level atoms, semiempirical models with ad-hoc velocities, hydrodynamical time-dependent models and a snapshot of a 3D MHD simulation. To such end, we included the action of the velocity fields on the atomic level polarization. Thus, we studied the impact of velocity gradients on the anisotropy of the radiation field, which controls the scattering polarization. We showed the effects of amplitude modulation, spectral shift and asymmetry that the vertical velocity gradients have on the zero-field linear polarization profiles; finally, we studied the emergent polarization in a forward scattering geometry including as well the Hanle effect produced by the magnetic field. 

Thus, we obtained the first tomographic view of a model quiet chromosphere that includes synthetic maps of linear polarization dominated by Hanle effect and of circular polarization dominated by Zeeman effect. We focused on the use of such maps to diagnose the spatial topology of the magnetic field, the thermodynamical state of the atmosphere and the vertical stratification of velocity. 
We also studied the relevance of dynamic and thermodynamic in the calculation of the chromospheric magnetic field orientation in the presence of the $90\deg$ and $180\deg$ ambiguities. Furthermore, we simulated synthetic observations by degrading our polarization maps, as the space telescope Solar-C would do with real observations, and we reconstructed them by following several methods (e.g., Principal Component Analysis). We found that Solar-C and the European Solar Telescope should be able to capture bevaviors similar to the ones simulated in this thesis for the Ca {\sc ii} IR triplet lines.

We dedicated a chapter to the tools and technical procedures developed in this thesis: the RT code; an adaptative method for numerical grids that improves the convergence of the RT calculation; a PCA code; a program to calculate response functions for chromospheric lines; and finally, some techniques for three-dimensional analysis and visualization.

 \clearemptydoublepage 

 \tableofcontents       
 \clearemptydoublepage  
 \mainmatter
 \clearpage
 \renewcommand{\thepage}{\arabic{page}}  
 \newpage

\chapter{Introduction}\label{cap:introduction}
In $1892$, Charles A. Young wrote \citep{wiki:2012aa}: 

``\textit{...This outer envelope...seems to be made up not of overlying strata
of different density, but rather of flames, beams and streamers, as
transient as those of our own aurora borealis. It is divided into two
portions...the outer portion...may almost, without exaggeration, be
likened to 'the stuff that dreams are made of', since it is chiefly
due to the 'corona' or glory which surrounds the darkened Sun during an eclipse... At its base, and in contact with the photosphere, is what resembles a sheet of scarlet fire... This is the
'chromosphere' ...}''

 These annotations already gave a clear qualitative description of the
outer layers of the solar atmosphere, containing also one of the first
scientific reports of what today is known as a chromospheric
emission. Beginning in the famous Indian eclipse of 1868 August 18, the
application of the yet novel spectroscopic visual techniques to the
Sun started to reveal several
crucial facts about its physical properties. They would also constitute the basis for
the flourishment of the present astrophysical spectropolarimetry.   

\begin{figure}[h!]
        \centering
        \begin{subfigure}{0.31\textwidth}
                \centering
                \includegraphics [width=\textwidth]{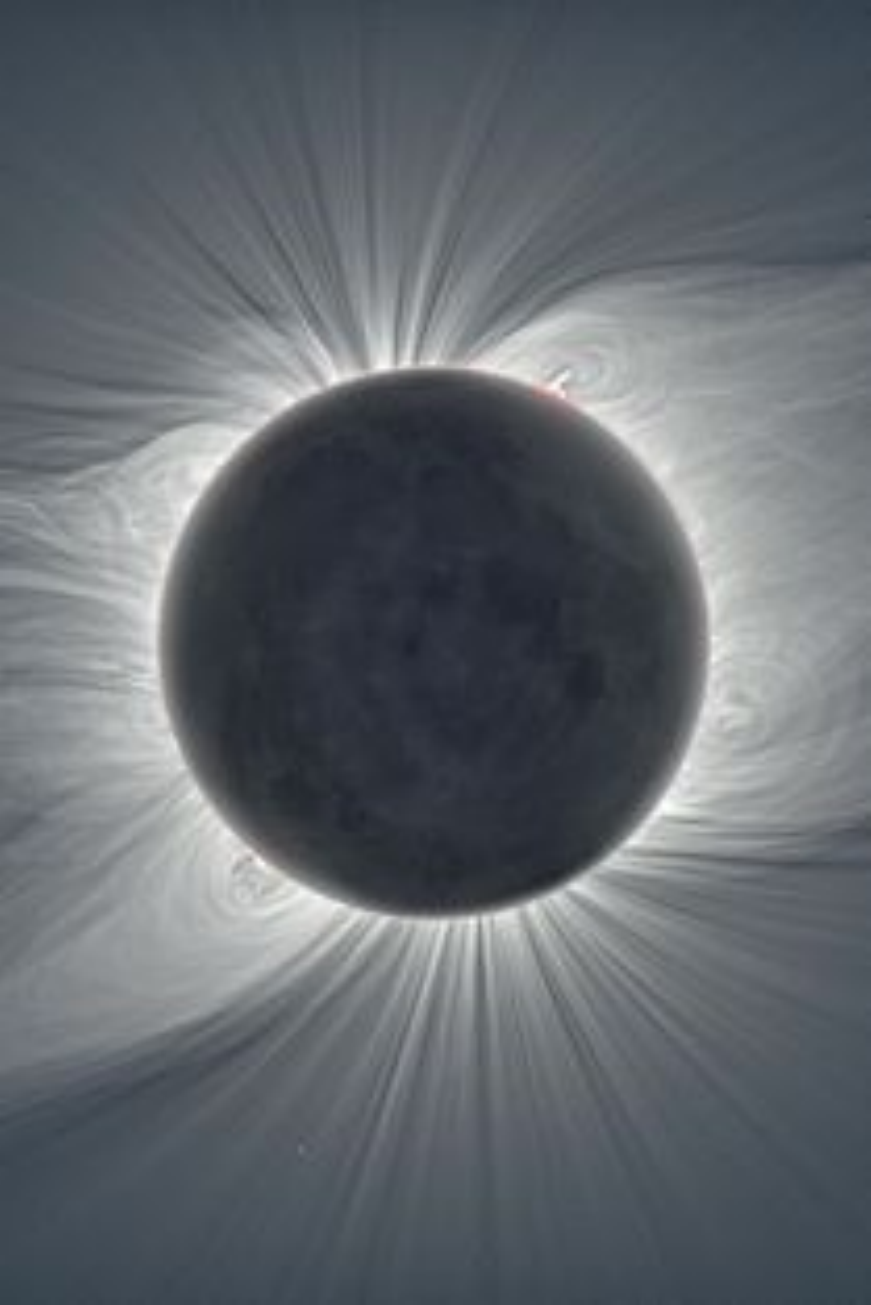}
              \end{subfigure}
        \begin{subfigure}{0.31\textwidth}
                \centering
                \includegraphics [width=\textwidth]{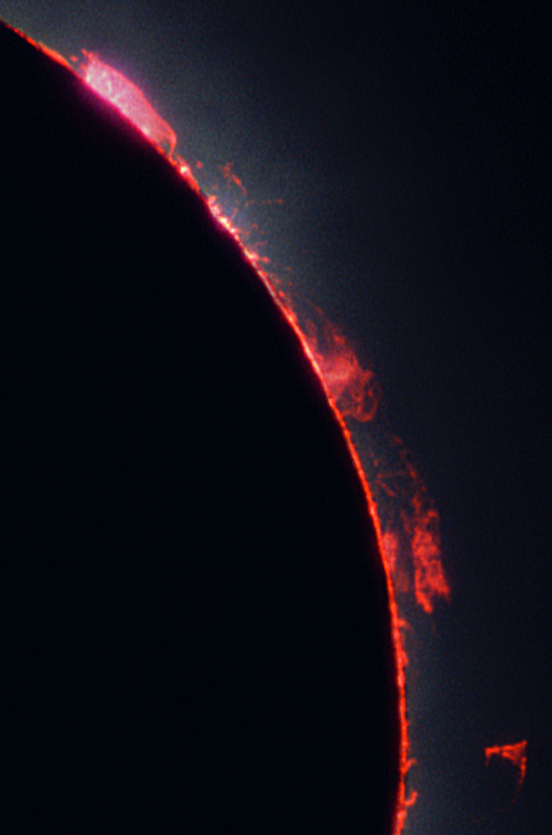}
              \end{subfigure}
        \begin{subfigure}{0.31\textwidth}
                \centering
                \includegraphics [width=0.98\textwidth]{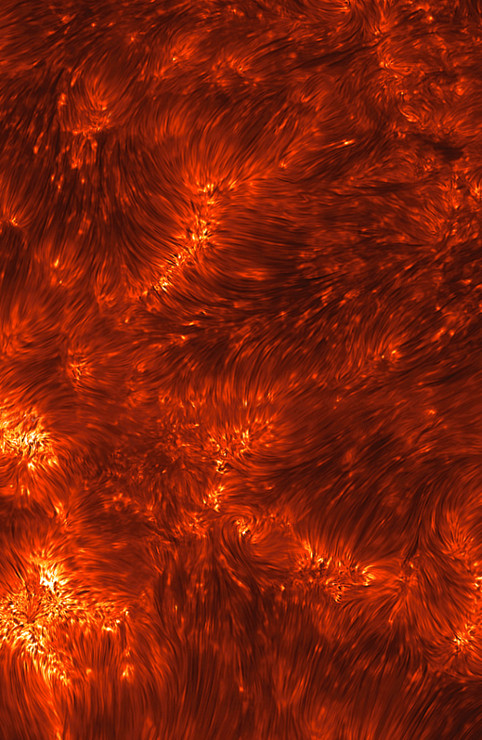}
              \end{subfigure}

        \caption{\textbf{Left}: Coronal magnetic fields ('\textit{the stuff that dreams are
            made of}') during the solar eclipse of 2008. \textbf{Center}:
          chromospheric emission at the solar limb during a total eclipse. \textbf{Right}:
          Fibrilar structures (spicules or mottles) in H$\alpha$. They
          trace network (brighter)
          and internetwork (darker) areas at the solar disk.}
                \label{fig:chromo}
\end{figure}

Irrespective of the scientific explanations we could give,
the vision of our moon exactly
matching the Sun's circumference will never cease to be amazing (Figure
\ref{fig:chromo}, left panel). But better eyes to observe it are always welcome.
Thanks to the instrumental works of Secchi in the 1860s and to the subsequent
establishment of a proper observational methodology, the
increasing interest and fascination of the scientists for the solar atmosphere
transformed it in a very attractive topic
of research. Besides the identification of the main outer regions of the Sun,
the incipient spectroscopy allowed their composition to be measured. Scientists
like Herschel, Rayet, and Janssen realized that the faint glow of the chromosphere was due to
an emission spectrum from hot, low density gases emitting at discrete
wavelengths, the
``scarlet fire'' being due to the strong Balmer H$\alpha$ emission
(Figure \ref{fig:chromo}, middle panel). Also
through spectroscopic methods, the discovery of the second most abundant element in the universe,
helium, was first done in emission lines seen in the solar
chromosphere during that Indian eclipse of 1868 (helium was not 
found on Earth until 1895!). 

At that time, the chromosphere could only be distinguished easily during a total
solar eclipse because it glows faintly relative to the
photosphere. But the invention of the first spectroheliograph by
Hale and Deslandres (1890) allowed the study of the solar disk chromosphere
at any time. It led Hale to reveal the ``chromospheric
network'' at various wavelengths in the Ca {\sc ii} H
and K lines and in H$\beta$, and showed that enhanced chromospheric
emission occurs in ``clouds'' or ``flocculi'' above photospheric
faculae \citep{Hale:1904aa}. These regions are overlaid and mixed with ubiquitous
hair-like fine structures, later termed ``jets'' or  ``spicules''
when they are seen at the limb, or ``mottles'' when seen on disk. Hale \citep{Hale:1909aa} also obtained the first spectroheliograms of the disk
in H$\alpha$, which revealed that ``\textit{...is clearly visible on the hydrogen
photographs. It is a decided definiteness of structure indicated by
radial or curving lines, or as some such distribution of the minor
flocculi as iron filings present in a magnetic field}''. Thus, together with the
discovery of the Zeeman effect in sunspots \citep{Hale:1908aa}, Hale had
confirmed a stunning and essential fact: the magnetic nature of
the Sun (figure \ref{fig:chromo}).

The magnetic field plays a
crucial role in the behavior of the solar atmosphere.
It is one of the three main drivers defining the chromosphere (together with dynamics and thermodynamics). Indeed, its importance in Astrophysics is universal. To
explain it, we can first consider the symmetry between electric and
magnetic fields in the
Maxwell equations describing the propagation
of electromagnetic waves. They are symmetric in their
interactions. On the contrary, there is a lack of symmetry between the
sources (charges) of the electric and magnetic fields in such
equations. Effectively, the matter throughout the cosmos is found to
consist only of electrically charged particles, i.e., electrons and
nucleons, with no indicia of scalar magnetic sources (magnetic monopoles).
On the other hand, since most of the gases in the universe are at least
partially ionized, there is an abundance of free electrons and
ions. Hence, a consequence of these facts is that an electric current density can be easily created by a very
weak electric field, quickly reducing to negligible
values any large-scale electric field in the reference frame of the moving plasma\footnote{Only in places without large electrical conductivity, as
the very good insulated regions of the planetary atmospheres, can
electric fields exist. This favours the emergence of life.}. In
other words, the abundance of free charges shortcircuits the electric
fields very fast, leaving the universe impregned only by the magnetic
field at large distances \citep{Parker:2007}. At the same time, charges in motion with respect to an external observer, are  
themselves sources of seed magnetic fields that are amplified by
rotation and convection in the stars (as stated by the induction equation in MHD
theory). That is one of the basis of the solar dynamo mechanism \citep{Charbonneau:2010ab}. The stellar magnetic fields are
then created and driven by organized macroscopic relative motions between
electric charges in the plasma\footnote{The magnetic field itself is a relativistic
  phenomenon. According to the special theory of relativity, the
  partition of the electromagnetic force into separate electric and
  magnetic components is not fundamental, but varies with the
  observational frame of reference: an electric force perceived by one
  observer may be perceived in a different frame of
  reference as a magnetic force. Special
  relativity combines the electric and magnetic fields into a rank-2
  tensor, called the electromagnetic tensor. Changing reference frames
  mixes these components. This is analogous to the way that special
  relativity mixes space and time into spacetime, and mass, momentum
  and energy into four-momentum. Interestingly, also the Stokes parameters
 (read further) form a Minkowskian four-vector}.
In the absence of motions, the most notable chromospheric
and coronal structures, such as those spicules or the longer ``iron
fillings'' described by Hale, would not exist.

In relation with the outer solar layers, it is believed that the dissipation of magnetic energy in
the $10^6$ K corona may be significantly modulated by the strength and
structure of the magnetic field in the chromosphere
\citep[e.g.,][]{Parker:2007}.
According to the variation of the magnetic field with height in the Sun, the chromosphere is
an interface layer lying
between the gas-dominated photosphere (where field lines are frozen in the plasma, dragged about by surface flows) and the field-dominated
corona (where the ionized plasma is forced to flow along the field lines). At such
extremes the field adopts the form of small-scale
intense flux tubes in the high-$\beta$ photosphere\footnote{The $\beta$ of the
  plasma is the ratio between the gas pressure and the magnetic
  pressure.} and produces loop-like structures in the low-$\beta$
corona. In the middle, magnetic
fields can be highly twisted and tangled, being expanded from below to fill the
chromosphere. 

\begin{figure}[h!]
                \centering
                \includegraphics [width=\textwidth]{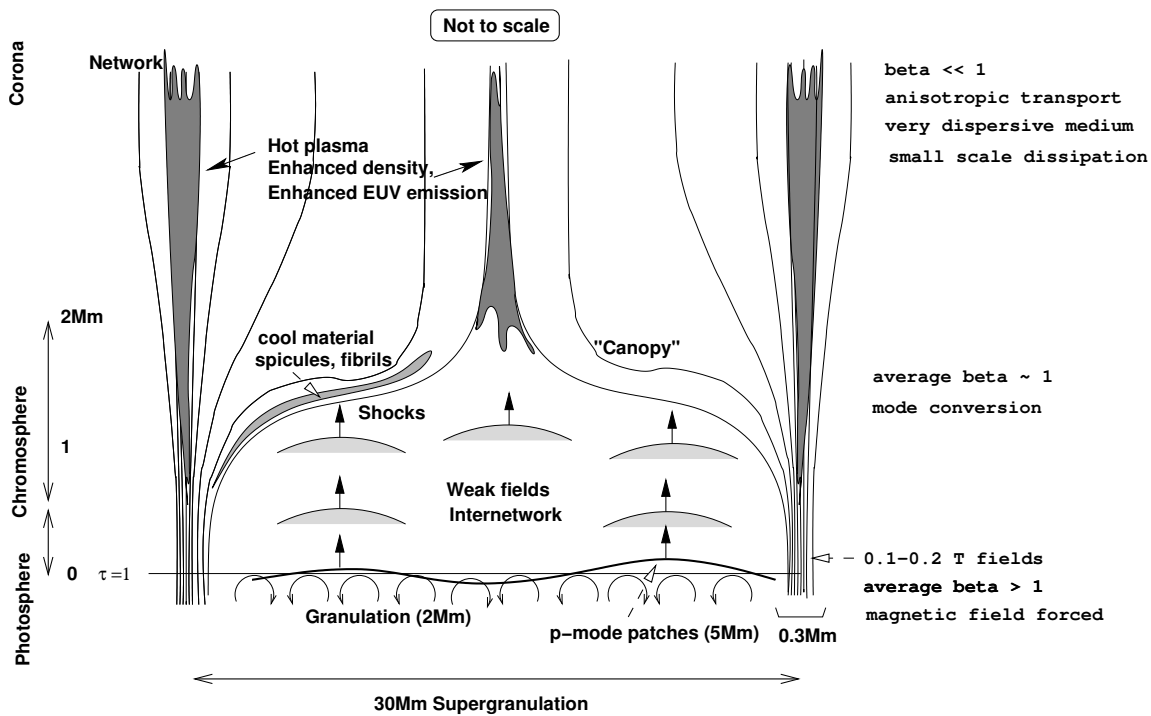}
        \caption{Schematic picture of the structure of the solar
          chromosphere in a quiet-Sun unipolar magnetic field region. From \cite{Judge:2006}.}
                \label{fig:chromosketch}
\end{figure}

The structure of the chromosphere is thus determined by the magnetic field,
while its dynamics is dominated by oscillations and flows
arriving amplified from the convective photosphere. Indeed, dynamics
is specially important in the solar chromosphere because of the much
larger velocities existing there in comparison with the photosphere. The combination of
intricate structure on small scales and fast dynamics make the
chromosphere one of the most defying regions for the comprehension of the solar
atmosphere. Some problems are the channelling of the highly
conducting and partially ionized plasma through the field lines, and the
changing of the force balance ($\beta$ parameter) within the
chromosphere, which leads to drastic variations in field
morphology and wave mode propagation. The magnetic field dramatically changes the ways energy can be
transported and dissipated, compared with the field-free case. 
Figure \ref{fig:chromosketch} sketches some of the complications introduced by magnetic
fields and dynamics \citep{Judge:2006}. 
In particular, the turbulent nature of the underlying photosphere will
inevitably lead to magnetic free energy (current systems) throughout
the entire atmosphere. It basically consists in very small scale
current sheets and dissipation regions below the current observable
scales \citep{Parker:1994aa}. Thus, the combination of magnetic fields
and dynamics, which includes shocks and turbulence, leads to
an atmospheric ``global warming''.

The temperature profile is the third distinctive attribute commonly used to define the
solar chromosphere. 
 In this region, most of the non-thermal energy that creates the corona and
the solar wind is released, with a heating rate requirement that is
between one and two orders of magnitude larger than in the corona \citep{Ayres:aa}. 
In quiet Sun regions, the chromosphere extends from the temperature minimum at
about $500$ km to the sudden steep increase around $2100$ km (the
transition region to the corona, where temperature changes from $10^4$ K
to $10^6$ K). The nature of the chromospheric temperature rise is still
unclear. Acoustic waves have long been proposed as the main energy source that
heats the quiet-Sun chromosphere. They steepen into shocks as they
propagate upward in the atmosphere, heating it as they dissipate. This
would produce a highly time-dependent heating \citep[e.g.,][]{Carlsson:2002}.

In the last instance, only one driver alters the
chromosphere at all scales: dynamics. Motions of charges generate and
sustain the
fields; macroscopic motions in plasmas can transport and distort the
fields, or can alter the optical properties of the fluid; and even temperature is
a proxy for microscopic motions. 
   
To
observationally understand the information that we receive from the
chromosphere, it is also important to discriminate what we are looking
at.
Since
the visual work of Secchi in the 1870s to the impressive satellite
pictures of today, the observational appearance of the chromosphere
has always shown the remarkable and beautiful fine
structure that seems to predominate (figure \ref{fig:chromo}, right
panel). It is then easy to imagine a chromosphere mostly composed by
those streamers of plasma (spicules) that Charles A. Young pointed out.

But, contrarily to that first impression, most of the mass
in the chromosphere have to be disposed in gravitationally stratified layers of plasma,
not in spicules defying gravity \citep{Judge:2010ab}.
We have here to distinguish between the fine-structured chromosphere and
the ambient chromosphere from which spicules originate. The
chromosphere fibrilar structure bears similarities in morphology and
dynamics to the overlying corona, being a kind of conspicuos interface
layer. When observing with broadband filters, the
instruments preferentially detect those bright, dynamic jets whose
line widths and Doppler shifts are sufficient to avoid the absorption
by the intervening material \citep{Judge:2010aa}. Thus, the
``non-spicule'' chromosphere, which in any case must
be present to account for material with significant opacity in the Ca
II lines \citep{Lites:1993aa}, cannot be easily seen in the
observations. Today we know that spicules have far smaller filling factor and density
(by several orders of magnitude) than the chromospheric
pool\footnote{They are important because they present a large areal
  interface to the corona, so having a great potential to supply large amounts of mass upwards, and
  channelling Alfvenic fluctuations that can release magnetic energy into the
  external layers \citep{de-Pontieu:2007aa}.}. 

A signature property of such ambient chromosphere pointed out before is its
geometrical extension
(near ten pressure scale heights), which is much larger than expected
in a hydrostatic atmosphere where gravity is balanced by pressure
gradients. During a long time the question was: which are the forces balancing the chromospheric stratification? 
In the past, the two competing solutions were
a hydrostatically stratified chromosphere supported on
radiation-pressure gradients \citep{Milne:1924aa}, and a similar model whose extra
support was given by
turbulence instead of radiation \citep{McCrea:1929aa}. It was not
until the middle of the twentieth century that the development
of the radiation transfer and spectral line
formation theory started to include the extreme departures from classical Local Thermodynamic Equilibrium (LTE) at the
very low densities and high temperatures of the chromospheric
regions. The solution inclined
favourably to Milne's model, and the discussion led to the
introduction of Non-LTE\footnote{Contrary to the photosphere, the atomic
  excitation of the chromospheric plasma strongly depends on
  a radiation field which does not correspond with local conditions
  but with the emission at distant points within the solar atmospheres.}
effects for explaining the observations. Years later, \cite{Athay:1961aa} concluded that, including NLTE effects, a hydrostatically
stratified distribution of plasma is in good agreeement with the limb observations made in
the continuum. They also showed that in NLTE,
the dependence of ionization equilibrium on electron pressure still naturally produces
higher ionization stages at higher layers in stratified media, as Saha
(1920) found assuming LTE\footnote{These facts seems to apply in most part of the 
chromosphere. Perhaps only in the highest chromospheric layers the spicules tend to appear
as a natural
consequence of the predominance of the magnetic field over the plasma
dynamics, which would channel and accelerate the particles along the
spicules altering the hydrostatic stratification along them.}.  
Several decades of research on
chromosheric lines have redounded in a well-established line formation
theory able to model chromospheric spectral lines under NLTE conditions
\citep[e.g., the monographs by
][]{Mihalas:1978,Cannon:1985,Rutten:2003ab}. 

Today, the large extent of the chromosphere is explained in terms of 
ionization of its dominant constituent, hydrogen. Given its large
ionization potential, hydrogen acts as a sponge that soaks up energy,
buffering the gas to some degree from local heatings (like acoustic
shocks) and moderating the temperatures \citep{Ayres:aa}. The
key point is that ionization frees electrons to feed regular and
continuous cooling by collisional excitation and subsequent
radiative de excitation of
abundant species such as Fe{\sc ii}, Mg{\sc ii} and Ca{\sc ii}. The
radiative cooling produced by those lines is a signature property of the chromosphere. This ``ionization valve''
works effectively along many scale heights because of the large dynamic range of the electron
fraction, $\mathrm{n_e/n_H}$. It is only $10^{-4}$ at the base of the chromosphere
(where the electrons are from singly ionized metals), but at the top
it approaches unity (hydrogen mostly ionized). This allows considerable
margin for the gas to balance heating even while the overall density
falls outward \citep{Ayres:1979aa}. Once the pool of neutral hydrogen is exhausted, the valve cannot
continue to balance the heating and a thermal escape
could propagate towards the corona. This scenario is appealing to
 explain the large extent and the ``thermostat''
role of solar-like chromospheres but, for the
moment, it is not enough to fully explain the so-called coronal
heating problem. To solve this and other various problems we have to
engage more pieces of the puzzle. How do we connect the physical
properties of the chromosphere with the observed spectral line radiation?
This thesis is about the answer to that question.

To validate and test the link between theory and observations,
a reliable diagnostic technique should be able to accomodate the
influence of magnetic field, temperature and velocity, while guided by
a detailed
line formation theory. 

On one hand, the scattering
in strong resonance transitions feeded by the hydrogen ionization becomes a dominant
energy transport mechanism (radiative
cooling). The nonlocality of such radiation fields, induced by the lower opacity in higher
layers, creates serious challenges for remote sensing
that have to be solved with the help of theoretical and numerical
approximations to the radiative transfer problem.

On the other hand, ``measuring" the chromospheric
magnetic field is notoriously difficult \citep[e.g., reviews
by][]{Casini:2007,Harvey:2009,Trujillo-Bueno:2010}. While
spectroscopic observations allow us to determine temperatures, flows
and waves, they do not provide any quantitative information on the
chromospheric magnetic field. To this end, we need to measure and
interpret the polarization that some physical mechanisms introduce in
chromospheric spectral lines. From Maxwell's electromagnetic wave theory it follows that the spectral radiation is characterized by its
intensity and also by its polarization, which is defined in the plane
perpendicular to the direction of propagation of the light ray
(the transversal plane). G. G. Stokes showed how
 intensity and polarization could
be described in a unified way by the Stokes\footnote{Apart from the polarimetry and the many contributions that Stokes did to
  science, some evidences suggest that he could have also been the
  first one (several years before Kirchhoff) in formulating
  the fundamental principles of spectroscopy, which allowed the
  identification of substances in the Sun and in the stars.} 4-vector. The
first vector component represents the ordinary intensity (I), the second
and third components (Q and U) describe linear polarization along two
reference directions in the transversal plane, while the fourth
component (V) relates to the circular polarization.
This is a very powerful way to
describe any partially polarized light beam while giving all the four vector
components in the same (intensity) units \citep{Born:1980}. In practice, this
representation implies that
spectropolarimetry is actually differential photometry. The
difference between the number of photons oscillating in one reference
direction along the transversal plane
and the number of photons oscillating perpendicularly in the same
plane gives Stokes Q and
U. Something similar holds for Stokes V, which is the difference
between right-handed and left-handed circularly polarized photons. Going from
spectroscopy to spectropolarimetry thus means an increment in the
dimensionality of information space from 1-D to 4-D.
Hence, in polarized radiative transfer (RT), instead of a scalar problem we
have to deal with the transfer of a 4-vector. This increases
the richness\footnote{Note that the information provided by the new dimensions
(Q, U, and V ) cannot be derived from Stokes I, but each independent
dimension
gives a different but complementary diagnostic window to the
universe.} but also the complexity of the physical situation. It is the price to pay for getting access to
stellar magnetic fields.

Historically, the discovery of the magnetic effects on the light was guided by
laboratory experiments. In $1896$,
Pieter Zeeman, disobeying the direct orders of his supervisor, used
laboratory equipment to measure the action of a strong magnetic field
on spectral lines. Thus, he discovered his homonymous effect,
consisting in the splitting of a spectral line into several
polarized components by the action of a static magnetic field. 

The circular and linear polarization signals that the Zeeman effect
can produce in a spectral line are caused by the
wavelength shifts between the so-called $\pi$ and $\sigma$ transitions of the
line (Zeeman splitting) induced by the presence of a
magnetic field. The amplitude of the circular polarization scales with
the ratio $\cal R$ between the Zeeman splitting and the Doppler line
width. The amplitude of the linear polarization scales with ${\cal
R}^2$ \citep[see][]{LL04}. Outside sunspots (where
$B\,{\lesssim}\,100$ G at chromospheric heights) ${\cal R}{\ll}1$, which
explains why it is so difficult to detect the linear polarization of the
Zeeman effect in a chromospheric line. Typically, only the Zeeman circular
polarization is detected, especially in long-wavelength chromospheric
lines such as those of the IR triplet of Ca{\sc ii} \citep[e.g.,
][Figure 3]{Trujillo-Bueno:2010}.  Then, the linear polarization observed
in quiet regions of the solar chromosphere has practically nothing to
do with the transverse Zeeman effect.

On the other hand, in 1922 Wood and Ellett published a paper
describing the effect of a magnetic field on the polarization of
resonance fluorescence radiation emitted by a cell of mercury vapor. It turned out that a constant
magnetic field of a few gauss was sufficient to depolarize the
scattered radiation. Wood and Ellett soon realized that this behavior could not be interpreted as a Zeeman effect since the Zeeman separation in such
fields was very small compared to the typical Doppler-broadened
linewidth of such radiation. In $1924$ Hanle gave a
classical explanation of the effect arguing that the external magnetic
field produces, due to the Lorentz force, a precession of the
atom electrons about the field direction. Such precession breaks the linear pattern of the oscillations re-emitted by
the atoms and leads
to a depolarized detected radiation. Attempts to
better understand the phenomenon were important in the subsequent development of quantum physics. 

It was through spectropolarimetry that measurements of stellar magnetic fields became possible: first, with the works
of Hale (remember, the spectroheliograph's inventor who detected
the Zeeman effect in intensity while studying sunspots), and
\cite{Babcock:1947aa}, who succesfully measured the
circularly polarized Zeeman components on stars; much later on, with the pioneering works of
\cite{Omont:1973aa, Bommier:1978} and \cite{Stenflo:1978aa},
the Hanle effect was studied in the context of the 
radiative transfer and measured in the solar atmosphere \citep{Stenflo:2013ab}. 

 Polarization is related to some symmetry breaking process. For
 instance, in the case
of the Zeeman effect, the spatial symmetry breaking induced by the
magnetic field is transferred to the radiation as an asymmetry between
the orientation sense of the circularly polarized blue and red sigma components. In a non-magnetized atmosphere,
the symmetry can be broken by a scattering process, depending on the angles
between the incident and scattered radiation. Similarly to the
light emitted from the Earth sky, which is linearly polarized by
molecular Rayleigh scattering, the solar spectrum is linearly
polarized by scattering processes in the Sun's atmosphere. 
In weakly magnetized regions, the linear polarization of chromospheric
lines is dominated by scattering processes. The
geometrical distribution of the solar atmospheric structures can
create, simply by the lack of symmetry in the illumination they produce, a phase
relationship between the excitation and emission processes in scattering
atoms. The quantum origin of
this polarization is the difference among the electronic populations
of sublevels pertaining to the levels of the spectral line under
consideration. This so-called atomic level polarization, which is
induced by the anisotropic illumination of the atoms, produces
selective emission and/or selective absorption of polarization
components without the need of a magnetic field
\citep[e.g.,][]{manso_trujillo03a,manso10}. The larger the anisotropy
of the incident radiation field the larger the induced atomic level
polarization and the larger can be the amplitude of the linear polarization
of the emergent spectral line radiation. In an optically thick plasma
like the solar atmosphere, the anisotropy of the radiation field
depends mainly on the spatial distribution of the physical quantities
that determine, at each point within the medium, the angular variation
of the incident intensity. Great attention has been paid to the
gradient of the source function \citep[e.g.,][]
{Trujillo-Bueno:2001aa,LL04} but, in a highly dynamic medium like the
solar chromosphere, the gradients of the macroscopic velocity of the
plasma may also play an important role in the modulation of the anisotropy. That is the central idea of
this thesis.

Those processes together with other processes modifying it (e.g.,
the Hanle effect and the collisions), can generate a net linear
polarization signal. Since the
emergent linear polarization has contributions from many scattering
angles, the measured quantity is a very small average. The small
amplitude of the scattering polarization signals delayed
the development of solar spectropolarimetry until technical
advances allowed the design of instruments with high polarimetric sensitivity.

\cite{Stenflo:1983aa} did the first survey of the scattering polarization throughout the solar
spectrum by measuring the linear
polarization at the solar limb. They covered from the far UV to the near
infrared, so revealing a new world of linearly polarized spectral
line signals that received the name of the \textit{second solar spectrum}. Such
signals respond to a rich variety of underlying physical mechanisms,
like the Hanle effect, what gives them a great potential
for field diagnosis. Since in the quiet chromosphere of the Sun the magnetic fields are weak
(which means a smaller Zeeman splitting) and
the spectral lines forming there tend to have larger thermal 
widths (decreasing the effective sensitivity to the Zeeman splitting), the Zeeman
effect shows a kind of blindness to the magnetism of these atmospheric layers. It leaves the
Hanle effect as a preferred mechanism to measure the magnetic field in
those quiet areas (which does not mean that it cannot be used in
active regions because its applicability depends on the spectral line considered).

The synthesis of the line scattering polarization requires several
ingredients, such as a well-established theory of line formation, a
precise characterization of many atomic processes, suitable iterative
methods of solution and reliable atomic and atmospheric models. Despite its intrinsic difficulty, there have been a good number of
studies contributing to the development and the establishment of this
reseach topic. A few representative examples can be
 the modelling of the Mg {\sc i}-b lines by
\cite{Trujillo-Bueno:1999,Trujillo-Bueno:2001aa}; the development of
the code HAZEL \citep{Asensio-Ramos:2008aa}
for the synthesis and inversion of Stokes profiles resulting from the
joint action of the Hanle and Zeeman effects in
 lines of neutral helium; the interpretation of the Ce {\sc ii} and
Ti {\sc i} lines by \cite{Manso-Sainz:2002a}; the modelling of the Ba
{\sc ii} 4554 {\AA} line by \cite{Belluzzi:2007}; and the work of \cite{Manso-Sainz:2003,manso10}, who successfully synthetized the
scattering polarization in the IR triplet lines of Ca {\sc ii} for the
first time. This latter example
is of especial relevance for this thesis because we focus on the same lines. The calcium IR triplet is a set of
subordinated chromospheric lines whose RT modelling requires taking
into account also the strong
resonance absorption in the H \& K lines of Ca {\sc ii}. Hence, their
scattering polarization is
directly sensitive to the chromospheric ``thermostat'' (radiative
cooling) and, as we
will see, also to the magnetic field and the dynamics. Being 
three lines with different heights of formation, they offer us a
tomographic heartbeat of the whole system photosphere $+$ chromosphere. 
The polarization of the
IR triplet of Ca {\sc ii} lines is a good choice to study 
the quiet Sun magnetism with the Hanle and Zeeman effects. They are suitable to evaluate the reliability of MHD models via spectral synthesis and comparison with
spectropolarimetric observations. 

Retrospectively, a rigorous quantum theory to describe polarization
and radiative transfer \citep[see ][]{LL04} together with the development of
high-sensitivity instrumentation have been important steps towards the understanding of the
solar chromosphere. However, to achieve a successful comparison
between observations and theory, much work is
still needed on both sides. For
example, despite the Sun's proximity, the
polarimetric accuracy needed to capture chromospheric vector fields is
not presently achievable at the desired fine spatial and temporal
scales (dividing the photons into space, time, frequency, and polarization states
quickly exhausts the supply). Innovations in this area are actively
being pursued, especially focused on larger telescope apertures both
on earth and in space. Examples are the European Solar
Telescope \citep[EST,][]{Collados:2013aa}, the Advanced Technology
Solar Telescope \citep[ATST,][]{Rimmele:2013aa} and the
Solar-C space telescope \citep{Shimizu:2011aa}. The spectropolarimetric technique has
been recently improved with the Zimpol 3
polarimeter \citep{Ramelli:2010aa}. On the theoretical side, the complexity
of the quantum theory for
treating partial redistribution effects \citep{Bommier:1997aa} or the
sophistication of the methods needed to carry out radiative transfer (RT) calculations in
3D models imply the need of doing approximations when
including scattering polarization \citep{Stepan:2013aa}. 

On the other hand, the research field of 3D radiation hydrodynamic
simulations of the solar atmosphere has 
reached a level of sophistication which is far beyond that of
idealised numerical experiments, and allows a direct confrontation
between models and real stars
\citep[e.g.,][]{Asplund:2000,Stein:1998aa,Leenaarts:2009}. By performing RT
calculations in such models, it is possible to
study, based on first principles, the
effect that the physical atmospheric properties produce on the emergent Stokes vector.

 In this thesis, we follow such strategy with emphasis on 
the radiative transfer problem with polarization. We pay
particular attention to the linear polarization generated by scattering
proceses with the aim of exploring
its potential application for deciphering the magnetism of the quiet
solar chromospheric regions. We try to improve the current theoretical diagnosis
capabilities based on
 radiative transfer, scattering polarization and Zeeman and Hanle
effects. Of particular interest is that we introduce and study the effects of dynamics on the
synthesis of the scattering polarization. Furthermore, we use
realistic 1D and 3D MHD models to synthetize temporal
series and tomographic spatial maps of the Stokes parameters in
the presence of shocks, magnetic fields and temperature gradients.

In Chapter \ref{cap:intro}, we review the theory of radiative transfer (RT) with polarization, the
numerical methods we have used and some general considerations about the
inclusion of macroscopic velocities in this problem. 

In Chapter \ref{cap:one}, we briefly present the computational methods
and computer programs that we have developed in the context of this thesis. The
chapter reports on seven tools: a RT code, a program to compute
response functions, a Principal Component Analysis program, two
interactive programs for visualization and a semi-empirical method that
facilitates the convergence of the RT problem. 

In Chapter \ref{cap:two}, we show some basic numerical experiments done for
understanding the effect of vertical velocity gradients on the
synthesis of the linear polarization in spectral lines. We present
also results for
non-magnetic semiempirical models. This chapter is an adapted version
of \cite{carlin12}. 

 In Chapter \ref{cap:three}, we calculate synthetic profiles
 using a more realistic dynamic chromospheric model (including shocks)
 for obtaining a temporal
 evolution of the linear polarization signals in a simulated solar-limb
 observation.  This chapter is an adapted version
of \cite{Carlin:2013aa}.

 In Chapter \ref{cap:four}, we show spatial maps of emergent Stokes profiles
 resulting from a snapshot of a realistic 3D MHD model in a disk-center
 observation, trying to relate the computed observables with the
 physical properties of the model chromosphere.

Finally, in Chapter \ref{cap:conc}, we summarize our conclusions and
discuss near-future directions of research.

\chapter{Spectral line polarization in stellar atmospheres}\label{cap:intro}
Stellar atmospheres are plasma regions of low density and high
temperature. They are constituted by a mixture
of many chemical elements in form of atoms, ions, free electrons and
molecules. Their physical conditions vary with the position,
generally more steeply along the vertical direction because of the
gravitational stratification. Due to the relatively low densities, the material
behaves as an ideal gas, whose state is determined by the particles
distribution (atomic populations) over all the free and bound energy
states accesible to the system. The calculation of the populations of
the atomic energy levels and
sublevels requires to consider the radiative and collisional processes producing atomic
transitions in each chemical species in the plasma. The collisional
processes are assumed to be isotropic and are described by the laws of statistical
mechanics. Being purely local interactions, this kind of transitions approach the atomic system to the
local thermodynamic equilibrium (LTE) with the surroundings, a
situation in which the matter and radiation are strongly coupled. On the contrary, radiative processes directly depend
on the (non-local) radiation field, which interacts with matter through radiative
excitations and photoionizations. These interactions detach the atomic
populations from their local thermodynamic values, making them 
sensitive to the physical conditions in distant regions (non-LTE;
hereafter NLTE). While LTE conditions are valid in
deep stellar atmospheric layers due to the higher densities and collisional
rates, the general case of NLTE must be accounted for when dealing with spectral
lines forming at higher layers. Thus, considering the external illumination
and all the microscopic processes
that alter the excitation state of the atoms, it is possible to characterize the macroscopic
behavior of each plasma element. In particular, it is our aim to focus
on the phenomena of scattering line
polarization and its modification by the action of weak ($B\leq 100$
G) magnetic fields (Hanle effect) in dynamic atmospheres. 

We will review in this chapter all the aspects related to the 
radiative transfer problem of spectral line polarization in weakly
magnetized atmospheres (the so-called radiative
transfer problem of the second kind). First, we
will explain how to quantify the excitation state of the plasma and the 
transfer of polarized light through it. We will
specify the radiative transfer coefficients by paying attention to the microscopic
processes defining them, which includes the Hanle and Zeeman effects produced by the
action of the magnetic field. We will present also the statistical
equilibrium equations. Later on, we will consider the
numerical methods employed for solving the ensuing NLTE radiative
transfer problem. In the
last sections we will discuss some
considerations to treat the
radiative transfer with macroscopic velocities, ending with
a somewhat detailed guide about the Hanle effect.

\section{Radiative transfer with polarization.}\label{sec:prt}
\subsection{Quantum mechanical description}
We refer to the plasma element as the smallest indivisible volume or resolution element of a
stellar atmosphere that is theoretically characterized when modelling the emergent spectral
radiation. 

We consider a plasma element composed by multi-level atoms
of the
 atomic species of interest, which is assumed devoid of hyperfine
 structure. In the
absence of interactions, the
independent particles of such a system are individually represented by a pure
quantum state. 
But, following statistical mechanics, the total ensemble of particles is
in a statistical mixture of states\footnote{The lack of information about the initial state
of the atomic subsystem due to its microscopic interactions avoids a
complete description based on a single pure state.} and, consequently,
it has to be described by the density operator \citep{ Bommier:1978, Blum:1981aa}
\begin{equation}\label{eq:densityop}
\centering 
\hat{\rho} \,= \,\sum_{\alpha} p_{\alpha} \vert\psi_{\alpha} \rangle \langle\psi_{\alpha} \vert,
\end{equation}
where $p_{\alpha}$ is the probability for the atoms to be in the pure
dynamical state identified by the vector $\vert\psi_{\alpha} \rangle$,
and where the sum is extended to all the pure states in which the
atoms can be found. The matrix elements of the density operator
(density-matrix elements) are evaluated on a given basis of the
Hilbert space associated with the quantum system. Such
density-matrix elements contain all
the accesible information about the system and its dynamical state.

The most natural basis in which the density-matrix elements can be defined
for an atomic system is the basis of eigenvectors of the total angular
momentum ${\mathbf J}$ of the atom. Each atomic
energy level is
identified by the set of integer (or half-integer) quantum numbers $\{J, M\}$ plus a set of
inner quantum numbers omitted for simplicity. Here, $J$ is the
angular quantum number of the energy level, while $M$ is the magnetic
quantum number, which is the
 eigenvalue of the projection of ${\mathbf J}$ along an
arbitrarily chosen quantization axis. According to the postulates of quantum mechanics,
the atomic system in a given level can occupy any of the $(2J
+1)$ possible $M$ magnetic substates ($-J \leq M \leq J$), which are degenerate if no
magnetic field is present.

On this basis, the general density matrix elements are then given by
\begin{equation}\label{eq:m_elements}
 \centering 
\bra{J M }\, \hat{\rho}\, \ket{J^{\prime} M^{\prime}} \,= \,\rho (J M, J^{\prime} M^{\prime}) ,
\end{equation}
being the diagonal elements proportional to the populations of
the corresponding magnetic sublevels. The off-diagonal components are the so-called
\textit{coherences} or \textit{phase relationships} describing the \textit{quantum
  interferences} that can exist between different magnetic
sublevels. Coherences between pairs of magnetic sublevels will be
assumed in this thesis to occur exclusively between sublevels of the
same $J$ level \citep[multilevel approximation; see ][]{LL04}.

Thus, the full description of an atomic system,  in the
general case in which polarization phenomena are accounted for,
requires the specification of a matrix for each energy level\footnote{
Instead of using one
quantity per atomic level to
describe the excitation state of the atoms in the non-polarized case, now $(2J+1)^2$
unknowns are needed. It is a considerable increase because it
applies to each energy level and at each position in the atmosphere.}. When
this matrix is not diagonal or when the diagonal elements are not equal, the atom is said to be polarized or to
show atomic level polarization. 

In particular, atomic polarization can be introduced
in the quantum system by any kind of external anisotropy to which the
atoms are sensitive (e.g., the incident radiation field). As a consequence,
the radiation re-emitted by a polarized
atomic system is, in turn, polarized. 

It is convenient to express the atomic density matrix in the spherical
tensor representation, obtaining the so-called multipole moments of the
atomic density matrix. It allows an easier
interpretation of the physical situation.
In the multilevel case, they are defined for each J-level as \citep{LL04}
\begin{equation}\label{eq:rhokq}
\centering 
\rho^K_Q (J) =\sum_{MM^{\prime}}(-1)^{J-M}\sqrt{2K+1}\left(
\begin{array}{ccc}
J & J & K \\
M & -M^{\prime} & -Q
\end{array}
\right)\rho(M,M^{\prime}),
\end{equation}
where the sum is extended to all possible values of $-J<M<J$, $K=0, ..., 2J$ and $-K\leq Q \leq K$, being 
the symbol between brackets a coefficient called $3j$-symbol
\citep[e.g.,][]{Brink:1968}.
 In this new basis, the overall population of each level J
is given by $\sqrt{2J+1}\rho^0_0$ and the population imbalances
between the corresponding magnetic sublevels are quantified by the
terms $\rho^K_0$. In particular, if $\rho^K_0$ with $K$ even is
nonzero, the system is said to be aligned (which produces linearly polarized radiation
), while the ones with $K$ odd quantify the atomic orientation (which
produces circularly polarized radiation). Finally, the quantum coherence
between pairs of magnetic sublevels are described
by the complex numbers $\rho^K_Q$ (with K and Q non-zero).
 \\
\subsection{The radiative transfer equation}
Consider a polarized electromagnetic wave
propagating through a plasma with a certain
refraction index. The refraction index of a medium is a complex
quantity that can be anisotropic, so showing a different value along
any of the three reference directions of space. Then, since the electromagnetic wave
oscillates in a plane, the two complex components of the electric
field $\varepsilon_a$ and $\varepsilon_b$ can
perceive a different refraction index. From a macroscopic point of
view, this simple idea explains the effects of
absorption, emission, dichroism and dispersion produced during the radiative
transfer of the electromagnetic wave. 
Thus, the total absorption is the ability
of the plasma of absorbing photons in any state of polarization
(intensity), and it is related to variations in the total modulus of the
refraction index. The emission is the opposite process in which the
atoms re-emit the energy absorbed in collisional and radiative processes. Dichroism is the ability of the plasma of absorbing
photons oscillating along a preferential direction (selective
absorption of polarization states) and it is connected with the
differential attenuation between the modulus of $\varepsilon_a$ and
$\varepsilon_b$ during the propagation. Finally,
anomalous dispersion is the ability of the plasma element to dephase
$\varepsilon_a$ and
$\varepsilon_b$, so changing the polarization state during the propagation.   

More specifically, the transfer of polarized light at frequency $\nu$ propagating along
the direction $\vec{\Omega}$ is described by the radiative
transfer equation (RTE), which gives the differential variation of the
Stokes vector inside each plasma element:
\begin{equation} \label{eq:basicrte}
\centering 
\frac{\rm{d}}{\rm{ds}}
\left( \begin{array}{c}
 I \\ 
 Q \\
 U\\
 V
\end{array} \right) 
=
\left( \begin{array}{c}
\epsilon_I \\ 
 \epsilon_Q \\
 \epsilon_U\\
 \epsilon_V
\end{array} \right) 
-
\left( \begin{array}{cccc}
 \eta_I & \eta_Q & \eta_U & \eta_V \\ 
 \eta_Q & \eta_I & \rho_V & -\rho_U \\
 \eta_U & -\rho_V & \eta_I & \rho_Q\\
 \eta_V & \rho_U & -\rho_Q & \eta_I
\end{array} \right) 
\left( \begin{array}{c}
 I \\ 
 Q \\
 U\\
 V
\end{array} \right) ,
\end{equation}
with $s$ the geometrical distance along the ray \citep{LL04}. The emission vector
and the absorption matrix in Eq. (\ref{eq:basicrte}) contain the radiative transfer
coefficients that quantify the emission
($\bm \epsilon$)\footnote{We will use the
  notation $\vec{\rm X}$ for physical vectors and $\mathbf{X}$ for formal
  vectors. The former  is reserved for vectorial
  magnitudes with three components in the ordinary space, such as
  the velocity or the magnetic field. The latter is for quantities
  that are better described by collections of points involving dimensions different
  than ordinary space. For instance, the Stokes parameters.},
total absorption ($\eta_I$), dichroism ($\eta_{Q,U,V}$) and dispersion
($\rho_{Q,U,V}$) at each spatial point within the model atmosphere
under consideration. Thus, the
entire atmosphere is modelled as a sucession of plasma elements that are
instantaneously intercepting the rays of light emitted by
the surrounding neighbors. 
\begin{figure}[h!]
\centering%
\includegraphics[width=0.55\textwidth]{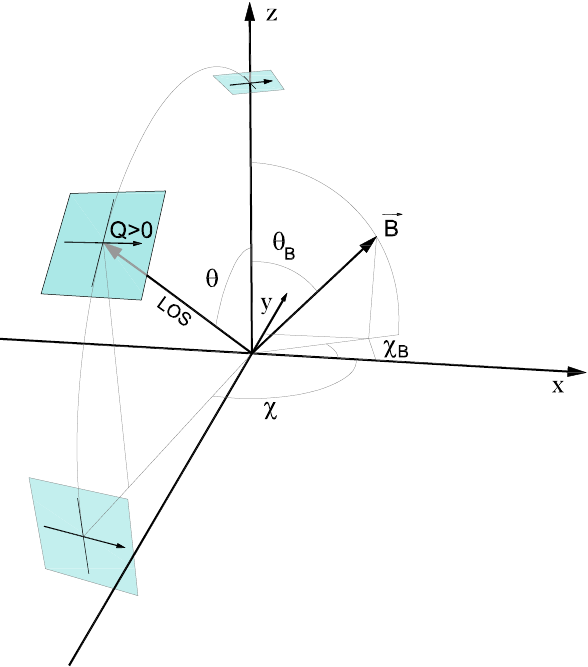}
\caption{Reference system (x,y,z) used to place the generic line of
  sight (LOS) and magnetic field vector ($\vec{{\rm B}}$). The vertical axis
  is the solar radial direction passing through the point
  considered. From the observer point of  view, the polarization is
  measured in the plane of the
  sky (coloured frames) containing the
  reference direction for $Q>0$.}
\label{fig:refsysA}
\end{figure}
We define the axis of reference assigned to a positive sign of Stokes Q
as the direction in the plane of the sky that is parallel to the limb
nearest to the scattering point
(e.g., in Fig. \ref{fig:refsysA}, such direction is parallel to the x axis
when $\chi=-\pi/2$). This also sets the same direction of
reference for other quantities related to geometrical tensors: the
radiation field components and the radiative transfer coefficients.  

\subsection{Line broadening mechanisms}\label{sec:spectral}
Before presenting the expressions for the total radiative coefficients, we need to complete the description of the plasma element specifying
the interaction mechanisms between the main atomic
species and the surroundings. These interactions are affected by
anisotropies, unresolved motions, quantum uncertainties, magnetic
fields and by the radiation field illuminating the plasma. All together define the spectral variation of the
absorption, emission and dispersion coefficients that characterize the
radiative transfer properties of bound-bound transitions. 

Since these properties are related to a complex refraction index, it is natural
that the spectral variations of the radiative coefficients are characterized by a complex line profile
$\Phi_{\nu}=\phi_{\nu}+i\psi_{\nu}$. The real part ($\phi_{\nu}$) describes the absorption and emission
properties whereas the imaginary part ($\psi_{\nu}$) specify the dispersion
effects in the polarization. They are probability distributions with
unit area whose functions are given by the complex Voigt profile\footnote{We
 will assume that the reader is familiarized with the expressions for the complex
 Voigt profile and the probabilistic distributions of Gauss, Maxwell 
and Lorentz.}: $H+iL$. Next, we will consider the treatment of $\phi_{\nu}$ but the
total dispersion profile follows from the same considerations with
just substituting the real part of the Voigt
profile H by the imaginary part L. 

The spectral variation of the total absorption and emission
coefficient $\phi(\nu)$ in a given spectral line is
obtained by normalizing in area the Voigt function $H(a,x)$. In a
weakly magnetized atmosphere with no macroscopic plasma
velocities, the
resulting profile is
\begin{equation}\label{eq:voigt}
\centering 
\phi(\nu-\nu_0)\, =\, \frac{H(a,x)}{\sqrt{\pi}\Delta \nu_D},
\end{equation}
with
\begin{equation}\label{eq:dampingpar}
\centering 
a\, =\, \frac{\gamma}{4\pi\Delta \nu_D},
\end{equation}
\begin{equation}\label{eq:redvar}
\centering 
x\, =\, \frac{\nu-\nu_0}{\Delta \nu_D},
\end{equation}
being $\gamma$ the total damping ``constant'', $\Delta \nu_D$ the total Doppler
width of the profiles and $\nu_0$ the line center frequency in the
atom frame. The Voigt
profile is the result of a convolution between a Lorentz profile
characterizing the radiative plus collisional broadening mechanisms and
the Gauss profile that describes the Doppler broadening. We have:

\begin{itemize}[leftmargin=0.23cm, labelsep=0.05cm]
\item Radiative broadening. 
The quantum uncertainty principle applied
  to the atomic energy levels limits the lifetime of the upper and
  lower levels of a transition. Thus, the infinitely sharp energy level
  is substituted by a statistical distribution function, the Lorentz
  profile, whose damping coefficient gives
  the radiative or \textit{natural} broadening of the level
  $\gamma_{\rm rad}=1/\Delta t$ (with $\Delta t$ the mean
  lifetime of the level). In general, a number of radiative
  transitions involving a
  level $i$ set its total natural damping to
\begin{equation}\label{eq:rad_damp1}
\centering
\gamma_i^{\rm{rad}}\, =\, \sum_{j< i}A_{i j} + \sum_{j> i}B_{i j}\bar{J}^0_0(\nu_{ij}),
\end{equation}
where the first sum accounts for the spontaneous emission rates and the
second sum for the absorption ones, being $A_{ij}$ and  $B_{ij}$
the corresponding Einstein coefficients and $J^0_0$ the
angle-averaged mean intensity. The induced emission to lower levels can be similarly
added \citep{Mihalas:1978}. 
The total radiative broadening of the transition
$\rm{u \rightarrow \ell}$  is
then given by $\gamma^{\rm rad}=\gamma^{\rm rad}_{\ell}+\gamma^{\rm
  rad}_u$\footnote{The total profile is a convolution
  between the Lorentz profiles of both levels. The convolution
  of two Lorentz profiles delivers a new Lorentz profile and the
  original broadening parameters add up linearly to give the resulting one.}. 
We consider electric dipole radiative interactions, which connects atomic energy levels such that $\Delta J=0,\,\pm1$ and
$\Delta M=0,\,\pm1$.

\item Collisional or pressure broadening. This broadening
  appears when the main species emitting
  the spectral line radiation are perturbed by elastic collisions with other surrounding
  particles. The usual assumptions for
  modelling the collisional broadening in spectral lines are that the
  collisions are instantaneous (impact
  approximation), that they are isotropic and that the atomic emissions before and after the
  collisions are totally uncorrelated \citep{SobelMan:1973aa}. In such case, the resulting
  spectral  broadening is (again) a Lorentz profile whose damping $\gamma^{\rm col}$
  can be simply added to $\gamma^{\rm rad}$ (by convolution of profiles), so giving the total
  $\gamma$ coefficient that enters in Eq. (\ref{eq:dampingpar}). Various
  collisional processes in turn contribute to $\gamma^{\rm col}$, being typically classified by the power index of the
  potential law that explains the collider interactions. Thus, we have the linear
  and quadratic Stark mechanisms ($\gamma^{\rm col}_2$ and $\gamma^{\rm col}_4$), resonance broadening ($\gamma^{\rm col}_3$) and Van der Waals broadening
  ($\gamma^{\rm col}_{6}$). Then,
\begin{equation}\label{eq:gamma_col}
\centering 
\gamma= \gamma^{\rm rad}+\gamma^{\rm col} \, =\, \gamma^{\rm rad} +
\sum_{i}\gamma^{\rm col}_i .
\end{equation}
Elastic collisions do not only broaden the profiles but also tend to eliminate
the phase correlations (coherences) between energy substates
$\ket{JM}$ and $\ket{JM^{\prime}}$, for what they are usually
referred to as depolarizing collisions. Elastic collisions do not change the
overall population of the level, but their
actual rates are necessary to treat transitions between sublevels in the polarized case \citep{Lamb:1971aa,Derouich:2003aa}. On the contrary, inelastic collisions
interchange energy between colliders, producing bound-bound
transitions among different energy levels.

\item Doppler broadening. Thermal motions below the plasma element
  scale produce spectral line broadening through the
  Doppler effect. The probabilistic distribution
  of velocities due to pure thermal motions is
  defined by the component form of a maxwellian
  distribution, which is
a gaussian. The Maxwell distribution is strictly valid under LTE
  conditions, but it is commonly used in most astrophysical applications. The spectral variation of the pure thermal broadening as
  seen by an stationary observer is then obtained as a
  convolution of a delta $\delta(\nu^{\prime}-\nu)$ that describes
  radiation emitted by a single
  particle (being $\nu^{\prime}\approx \nu-v_{\rm
    LOS}\nu_0/c$ the Doppler-shifted frequency emitted by the
  particle) with the
  Maxwell distribution that characterizes the thermal motions of the
  emitting/absorbing atoms. On the other hand, it is usual to also assume here the presence of
  microturbulent motions when modeling the observed spectral line
  radiation using one-dimensional models of stellar atmospheres. The microturbulent
  broadening is given by non-thermal
``turbulent'' velocities ($v_{\rm micro}$) but are also assumed to have a random nature
under the resolution element. The
total spectral profile produced by both Doppler broadening mechanisms is other
Gaussian whose total Doppler width is\footnote{The resulting profile
  is then a convolution of two Gaussian contributions, which is another
  Gaussian whose Doppler width is the geometrical sum of the contributing Doppler
  widths.}
\begin{equation}\label{eq:dnud_def}
\centering 
\Delta \nu_D \, =\, \frac{\nu_0}{c}\sqrt{\frac{2kT}{m}+v^2_{\rm micro}}, 
\end{equation}
with $m$ the atomic mass of the main species. The microturbulent
velocity $v_{\rm micro}$ is an ad-hoc fitting
parameter that was introduced in the past to correct for
deficiencies in plane-parallel modeling.  
\end{itemize}

Other two ``broadening'' mechanisms affecting the spectral variation of the profiles are:

\begin{itemize}[leftmargin=0.23cm, labelsep=0.05cm]
\item Statistical redistribution in frequency. Thermal motions in the
  stellar atmosphere produce
  complete redistribution in frequencies (CRD) within the Doppler Gaussian core of the
  emerging line profiles\footnote{If the scattering is
    coherent in the frame of the atom (no elastic collisions),
    frequency redistribution (in the
    observer frame) occurs
    over a range of $3\Delta \nu_D$ around the line center.}. However, toward the line profile wings (lorentzian region), the
  probabilistic distribution of frequencies is instead controlled by the radiative and collisional
  elastic rates. If the former rates dominate
  ($\rm{\gamma^{rad}>\gamma^{col}}$) at some height, the electrons leave their 
  atomic levels at exactly the same energy they arrived in, so maintaining the
memory of the previous process (coherent
  scattering). But, as the elastic collisional rates increases
  downward in the atmosphere with the perturber density, they
  are able to produce a frequency reshuffling at those layers before radiative transitions
  occur. Thus, it approaches CRD in the spectral line wings. In intermediate layers where
  $\rm{\gamma^{rad} \approx\gamma^{col}}$, scattering is neither
  coherent nor completely redistributed, which is known as partial frequency
  redistribution or PRD. Both coherent scattering and complete redistribution
  can be equivalently treated, frequency by frequency, just correctly modelling the
  local processes that build the line profile. However, in PRD, the
  probability of scattering photons from one  frequency to another is
  sensitive to the non-local monochromatic radiation field $J_{\nu}$
  reaching each plasma element, which must be accounted for. In some cases, the lines affected by PRD exhibit
 extense line wings with challenging and complicated polarization
 patterns not fully understood so far. That is not the case of the IR
 triplet of the Ca {\sc ii} lines, which are well described by CRD.

\item Zeeman splitting. Considering a magnetically sensitive transition
  that connects an upper
  energy level $\rm{J_u}$ with a lower level $\rm{J_{\ell}}$, the
  presence of a magnetic field produces an energy splitting in the
  magnetic sublevels. This changes the spectral profiles. In general, the radiation emitted by the
  various transitions between the upper sublevels ${\rm M_u}$ and the
  lower ones ${\rm M_{\ell}}$ is usually referred to as Zeeman
  components. They are characterized by well- defined polarization properties that depend
on the inclination $\hat{\theta}$ and azimuth $\hat{\chi}$ angles
of the magnetic field vector measured in the reference frame of the ray with direction $\vec{\Omega}$. For the electric dipole mechanism, the only allowed
  transitions are those with $\rm{\Delta M=M_u-M_{\ell}=0}$ ($\pi$ components) and those
  with $\Delta M=\pm 1$ ($\sigma_{\rm b}$ and $\sigma_{\rm r}$
  components, respectively). To specify the Zeeman profiles, we assume an isolated
  spectral line, no macroscopic
  velocity and a magnetic
  field weak enough for the Zeeman regime to hold.
 Furthermore, atomic polarization between magnetic sublevels
  is neglected in both levels of the transition.  Then, the total line
  profiles are \citep[e.g., ][]{LL04}

\begin{subequations}\label{eq:zeeeqs}
\begin{align}
\phi_{I} (\nu, \vec{\Omega})&={\displaystyle\frac{1}{2}\left[ \phi_0 \sin^2 \hat{\theta}+\frac{\phi_{-1}+\phi_1}{2} \left(1+\cos^2 \hat{\theta} \right)\right]}, \label{eq:zee1}
 \\
\phi_{Q} (\nu, \vec{\Omega})&= {\displaystyle
\frac{1}{2}\left[\phi_0 -\frac{\phi_{-1}+\phi_1}{2}\right]
\sin^2 \hat{\theta} \cos 2 \hat{\chi} }  , \label{eq:zee2} 
\\
\phi_{U} (\nu, \vec{\Omega}) &= {\displaystyle 
\frac{1}{2}\left[\phi_0 -\frac{\phi_{-1}+\phi_1}{2}\right]
\sin^2 \hat{\theta} \sin 2 \hat{\chi}}, \label{eq:zee3}
\\
\phi_{V} (\nu, \vec{\Omega}) &= {\displaystyle 
\frac{1}{2}\left[\phi_1 - \phi_{-1}\right] \cos{\hat{\theta}} } , \label{eq:zee4}
\end{align}
 \end{subequations}
where
\begin{equation}\label{eq:phicomp}
\centering 
\phi_q =\sum_{M_{\ell}M_u} 3 \left(
\begin{array}{ccc}
J_u & J_{\ell} & 1 \\
-M_u & M_{\ell} & -q
\end{array}
\right)^2\phi(\nu-\nu_{J_u M_u, J_{\ell} M_{\ell}}) \qquad (q=-1,0,-1)
\end{equation}
describe the superposition of Zeeman components. Each $\phi(\nu-\nu_{J_u
  M_u, J_{\ell} M_{\ell}})$ is a profile like Eq. (\ref{eq:voigt})
evaluated around its corresponding Zeeman frequency 
\begin{equation}\label{eq:kele}
\centering 
\nu_{J_u  M_u, J_{\ell} M_{\ell}} = \nu_0 + \nu_L (g_u M_u - g_{\ell} M_{\ell}), 
\end{equation}
with $\nu_0$ the central wavelength of the transition and $\nu_L$ the
Larmor frequency ($\nu_L [s^{-1}] =1.3996 \,\times \,10^6 \,
B[G]$). In wavelength units, the Larmor frequency becomes the
Zeeman splitting $\Delta\lambda_B=\lambda^2_0 \nu_L / c$. When the
Zeeman splitting is of the same order as the thermal Doppler width of the
profiles, the polarization signals are in \textit{Zeeman regime}. If
the Zeeman splitting is small compared to the thermal width (e.g., due to weak
  magnetic fields or in some lines at
  optical wavelengths), the \textit{weak-field regime} holds. In both cases the Zeeman
  components superpose producing a broadened line profile (magnetic broadening).
\end{itemize}
\subsection{Radiative transfer coefficients}
The coefficients appearing in Eq. (\ref{eq:basicrte}) are the
important connection between the \textit{micro} and the \textit{macro} state of the
plasma. They have two
contributions: continuum and
line processes. 
In the solar atmosphere,
the continuum opacities and emissivities
are due to free-free and bound-free
transitions, Thompson scattering and Rayleigh scattering. 

On the other hand, the line terms are due to bound-bound transitions
in the atomic species under consideration. Thus, line emission is described by the quantities
$\epsilon^l_I,\epsilon^l_Q,\epsilon^l_U$ and $\epsilon^l_V$ in terms
of the atomic density matrix elements and the emission profile $\phi_{\nu}$. Namely, in the case of
weak magnetic field and no stimulated emission, the line emissions in
I, U and Q  for a
transition $u\rightarrow {\ell}$ are 
\citep{manso10}:
\begin{subequations}\label{eq:emi_iqu}
\begin{align}
\epsilon^l_I(\nu,\vec{\Omega}) = \, \epsilon^0 \rho^0_0 +
\epsilon^0\omega^{(2)}_{J_u J_{\ell}} \sqrt{3} &\Big\{
\frac{1}{2\sqrt{6}}\Big(3\mu^2-1 \Big)\rho^2_0  - \mu\sqrt{1-\mu^2}\Big(\cos{\chi} {\rm Re}[\rho^2_1]
-\sin{\chi} {\rm Im}[\rho^2_1]\Big)\nonumber\\
& +\frac{1}{2}\Big (1-\mu^2 \Big) \Big (\cos{2\chi}{\rm Re}[\rho^2_2]
-\sin{2\chi}{\rm Im}[\rho^2_2]\Big) \Big\},
\label{eq:ei}
\end{align}
\begin{align}
\epsilon^l_Q(\nu,\vec{\Omega}) = \,-\epsilon^0\omega^{(2)}_{J_u J_{\ell}} \sqrt{3}&\Big\{
\frac{3}{2\sqrt{6}}\Big (\mu^2 -1\Big)\rho^2_0  -\mu\sqrt{1-\mu^2}\Big (\cos{\chi} {\rm Re}[\rho^2_1]
-\sin{\chi} {\rm Im}[\rho^2_1]\Big)\nonumber\\
& -\frac{1}{2}\Big (1+\mu^2 \Big) \Big (\cos{2\chi}{\rm Re}[\rho^2_2]
-\sin{2\chi}{\rm Im}[\rho^2_2]\Big) \Big\},
\label{eq:eq}
\end{align}
\begin{align}
\epsilon^l_U(\nu,\vec{\Omega}) = \, -\epsilon^0\omega^{(2)}_{J_u
J_{\ell}} \sqrt{3} &\Big\{ \sqrt{1-\mu^2}\Big (\sin{\chi} {\rm
Re}[\rho^2_1]+\cos{\chi} {\rm Im}[\rho^2_1]\Big) \nonumber\\
& +\mu \Big (\sin{2\chi}{\rm Re}[\rho^2_2]+\cos{2\chi}
{\rm Im}[\rho^2_2]\Big)\Big\},
\label{eq:eu}
\end{align}
\end{subequations}
being $\mathrm{\mu=\cos(\theta)}$ and $\{\theta,\chi\}$ the
inclination and azimuth of the ray $\vec{\Omega}$ with respect to the
local solar vertical
(quantization axis). The direction of reference for $Q>0$ is parallel
to the nearest solar limb from the observed point. All the
density-matrix components in these expressions correspond to the upper
level of the transition. The $\omega^{(2)}_{J_u J_{\ell}}$ coefficients were introduced by
\cite{Landi-DeglInnocenti:1984} and
$\epsilon^0=(h\nu/4\pi)A_{u{\ell}}n_{\rm at}\sqrt{2J_u+1}\rho^0_0(J_u)
\phi_{\nu} $ is calculated with the Voigt profile $\phi_{\nu}$ and with
$n_{\rm at}$ the total number of atoms of the considered species per unit volume. 

 Since in this thesis we assume a negligible contribution of atomic
 polarization to Stokes V, the corresponding line emission coefficient
 is dominated by the Zeeman effect. Thus, under the same previous assumptions: 
\begin{equation}\label{eq:ev}
\centering 
\epsilon^l_{V} (\nu, \vec{\Omega}) = \epsilon^0 \phi_V(\nu,\vec{\Omega}),
 \end{equation}
where $\phi_V$ was defined in Eq. (\ref{eq:zee4}). We remark again
that the angles ($\hat{\theta},\,\hat{\chi}$) appearing in such
equation are measured with respect to the magnetic field direction. 
The corresponding absorption coefficients $\eta^l_I,\eta^l_Q,\eta^l_U, \eta^l_V$ have
identical expressions to the emission ones, but changing
$A_{u{\ell}}\leftrightarrows B_{{\ell} u}$ and $u\leftrightarrows
{\ell}$ in all the subscripts (including the ones in the expression of
$\epsilon^0$). Furthermore, the
$\rho^K_Q$ elements become the ones of the lower level of
the transition.

The anomalous dispersion coefficients $\rho_Q,\rho_U, \rho_V$  are given by the same
equations than the $\eta^l_Q,\eta^l_U, \eta^l_V$, respectively, but
substituting the profiles $\phi_{\nu}$ (appearing in $\epsilon^0$ and in the Zeeman
components of Eq. (\ref{eq:phicomp})) by the
normalized anomalous dispersion profile $\psi_{\nu}$, introduced in
Sec. \ref{sec:spectral}. Considering the total complex profile $\Phi_{\nu}$
(Sec. \ref{sec:spectral}), the absorption and dispersion coefficients
can be seen as real and imaginary part of the same complex coefficient
$X_k=\eta_k+i\rho_k$ with $k \equiv Q,\,U,\,V$. 

The source function $S_{\nu}=\epsilon_I/\eta_I$ is an important
quantity related to the intensity coefficients that describes the
propagation of the intensity in the plasma. Under LTE conditions, it equals the Planck
function. 

\subsection{Statistical equilibrium equations}
\begin{figure}
\centering
\includegraphics[scale=0.8]{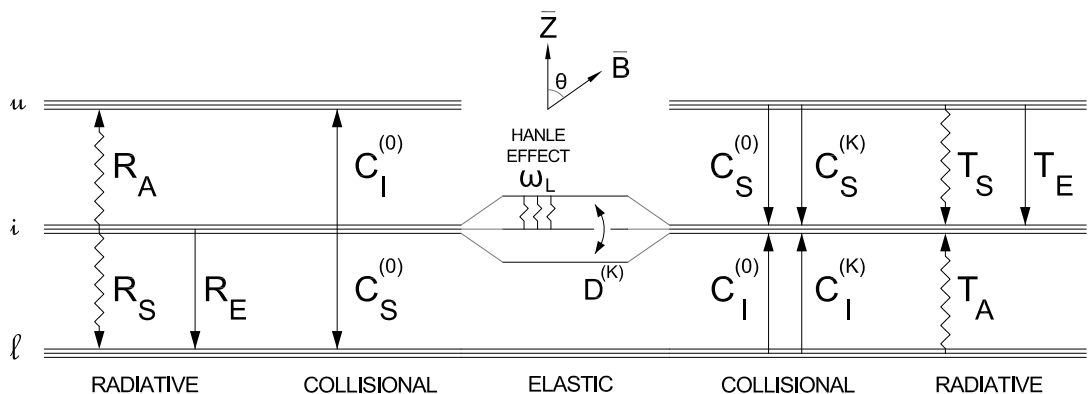}
\caption{Schematic representation of the radiative, collisional and
  magnetic processes included in the rate equations of a given level
  $i$. All the upper and lower levels are represented by $u$ and
  ${\ell}$. On the right, the processes populating $i$
  (transfer rates). On the left, the processes depopulating $i$
  (relaxation rates). In the middle, the processes that do not change
  the overall population of $i$. \label{fig:see}}
\end{figure}

In the case of LTE, the populations of the atomic energy levels are
given by the Saha-Boltzmann distributions \citep{Mihalas:1978}. They are
totally determined by the local conditions of the plasma (basically,
the density and a common temperature for all the particles) resulting in magnetic
energy sublevels that are equally populated (hence, without atomic
level polarization).
When the spectral line is formed under NLTE conditions, the energy level and
sublevel populations are not dominated by the collisional rates but by the radiative
transitions. The radiation field interacts with matter through
radiative excitations, photoionizations and their inverse processes
(radiative desexcitations and recombinations), which strongly affects the
atomic level populations especially in the outer
atmosphere. 
To determine the resulting excitation state of the atoms in the
plasma element, we have to solve the rate equations for the atomic
density matrix corresponding to each level $i$. There is a rate equation per
multipolar component of the density matrix. The rate equations
account for the time evolution of the atomic system by specifying
all the processes (see Fig. \ref{fig:see}) that produce trasitions between energy
states. In the case of a multilevel atom without hyperfine structure,
and neglecting coherence between sublevels of different levels, the rate
of change of the density matrix element
$\rho^K_Q(J_i)$ in the solar vertical frame reads \citep{LL04}:

\begin{align}
\begin{split}\label{eq:generalsee} 
\frac{\rm d}{{\rm d}t} \rho^K_Q(J_i) = -i\omega_L g_{J_i}
\sum_{Q^{\prime}} K^K_{QQ^{\prime}} &\cdot \rho^K_{Q^{\prime}}(J_i) -D^{(K)}(J_i)\cdot \rho^{K}_{Q}(J_i)\\
  + \sum_{J_u} \sum_{K_u Q_u} \rho^{K_u}_{Q_u}(J_u) \cdot T_E(J_u;K_uQ_u \rightarrow J_i;K&Q) + \sum_{J_l} \sum_{K_l Q_l} \rho^{K_l}_{Q_l}(J_l) \cdot T_A(J_l;K_lQ_l \rightarrow J_i;KQ) \\
 + \sum_{J_u} \sum_{K_u Q_u} \rho^{K_u}_{Q_u}(J_u)  \cdot T_S (&J_u ; K_u Q_u \rightarrow J_i ; KQ) \\
  + \sum_{J_u} \sqrt{\frac{2J_u+1}{2J_i+1}} \rho^{K}_{Q}(J_u) \cdot C^{(K)}_S(J_u \rightarrow&J_i) + \sum_{J_l}\sqrt{\frac{2J_l+1}{2J_i+1}} \rho^{K}_{Q}(J_l) \cdot C^{(K)}_I(J_l \rightarrow J_i) \\
 - \sum_{K^{\prime} Q^{\prime}}  \rho^{K^{\prime}}_{Q^{\prime}}(J_i)\cdot R_E(J_i;KQ,K^{\prime}Q^{\prime} \rightarrow &J_l) - \sum_{K^{\prime}Q^{\prime}} \rho^{K^{\prime}}_{Q^{\prime}}(J_i)\cdot R_A(J_i;KQ,K^{\prime}Q^{\prime} \rightarrow J_u) \\
  - \sum_{K^{\prime} Q^{\prime}} \rho^{K^{\prime}}_{Q^{\prime}}(J_i)\cdot R_S&(J_i;KQ,K^{\prime}Q^{\prime}\rightarrow J_l ) \\
 - \rho^{K}_{Q}(J_i) \cdot \bigg[ \sum_{J_u} C^{(0)}_I(J_i & \rightarrow  J_u )+\sum_{J_l} {C^{(0)}_S }(J_i \rightarrow J_l ) \bigg] ,
\end{split}
\end{align}
where we have included the effect of elastic and
inelastic collisions assuming they are isotropic and
that the impact approximation is valid.
The first term in the r.h.s. of this equation accounts for
the effect of the magnetic field (Hanle effect, see Sec. \ref{sec:introhanle}). The
second term shows the action of the depolarizing
collisions through the elastic collisional rate $D^{(K)}$. Both terms
affect only the population imbalances inside the same energy level,
either by the coherence or by population redistributions
between energy sublevels. 

In general, the rate $X(J; KQ \rightarrow J^{\prime};K^{\prime}Q^{\prime})$
represents the probality that a transition carries atomic coherence from
the level $J$, where is described by
the multipole $\rho^K_Q$, to
the level $J^{\prime}$, where it generates
$\rho^{K^{\prime}}_{Q^{\prime}}$. The quantities $C_I$ and $C_S$ are the inelastic
and superelastic collisional rates, respectively. The
radiative rates are divided in the (populating) transfer rates
$T_A$, $T_E$, $T_S$,
and the (depopulating) relaxation rates
$R_A$, $R_E$, $R_S$, which together describe the effects of absorption (A), spontaneous emission (E) and
stimulated emission (S).  
The explicit expressions of
all the rates can be found in Chapter 7 of 
\cite{LL04}. 

The whole system of $\rm{N_{KQ}}$ equations (one per unknown
$\rho^K_Q$) at a given position in the atmosphere can be written in
matrix form as:
\begin{equation}\label{eq:seesystem}
\centering 
{\bf A}({\bm \rho}) \cdot {\bm \rho}={\bf f} , 
\end{equation}
where ${\bf A}$ is a  $\rm{N_{KQ} \times N_{KQ}}$  matrix containing
all the transition rates, ${\bm \rho}$ is a vector of length $\rm{N_{KQ}}$ containing the
$\rho^K_Q$ elements for all the levels in the atomic model and
$\bf{f}$ is a vector with zeros everywhere (assuming statistical
equilibrium; i.e., $\frac{\rm d}{\rm dt}\rho^K_Q(J_i)=0$). Note that
${\bf A}$ implicitly depends on the unknown $\rho^K_Q$ elements
through the radiation field and that such a system of equations is highly non-linear. 

The resulting set of equations is not
linearly independent. To close the system, we have to substitute one of the
equations, typically that of the ground level, by the equation of 
conservation of particles (the trace equation in the density matrix
formalism). It establishes that the overall
population of the atomic model considered is conserved: 
\begin{equation}\label{eq:conserva}
\centering 
\sum_i \sqrt{2J_i+1}\rho^0_0(J_i)=1
\end{equation}
Thus, one of the zeros in $\mathbf{f}$ must be substituted by $1$.

The combination of the SEEs with the RTE poses the radiative transfer
problem. The input of the former
(the radiation field tensor components) depends on the output of the
latter (the Stokes vector) and viceversa, which implies that
an iterative method is needed to find the solution. Once the
self-consistent solution for the $\rho^K_Q(J)$ elements is found, we
can compute the emergent Stokes profiles for any desired line of sight.

\subsection{Radiation field.}\label{sec:radfield}
In concordance to the previous treatment of the density matrix, the radiation field illuminating each
plasma element is also expressed in the spherical tensor
representation. Thus, we obtain the tensor components $J^K_Q$, having
$-K\leq Q \leq K$ with only three possible ranks $K=0,1, 2$
\citep{LL04}. They are integrals over frequency and angle of the Stokes
vector components $I$, $Q$, $U$ and $V$.
In particular, the even $K$ components are the main excitation sources for the
scattering polarization. The explicit expressions for $K=0$ and $K=2$ are\footnote{The components with
  $Q<0$ are obtained with the conjugation property $J^K_{-Q}=(-1)^Q[J^K_Q]^*$ (with $*$ the symbol
  for complex conjugation).}
\begin{equation}
\centering 
{\bar{J}^0_0}(z)=\displaystyle{ \int {\rm d}{\nu} 
\oint \frac{{\rm d} \vec{\Omega}}{4\pi} \phi_{\nu} {{I_{\nu
      \vec{\Omega}}}}} ,
\end{equation}
\begin{equation}
\centering 
{{\bar{J}^2_0}}(z)=\displaystyle{\int {\rm d}{\nu} 
\oint \frac{{\rm d} \vec{\Omega}}{4\pi} \phi_{\nu}
\frac{1}{2\sqrt{2}} \left[(3\mu^2-1){{I_{\nu \vec{\Omega}}}}+
3(1-\mu^2){Q_{\nu \vec{\Omega}}}\right]},
\end{equation}
\begin{equation}
\centering 
{{\bar{J}^2_1}}(z)=\displaystyle{\int {\rm d}{\nu} 
\oint \frac{{\rm d} \vec{\Omega}}{4\pi} \phi_{\nu}
\frac{\sqrt{3}}{2} {e}^{i\chi}\sqrt{1-\mu^2} \big[-\mu ({I_{\nu
    \vec{\Omega}}}-{Q_{\nu \vec{\Omega}}} )+iU_{\nu \vec{\Omega}}\big]},
\end{equation}
\begin{equation}
\centering 
{{\bar{J}^2_2}}(z)=\displaystyle{\int {\rm d}{\nu} 
\oint \frac{{\rm d} \vec{\Omega}}{4\pi} \phi_{\nu}
\frac{\sqrt{3}}{2} {e}^{i 2\chi}\ \big[\frac{1}{2} ({1-\mu^2}) {I_{\nu
    \vec{\Omega}}}+\frac{1}{2} ({1+\mu^2}) {Q_{\nu
    \vec{\Omega}}}+i \mu U_{\nu \vec{\Omega}}\big]}.
\end{equation}
As usually along this thesis, the reference direction for $Q>0$
(defined in the plane of the sky) is chosen
parallel to the solar limb that is nearest to the local vertical (z
axis)\footnote{This reference is sometimes referred to as the direction
  perpendicular to the plane containing the line of sight
  ($\vec{\Omega}$) and the local solar vertical (z axis).}.
 The anisotropy factor for each spectral line transition is defined as:
\begin{equation}\label{eq:aniso}
\centering 
w_{\mathrm{line}} (z)={\sqrt{2}} \frac{\bar{J}^2_0(z)}{\bar{J}^0_0(z)}.
\end{equation}
It ranges from $w_{\mathrm{line}}=-0.5$ (azimuthally
independent incident radiation
field contained in the horizontal plane) to $w_{\mathrm{line}}=1$ (collimated vertical
beam), vanishing at the bottom of the atmosphere where the radiation field is
unpolarized and isotropic. The dominating factor $(3\mu^2-1)I$ in the definition of $\bar{J}^2_0$ makes that the rays 
with $54.73\deg<\theta< 125.27\deg$ (which are mainly horizontal) contribute always negatively to the
anisotropy factor, while radiation coming from other directions
(that is, mainly vertical) always contributes positively. The
anisotropy factor is thus
especially sensitive to temperature gradients \citep[][]{Trujillo-Bueno:2001aa,LL04}.

The $J^2_Q$ components with $Q\neq 0$ measure the breaking of the
axial symmetry of the radiation field.
In plane-parallel atmospheres only the
magnetic field can break the symmetry of the radiation field and generates such components.

\section{Numerical methods for polarized radiative
  transfer.}\label{sec:numerical} 

\subsection{Integration of the transfer equation.}\label{sec:formalsolver} 
The formal solver used in
our calculations is based on a
short-characteristics scheme \citep{kunasz-auer}, which allows the numerical integration
of the RT equations along ray paths between neighbouring points. 

The RTE (see Eqs. (\ref{eq:basicrte})) in compact form can be written as
\begin{equation} \label{eq:matrixrte}
\centering 
\frac{\rm{d}}{\rm{ds}} {\bf I} = {\bm \epsilon}- {\bf K I},
\end{equation}
where s is the geometrical distance along the ray and ${\bf K}$, ${\bm \epsilon}$ and ${\bf I}$ are the propagation
matrix, the emission and the Stokes vectors, respectively. Making a
change of variable to $\tau=\int \eta_I {\rm d} s$ and formally integrating
between consecutive points along the ray under consideration we
get \citep{Rees:1989}:
\begin{equation} \label{eq:solformal}
\centering 
 {\bf I}_{\rm O} =  {\bf I}_{\rm M} \,{\rm e}^{-\Delta\tau} +\int^{\tau_{\rm M}}_{\tau_{\rm O}} \left[
   \frac{1}{\eta_I}{\bm \epsilon}-\left( \frac{1}{\eta_I}{\bf K}- {\bf
       1} \right){\bm I} \right] {\rm e}^{-(\tau-\tau_{\rm O})} {\rm d}\tau,
\end{equation}
where $ {\bf 1}$ is the $4\times 4$ identity matrix, $\Delta\tau$
is the optical thickness between M and O and $\tau$ is the optical
path along the ray.

For a given direction and frequency, the optical depth
discretization is obtained from the geometrical depth grid, taking
into acount its definition and assuming an exponential dependence of
$\eta_I$ with \textit{z}. If we assume a parabolic variation of
$\mathbf{S}={\bm \epsilon} / \eta_I$ between three successive points M, O and P along the
ray, while assuming that $({\bf K}/\eta_I-1){\bf I}$ varies linearly between M and
O, then we can perform the integral in Eq. (\ref{eq:solformal}) and obtain
the Stokes vector at point O with \citep{Trujillo-Bueno:2003} 
\begin{equation} \label{eq:idelopar}
\centering 
 [\mathbf{1}+\Psi^{\prime}_{\rm O} \,\mathbf{K}^{\prime}_{\rm
   O}]\mathbf{I}_{\rm O} =   \, [{\rm e}^{-\Delta\tau_{\rm M}}\mathbf{1}+\Psi^{\prime}_{\rm M}\, \mathbf{K}^{\prime}_{\rm M}]\, \mathbf{I}_{\rm M} 
 +\Psi_{\rm M} \, \mathbf{S}_{\rm M}+\Psi_{\rm O} \,\mathbf{S}_{\rm O}+\Psi_{\rm P}, \,\mathbf{S}_{\rm P},
\end{equation}
where $\Psi_{M, O,P}$ are numerical coefficients that depend on the
optical distances $\Delta \tau_{\rm M}$ between M and O, and $\Delta
\tau_{\rm P}$
between O and P. Similarly, $\Psi^{\prime}_{\rm M, O}$ only depend on
$\Delta \tau_{\rm M}$. Explicit expressions for these coefficients are given
in \cite{kunasz-auer}. This generalization of the
short-characteristics method to the polarized case is called
DELOPAR. To apply it, we start at one atmosphere's boundary and
successively calculate $\mathbf{I}_{\rm O}$ at each grid point, proceeding
along any given ray direction until the opposite boundary. 

Finally, to obtain the radiation field tensors (see Section \ref{sec:radfield}) we have to integrate
numerically the calculated
Stokes parameters for all the points in the frequency and angular
discretization. For the frequency quadrature we use a trapezoidal
rule over the absorption profiles. For the polar angles we use a
Gaussian quadrature in inclination and an equally spaced trapezoidal
rule for the azimuth. Formally, we can write the radiation field tensor components as
\begin{equation}\label{eq:numethod1}
\centering 
\mathbf{J^{K}_Q}=\mathbf{\Lambda}[{\bm\rho}^{\bf K}_{\bf Q}] +{\bf T^K_Q} ,
 \end{equation}
where $\mathbf{\Lambda}$ is an operator\footnote{To illustrate
  the following concepts we symplify for the moment the notation of the $\Lambda$
operator. As will be detailed later, it is different at each spatial
grid point but also when connecting different tensor components.} giving the response of the
radiation field to perturbations in the density-matrix elements and
the $\bf{T^K_Q}$ vector gives the contribution of the boundary conditions to the
radiation field at the spatial points considered.

\subsection{Statement of the iterative problem.}\label{sec:iterative} 
In general, to solve the RT problem with polarization it is 
necessary to calculate the self-consistent values for the $\rho^K_Q$ multipolar
components of the density matrix at each spatial point $i$ in the
model. The main problem is that such components are coupled in a
highly non-linear and non-local way through the radiation field. The
strategy is to start from an estimation of the solution and to perform
iterative corrections to that estimation until arriving to self-consistent values.

The procedure is as follows. Given an initial guess $\rho^{K\, \rm{old}}_Q(i)$ at each depth,
the radiative transfer equations are integrated to obtain the Stokes
parameters (at each depth and for any angle and frequency), which gives
the radiation field tensor components $J^{K\, \rm{old}}_Q(i)$ at each
of the $\mathrm{N_z}$ heights of the model atmosphere under
consideration. From Eq. (\ref{eq:numethod1}), it can be expressed as
\begin{equation}\label{eq:numethod2}
\centering 
\mathbf{J^{K\, \rm{old}}_Q}=\mathbf{\Lambda}[{\bm\rho}^{{\bf K}\,
  \rm{old}}_{\bf Q}] +{\bf T^K_Q} ,
 \end{equation}
Once the $J^{K\, \rm{old}}_Q(i)$ are estimated, the transfer rates can be
calculated and the algebraic system
formed by the SEEs (Eq. \ref{eq:seesystem}) is linearized and solved as
\begin{equation}\label{eq:numethod3}
\centering 
{\bf A}^{\rm old} \cdot {\bm \rho}^{\rm new}={\bf f} 
\end{equation}
 to obtain the new corrected elements $\bm{\rho}^{\rm{new}}={\bm
   \rho^{\rm{old}}}+\delta {\bm \rho}$. We remark that $\bf{A^{\rm
     old}}$ depends on the ${\bm\rho^{\rm{old}}}$ values. 

 If the initial guess is not the exact solution, then $\delta {\bm \rho}
 \neq 0 $ and Eq. (\ref{eq:numethod3}) will
 have a certain residual error. The objective is to find the
 $\bm{\rho}^{\rm{new}}$ values (equivalently, the
 corrections $\delta {\bm \rho}$) that leads to precisely fulfill
 Eq. (\ref{eq:seesystem}). The linearization makes the
 solution at each step to never be exact, but successive iterations reduce the errors to the desired
 small size as $\delta {\bm \rho}\rightarrow 0$ \citep{Hubeny:1992aa}.

 The different iterative
 methods applied to the RT problem are distinguished by the way they
 approximate  in Eq. (\ref{eq:numethod3}) the $\bm{\rho}^{\rm{new}}$ values from the
 $\bm{\rho}^{\rm{old}}$ by modifying $\bf{A^{\rm old}}$. The simplest method is the $\Lambda$-iteration, which consists in introducing the 
$\bm{\rho}^{\rm{old}}$ values again into
Eq. (\ref{eq:numethod2}) to build $\bf{A^{\rm old}}$, thus using only
the $\rho^K_Q$ values of the
previous iterative step. The convergence rate of this method is very
poor in optically thick atmospheres because the numerical information
is propagated through the spatial grid along one photon mean free path per
iteration, which takes many iterations
to radiatively connect all the points in the atmosphere.

In the following, we specify the conceptual strategies used by more
sophisticated methods to estimate the new  $\rho^K_Q$ values. Basically, we require a formal solver
 to integrate the RT equations 
 and a suitable linearization of the SEE
 that guarantees convergence in the iterative process.
 \subsection{Iterative scheme.}\label{sec:mali} 
The linearization of Eqs. (\ref{eq:seesystem}) can
be achieved by several methods. The one we use in our calculations is based on two
techniques: operator splitting \citep{Cannon:1973} and
preconditioning \citep{rybicki-hummer,Socas-Navarro:1997}.

Applied to the polarized case, the operator splitting strategy
rewrites the formal solution of the RT equation given by Eq. (\ref{eq:numethod1}) as:
\begin{equation} \label{eq:op_splitting}
\centering 
\mathbf{J^{K}_Q}=\mathbf{\Lambda}^{\bf *}[{\bm\rho}^{{\bf K}}_{\bf Q}] +(\mathbf{\Lambda}-\mathbf{\Lambda}^{\bf *})[{\bm\rho}^{{\bf K}}_{\bf Q}] +{\bf T^K_Q} ,
\end{equation}
where $\mathbf{\Lambda}^{\bf *}$ is an approximation to the full
operator $\mathbf{\Lambda}$. This splitting will allow the substitution of \textit{some}
$\rho^{K \, {\rm old}}_Q$ values by their implicit ``new'' values in the equations, which
approaches the final solution at a higher convergence rate. A good
choice for the approximate
operator $\Lambda^{*}$ is the diagonal of the full $\Lambda$ operator (so making it \textit{local}) because it
is easy to obtain and to invert. This is the extension
of the Accelerated $\Lambda$-iteration (ALI\footnote{Sometimes known as Jacobi iteration in the astrophysics
literature; as MALI, when applied to
multilevel systems; or as DALI, when applied to the density matrix
formalism in the polarized case.}) method to the polarized case \citep{Trujillo-Bueno:1995}. 

The next step is to approximate the dominant tensor component
$J^0_0(i)$ by considering
that, for each \textit{i}-th
spatial point, it can be obtained using the $\rho^{K \, {\rm old}}_Q$
values at all grid points but only $\rho^{0\,{\rm new}}_0(i)$ at point
\textit{i}. From Eq. (\ref{eq:op_splitting}), this can be expressed as
\begin{equation} \label{eq:op_splitting2}
\centering 
J^{0}_0(i)=J^{0\,{\rm old}}_0(i) + \Lambda(i,i)[\rho^{0}_0(i)^{
  \rm{new}} -\rho^{0}_0(i)^{
  \rm{old}}],
\end{equation}
where the numerical values of the r.h.s. are known except
for $\rho^{0}_0(i)^{ \rm{new}}$, which is the implicit unknown to be
calculated in the current iterative step. Here, the $\rho^0_0$ values correspond to the
upper level of the transition.

Substitution of
Eq. (\ref{eq:op_splitting2}) into the SEE yields a
non-linear system of algebraic equations. This non-linearity is due to
terms of the form $\rho^K_Q({\ell}) J^0_0$ that are tied to the
absorption rates pumping population and coherence from each lower level $\ell$. In order
to linearize such terms we calculate them at each spatial point $i$ making \citep{Trujillo-Bueno:2003}
\begin{equation} \label{eq:preconditioning}
\centering 
\rho^K_Q(J_{\ell}) J^{0}_0 \rightarrow \rho^{K}_Q(J_{\ell})^{ \rm{new}} J^{0\,{\rm old}}_0 + \Lambda(i,i)[ \rho^{K}_Q(J_{\ell})^{\rm old}\rho^{0}_0(J_{ u})^{
  \rm{new}}-  \rho^{K}_Q(J_{\ell})^{\rm new}\rho^{0}_0(J_{u})^{\rm{old}}  ],
\end{equation}
again with all the quantities evaluated at point $i$, the $\rho^0_0$
components corresponding to the upper level of
the transition and $\rho^K_Q$ to the lower one. This is the
preconditioning scheme we use in our calculations. It allows to transform the
statistical equilibrium equations given by
Eqs. (\ref{eq:seesystem}) in a linear system at each
iterative step. 

In summary, the combination of preconditioning and operator splitting
achieves linearity building ${\bf A^{\rm
    old}}$ with a strategical combination\footnote{Avoiding
multiplications of two ``new'' values.} of ``new'' and ``old''
 terms indicated by
Eq. (\ref{eq:preconditioning}). With respect to $\Lambda$ iteration, the only extra calculation
in the ALI method are the local elements $\Lambda(i,i)$, which
can be efortlessly obtained using a
formal solver based on short-characteristics.
\subsection{Evaluation of \texorpdfstring{$\Lambda^{*}$}{L*}}\label{sec:calcop} 
We can detail Eq. (\ref{eq:numethod1}) as
\begin{subequations}\label{eq:formalsol1}
\begin{align}
  0\rightarrow&\quad  \mathbf{J^0_0}=\mathbf{\Lambda}_{00}[{\bm\rho}^{\bf 0}_{\bf
      0}(u)] +\mathbf{\Lambda}_{01}[{\bm\rho}^{\bf 2}_{\bf
      0}(u)] + \cdots +\mathbf{\Lambda}_{\rm  0N}[{\bf \hat{{\bm \rho}}}^{\bf 2}_{\bf
      2}(u)]  +{\bf T}^0_0\label{eq:forsol1}
 \displaybreak[0] \\
 1\rightarrow&\quad     \mathbf{J^2_0}=\mathbf{\Lambda}_{10}[{\bm\rho}^{\bf 0}_{\bf
      0}(u)] +\mathbf{\Lambda}_{11}[{\bm\rho}^{\bf 2}_{\bf
      0}(u)] + \cdots +\mathbf{\Lambda}_{\rm 1N}[{\bf \hat{{\bm \rho}}}^{\bf 2}_{\bf
      2}(u)]  +{\bf T}^2_0\label{eq:forsol2}
 \displaybreak[0] \\
\beta\rightarrow&\quad  \vdots \nonumber
 \displaybreak[0] \\
 N\rightarrow&\quad  \mathbf{\hat{ J}^2_2}=\mathbf{\Lambda}_{\rm N0}[{\bm\rho}^{\bf 0}_{\bf
      0}(u)] +\mathbf{\Lambda}_{\rm N1}[{\bm\rho}^{\bf 2}_{\bf
      0}(u)] + \cdots +\mathbf{\Lambda}_{\rm NN}[{\bf \hat{{\bm \rho}}}^{\bf 2}_{\bf
      2}(u)] +{\bf \hat{T}}^2_2 \label{eq:forsol3}
 \end{align}
 \end{subequations}
where ${\bf J^K_Q}$, ${\bf T^K_Q}$ and ${\bm \rho}^{\bf K}_{\bf Q}$ are formal
vectors with $N_z$ components, one per spatial grid
point, whose imaginary parts are 
indicated with the hat ( $\hat{}$ ); and where $N+1$ is the total number of multipole components different
from zero in the problem considered (6 in our most general case). Each $\Lambda_{\alpha \beta}$ are $N_z \times N_z$ operators
with elements $\Lambda_{\alpha \beta}(i,j)$. Their expressions in terms of $\Psi$, $\Psi^{\prime}$ and
$\rho^K_Q(\ell)$ multipoles are given in \cite{Manso-Sainz:2002}.

In the standard methods, only the calculation of a few $\Lambda_{\alpha \beta}(i,j)$ elements is
necessary. We show the Eqs. (\ref{eq:formalsol1}) just to illustrate how the 
calculation of the required operator elements can be done at the same
time the formal solution of the Stokes vector is computed. Namely, to
obtain an specific $\Lambda_{\alpha \beta}(i,j)$ element, we have to
calculate the $j$-th
component of the ${\bf J^K_Q}$ multipole in the equation $\beta$ after
making ${\bf T^K_Q}=0$ and taking equal to zero all the ${\bm \rho^K_Q}$ vectors except for the $i$-th
component of the ${\bm \rho}^{\bf K}_{\bf Q}$ multipole corresponding to the
$\alpha$ position, which has value unity. The result is that the
$\Lambda_{\alpha \beta}(i,j)$ values can be
obtained at the same time and following similar operations than when
calculating the radiation field tensors with the formal solution.

To understand how the different iterative methods work, we can also make
explicit the action of a given $\Lambda_{\alpha \beta}$
operator. Take for instance $\Lambda_{00}$. In general, that component operates as
\begin{subequations}\label{eq:FSdesglosada}
\begin{align}
 {\displaystyle  J^0_0(i)}=\Lambda_{00}(i,1)[\rho^0_0(1)]^{\rm
   A}+&\cdots+\Lambda_{00}(i,i-1)[\rho^0_0(i-1)]^{\rm A}\\[1.4em]
+\Lambda_{00}(i,i)&[\rho^0_0(i)]^{\rm  B}\\[1.4em]
+\Lambda_{00}(i,i+1)[\rho^0_0(i+1)]^{\rm C}&+\cdots+\Lambda_{00}(i,N_z)[\rho^0_0(N_z)]^{\rm C}\\[1.4em]
+ &T^K_Q   ,
\end{align}
\end{subequations}
where all the letters in brackets especify a spatial position and the superscripts A,B and C can be ``old'' or ``new'' depending
on the iterative scheme followed. Thus, in the simplest case of $\Lambda$-iteration they are $A=B=C=\rm{old}$, which means that $J^K_Q=J^{K\,
  {\rm old}}_Q$ for any K and Q. This can be interpreted as if we were solving
the SEE with a radiation field
that does not react to the corrections in the populations because the operator
${\bf A}$ of Eqs. (\ref{eq:seesystem}) was exclusively built with
the old population estimation. 
In the ALI method used in this thesis, $A=C=$old but
$B={\rm new}$, which means that $J^0_0$ is only affected by the local
corrections in the new populations. 
In the case we choose $A=B=\rm{new}$ with $C={\rm old}$, we would be updating the
contributions given by all the spatial points that were previously
considered when performing the formal solution of the RT equation
along a ray. In other words, we use an approximate
triangular operator ${\bm \Lambda^*}$ that contains more
information about the radiative couplings between points in the
atmosphere than the diagonal approximate ${\bm \Lambda^*}$ operator. 
This gives a faster iterative method known as
Gauss-Seidel and SOR, which can be also efficiently
solved \citep{Trujillo-Bueno:1995}. 
The main drawback of these operator splitting methods is the
deterioration of the convergence rate when the spatial resolution of
the grid is refined. In the limit of an infinitely fine grid, all such
methods will converge as slow as the $\Lambda$-iteration method. For
iterative methods whose convergence rate is insensitive to the grid
size consult \cite{Fabiani-Bendicho:1997}.
 
\section{Line formation in moving atmospheres}\label{sec:rtmoving} 
The existence of systematic velocity fields in solar and stellar
atmospheres is well-documented by abundant observational
evidence. Plasma motions appear to be present on all scales, from
microscopic thermal to macroscopic motions (i.e., nonthermal
velocities that are coherent over distances much larger than a
particle mean-free path). Indeed, the consideration of systematic motions is
essential to explain the observations in many astrophysical contexts: pulsating stars, hot stars with fast stellar
winds, expanding nebulae, supernovae and fast changes
in solar prominences are good examples. In particular, the increasing
interest in the highly inhomogeneous and dynamic solar chromosphere has served to highlight 
plasma dynamics as a key point to understand the outer solar layers. 
 
The modelling of radiative transfer in dynamic atmospheres is more complicated than in
static atmospheres. On one hand, the presence of
velocity gradients makes the
opacity and emissivity angle-dependent, which may have a significant
impact on the radiation field anisotropy. On the other hand, any of the existing methods to solve the RTE add higher levels of
complexity in dynamic atmospheres, with significant increase of the computational
demand, either because the RTE adopts a more involved form to treat the
problem in a comoving frame,  or due to a meaningful
increment in the numerical resolution \citep{Mihalas:1978}. Such drawbacks are more severe when
including the polarization. 

Most of the theoretical works done so far 
have only considered polarized light in static
atmospheres, being however unbalanced by the large number of
observational investigations studying it in dynamic media.
On top of that, most of the computational tools and theoretical investigations in moving atmospheres
only consider the effect of the velocities in the 
framework of the Zeeman line-formation theory
\citep[e.g., ][]{Stenflo:1994,Socas-Navarro:2000,Uitenbroek:2011aa}, sometimes even without
putting an special emphasis on the analysis of dynamic effects. Other
authors treat the problem in very schematic situations \citep{2011IAUS..274..291M,
  Landolfi:1996aa}. With respect to the scattering
polarization, the work of \cite{Nagendra:1996aa} accounts for the
impact of macroscopic velocities in moving media using the two-level
atom approximation for describing resonance lines.

\subsection{Effect of the velocity on the radiative transfer.}\label{sec:rtvel}
Although in many physical
situations the RTE can be
formulated as in the static regime, in many others a full dynamic
treatment is required. For instance, the influence of
bulk motions is small for the continuum, but can be
important in spectral lines because even small Doppler
shifts can cause large changes in opacity (leading to significant variations in the radiation received by a
stationary observer). In general, the effects acting on the
emitted line radiation due to systematic velocities
 in a stellar atmosphere are: Doppler shifts, producing variations of the line
profiles with frequency; aberration of photons, producing variations of the
line profiles with the ray direction; and advection, producing variations of
the line profiles with the distance along the line of sight. 

The numerical methods required for solving the
general line-formation problem in moving media are reviewed in      
\cite{Mihalas:1986aa}. Nowadays, the discussion is still around techniques and valid approximations that can be
applied in different situations to solve the RTE for the intensity. In a nutshell, there are four essential ways of treating radiative transfer in
atmospheres with velocity gradients: (i) Monte Carlo methods \citep{Bernes:1979,Lucy:2005aa};
(ii) Sobolev or supersonic approximations \citep{Sobolev:1958,Sobolev:1960}; (iii) observers' frame
methods; and, (iv) comoving frame methods \citep[][]{Mihalas:1978}. 
The first method is not suitable for treating optically thick atmospheres
. The relatively low solar velocity fields, with outflows not much larger than
the microturbulent velocities (in comparison with extreme events), also excludes the Sobolev
approximation.
 
The preference between the observer's and the comoving frame methods
in points (iii) and (iv) will depend on various issues. For instance, the application of the comoving frame method has the advantage that
the opacity and emissivity coefficients are not affected by motion and
may be treated as in the static case (i.e., since they are isotropic for
the static case, then they are isotropic also in a local reference
frame in a moving atmosphere). Other important point is that the comoving frame method admits the simplification
of using angle-averaged redistribution functions in calculations where PRD is required. However, the RTE becomes
more complicated due to an additional term containing the frequency
derivative of the intensity. Lorentz transformations have also to be
considered to transmit the information between different comoving
frames. Finally, the comoving-frame transfer equation is usually solved
using Feautrier variables, which furthermore limits the applicability of this method to
monotonic velocity fields. Very recently, we have been aware of
improved (and more sophisticated) methods to solve the RTE in the comoving frame that allow the
treatment of arbitrary velocity fields \citep{Baron:2012aa}, but they do
not represent a significant advantage with respect to the
computational effort of doing it in the observer's frame.

Thus, although the solution of the RTE in the
comoving frame would require much less angle-quadrature points when
macroscopic velocities are present \citep{Mihalas:1978}, we treat the
radiative transfer in the observer's frame because it is better
suited to the non-monotonic velocities that are present in the
chromosphere. Other authors have used this
strategy for solving the problem under the Zeeman
line-formation theory in the absence of atomic polarization
\citep[e.g., ][]{Stenflo:1994,Uitenbroek:2002}.
In the observer's frame, the opacity and emissivity of the material, as
seen by a stationary observer, become angle-dependent but the formal
structure of the RTE is maintained. Consequently, it is also a suitable
method for including the effect of the macroscopic velocity in a previously
existing code, as it has been the case in this thesis. The details of the implementation are left for the next chapters.

\begin{itemize}
\item Frequency redistribution in dynamic cases:
\end{itemize}
An issue that will trascend our results is how they could be affected by PRD effects in dynamic situations. 
The investigation of PRD in the
polarization is presently limited by the applicability of the
theoretical framework developed so far. The quantum theory of spectral
line polarization \citep{LL04} that we follow in
our investigation is based on the hypothesis that the pumping
radiation field has no spectral structure across intervals
smaller than the frequency separation between interfering atomic
levels (flat-spectrum approximation), which is equivalent to the CRD
approximation. 
Despite its limitations, this
theory represents the most robust quantum approach to the physics of
polarization developed so far and it even has demonstrated to be useful in
situations out of their applicability regime. 


In absence of macroscopic motions, CRD is a good
approximation for weak lines whose extreme wings do not require
extensive line transfer. On the contrary, the modelling of very strong
(typically resonant) lines should be done
in principle using PRD. For CRD to be valid across the whole line profile, the
intrinsic radiative rates (most importantly $A_{u{\ell}}$) must be low enough,
in such a way that collisional rates overcome them in middle
layers\footnote{That is not necessary for radiation forming at higher
  layers because it is already redistributed due to thermal
  motions. See Section \ref{sec:spectral}.}. On the contrary, if the radiative transition probabilities are very large
(as in resonant lines), collisional damping will no longer be an efficient frequency mixer already from
deeper layers, leaving an intermediate extended region in which
PRD holds. Even in that case, the intensity profiles of solar resonant lines
(e.g., Mg {\sc ii} or Ca {\sc ii} H and K) calculated with PRD and
with CRD have almost identical
absorption cores\footnote{The 
effect of modelling resonant lines in CRD is that the wings are darker than in PRD because the PRD
line source function (which is nearly pure coherent scattering in the
wings) uncouples from the thermal source function deeper in the
atmosphere and, as a consequence, is smaller than in CRD.}.
In typically subordinated lines whose lower atomic energy levels
are not connected to the fundamental levels (e.g., the Ca {\sc ii} IR triplet lines), CRD can be
safety applied in static media \citep{Uitenbroek:1989aa}. 

On the other hand, for flow speeds on the order
of the mean thermal speed\footnote{For
flows in which typical speeds are much larger than the mean thermal
speed, CRD and coherent scattering lead to essentially identical results \citep{Lucy:1971aa,Mihalas:1986aa}.},
velocity gradients could cause the line center intensity to depend more
strongly on frequency and direction than in comparable static media
because they tend to assymetrize the Maxwell distributions describing the
thermal motions at the core. Hence, the CRD assumption would deteriorate
\citep{Hummer:1968aa}. 
On the basis of previous works, we expect the difference between PRD
and CRD profiles \textit{of resonant lines} to be enhanced as the ratio of
systematic to turbulent velocities decreases and as the line formation
region is larger, showing an effectively thick chromosphere \citep{Drake:1983aa}. In next chapters, we deal
with atmospheres whose maximum velocities ($\sim 0-25\,\rm{ km\,s^{-1}}$) are comparable to the mean
turbulent velocity ($\sim 3\,\rm{ km\,s^{-1}}$). However, we are interested in the 
polarization amplitudes of non-resonant lines at their cores, where 
the PRD effects should be notably less limitating. Indeed, the exact line profile at the
wings is not relevant to our aims because their physics is actually
very different from the physics of the core, which is the true
recorder of the Hanle and Zeeman effects and where the observational
Ca{\sc ii} IR triplet signals show a measurable polarization.

 
\subsection{Effect of the velocity in the SEE.}\label{sec:seevel}
Being a conservation law applied to a volume element, the statistical equilibrium
equations depend on the plasma velocities. For instance, the $\rm{\rho^0_0}$ component of
each atomic level J satisfy a rate equation of the form
\begin{equation}\label{eq:ssev}
\centering 
\frac{\partial\rho^0_0(J)}{\partial t}+\nabla\left[ \vec{v}\rho^0_0(J)\right]=\,
  \mathcal{P}(J\rightarrow J^{\prime})-\mathcal{P}^{\prime}(J\rightarrow J^{\prime}) 
\end{equation}
where the r.h.s contains the variations of
$\rm{\rho^0_0}$ due to radiative and collisional
transitions (both included in the generic processes $\mathcal{P}$ and $\mathcal{P}^{\prime}$) that populate
$(J^{\prime}\rightarrow J)$ and depopulate
$(J\rightarrow J^{\prime})$ the level $J$. In static cases, the
l.h.s. is zero because statistical equilibrium is assumed. In
dynamic but non-relativistic calculations, the atmosphere can still be considered stationary ($\partial
/\partial t=0$) and the advection term is usually 
neglected. 

The advection term accounts for the amount of material that
enters or exits the fluid element volume considered. The rate at which the non-relativistic material can enter
or leave the volume element in the simulation is assumed slow
compared with that at which the atomic system reaches the statistical equilibrium
(which is a fast microscopic process following the atomic transition rates). The conclusion is that statistical equilibrium can be
safety applied.   
 \subsection{Effect of the velocity on the atomic density matrix.}\label{sec:atomvel}

Due to the Doppler effect, the radiation field experienced by an atom depends
on its velocity $\vec{v}$. Consequently, the density matrix will
also depend on it \citep{LL04}. Being $\rho^K_Q(J;\vec{v})$ the spherical
statistical tensor of the atomic system dependent on $\vec{v}$, and
$f(\vec{v})$ the velocity distribution function of the atoms in a
given point of the atmosphere, a complete statistical description of
the atomic collectivity is given by the product
$f(\vec{v})\rho^K_Q(J;\vec{v})$. Such quantity is the so-called
velocity-space density matrix.

This formalism is presently being developed for the case
of a two-level atom with infinitely-sharp levels and it has shown to
be equivalent to the redistribution matrix formalism for treating PRD
effects \citep{Belluzzi:2013aa}. Indeed, it constitutes a more general approach to such
a problem because it is able to account for the phenomenon of pure
Doppler redistribution including broadening collisions and
lower level polarization in multilevel systems. Consequently, the velocity-space
density matrix formalism seems to be also a likely solution for
describing the macroscopic velocity effects in PRD-affected lines.



\section{The Hanle Effect in the solar atmosphere}\label{sec:introhanle}
\subsection{Description}\label{sec:hanle_description}
\begin{figure}[h!]
        \centering
                          \fbox{
        \begin{subfigure}[b]{0.25\textwidth}
                \centering
                \includegraphics [width=\textwidth]{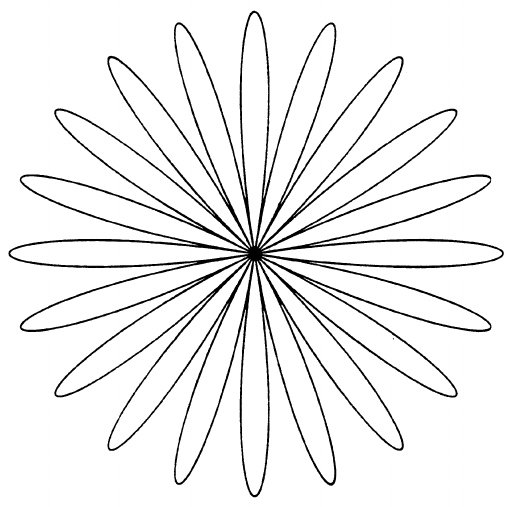}
        \end{subfigure}
        \begin{subfigure}[b]{0.25\textwidth}
                \centering
                \includegraphics [width=\textwidth]{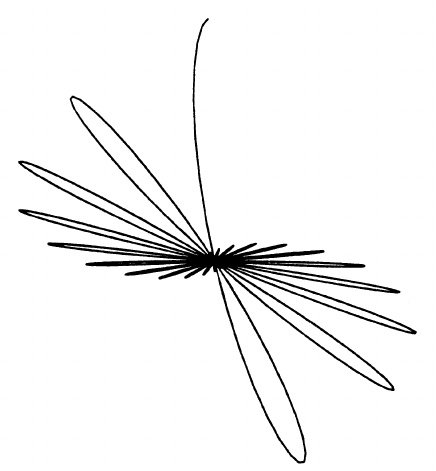}
        \end{subfigure}}
        \caption{Rosette pattern describing the motion of an undamped charged oscillator
          in a magnetic field (left) and the pattern obtained when
          taking radiative decay into account (right).}
\label{fig:rosette}
\end{figure}

The Hanle effect can be understood classically \citep{Hanle:1924} with
the oscillator model for the atomic electrons. In this picture,
the magnetic field makes the oscillating electrons to precess around the magnetic
field direction, producing a rosette pattern
(Fig. \ref{fig:rosette}, left drawing). There is a competition between the radiative damping
rate of such oscillator and the rate of Larmor precession, which is
proportional to the magnetic field strength \citep{Stenflo:1994,Trujillo-Bueno:2001aa,LL04}. Thus, when the former
prevails, the Hanle effect can leave its
fingerprint on the Stokes parameters (Fig. \ref{fig:rosette}, right drawing). From the point of view of the observer, the
action of the Hanle effect is a modification of the linear
polarization, measured by Stokes
Q and U. When the Zeeman splitting is negligible compared with the
line width\footnote{In this regime, the Hanle effect can be
  fully operative, well as saturated Hanle effect or, when the field
  is even smaller, as lower-level or upper-level Hanle effect. But for
strong magnetic fields, the transversal Zeeman effect dominates the linear
polarization through Eqs. (\ref{eq:zee2}) and  (\ref{eq:zee3}).}, the transverse
Zeeman effect signals (affecting Q and U) are of second order, in such
a way that Q and U 
only measure the scattering polarization and its modification by the Hanle effect. 

In the following, we introduce how the orientation and strength of a
magnetic field in a solar plasma element produces the Hanle effect
in the quantum theory.
 Consider a
spectral line whose upper and/or lower levels can be polarized
($\mathrm{J_{\ell,u} > 1/2}$). Then, the Hanle effect consists in the
modification of such atomic polarization through a
magnetic field whose Larmor frequency ($\mathrm{\omega_L=8.79
  \times 10^6 B}$) is comparable to the inverse of the radiative
lifetime of the energy levels of the transition. This defines one
critical magnetic field strength $B_H$ per polarized level, given in gauss by
\begin{equation}\label{eq:hanlecri}
 \centering 
     8.79 \times 10^6 \,g_L \, B_H \sim 1/\mathrm{t_{life}},
\end{equation}
where $\mathrm{g_L}$ is the Land\'e factor and
$\mathrm{t_{life}}$ (in seconds) can be approximated with
$\mathrm{1/A_{u\ell}}$ for the upper level and with
$\mathrm{1/(B_{\ell u} J^0_0)}$ for the lower one. As the Larmor
frequency in wavelength units is the Zeeman
splitting, Eq. (\ref{eq:hanlecri}) is equivalent to say
that the Hanle effect operates when the Zeeman splitting is comparable
to the \textit{natural} width of the energy levels.
Note that the Zeeman effect produces clear polarization signals when the Zeeman splitting is comparable to the
\textit{thermal} width of the line profiles.

The above considerations suggest that the
Hanle effect can be seen as a result of the coupling between the
magnetic field with Larmor frequency and the radiative rates. In the quantum approach, this
coupling alters the
quantum coherence between pairs of magnetic sublevels and modifies the atomic level polarization. The direct action of the magnetic field on the
density-matrix multipoles is introduced in the rate
Eqs. (\ref{eq:generalsee}) through the 
magnetic term, which also accounts for the geometry of the physical
situation. Geometry is a essential piece
of the Hanle effect. The Hanle magnetic term is \citep{LL04} 
\begin{equation}\label{eq:magterm}
\centering 
-i\omega_L g_{L}\sum_{Q'}\mathcal{K}^K_{QQ'}\rho^K_{Q'}(J).
\end{equation}
Choosing the quantization axis along the vertical, the only non-zero
\textit{magnetic kernels} for $K=0,1,2$ can be expressed as
\begin{subequations}\label{eq:kernelitos}
\begin{align}
\mathcal{K}^K_{QQ}&=\, \displaystyle Q \,\cos{\theta_B} \qquad &(\Delta
Q=0),
\displaybreak[0] \\
\mathcal{K}^K_{QQ'}&=\, \displaystyle \sqrt{\frac{2K-1-|Q \cdot Q^{\prime}|/K}{2}} \cdot\sin{\theta_B}\cdot
e^{\mathrm{ i} (Q^{\prime}-Q) \chi_B}\qquad &(\Delta Q=\pm 1),
\end{align}
\end{subequations}
where the angles define the orientation (inclination $\theta_B$ and
azimuth $\chi_B$) of the magnetic field in the 
reference frame of the solar vertical passing through the scattering point.
Eqs. (\ref{eq:magterm}) and (\ref{eq:kernelitos}) indicate that the
magnetic field can create or destroy coherences with a certain $Q \neq 0$ by changing and dephasing
any multipolar component with $Q'= Q \pm 1$, but it cannot modify the
overall population of a level by itself. Similarly, those equations indicate
that population imbalances $\rho^K_0$ can be created from coherence terms with $Q=\pm 1$.
This way, the magnetic field transfers the
information about its orientation and strength to the quantum
system. The Hanle effect becomes important when the orientation of the magnetic
field produces a significative symmetry breaking in the atmosphere
($\theta_B \rightarrow 90\deg$)
and when the field strength is near the critical
values for the energy levels (Hanle regime). 
  In other cases, coherences vanish and the magnetic kernel in the SEE
  cancels out.



\subsection{A hands-on guide to the Hanle equations}\label{sec:hanle_walk}
Next, we underline the physics of the Hanle effect and give some technical
details and approximations to work with it.
    
\begin{figure}[h!]
\centering%
\includegraphics[width=0.6\textwidth]{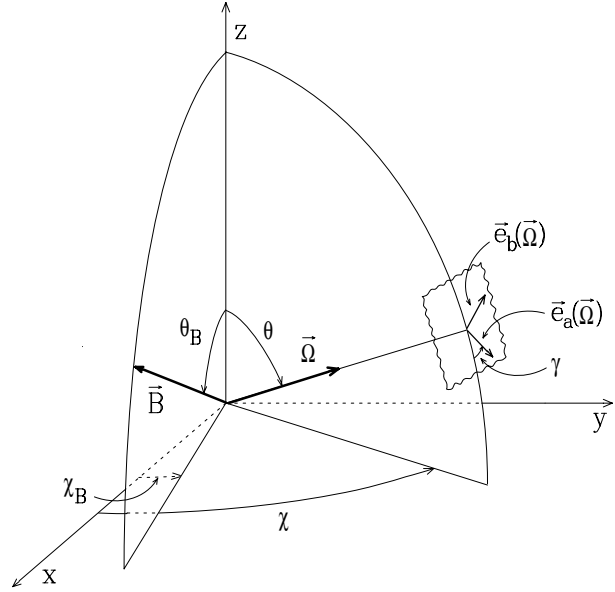}
\caption{Reference system (x,y,z) used to place the generic line of sight
  ($\vec{\Omega}$) and magnetic field direction ($\vec{{\rm B}}$). The vertical axis
  is the radial direction of the Sun through the point being
  considered. The reference direction set by the observer in the plane
  of the sky are given by the unit vectors ${\rm \vec{e}_a}$ and ${\rm \vec{e}_b}$. The
  positive Stokes-Q direction is defined in that plane by
  $\gamma$. The magnetic field reference frame is similar to the one in $\vec{\Omega}$, but
  placed along $\vec{\rm B}$ and having $\gamma_{\rm B}=0$ for simplicity. }
\label{fig:refsys}
\end{figure}
\subsubsection{ Eddington-Barbier approximation}
We start estimating the linear polarization at line
center \citep{Trujillo-Bueno:2003c}. We consider a
strong spectral line not affected by stimulated emission and emerging from a plane-parallel
atmosphere that is assumed to be well described by the Eddington-Barbier
approximation. Then, its fractional linear polarization X$/$I
(with X$=$Q or X$=$U) is given by\footnote{see also
Appendix \ref{app:C} for an alternative derivation}:
\begin{equation}
\centering 
\frac{X}{I} \simeq \,\frac{\epsilon^l_{X}}{\epsilon^l_I}-\frac{\eta^l_{X}}{\eta^l_I} ,
\label{eq:basicEB}
\end{equation}
where all the terms in the r.h.s. have to be evaluated at $\mathrm{\tau^{LOS}=1}$.
Afterwards, this approximate formula will give us some useful expressions.
\subsubsection{No Zeeman-splitting case}
 In the Hanle regime, the Zeeman splitting is small in
comparison with the spectral line width, so the absorption and
emission profiles do not depend on the magnetic quantum numbers
[$\mathrm{\phi=\phi(\nu-\nu_{u{\ell}})= \phi(J_u,J_{\ell})}$]. Then,
Eq. (7.16e) of \cite{LL04} can be written as\footnote{We follow the standard notation for the following quantities: the Planck constant (h), the number density of atoms of the
  considered species (N), the upper and lower level atomic quantum numbers
  ($\alpha_u J_u$ and $\alpha_{\ell} J_{\ell}$), the Einstein
  coefficients ($A_{u{\ell}}$ and $B_{{\ell} u}$), and the
  numerical coefficients $\omega^{(K)}_{J_u J_{\ell}}$ (\cite{Landi-DeglInnocenti:1984}). The latter
  can be calculated with
\begin{equation*}
\omega^{(K)}_{J_u
  J_{\ell}}=\frac{\sixj{1}{1}{K}{J_u}{J_u}{J_{\ell}}{}}{\sixj{1}{1}{0}{J_u}{J_u}{J_{\ell}}{}},
\end{equation*}
where the symbols in brackets are called 6-j symbols
\citep{Racah:1942} and take into account the coupling of electronic
angular momentum. Analytic expressions of the 6-j symbols are
found in Eqs. (2.36) of \cite{LL04}.} :
\begin{equation}
\centering 
\epsilon_i(\nu,\vec{\Omega}) = \, \frac{h\nu}{4\pi}N
\sum_{\alpha_{\ell} J_{\ell}}\sum_{\alpha_u J_u}
\sqrt{2J_u+1}A_{u{\ell}}\sum_{KQ}\omega^{(K)}_{J_u J_{\ell}} \mathcal{T}^K_Q(i,\vec{\Omega})\rho^K_Q(J_u)\phi(\nu-\nu_{u{\ell}}).
\label{eq:step1}
\end{equation}

Hereafter, we will refer to the emission
 vector components and propagation matrix
 elements as $\mathrm{\epsilon_i}$
 and  $\mathrm{\eta_{i}}$, being $\mathrm{ i=0,1,2}$ for Stokes I, Q
 and U. Similar expressions for the absorption coefficients
$\mathrm{\eta_i(\nu,\vec{\Omega})}$ are obtained from the previous equation with the changes
$\mathrm A_{u{\ell}}\leftrightarrows B_{{\ell} u}$,
  $J_{u}\leftrightarrows J_{\ell}$ and
$\mathrm \omega^{(K)}_{J_uJ_{\ell}}\leftrightarrows\omega^{(K)}_{J_{\ell}J_u} \cdot (-1)^K $. In Eq. (\ref{eq:step1}), the tensor $\mathcal{T}^K_Q$
gives the geometrical transformation that projects the multipolar components $\rho^K_Q$ from the quantization axis to the LOS
reference system, thus expliciting the polarized emission of the
media along the considered ray. We always have to choose a quantization axis for the
 angular momentum (typically the local solar vertical or the magnetic
 field direction). It has associated the reference system in which
 $\mathcal{T}^K_Q$ and $\rho^K_Q$ are defined.
 Besides that, the previously assumed independence of the spectral profiles on the magnetic
 quantum numbers makes the emissivity and absorption coefficients invariant under rotations of the reference
 frame. It means that in Eq. (\ref{eq:step1}) we can express  $\mathcal{T}^K_Q$ and $\mathrm{\rho^K_Q}$
 in any system (the same for both). Apart from this, we must also
 choose the observer reference system by setting the LOS
 ($\vec{\Omega}$) and the angle $\gamma=\gamma_{\Omega}$ that defines the 
 direction for positive Q (see Fig. \ref{fig:refsys}).
     
\subsubsection{Non-blended spectral line}

If the
spectral lines of the atomic system do not overlap, there is only one transition involved at line center (summations
in $\alpha J$ disappear from Eq. \ref {eq:step1}). Then, making the substitution
$\epsilon^0=\frac{h\nu}{4\pi}N \sqrt{2J_u+1}A_{u{\ell}}\phi(0)$, we get:
\begin{equation}
\centering 
\epsilon^l_i(\vec{\Omega}) = \, \epsilon^0\sum_{KQ}\omega^{(K)}_{J_u J_{\ell}}
\left[\mathcal{T}^K_Q(i,\vec{\Omega})\right]^{\vec{z}}   \left[\rho^K_Q(J_u) \right]^{\vec{z}},
\label{eq:step2}
\end{equation}
 where the superscript on the square brackets indicates that we are
 evaluating those quantities in the reference frame of the local
 vertical. The
advantage of setting the quantization axis along the vertical is that the
angles of the geometrical tensors are referred to a common and invariant direction, which facilitates the
calculation of the radiation field components ($\bar{J}^K_Q$) independently on the
magnetic field direction (the magnetic field can be differently oriented
from point to point).
Eq. (\ref{eq:step2}) is the simplest general expression for the Hanle effect that allows
us to derive other published in literature. It shows clearly
that the emission of polarized light in the Hanle
regime is directly proportional to the atomic density matrix elements projected to
our line of sight by the tensor $\mathcal{T}^K_Q$. If the illumination
at the atmosphere boundaries is not
polarized and there are no sources of atomic orientation, the components
with odd $K$ are zero in the summation. However, an important point is that the mere existence of a limb darkening at every point of
the solar atmosphere generates atomic
alignment and terms with even K that
contribute to the emisivity in Q and U. Such $\rho^K_Q$ elements are tipically
modified by the magnetic kernel in the SEE, following Eq. (\ref{eq:magterm}). That is
the essence of the Hanle effect on the scattering polarization.

To explicit the dependence of the polarized emission on the
coherences we recall Equations (\ref{eq:emi_iqu}). They are
derived from Eq. (\ref{eq:step2}) with $i=0,1,2$, 
$K=0,2$ and $-K<Q<K$ \footnote{To perform the summation
 apply that $\rho^K_{-Q}=(-1)^Q{\rho^K_{Q}}^*$.}, after choosing a reference direction for
$Q>0$ that is parallel to the limb nearest to the scattering point
(that is, with $\gamma=\pi/2$). The setting of a reference direction simplify the
geometrical tensors\footnote{In such case, the
  $\mathcal{T}^K_Q(i)$ components for $i=0,1,2$ (I,Q,U) used are:
\begin{align}
\mathcal{T}^0_0(0,\vec{\Omega})=1 \qquad  ;& \qquad
\mathcal{T}^2_0(0,\vec{\Omega})=\frac{1}{2\sqrt{2}}(3\mu^2 -1) 
\nonumber \\ 
\mathcal{T}^2_0(1,\vec{\Omega})=\frac{3}{2\sqrt{2}}\mu^2 \qquad  ; \qquad
\mathcal{T}^2_1(1,\vec{\Omega})=\frac{\sqrt{3}}{2}&\mu\sqrt{1-\mu^2}\,{\rm e^{i\chi_B}}\qquad; \qquad  
\mathcal{T}^2_2(1,\vec{\Omega})=\frac{\sqrt{3}}{4}(1+\mu^2) \, {\rm e^{i2\chi_B}}
\nonumber\\
\mathcal{T}^2_1(2,\vec{\Omega})=\rm{i}\frac{\sqrt{3}}{2}\sqrt{1-\mu^2}\,{\rm
  e^{i\chi_B}} \qquad  ;& \qquad
\mathcal{T}^2_2(2,\vec{\Omega})=\rm{i}\frac{\sqrt{3}}{2}\mu \,{\rm
  e^{i2\chi_B}} 
 \nonumber
\end{align}} \citep[from Table 5.6 in ][]{LL04}. For
instance, the emissivity in Stokes U:
\begin{align}
\epsilon^l_2(\vec{\Omega}) &= \, -\epsilon^0\omega^{(2)}_{J_u
J_{\ell}} \sqrt{3} \{ \sqrt{1-\mu^2}(\sin{\chi} {\rm
Re}[\rho^2_1]+\cos{\chi} {\rm Im}[\rho^2_1]) \nonumber\\
& \qquad \qquad \qquad+\mu (\sin{2\chi}{\rm Re}[\rho^2_2]+\cos{2\chi}
{\rm Im}[\rho^2_2])\} ,
\label{eq:step3}
\end{align}
with $\chi$ and
$\theta=\cos^{-1}(\mu)$ the azimuth and inclination of
the LOS in the reference frame of the local vertical. In particular, the emission in U vanishes when coherences
are zero. However, Stokes Q also depends on
the non-vanishing $\rho^2_0$ components excited by
the radiation field anisotropy.

The explicit angular dependence set by the tensors $\mathcal{T}^K_Q$ 
always measures rotations from the
reference frame in the quantization axis ($\vec{\mathrm{z}}$ or
$\vec{\mathrm{B}}$, usually) to the one in the LOS (see Fig. \ref{fig:refsys}). On the other hand,
 any spherical tensor can be expressed along a different quantization axis
 by following the corresponding
rotation law. Namely, we can pose the $\rho^K_Q$ elements
in the $\vec{\mathrm{z}}$ system as a combination of
  the ones in the $\vec{\mathrm{B}}$ system with:
\begin{equation}
\centering 
[\rho^K_Q(J)]^{
    \vec{{\rm z}}}=\sum_{Q^{\prime}}[\rho^K_{Q^{\prime}}(J)]^{\vec{{\rm B}}}\mathcal{D}^K_{Q^{\prime}Q}(R)^*,
\label{eq:law1}
\end{equation}
where the matrices $\mathcal{D}^K_{Q^{\prime}Q}$ carry out the
rotations given by the Euler angles $R=(\alpha\beta\gamma)$ going from
the magnetic field to the solar vertical reference frame. The
advantage of applying this relation in order to express a given equation
as a function of the $\rho^K_Q$
elements of the magnetic field frame is that they simplify in certain
circumstances (see below). 
\subsubsection{Hanle effect in the saturation regime}
In saturation, the magnetic field strength is very intense in
comparison with the critical Hanle field of the atomic levels
($B>>B_H$). Thus, the separation
in energy between them destroy all coherences
 \textit{in the
magnetic field reference frame} ($[\rho^K_Q]^{SAT}=\rho^K_Q\cdot\delta_{Q0}$). In this case, the
polarization of the emergent spectral line radiation does not depend on the field strength but only on its
orientation, which still plays an active role in the radiative
transfer through the radiative coefficients. 
The emission
(and similarly the absorption) coefficients can then be particularized from
Eqs. (\ref{eq:emi_iqu}) (or Eq. (\ref{eq:step2})) by conveniently expressing the components
$\rho^K_Q$ as in Eq.(\ref{eq:law1}).  
 Following it, we write the rotation
going from the $\vec{\mathrm{B}}$ to the $\vec{\mathrm{z}}$ frame as
$R_1=(0,-\theta_B,-\chi_B)$. Recalling the saturation
assumption, the coherence terms vanish in the $\vec{\mathrm{B}}$ frame and the
only relevant transformations are thus given by
$[\rho^2_Q]^{\vec{\mathrm{z}}}=[\rho^2_0]^{\vec{\mathrm{B}}}
\mathcal{D}^K_{0Q}(R_1)^*$, where:
\begin{align}
\mathcal{D}^2_{00}(R_1)^* &=\frac{1}{2}(3\cos^2{\theta_B}-1)\\
\mathcal{D}^2_{01}(R_1)^* &= \sqrt{\frac{3}{2}}\sin{\theta_B}\cos{\theta_B}
{\rm e^{i\chi_B}} \nonumber \\
\mathcal{D}^2_{02}(R_1)^* &= -\sqrt{\frac{3}{8}}\sin^2{\theta_B}{\rm
  e^{-i\chi_B}}. \nonumber
\end{align}
Substituting the so given $\rho^2_Q$ in Eq. (\ref{eq:step3}) we
obtain:
\begin{align}
\epsilon^l_2(\vec{\Omega}) & \propto \, - \Big[
\sqrt{1-\mu^2}\sin{(\chi-\chi_B)}\sin{(2\theta_B)}
+\mu\sin{[2(\chi-\chi_B)]}\sin^2{\theta_B} \Big]\cdot [\rho^2_0]^{\vec{\mathrm{B}}},
\label{eq:step4}
\end{align}
where the alignment in the magnetic case can be related to the alignment in the
non-magnetic case as\footnote{This equality results from comparing the
  statistical equilibrium equations (SEE) for a two
level atom in the magnetic and non-magnetic case. As the SEE are
formally invariant, such expression must hold.}
\begin{align}
[\rho^2_0]^{\vec{\mathrm{B}}}=& \, \frac{1}{2}(3\cos^2{\theta_B}-1)\cdot[\rho^2_0]^{\mathrm{B=0}} ,
\label{eq:ajaja}
\end{align}
 Following similar steps for the emission and absorption terms in $i=0,1,2$, we can evaluate the
Eq. (\ref{eq:basicEB}) for Q$/$I and U$/$I. In the weak-anisotropy regime
($\rho^2_0<<\rho^0_0$) holding in stellar atmospheres we can finally write:
\begin{subequations}\label{eq:step5}
\begin{align}
\frac{Q}{I} \simeq& \, \frac{3}{8\sqrt{2}} \Big[
(1-\mu^2)(3\cos^2{\theta_B}-1) - 2\mu\sqrt{1-\mu^2}\sin{(2\theta_B)}\cos{(\chi_B-\chi)}\nonumber \\
&+(1+\mu^2) \sin^2{\theta_B}\cos{[2(\chi_B-\chi)]} \Big]\cdot (3\cos^2{\theta_B}-1) \cdot\mathcal{F} \label{eq:step5a}
\displaybreak[0] \\
\frac{U}{I} \simeq &\, -\frac{3}{4\sqrt{2}} \Big[\sqrt{1-\mu^2}\sin{(2\theta_B)}\sin{(\chi_B-\chi)}\nonumber \\
&-\mu\sin^2{\theta_B}\sin{[2(\chi_B-\chi)]} \Big]\cdot (3\cos^2{\theta_B}-1) \mathcal{F}, \label{eq:step5b}
 \end{align}
 \end{subequations}
where $\mathcal{F}=\omega^{(2)}_{J_u J_{\ell}} \sigma^2_0(J_u)
-\omega^{(2)}_{J_{\ell} J_u} \sigma^2_0(J_{\ell})$ is the non-magnetic
contribution of the fractional atomic alignment
($\sigma^2_0=\rho^2_0/\rho^0_0$) generated in the levels of the
transition. The evaluation of Eqs. (\ref{eq:step5}) for
different configurations of magnetic field and LOS is an interesting
exercise that shows the behavior of the scattering polarization in
every case. 

Particularizing to the configuration in which the LOS and the magnetic field
vector are contained in perpendicular planes ($\chi_B-\chi=\pi/2$), we obtain the Eqs (3) and (4)
of \cite{Trujillo-Bueno:2010}. If, instead, we particularize to the \textbf{forward
scattering case} ($\mu=1$) with arbitrary magnetic field direction, we obtain: 
\begin{subequations}\label{eq:step6}
\begin{empheq}[box=\fbox]{align}
\frac{Q}{I} &\simeq \, \frac{3}{4\sqrt{2}}\cdot\sin^2{\theta_B}\cdot(3\cos^2{\theta_B}-1)\cos{[2(\chi_B-\chi)]}\cdot\mathcal{F} \label{eq:step6a}
\displaybreak[0] \\
\frac{U}{I} &\simeq \,\frac{3 }{4\sqrt{2}}\cdot\sin^2{\theta_B}\cdot(3\cos^2{\theta_B}-1) \sin{[2(\chi_B-\chi)]}\cdot\mathcal{F}, \label{eq:step6b}
 \end{empheq}
 \end{subequations}
Remind that all the terms
in these approximated expressions have to be evaluated at $\mathrm{\tau^{LOS}=1}$. Apart from the effect of the limb-darkening induced anisotropy on the
linear polarization (contained in $\mathcal{F}$), Eqs. (\ref{eq:step6}) isolate 
other geometric effects in the polarization of the saturated Hanle
effect. Note that in forward scattering both Stokes parameters are basically equivalent
in their geometrical dependence, having the same maximum and minimum values. Furthermore, a vertical magnetic field makes
zero the forward scattering Stokes parameters because nothing breaks the
axial symmetry around the LOS when considering a plane-parallel atmosphere.

Being in forward scattering, we also particularize Eqs. (\ref{eq:step6}) to have
$\chi_B-\chi=\pi/2$, which is equivalent to align the positive-Q
direction along the projection of the magnetic field onto the plane of the sky (because
$\gamma$ was set to $\pi/2$). Then, we obtain
\begin{align}
\frac{Q}{I} &\simeq \, -\frac{3}{4\sqrt{2}}\sin^2{\theta_B}\cdot(3\cos^2{\theta_B}-1)\cdot\mathcal{F},
\label{eq:step7}
\end{align}
 with $U=0$.
On the other hand, dividing Eqs. ({\ref{eq:step6}) between them, we arrive to\footnote{ See
    \cite{Collados:2003aa} for an observational application of this equation.}
\begin{align}
\frac{U}{Q} &= \, \tan{[2(\chi_B-\chi)]}.
\label{eq:step8}
\end{align}
If we set the positive Stokes Q along
  the x axis (see Fig. \ref{fig:refsys}), then $\chi=-\pi/2$. In this configuration, the azimuth of the magnetic field defines
the direction of maximum polarization for the emerging light
beam (such maximum is given by Eq. (\ref{eq:step7})). We will take
advantage of this fact in Chapter \ref{cap:four}. 
\subsubsection{Zero-field regime}
Finally, in the zero-field regime and for any line of sight, the result would be simpler. 
Repeating all the procedure, we obtain\footnote{Expression
  appearing for instance in
   \cite{Trujillo-Bueno:2003b,Trujillo-Bueno:2010}. See also the Appendix for an alternative derivation.}: 
\begin{align}
\frac{Q}{I} &\simeq \, \frac{3}{2\sqrt{2}}(1-\mu^2)\mathcal{F},
\label{eq:step9}
\end{align}
with $U=0$, and always with $\mu=\cos{\theta}$ and the positive-Q reference
being parallel to the limb nearest to the scattering point. Here, the
polarization is zero in forward scattering and maximum when looking at the solar limb.

Note that in any situation the anisotropy of the radiation field is always
affecting the microscopic system in the plasma element through the
factor $\mathcal{F}$. However, the
expression of such atomic polarization into a measurable radiative
quantity depends on the observer point of view\footnote{If we observe
  an homogeneuos non-magnetic atmosphere at disk center (forward
  scattering geometry) Stokes Q and U are always zero. Irrespectively
  of the illumination received at each point along the atmospheric
  vertical, different thermodynamical stratifications alway give zero
  linear polarization. If there is an inclined magnetic field, the
  symmetry breaking produces a linear polarization signal, not only
  giving information about the magnetic orientation and strength, but
  also about the anisotropy of the radiation field. Thus, the
  previously hidden atomic polarization is revealed and the differents MHD stratifications can be distinguished.}. 
Hereafter, in all the subsequent chapters, the quantization axis for
the angular momentum will be along the local solar vertical.

\subsection{Hanle effect at work: two examples.}\label{sec:hanle_examples}

The cases in which the Hanle effect is commonly employed to measure
solar magnetic fields are illustrated here in two particular geometries:
\begin{itemize}
\item  Observation near the solar limb (typically $\mu=0.1$). 
Here, the non-magnetic signals are given by Stokes Q due to pure
scattering (i.e., to the atomic polarization caused by the anisotropy of the radiation field). If a magnetic
field is present, the Hanle effect produces depolarization. To measure
the magnetic field we need to know the theoretical non-magnetic polarization amplitudes as a reference. We
could also calibrate theoretically the relative
response that similar lines have to the magnetic field 
and use the calibration for real measures (line-ratio technique). 
In the case of the Ca {\sc ii} IR triplet lines that can be observed at the solar
limb (Fig. \ref{fig:stenflo}), we expect that the relative
polarization amplitudes observed are due to a depolarizing magnetic field. This case will
be treated in chapter \ref{cap:three}.

\begin{figure}[htb!]
\centering%
\includegraphics[scale=1.0]{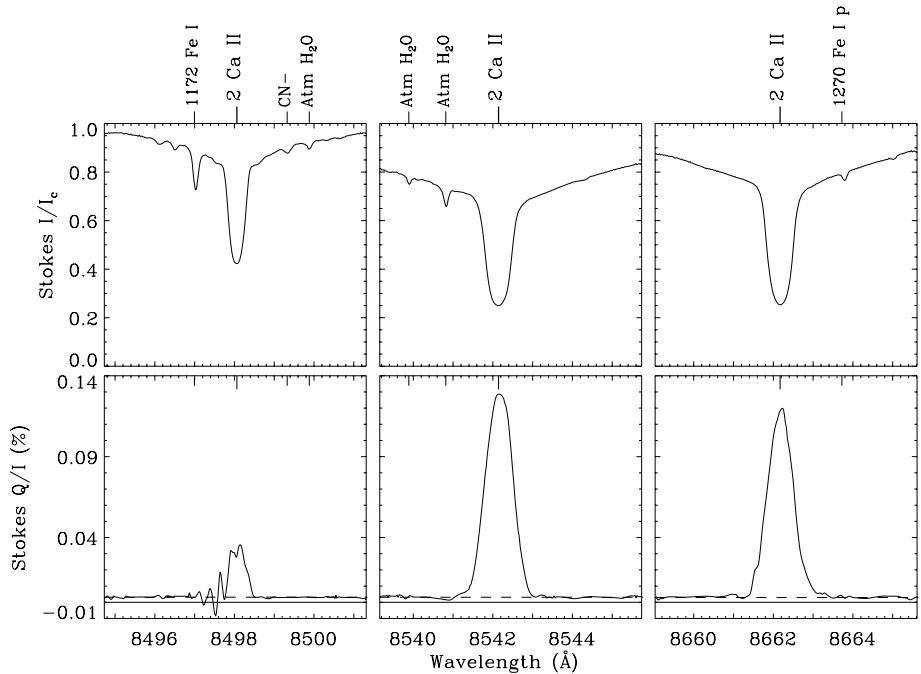}
\caption{Recordings (November 17, 1994 ) of the spectral regions around the three IR lines
  of Ca {\sc ii} made with ZIMPOL near the solar north pole limb by
  \cite{stenflo00b}. The level of the continuum polarization is represented by the horizontal dashed lines. }
\label{fig:stenflo}
\end{figure}

\item Forward scattering at disk center ($\mu=1$). In this case, the non-magnetic
signals are zero because the observer is looking along a symmetry
axis (assuming a plane-parallel atmosphere). If the field is purely vertical nothing changes. If the field is
inclined, Hanle-induced polarization appears. The more horizontal the
magnetic field is, the larger the polarization. This case will be
treated in detail in Chapter \ref{cap:four}.

\begin{figure}[htb!]
\centering%
\includegraphics[scale=0.4]{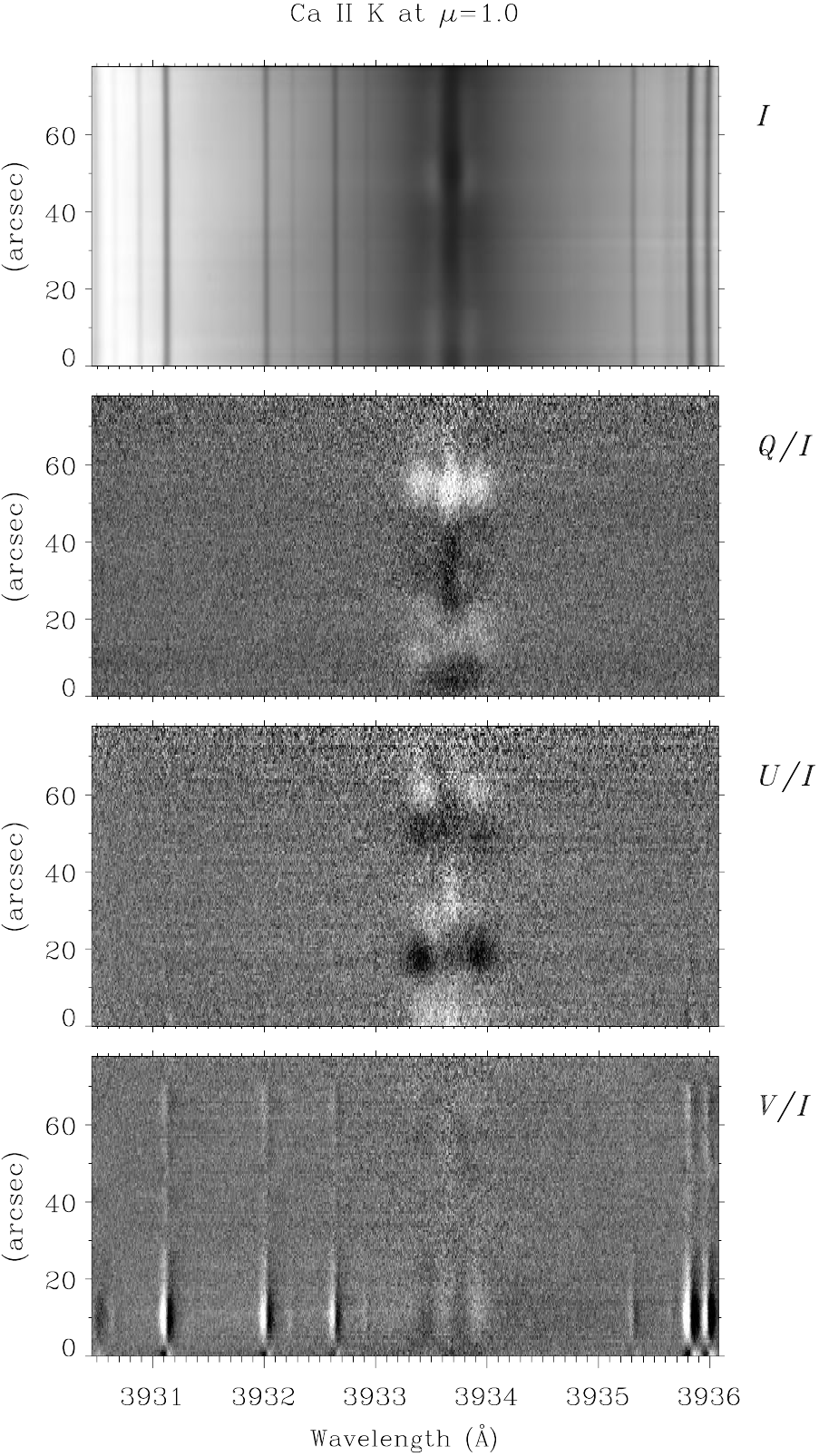}
\caption{Spectropolarimetric observation of the K-line of Ca{\sc ii}
at solar disk center in quiet regions. Obtained with ZIMPOL at the McMath Telescope in Kitt
Peak Observatory  by \cite{Stenflo:2006aa}. }
\label{fig:klinestenflo}
\end{figure}


\end{itemize}

\clearpage{}

 
\clearemptydoublepage  
%
\chapter{Developed tools and methodology.}\label{cap:one}

In this chapter, we describe some computer
programs, procedures and techniques that we have developed to treat the
radiative transfer (RT) of scattering polarization in dynamic solar
models, including software that we have found very useful to analyze and understand the results of our investigations. The most
important tools to mention are: the RT code that we have used
to synthetize the scattering
polarization in the presence of velocity gradients; an iterative
computer procedure to suitably redistribute the nodes of the numerical spatial
grid and achieve convergence of the RT problem in atmospheric models
with strong height variations; a code to calculate
response functions of the Stokes parameters; a code to perform
Principal Component Analysis in measured and theoretical Stokes profiles; a
Python-based software to interactively analyze the atmospheric models together with
the resulting spectropolarimetric maps; and finally, a series of
codes and plotting routines that use
the Mayavi package to visualize and analyze 3D models
of the solar atmosphere interactively.

\section{A RT code for dynamic scattering polarization.}\label{sec:traviata}	

\subsection{Traviata.}\label{sec:trav1}	
Traviata is a multilevel RT code for the synthesis of the spectral line
polarization that is due to atomic polarization and the Hanle effect in weakly
magnetized stellar atmospheres
\citep{Manso-Sainz:2003}. The relevant equations that it implements
have been already described in Chapter \ref{cap:intro}.
A very significant part of this thesis work is based on the improvements applied to
Traviata. 
The following is a brief summary of the
capabilities of the original program.

The physical assumptions in which the calculations were based are:
\begin{enumerate} 
\item Plane-parallel geometry of arbitrary optical thickness.
 \item Complete redistribution in frequency. 
\item NLTE.
\item Atomic levels can be polarized. Populations imbalances and
  coherence between different magnetic energy
  sublevels of the same J level can exist. The
  magnetic sublevels of different J levels do not interfere (multilevel
  approximation).
\item Magnetic fields are weak enough for the splitting of the
  absorption profiles to be negligible (much lower than the
  thermal width of the spectral line).
 \item Impact approximation for collisions.
\item Stimulated emission is neglected.
\end{enumerate}
To solve the RT of the second kind, Traviata follows the following
numerical approach
\citep{Manso-Sainz:2002, Manso-Sainz:2003, manso10}: 
\begin{enumerate}
\item The equations are formulated within the spherical tensors representation of the density matrix and radiation
field tensor. There are $\rm{(2J + 1)(J + 1)}$ unknowns
$\rm{\rho^K_Q}$ with K even\footnote{Components with K odd are zero
  because it is assumed that no natural sources of atomic orientation
  are present in the atmosphere and the boundaries illumination is not
circularly polarized.}, for
each level with integer angular momentum J , and $\rm{(2J + 1)J}$ unknowns
for each level with semi-integer angular momentum J .

\item The radiative transfer equations for the Stokes parameters are
integrated along short characteristics using the quasi-parabolic DELO
method (DELOPAR; see Section \ref{sec:formalsolver}).

\item The iterative corrections for the unknowns (the statistical
  tensors $\rm{\rho^K_Q}$) are calculated applying a suitable
  generalization to the multilevel atom case of the Jacobian iterative
  method described in \cite{Trujillo-Bueno:1999aa}. This implies
  writing down the statistical equilibrium equations taking explicitly
  into account the contribution of the diagonal components of the
  $\Lambda$ operator and linearizing according to
  Eq.(\ref{eq:preconditioning}) and the procedure in
  Sec. \ref{sec:mali}.

\item The iterative scheme can be speeded up through polynomial Ng-acceleration \cite{Auer:1987}.
\end{enumerate}
When using semiempirical models of the solar atmosphere, the number density of the ion under consideration is first computed in
NLTE. Having the overall
population of the ion and after reading the
model variables, Traviata computes the background continuum opacities.
The first initialization is done calculating the population of
the levels in LTE. In a second step, they are then used as a guess to
obtain the population of the levels in the standard NLTE case without polarization.
Afterwards, it solves the SEE  at each point in the atmosphere for the multipolar components of the atomic
density matrix plus the conservation of particles equation (Eq. (\ref{eq:conserva})) to
calculate the full excitation state of the atomic system. Next, it calculates the
radiation field at each point in the model atmosphere for each 
allowed radiative transition in the model atom applying the formal integration
of the radiative transfer equations for the Stokes parameters (DELOPAR method). The radiation field tensors $\rm{J^K_Q}$ are calculated
and used to obtain the new excitation state of the atomic system and
so on, iteratively, until reaching the final solution for the
$\rho^K_Q$ values.  

\subsection{The new version of Traviata}\label{sec:dynamic_traviata}	
We have extended Traviata to investigate the effect of macroscopic vertical velocities
on the linear polarization produced by scattering processes in
spectral lines, as well as on the circular polarization profiles
caused by the Zeeman effect. Maintaining the previous characteristics of
the code, we have implemented the following improvements:

\begin{enumerate}

\item  Inclusion of the effect of vertical (non-relativistic) velocities in the RT. 

The RT problem considered by Traviata is posed and solved in the
observer's frame (see Sec. \ref{sec:rtvel}). In this strategy, there is a unique reference frame
 fixed in space and all the motions in the plasma are related to
 it. Thus, each plasma element receives an spectral illumination that
 depends on the inclination of the rays but also on the relative
 motion with respect to other plasma elements that shift the line
 profiles by the Doppler effect (more details in Chapter
 \ref{cap:two}). In this case,
  the absorption, emission and dispersion line profiles depend on position, 
  frequency, spectral
 transition and, additionally, on the inclination of the
 rays of the angular quadrature. It extends to associated quantities such as the absorption, emission and
  dispersion coefficients. The line profiles are calculated for each
spatial point $i$ by shifting the static profiles (see Sec. \ref{sec:spectral})
 with the Doppler velocity $\mu_k V_i$, where $\mu_k=\cos(\theta_k)$,
 $\theta_k$ is the inclination of the ray $k$ in the angular
 quadrature, and $V_{i}=(\nu_0 v_i/c)/\Delta \nu_{Di}$ is the Doppler
 shift, in Doppler width units, produced by
 the vertical velocity $v_i$ at point $i$. For instance, the dynamic absorption profile
 would be $\phi^{\prime} = \phi({\rm x}-\mu_k {V_i})$ with ${\rm x}$ the Doppler frequency
  axis. Upward velocities gives
 blue shifts. For a given model stratification, the maximum effect of the velocity
 on the profiles occurs at the height where the Doppler velocity is $V_{\rm max}={\rm Max}[{\nu_0} \bm{v}/(c \bm{
    \Delta \nu_D})]$.

\item Redefinition of variables and numerical grids.
\begin{itemize}[leftmargin=0.23cm, labelsep=0.05cm]
\item The frequency axis is common and
  fixed along the whole calculation\footnote{One advantage of working in the observer's frame
  is that it is not necessary to perform interpolations or involved transformations between
  different points.}.
 The frequency axis is symmetric with respect to line center and has
 now double extension in wavelength than in
  the static case in order to cover the asymmetries and Doppler shifts
  produced by motions in the profiles.
 
\item The extension of the frequency axis is larger enough for the line
  coefficients to have a smooth transition to the continuum even in the
presence of large velocities. This avoids imprecisions in
  the normalization of the profiles to unit area.

\item The line core region is adaptively built with a resolution
  notably larger than in the wings to correctly sample the features of
  the emergent moving profiles. The
  wings resolution is smaller and quadratically decreasing towards the
  continuum. The minimum frequency resolution at the
  core must give 2 points per \textit{minimum} Doppler width\footnote{As the Doppler width varies significatively with temperature,
  we distinguish between quantities measured in \textit{minimum}
  Doppler units or in \textit{maximum} Doppler units. Thus, the
  maximum Doppler width unit is given by the temperature at the hottest
    layer and the minimum Doppler width unit is given by the temperature minimum.}
  $\Delta \nu_D^{\rm min}$
  in the
  atmosphere ($\Delta
  {\rm x}=1/2 $, in minimum Doppler units). 
Furthermore, the cutoff frequency ${\rm x_c}$ setting the end of the core is
dictated by the maximum expected Doppler shift
$\delta {\rm x}^{\rm{max}}=\nu_0 v^{\rm max}/(c \Delta \nu_D^{\rm min})$
(expressed here in minimum Doppler
units). Then, the core width in Traviata is set to the most restrictive case:
\begin{equation}\label{eq:core}
{\rm x_c} =\delta \rm{x^{ max}}+ N^0_c 
\end{equation}
where everything is in minimum Doppler units and $N^0_c$ is a safeguard
constant taking into account an extra width due to the large profiles
width at the hottest
layers. 
 
\item The angular and spatial numerical grids are heavily restricted by the
  presence of velocities and the maximum Doppler shift in the model stratification\footnote{Note that
    the maximum Doppler velocity $V_{\rm max}$ defined before and the maximum
    velocity Doppler shift
   $\delta {\rm x}_{\rm{max}}$ (given above in arbitrary units) are not the same quantity. The
    subtle difference is that the former includes the Doppler width, so giving
  the maximum shift relative to the actual profile width. But the latter
   gives the maximum Doppler frequency shift in the atmosphere expressed in
  Doppler units (that can be maximum Doppler units, minimum Doppler
  units or any other unity).}. If we want
  the spectral variation of the profiles to be well sampled, the
  change in Doppler velocity
  $\Delta(\mu_k V_i)=\Delta\mu_k \cdot V_i+\Delta V_i \cdot\mu_k$ in
  our spatial and angular grid has to be
  small enough (i.e., of the order of half Doppler width)
  everywhere, not only between
  adjacent points along the same ray but also between adjacent points along the
  same height. 

On one hand, this sets a constraint for the angular grid that can be
evaluated as:

\begin{equation}\label{eq:angular_con}
    |\Delta \mu_{\rm max} V_{\rm max}| \lesssim \frac{1}{2},
\end{equation}

 where $\Delta \mu_{\rm max}$ gives the maximum allowed angular step
 in the quadrature.
On the other hand, the spatial grid is constrained by:

\begin{equation}\label{eq:spatial_con}
\centering
\Delta V_{i}=|V_{i}-V_{i-1}|\lesssim 1/2,
\end{equation}
which limits the maximum Doppler velocity gradient between
two adjacent layers $i$ and $i-1$ in the atmosphere.

\item Once the iterative problem is solved for the atomic density-matrix elements, the emergent Stokes vector
  is calculated using a very fine frequency grid for capturing with
  precision all the
  spectral features. The maximum number of spectral grid points is
  limited to $191$ during the  iteration and to $505$ in the last
  formal integration of the RT equation.
 
\end{itemize}

\item Inclusion of the variation of the magnetic field with height. 

In semiempirical models like FALC \citep{Fontenla:1993}, there is no description of the magnetic
field. Usually, in this case the magnetic field is added ad-hoc with
constant parameters along the whole atmosphere. For Traviata to be
able to compute realistic models showing variations of the magnetic
field strength and orientation with height, we have also extended the
variables to consider such variations.

\item Calculation of Stokes V in the Zeeman regime (optional choice). 
\begin{itemize}[leftmargin=0.23cm, labelsep=0.05cm]
  \item Under the
  assumption of the weak field regime, Stokes V is due to the longitudinal
  Zeeman effect while Stokes Q and U are due to scattering and Hanle
  polarization, being the linear signals of the transverse Zeeman
  effect a second-order contribution. In such a case, the radiative
  transfer equations for Stokes Q and U decouple from the RT equation
  for Stokes V to a good level of approximation. 

    Dynamic Traviata solves the iterative problem for the populations of the energy
    sublevels using the transfer equations in the scattering and
    Hanle regime. At the end, when the atomic density multipoles
    are solved, the total level populations are used to solve the Zeeman
    transfer equation independently for I and V along the observer LOS:

\begin{equation} \label{eq:zeerte}
\frac{\rm{d}}{\rm{ds}}
\left( \begin{array}{c}
 I \\ 
 V
\end{array} \right) 
=
\left( \begin{array}{c}
\epsilon_I \\ 
 \epsilon_V
\end{array} \right) 
-
\left( \begin{array}{cc}
 \eta_I  & \eta_V \\ 
 \eta_V  & \eta_I
\end{array} \right) 
\left( \begin{array}{c}
 I \\ 
 V
\end{array} \right) .
\end{equation}
The dispersion or magneto-optical terms have been neglected in the
calculation of Stokes V because 
their contributions to the circular polarization are insignificant in
weakly magnetized atmospheres. To solve the system of equations we transform it in two independent equations:

\begin{equation} \label{eq:zeerteplus}
\frac{\rm{d}}{\rm{ds}}
\left( \begin{array}{c}
 I^{+} \\ 
 I^{-}
\end{array} \right) 
=
\left( \begin{array}{c}
\epsilon^{+} \\ 
 \epsilon^{-}
\end{array} \right) 
-
\left( \begin{array}{c}
 \eta^{+} I^{+} \\ 
 \eta^{-} I^{-}
\end{array} \right),
\end{equation}
creating the variables $I^{\pm}=I \pm V$, $\epsilon^{\pm}=\epsilon_I \pm
\epsilon_V$ and $\eta^{\pm}=\eta_I \pm
\eta_V$. Both equations are solved along the LOS using a parabolic
short-characteristic method to get the emergent $I^{+}$ and $I^{-}$
profiles for each frequency and spectral transition. The solution for
the original
variables is then obtained with $I=(I^{+}+I^{-})/2$ and
$V=(I^{+}-I^{-})/2$. There is a slight inconsistency in
calculating again the intensity with Stokes V while neglecting Q and U, but it is
justified by the uncoupling between linear and circular polarization
in weak magnetic fields. The intensity profile resulting from the scattering
problem is almost the
same as the one obtained from the Stokes V calculation, being in most cases
undistinguishable except for a negligible (magnetic) broadening in
some Zeeman signals.

  \item  The Stokes V
  line radiative coefficients are calculated neglecting atomic orientation. 
This implies that they can
be calculated at each point from
  Eqs. (\ref{eq:emi_iqu}) and (\ref{eq:ev}), where the spectral
  profiles for the Stokes V coefficients are given by
  Eq. (\ref{eq:zee4}). We repeat it here for the sake of clarity:
\begin{equation}\label{eq:v_repeat}
 \centering
 \phi_{V} (\nu, \vec{\Omega})
  =\frac{1}{2}\left[\phi_1 - \phi_{-1}\right] \cos{\hat{\theta}},
\end{equation}
with $\hat{\theta}$ the direct angle between the local magnetic field
vector and the
  direction $\vec{\Omega}$ of the given emerging ray passing through
  the considered plasma
  element. Working in the reference frame of the solar vertical, this angle can be calculated
  locally at every point in the atmoshere with:
\begin{equation}\label{eq:v_angle}
\centering
\cos{\hat{\theta}}= \cos{\theta}\cos{\theta_B}+\sin{\theta}\sin{\theta_B} \cos{(\chi_B-\chi)} ,
\end{equation}
with $\theta$ and $\chi$ the angles defining the LOS.

\item With these modifications Traviata gives the linear polarization
  signals produced by scattering polarization and Hanle effect, while
  additionally delivers the Stokes V profiles resulting from the Zeeman effect, always
  considering the velocity and
  the magnetic field gradients. 
\end{itemize}

\item Automatization for working with a large number of models.
  
  The code is launched from an IDL program that provides all the
  initialization settings. This program is adapted to the
  particular model atmosphere under consideration. For
  the case of a 3D data cube, it is considerably
  sophisticated because a strategy
  for running each model column in a specific order and with an
  adaptative initialization of the density-matrix elements (see next
  point) is needed. The automatization also includes the management and listing
  of models with convergence problems that have to be recomputed with other
  interpolation grid, with other parameter settings or with other
  initializing populations. Finally, the automatization includes the
  compilation process (which is now more complex due to a
  re-modularization and re-organization of the subroutines) and the
  possibility of reading and representing the outputs with additional programs.

\item  Initialization of the populations. 

The code has 4 execution modes. In the first mode the code reads a
file with a given solution of the $\rho^K_Q$ elements and directly
integrate the RTE to obtain the emergent Stokes vector for one LOS. In
the second mode Traviata solves the full RT problem by initializing
the iterative process with the overall populations given by the
atmospheric model (when they are available). In
the third mode, the code solves the problem initializing with the $\rho^K_Q$ elements that
were previously obtained for an adjacent column in the model. This can save some
time when running a dataset whose consecutive models in the series are
similar. Finally, the fourth mode is the standard one in which the
initialization is done in two steps; first, the populations in LTE are
obtained, and then such populations are used as inputs
to solve the NLTE populations that finally initializes the iterative problem.    

\item Parallelization.

The restrictive requirements needed to include the effect of the
velocity and the aim of computing a large number of
models required the code to be parallelized. The slowest operation is, with difference, the
application of the formal solver for each ray and frequency to
obtain the radiation field tensors. To accelerate it, we have
implemented a parallelization based on OpenMP directives acting
directly on the radiative transfer loop.

OpenMP supports shared memory multiprocessing programming
in Fortran on most processor architectures and operating
systems. It is an implementation of multithreading, a method of
parallelizing by which a master thread (a series of instructions
executed consecutively) forks a specified number of slave threads that
solve a task in individual parts. The threads then run independently, with
the runtime environment allocating them to different
processors. Scalability is limited by memory architecture. 

OpenMP puts a heavy demand on stack memory\footnote{In most modern
  computer systems, each thread has a reserved region of memory
  referred to as its stack. When a function executes, it may add some
  of its state data to the top of the stack. Stack-based memory allocation is very simple and typically faster than heap-based memory allocation (also known as dynamic memory allocation). Another feature is that memory on the stack is automatically, and very efficiently, reclaimed when the function exits, which can be convenient for the programmer.}  
, which has to be considered when programming. 
In this respect, we have concluded that the stack memory has to be free
enough for OpenMP to use it efficiently. Indeed, stack
memory has to be usually enlarged by user commands before running Dynamic Traviata because the
increment in the number of variables and their sizes can easily produce a stack
exhaustion. Besides that, the use of heap arrays can produce
significative performance penalties in codes running OpenMP. Again the solution is to avoid heap
allocation in favor of stack memory, which furthermore has a faster access. 

The performance improvement reached after the parallelization when
executing the code with 8 processors gives a speed up factor of 7,
approximately. When a dataset of
thousands of models is considered, it represents a difference
between $5$ days and more than 1 month of computing time. 
\end{enumerate}

\subsubsection{Computational time}

To give an idea of the increase in the computing time
when including velocities in the RT problem with
polarization, we made a simple estimation. Consider ${\rm N_{\rm atm}}$ to be
the number of points in the atmosphere where the specific intensity
has to be calculated (i.e., the number of layers ${\rm N_z}$ in the
planeparallel atmosphere). Let ${\rm N}_{\alpha}$ be
the number of quadrature angles chosen for the numerical calculation of the
radiation field tensor components and ${\rm N}_{\nu}$ the number of
frequencies at which the RT equation has to be solved. Let ${\rm N_{ it}}$
be the number of iterations needed for reaching the desired
precision, which in MALI (and by
extension in DALI) methods is ${\rm N_{ it} \propto N_{z}}$
\citep{Trujillo-Bueno:1995}. Since with the short-characteristics
method the computational time needed for doing a formal solution scales
linearly with ${\rm N_{\rm atm}}$, ${\rm N}_{\nu}$ and ${\rm N_{
    \alpha}}$, the total computational time ${\rm T_{cpu}}$ is 

\begin{equation}\label{eq:nlaw1}
\centering
    \rm{T_{\rm cpu}} \propto {\rm N_{\nu}N_{\alpha} N_{ atm} N_{it}}= {\rm N_{\nu}
      N_{\alpha} N_{z}^2}.
\end{equation}

In the estimation for the dynamic case, the main difference with the
static case is the larger number of points due to the more exigent
grid requirements. Thus, on the empirical basis of the experience with
different models, we point out that:

i) due the new requirements in 
resolution and extension of the frequency grid, the number of
frequencies in the problem with chromospheric velocities has to be $\approx 4$ times larger
than in a static problem with similar characteristics. 

ii) the number of vertical points in the spatial grid is increased
around a
factor $1.5$ with respect to the static case. In average, for the
chromospheric models we have used, it usually
requires between $13$ and $15$ points per decade approximately.

iii) the number of points in the angular quadrature for the
inclination is usually a few times larger than in the static
case. Typically, between $20$ and $40$ points for the whole
inclination range, depending on the maximum
velocity gradient. If a normal static resolution is about $8$
points in inclination, we evaluate the average increment in the dynamic case in a factor $4$.
Then, we obtain
 
\begin{equation}\label{eq:nlaw2}
\centering
    \rm{T_{\rm cpu}} \propto  {\rm 4 N_{\nu}
      4 N_{\alpha} (1.5 N_{z})^2}= 36 \,{\rm N_{\nu}
      ( N_{\alpha}  N_{z})^2},
\end{equation}
 which is $36$ times more than the static case. This estimation does
 not take into account the improvement in the convergence rate
 achieved by the Ng method. Estimating it in a factor $2$, the final difference between static and dynamic
 problems is a factor $18$ in computacional time. 

Accounting for the parallelization of the RT loop and using $8$
processors, we reduce that time in a factor $7$. Then, the dynamic
problem is reduced to have a computing time of the same order of magnitude as the
static one. 
\subsubsection{Testing the code}
We put the new version of the code under scrutiny applying a set of different tests to confirm
its functionality. We reproduced all the results of \cite{manso10} in
the static case. We did calculations with a constant ad-hoc velocity
field, obtaining the same results but shifted in wavelength. To
test V$/$I in the presence of velocity gradients and magnetic field gradients we compared the outputs
with the results of the RH code \citep{Uitenbroek:2001}. The
agreement in the Ca {\sc ii} IR triplet lines were satisfactory. The 
effect of increasing velocity gradients on the linear polarization signals
are detailed along this work. In all cases,
we always put especial care in setting a suitable number of points in
all numerical grids, following the criteria explained in this section
to avoid numerical errors due to lack of resolution. 


			
\subsection{A FEM-based optimization method for the spatial grid.}
\label{sec:fem}

In a finite element method (FEM), the spatial domain of a certain computational problem is divided in cells (finite elements). In their interior, the
solution is assumed to have a certain simplified variation that is parameterized by some free parameters
(\citeauthor{Suli:2012} \citeyear{Suli:2012}, \citeauthor{Guo:1986} \citeyear{Guo:1986}). Using the calculus of
variations, the method minimizes a local error function (objective function) defined for
each cell as a measure of quality of the grid,  thus obtainig a precise
solution for the problem. These methods are tipically used in boundary
value problems characterized by a set of differential equations and
boundary conditions. Among the different variants, one strategy for minimizing the objective function is to adjust the position of the grid nodes, then obtaining the optimal grid that produce a stable solution for the differential
equations \citep{Diaz:1983}.

In our case, the boundary value problem is to solve the RTE, which is
the equation governing the propagation of light (Stokes vector) along
each ray as well as the origin of instabilities during the
iteration. It would have to be solved for each step of the iterative
problem because the local values of the propagation matrix change
accordingly with the density matrix values resulting from the solution
of the SEE. Thus, we should incorporate the corresponding finite
element equations inside the RT code, changing the discretization of
the grid step by step during the iteration. Other theoretical requirement is that the boundary value problem should be well-posed\footnote{ The mathematical term ``well-posed problem'' stems from a definition given by Jacques Haddamard \citep{Parker:2013}. He believed that mathematical models of physical phenomena should have these properties:\\
  1) A solution exists.\\
  2) The solution is unique. \\
  3) The behavior of the solution hardly changes when there is a
  slight change in the initial condition (topology).}. This seems to be confirmed by numerical experimentation but, to the best of our knowledge, it is not mathematically demonstrated in RT schemes where the set of SEE are coupled with the RTE.

  Instead of dealing with the discretization problem at each iteration, we did not include
the finite element method in the RT problem, but simply
applied the FEM theory for redistributing the nodes in such a way that
some basic aims were fulfilled. Namely, to generate a new starting spatial grid that: (a) produces softer variations of the physical quantities along
the model; (b) improves the convergence guarantee; (c) minimizes the number
of grid points; and (d) requires a minimum effort. Thus, we developed a very simple and fast
algorithm that succesfully creates the adaptative grid by following
the next scheme.

For each vertical extension in the atmospheric model, we construct the
finite element approximation of this problem by subdividing the
normalized spatial domain $\Omega=[0,1]$ into $N$ subintervals or
finite elements $[x_k,x_{k+1}]$, $k=0,...,N-1$, by the set of nodes
$\lbrace x_0=0 <x_1 <\cdots<x_N \rbrace$, where $h=1/N$, with $N\ge 2$. 

We now choose an atmospheric quantity $u$ that will help in our method for
redistributing the nodes. Namely, we chose the vertical velocity (in
Doppler units) because the formal solution is very sensitive to its
variations due to the frequency shifts it introduces in the absorption
profiles (Sections \ref{sec:numeric} and \ref{sec:me}). Furthermore,
when expressed in Doppler units, it is a measure of the importance of
the macroscopic velocity with respect to the microscopic (thermal and
microturbulent) velocities, giving account for the relative influence
of vertical motions on the profiles. Indeed, if the Doppler velocity
presents larger gradients, the RT calculation will be more unstable and error prone. 

In the standard method, we should first define local piecewise finite
element functions acting as a base in a certain space. And then $u$
should really be a norm\footnote{ A norm is a function that assigns a strictly positive length or size to each vector in a vector space. The vector norm is the generalization to abstract vector spaces of the vector modulus in euclidean spaces.} measuring the interpolation
error between such local basis functions and the real solution. But we seek simplification.
Our solution is not the solution to the boundary value problem, but
the distribution of nodes in the physical magnitudes that will give us
a quicker convergence. Then, we redefine the problem in two statements :  1) find the
location of the nodes $x_k$ in order to equalize a quantity $f_k(u_k)$, under the geometry constraints that the
nodes must maintain a relative location strictly increasing inside the
domain $(0,1)$;
 2) Find the quantity $f_k(u_k)$ that roughly bounds the numerical interpolation error of $u_k$ along height.  After some experimentation, we define $f_k$ as:
\begin{equation}\label{eq:fk}
 \centering
    f_k = \, \alpha \hat{h}_k + \beta \frac{\partial
      \hat{u}_k}{\partial x} +\gamma \frac{\partial^2
      \hat{u}_k}{\partial x^2},
\end{equation}
with $\alpha,\beta,\gamma$ being free parameters weighting the length
of each element ($h_k=x_{k+1}-x_k$) and the first and second local
derivatives of $u$, respectively. The hat $\hat{}$  indicates that the quantities are
normalized to their maximum value in the whole domain $\Omega$. This
is done for making them comparable in magnitude. 

Equalizing the quantity $f_k$ given by Expression (\ref{eq:fk}) in each finite element, we are requiring the placing of more discretization nodes
where the first and second derivatives are larger, so covering zones
of $u$ with large gradients (first derivative) and with pronounced
curvature (second derivative). To avoid large distances between nodes
when they move to fill the regions with large derivatives, the expression also depends on $h_k$, which acts as a binding force (if $u$ were constant, the nodes would be
equally spaced). The formula has an empirical justification because it gives good results with difficult models that did not converge easily (or at all) with other spatial grids. Its success can be also understood if we consider that the redistribution of nodes is better suited to the quasi-parabolic variation of the source function assumed by the DELOPAR method (Sec. \ref{sec:formalsolver}). Imagine, for example, a grid point having its left-side neighbour very close and its right-side neighbour significantly apart, covering the three a region with a large gradient in velocity. In such situation there are significant differences between the optical depth increments, which leads to a source function being far from the parabolic variation that DELOPAR expects (hence, with larger error gradient) to transfer the Stokes vector. Instead of changing the DELOPAR formal solver, we have adjusted the nodes to allow DELOPAR to produce accurate results. 
 
Technically, we approximate the derivatives with their values at the geometrical
center of each element (the seminodes $x_{k+1/2}=(x_k+x_{k+1})/2$)
using a standar centered finite difference scheme:

\begin{subequations}\label{eq:finitediff}
  \begin{align}
    \frac{\partial u_k}{\partial x} &= \frac{\partial u}{\partial x}\bigg|_{k+1/2} = \, \frac{u_{k}-u_{k+1}}{h_k},\\
    \frac{\partial^2 u_k}{\partial x^2} &= \frac{\partial^2
      u}{\partial x^2}\bigg|_{k+1/2} = \,
    \frac{u_{k}+2u_{k+1/2}+u_{k+1}}{h^2_k}
  \end{align}
\end{subequations}

Finally, the nodes can be equally redistributed in $f_k$ using the
expression of a mass center for each pair of finite elements at both sides of a
node :
\begin{equation}\label{eq:masscenter}
\centering
    x^{\nu+1}_k = \, \frac{\underset{i=k-1,k}{\sum}x^{\nu}_i
      f^{\nu}_i}{\underset{i=k-1,k}{\sum} f^{\nu}_i} ,
\end{equation}
where $x^{\nu+1}_k$ is the new location of the k\textit{th} node, and
$x^{\nu}_i$ is the center of the corresponding finite element in the
$\nu$\textit{th} iteration. The boundary nodes are fixed. 

Following this procedure, the
grid points are automatically repositioned in stable places after a short
iterative process. The effect is summarized in Fig
\ref{fig:adapgrid}. We start considering the vertical velocity sampled
at an initial spatial grid (lower panel). Note the large number of
points where there are almost no variations in the velocity, being
scarce where gradients are larger, typically in higher
regions. Following Eq. \ref{eq:masscenter}, we illustrate the
iterative calculation of a new grid, which is guided by the weighting
function $f_k$ (middle panel). Note how the grid nodes are shifted to
the upper part of the atmosphere, and that still the lower parts are
well covered, which is important because most of the plasma mass is
there. The final product is a grid that softly samples the Doppler
velocity (upper panel). The expressions for $f_k$ could be better, so
it would be interesting to investigate other similar possibilities
(e.g., using the coefficients of a series expansion).

\begin{figure}[htb!]
\centering%
\includegraphics[scale=0.60]{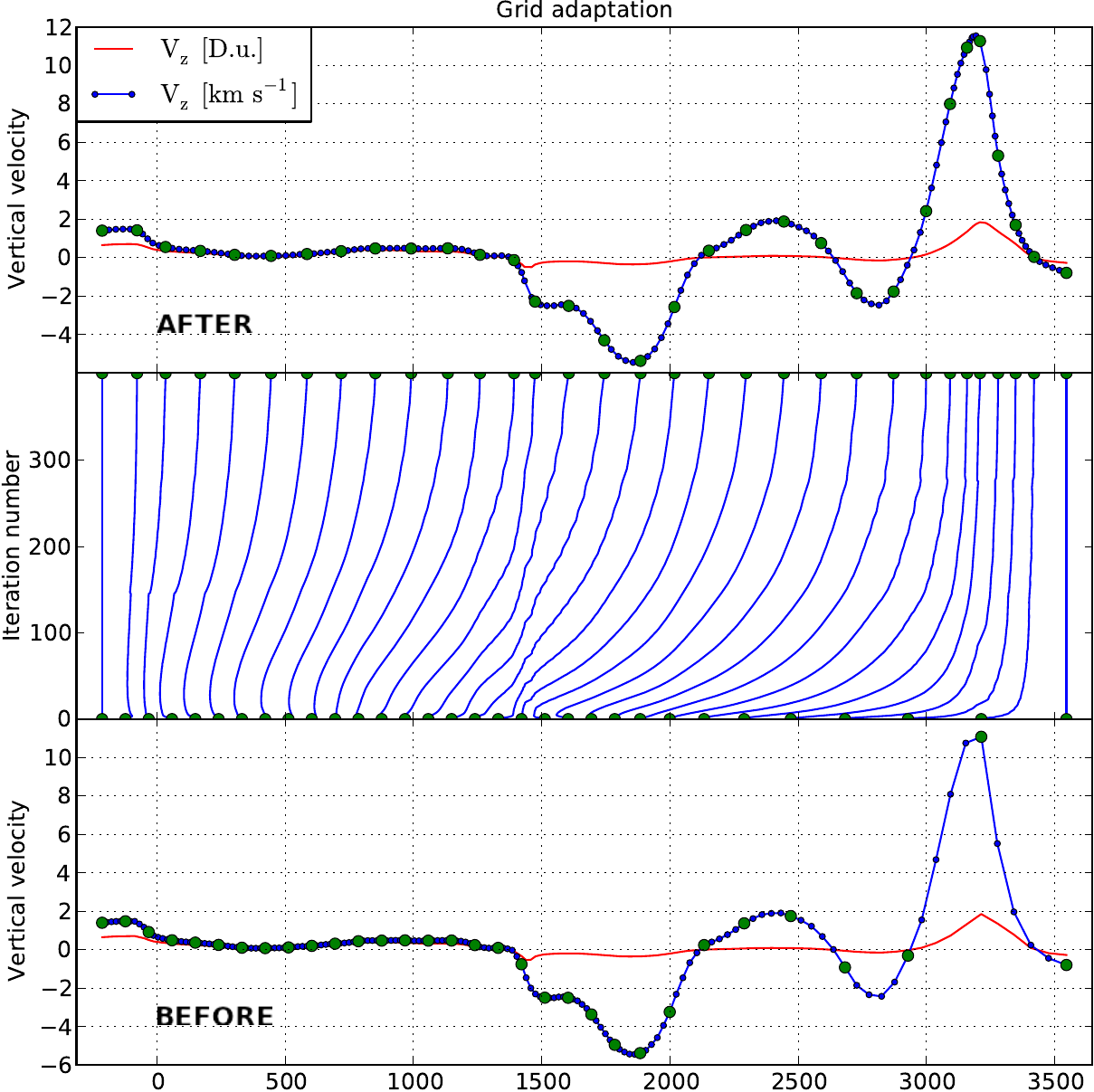}
\caption{Blue lines in the bottom/top panel illustrate the vertical velocity before/after the grid points reallocation. The red lines represent the corresponding Doppler velocity ($\mathrm{\varpropto v_z/T^{1/2}}$), which is used to shift the nodes. We have selected $1$ grid point every $5$ (see green dots) to illustrate how they are reallocated (middle panel).  }
\label{fig:adapgrid}
\end{figure}
The advantages of this strategy are: 

\begin{list}{}{}
\item 1) The computational
implementation is easy and fast.
\item 2) The procedure is
flexible because $f_k$ can be easily redefined and adjusted with the
weights $\alpha,\beta, \gamma$ .
\item 3) The new grid produces a smooth
variation of the physical quantities from node to node, which in general tends to diminish the
corresponding numerical error gradient.
\item  4) The radiative transfer equation requirements for the velocity gradients (explained in Section \ref{sec:numeric}) are easily accounted for by the parameter $\beta$.
\item 5) The better distribution of points allows a significant reduction of the total number of nodes without losing numerical precision in the RT solution.
\end{list}

The final income is that we can solve the \textit{impossible-to-converge} atmosphere models in less time and with less effort. 

When applied in combination with a
short-characteristic method, the results given by the redistribution
method depend on the spatial resolution. Taking that into account, before redistributing the
grid we interpolate to a number of points that depend on the
atmospheric extension in order to get a certain number of points per decade. After choosing the overall grid points, we redistribute them.

 Just to put an example of the application of this method for different number of grid points, we illustrate the evolution of the
 Maximum Relative Change during the RT iteration in 2 different
 models (Fig. \ref{fig:mre}). Depending on the number of points per
 decade\footnote{We refer to the number of points per decade of
   variation in the optical depth for the K line. We use it as a
   measure of the resolution of the model.} (p/dec), the
 redistribution of nodes has a better or worse effect on the
 convergence, in comparison with the case of no-redistribution and
 high resolution (red lines in Fig. \ref{fig:mre}). In general, $10$
 p/dec seems to be insufficient for most of the expanded models,
 appearing a delay in convergence (blue line, upper panel,
 Fig. \ref{fig:mre}) and in many cases a lack of it. Note that these
 models can have shocks in higher layers, which requires more
 discretization points. On the contrary,
 in compressed atmospheres the redistribution with $10$ p/dec can give the
 best results (blue line, lower panel,
 Fig. \ref{fig:mre}). Redistribution with a very large number of
 points is inefficient in most cases (orange lines very near
 the red ones). It happens because 
 redistribution plays a minor role against the decreased convergence
 efficiency of the DELOPAR method in very resolved grids. A trade off
 is represented by the green lines, corresponding to the cases with
 redistribution of $14$ p/dec. With this choice, we have solved the
 convergence problems in many highly variable models redistributing
 the grid points and using the less
 number of them possible. Tuning this procedure for any arbitrary column, we can obtain an improvement in the computational time.

\begin{figure}[htb!]
\centering
\includegraphics[scale=0.55]{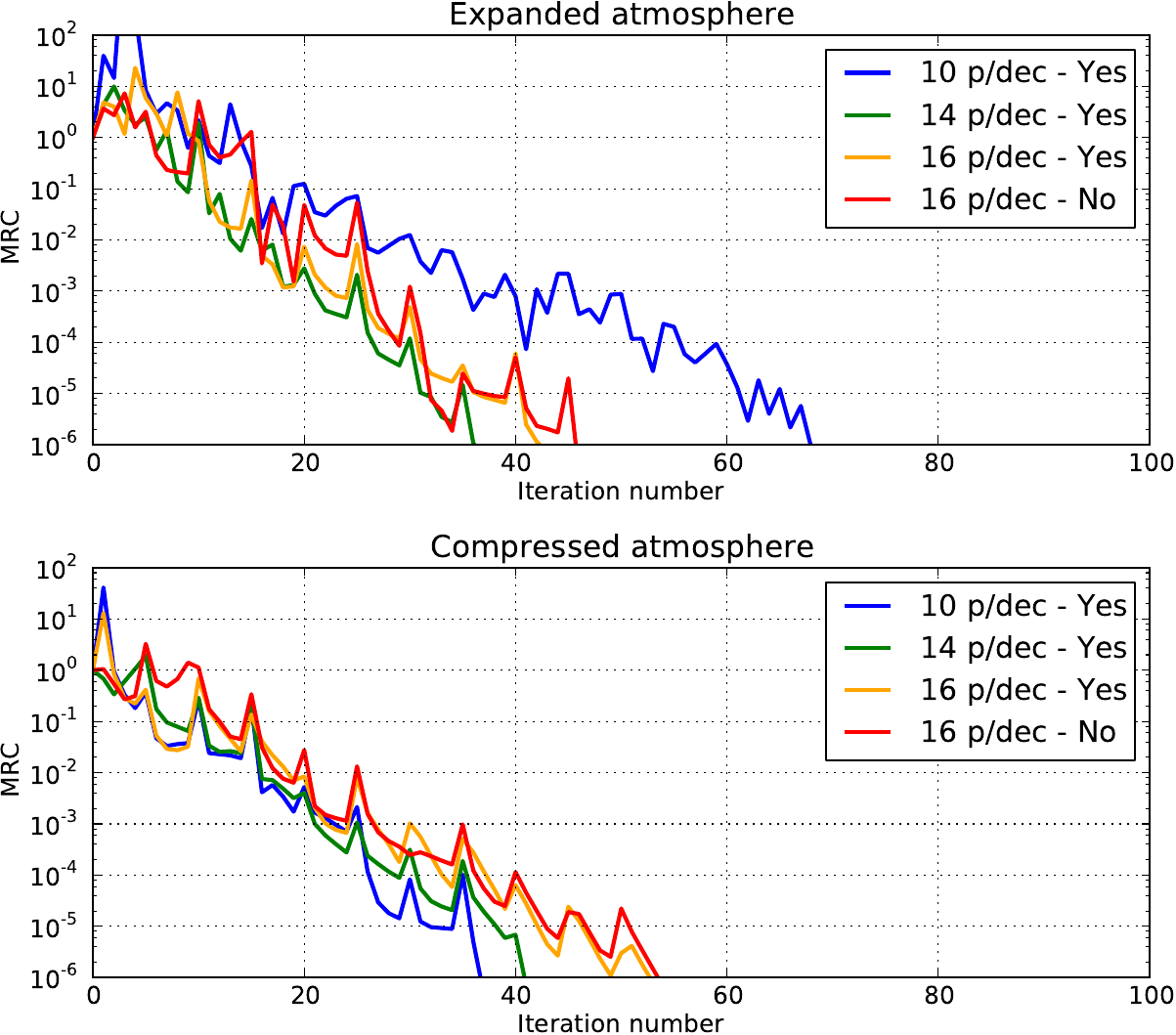}
\caption{Comparison of different spatial grid interpolations for a typical expanded model (upper panel) and a typical compressed one (lower panel). The no-redistributed grid with large resolution corresponds to the red lines. The other lines are for a varying number of redistributed grid points.}
\label{fig:mre}
\end{figure}
\section{Response functions in chromospheric solar models.}\label{sec:RF}	
The response function $R^{F,X}_{\lambda}$ measures the sensitivity of a certain spectral
quantity $F_{\lambda}$ to local perturbations of the physical property
$X(z)$ in the model atmosphere under consideration
\citep{Landi-DeglInnocenti:1977aa,Ruiz-Cobo:1994aa} . It is defined using the following integral expression \citep{Uitenbroek:2006aa}

\begin{equation}\label{eq:rf1}
    F_{\lambda} = \int^{z_0}_{-\infty}R^{F,X}_{\lambda}(z)  X(z) dz.
\end{equation}
 $F_{\lambda} $ can be any function of the emergent Stokes parameters obtained from the model
atmosphere. Equation (\ref{eq:rf1}) tell us that the response functions (RFs)  behave
like partial derivatives of the spectral profile ($ F_{\lambda}$, or a
Stokes profile in particular) with respect to a given
atmospheric physical quantity at a given depth of
the atmosphere. Thus, they provide a measure of the sensitivities of the observed
spectral quantities to the physical magnitudes characterizing the
state of the atmosphere. More precisely, they specify the heights at
which a perturbation produces alterations in the spectral quantity. This is very important because, under strong non-LTE conditions (i.e.,
 when the level populations in a transition are
 controlled by radiative processes or scattering),
 the source function at higher and less dense layers can be controlled by
 photons created much deeper than $\tau=1$ in the atmosphere or
 in hot layers well above. As the RFs account for it, they consequently provide the most reliable
estimate of the heights of formation for a given model.

 Due to the linearity of the integral, if the emergent
$F_{\lambda}$ is recalculated after perturbing $X$ with $\Delta X$,
the variation in $F_{\lambda}$ is :

\begin{equation}\label{eq:rf2}
    \Delta F_{\lambda} = \int^{z_0}_{-\infty}R^{F,X}_{\lambda}(z) \Delta X(z) dz,
\end{equation}

The application of a suitable perturbation allows us to calculate the
response function
$R$. Following \cite{Fossum:2005aa}, we perturb with $\Delta X
(z^{\prime}) = x(z^{\prime}) H(z^{\prime} - z)$, using the step
function $H$, which is $0$ for $z^{\prime} > z$ and
$1$ for $z^{\prime} \leq z$. Substitution of the perturbation
into Equation (\ref{eq:rf2}) and subsequent differentiation yields:
 
\begin{subequations}\label{eq:rf4}
\begin{align}
    \Delta F^z_{\lambda} &= F^z_{\lambda}-F_{\lambda}= 
    \int^{z}_{-\infty}R^{F,X}_{\lambda}(z^{\prime}) x(z^{\prime})
    dz^{\prime},\label{eq:pepe}\\
R^{F,X}_{\lambda}(z) &= \frac{1}{x(z)}
\frac{\mathrm{d}}{\mathrm{d}z} (\Delta F^z_{\lambda}). \label{eq:pepe2}
\end{align}
\end{subequations}

Following the previous equations, we have developed a program that uses Dynamic
Traviata to calculate RFs of the polarization profiles in
chromospheric dynamic models. The numerical evaluation of RFs can be done following these steps:
\begin{enumerate}
\item Synthesis of the desired spectral quantity $ F_{\lambda}$ in a
  given unperturbed model atmosphere.

\item Considering a height $z$: perturbation of the physical quantity of interest (temperature, density, magnetic field, velocity,...) from the lowest height of the
atmosphere to the height $z$ (step function) and synthesis (including calculation of the 
atomic density matrix) of the corresponding  emergent perturbed spectrum $F^z_{\lambda}$.

\item Calculation of the difference $\Delta
  F^z_{\lambda}$ between both spectra. 

\item Repetition of steps 2 and 3 for other height $z$, starting from
  the bottom upward
  until the last RT calculation, which is finally done perturbing all points
  in the atmosphere. 
\item Application of Equation (\ref{eq:pepe2}).
\end{enumerate}
\begin{figure}[h!]
        \centering
        \begin{subfigure}[b]{0.49\textwidth}
                \centering
                \includegraphics [width=\textwidth]{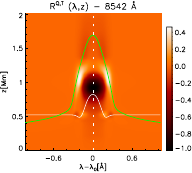}
                \label{fig:rf_T1}
        \end{subfigure}
        \begin{subfigure}[b]{0.49\textwidth}
                \centering
                \includegraphics [width=\textwidth]{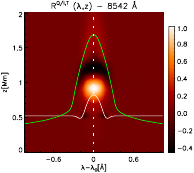}
                \label{fig:rf_T2}
        \end{subfigure}
        \caption{Response functions to temperature of Q (left) and
          Q$/$I (right) for the $8542$ {\AA } line. The green line
          traces the heights of optical depth unity calculated along the line of
          sight ($\mu=0.1$) for each wavelength. The white line is the
        Q$/$I profile in arbitrary units. The RFs are normalized.}
\label{fig:rf_Q_T}
\end{figure}
Thus, for a model atmosphere sampled at $N_z$ heights, the previous method complete $N_z+1$
full radiative transfer calculations, which can be very expensive
depending on the model. An illustrative example is shown in Figure
\ref{fig:rf_Q_T} assuming a semiempirical FAL-C model. 
\begin{figure}[h!]
        \centering
        \begin{subfigure}[a]{\textwidth}
                \centering
                \includegraphics [width=\textwidth]{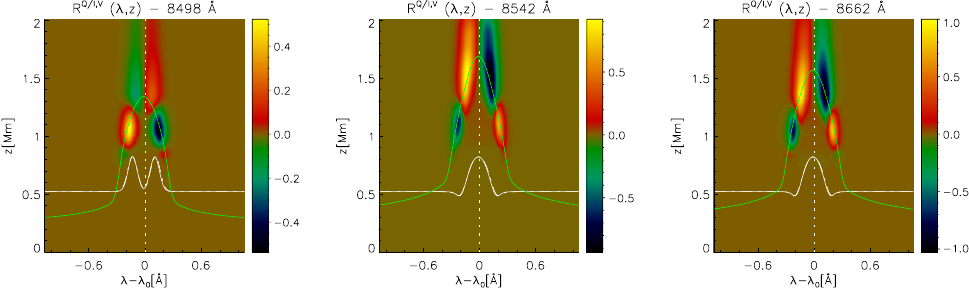}
              \end{subfigure}
        \begin{subfigure}[b]{\textwidth}
                \centering
                \includegraphics [width=\textwidth]{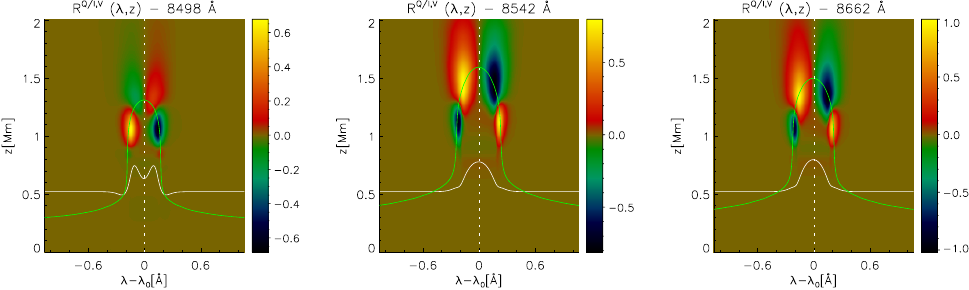}
              \end{subfigure}
        \caption{\footnotesize{Response functions to macroscopic velocity for Q$/$I in FAL-C model with constant microturbulent velocity (upper panel)
          or with the original height-dependent microturbulent velocity of the model. The
          unperturbed macroscopic velocity is zero and the pertubation is
          constant and positive (hence, upward).} }
                \label{fig:rf_micro}
\end{figure}
\begin{figure}[h!]
                \centering
                \includegraphics [width=\textwidth]{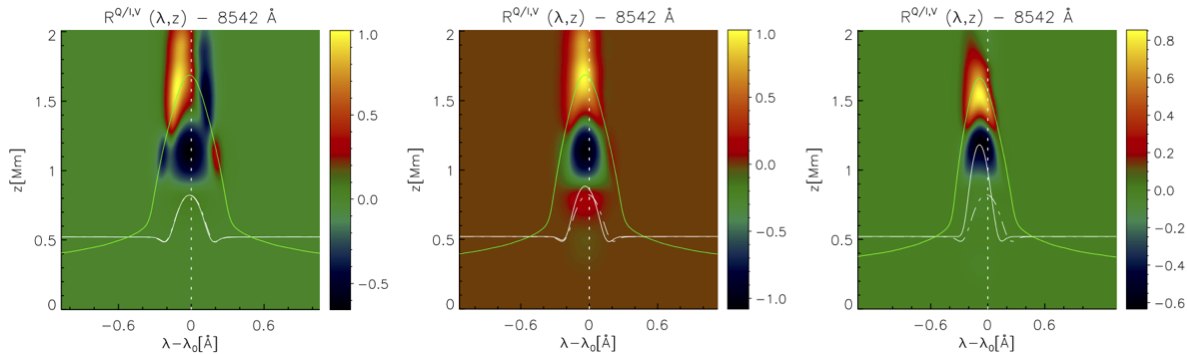}
        \caption{\footnotesize{Response functions to the velocity gradient of Q$/$I
          using the FAL-C
          model with three ad-hoc velocity fields of the form $
          v_z= 0.05 \, v_{\rm max} \, (z[{\rm Mm}]+0.1)$
          being $ v_{\rm max}=1$, $10$, $30$ $\rm km\,s^{-1}$ from left
          to right, respectively.}}
                \label{fig:rf_vgrad}
\end{figure}
It is the response function to temperature in Stokes
Q (left) and in Stokes Q$/$I (right) for a same static chromospheric
model. It can be seen how the line core is sensitive to perturbations
in temperature at heights far below $\tau^{\rm LOS}_{\nu}=1$ (green
line). For the $8542$ {\AA } line of Ca{\sc ii} formed in this model, the largest response
takes place where the chromospheric temperature gradients increase above the minimum of
temperature region.
The response functions can be very useful to understand the line
formation in a given model atmosphere and to characterize such a model. For instance, in Figure \ref{fig:rf_micro} we show the difference between
calculations in the FAL-C model with variable (lower panels) and constant (upper panels) microturbulent velocity for the three lines of the triplet, which
gives a quantitative grasp about the effect of this parameter in the radiative transfer. In
this example, the response functions are computed for perturbations
in bulk (macroscopic) velocity. Note that the velocity has a particularity: if we want to perturb a static
model, we cannot add a perturbation being a multiplicative factor from
the unperturbed model (because $v=0$). Instead, we have to consider a somehow arbitrary
velocity perturbation. In our case we have simply \textit{added} a
constant velocity perturbation, the same for all the heights.   

Although the RFs depend on the model considered, it is possible to
find similarities and patterns appearing in calculations from
different atmospheric models. Figure \ref{fig:rf_vgrad} is an example showing some
common features in the RFs for three models under changes in the velocity
gradient. 

\section{Tools for Principal Component Analysis.}\label{sec:PCA}	
Principal Component Analysis (PCA) is a statistical
technique based on pattern recognition that is used for dimensionality
reduction and signal processing. It has been applied in diverse fields of
scientific research, from human face recognition \citep{Turk:1991aa} to
the classification and inversion
of stellar spectra \citep{Lopez-Ariste:2001aa,Socas-Navarro:2001}. These techniques are based on the intuitive concept that a
certain natural phenomena manifests itself through a corresponding finite set
of basic recognizable patterns. In particular, PCA assumes that any other measurement
resulting from a particular realization of that natural
phenomenon can be modelled by decomposing the measurement into a
linear combination of orthogonal basic patterns. Finding the most
representative ones, we will be able to explain observations based on
the same phenomenon and
to discriminate it from the influence of other phenomena (e.g., observational noise).

Next, we explain the basic operations defining PCA applied to
spectral profiles \citep{Rees:2000}. We have developed a computer code that
perform these operations and we have used it to analyze RT observations and theoretical results.

The application of PCA starts with the creation of an initial database of $M$
model profiles,
$\phi(\lambda)$, covering in a statistical sense all possible physical
states under which the measured profiles $\psi(\lambda)$ can
form. Each physical state represents a model profile in the
database. The model profiles can be synthetic when obtained from a physical
model of the phenomenon. But they can also be obtained from a
representative set of observations, so forming an observational database.

If $N$ is the number of sampling wavelengths for the
model and measurement profiles, we first construct the $N\,\times
\,M$ database matrix [step $1$] 
\begin{equation}
  \label{eq:pca_1}
  D_{ij}=\phi_{j}(\lambda_i)\, ,\quad  i=1,\ldots, N\,;\quad
  j=1,\ldots ,M
\end{equation}
and the associated $N\,\times
\,N$ correlation matrix [step $2$],
\begin{equation}
  \label{eq:pca_2}
  \bf{C}= \bf{D}\bf{D}^{\rm T}.
\end{equation}
Next, we have to solve the eigenvalue problem for the correlation
matrix [step $3$],
\begin{equation}
  \label{eq:pca_3}
  \mathbf{C \, f}_i= e_i \mathbf{f}_i \, , \qquad i=1,\ldots , N
\end{equation}
using the method of Singular Value Decomposition (SVD). The
$N$-dimensional singular vectors $\mathbf{f}_i$ (\textit{principal components})
represent a set of $N$
eigenprofiles in terms of which the profiles $\phi_j(\lambda)$ can be
reconstructed exactly. The
eigenvalues $e_i$ in Eq. (\ref{eq:pca_3}) are ordered according to
decreasing norm (hence, of decreasing significance), and they estimate
the signal variance captured in the individual principal components. Usually, the
first $n$ eigenprofiles, with $n\,<< \,N$, form a basis that is sufficient to
reconstruct all the model profiles within the typical noise of the
observations. The cumulative fractional variance
  explained by those few eigenvectors can be calculated with [step4]:
\begin{equation}
  \label{eq:pca_4}
  100\cdot \frac{\sum^{n}_{k=1} e_{k}}{\sum^{N}_{i=1} e_i}.
\end{equation}

The basis represents the orthogonal ``universal'' patterns in which
all profiles corresponding to the physical model adopted can be decomposed within
the observational noise. These eigenprofiles are stored in the $n\, \times\, N$ matrix
$\mathbf{B}$ [step $5$]. 

If the initial database is
actually representative of the reality (either because the
physical model is correct or because the observations taken to build
the database are statistically well chosen), it is then
unnecesary to use more than the PCA basis to approximate
any other measured profile.

Thus, we can project any given measured $\psi_j(\lambda)$ on the basis of eigenprofiles $\mathbf{f}_k$,  obtaining the
corresponding projections $e^{j}_{k}$ with $k=1 \ldots n$ [step $6$]. Such
projections allow us to reconstruct
the observations in the basis subspace by doing [step $7$]:
\begin{equation}
  \label{eq:pca_5}
  \bm{\psi^{\prime}}_j= \sum^{n}_{k=1} e^j_k \mathbf{f}_k, 
\end{equation}
\begin{figure}[h!]
        \centering
        \begin{subfigure}[a]{0.48\textwidth}
                \begin{center}
                  \quad
                  \fbox{\includegraphics[width=3.5in]{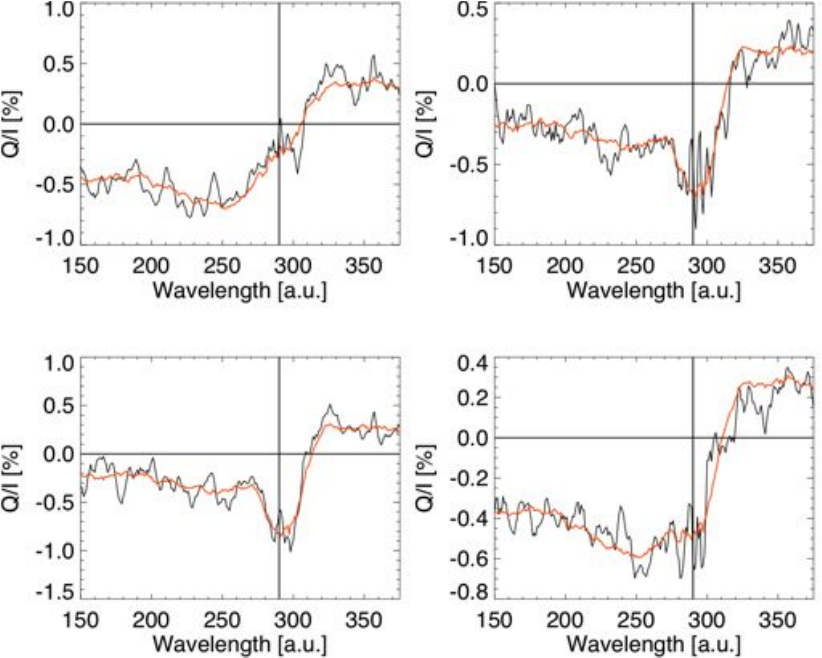} }
              \end{center}
        \end{subfigure}
\quad
    \begin{subfigure}[b]{0.48\textwidth}
                \begin{center}
                  \qquad \quad
                  $\begin{array}{c}
                  \includegraphics[width=1.7in]{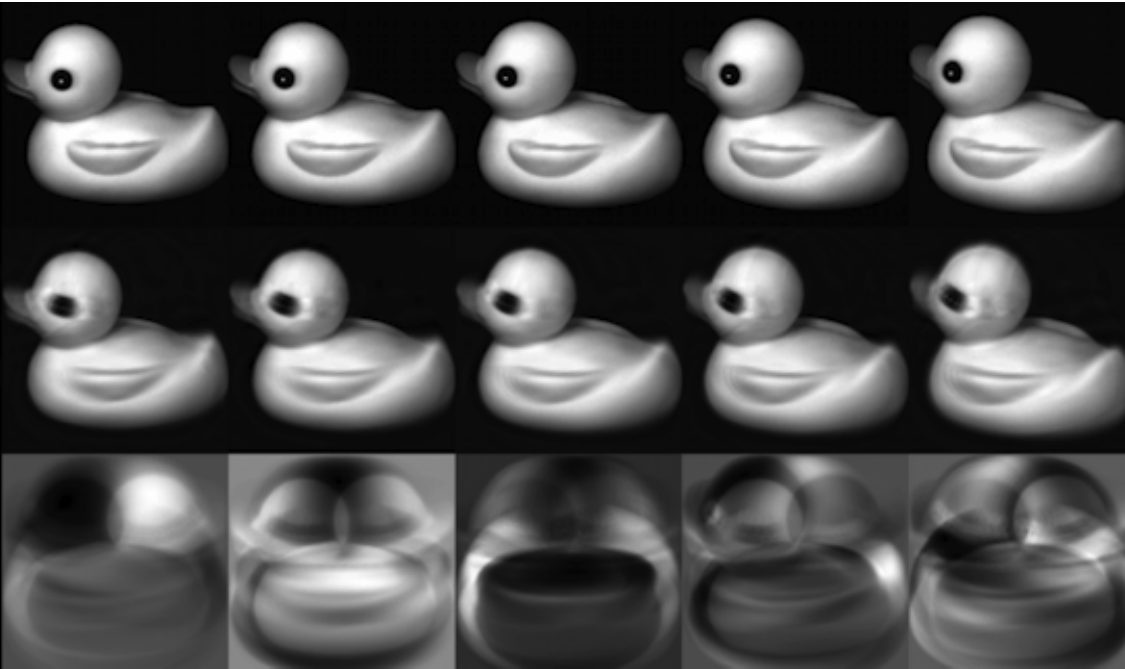} \\
                  \includegraphics[width=1.9in]{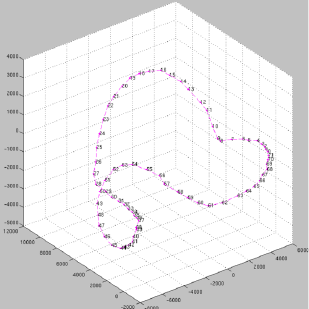} 
                \end{array}$
              \end{center}
    \end{subfigure}
        \caption{Left: Examples of four PCA-reconstructed linear
          polarization profiles. Right upper picture: some
          measurements (upper row ducks) with their corresponding
          reconstructed images (middle row ducks) after decomposing
          the former set in the first five eigenvectors (lower row) of the
          PCA basis. Right lower picture: path followed by the projected
          components in the space formed by the first three eigenvectors.   }
                \label{fig:pcaboth}
\end{figure}

where $ \bm{\psi^{\prime}}_j \sim   \bm{\psi}_j$, differing in an
error of the order of
the observational uncertainty\footnote{The error can be estimated from
  $ ||\bm{\psi^{\prime}}_j - \bm{\psi}_j||^2=\sum^{N}_{i=n} (e^j_i)^2$.}.  The finite size of the initial database implies a discrete coverage of
the physical parameter space, which is the main source of numerical
error in the reconstruction process.

Before performing the previous operations, our program statistically
standardizes the set of model profiles to create the initial
database. This is a common operation in PCA, consisting in normalizing and
regularizing the mean and the statistical deviation of the profiles to
be able to explain the shape of the profiles without worrying about
scale factors. The program can be used to create the basis from observations
of Stokes profiles or from the theoretical results of our
investigation. To test it, we applied the code to spectropolarimetric
observations to clean them from the noise (see figure \ref{fig:pcaboth}) and we reproduced some results
of \cite{Rees:2003aa}.

\section{The Solar Inspector}\label{sec:chromocube}	
To understand the results of a radiative transfer investigation it is
necessary to carry out a combined analysis of both the emergent radiative
quantities (e.g., Stokes profiles) and the physical properties of the atmospheric model. In order to facilitate the
analysis of
the 3D models described in Chapter \ref{cap:four}, we have developed a
visual interactive application (Solar Inspector). It shows the Stokes parameters of all
the considered spectral lines in combination with different maps and
with the stratifications of temperature, vertical velocity and magnetic
field intensity. The Stokes parameters are the result of solving the
RT problem as explained in Chapter \ref{cap:four}. In total, there
are $63$
different maps. Some of them correspond to temperature, magnetic field inclination,
velocity, etc. evaluated
at line-center optical depth unity for every spectral line of the Ca {\sc ii}
IR triplet. Others are maps of Stokes profiles (or derivatives
quantities) for every spectral
line. 

The maps correspond to the same atmospheric area and are
sorted out in a specific order that allows easy and useful comparisons. The user can
quickly alternate between them (see Figure
\ref{fig:chromocube}) using the mouse
wheel. With just a click the
scroll step is modified, changing the adjacent maps to be
compared. This simple strategy gives, for instance, a quick matching between features for
the same map in different spectral lines or between different maps of Stokes
parameters for the same spectral
line. Clicking in the pixels of the maps, the corresponding atmospheric
stratifications (regions A, B and C in Figure \ref{fig:chromocube}) and the emergent
Stokes profiles (panels on the right in Figure \ref{fig:chromocube})
for that pixel are showed. Clicking on the different
regions of the screen we can select
different operations, such as overlapping Stokes profiles
from different pixels to be compared or saving the actual 
image to a pdf file.      
  
The program is built in Python by embedding
\textit{matplotlib}\footnote{Matplotlib is a high-quality plotting package for
  Python that allows to easily visualize any kind of scientific data, giving a great control
  to the user.} plots into wxPython
GUIs\footnote{Graphical User Interface: it is the software that serves
  as user interface to communicate with the computer operative system
  (creating the windows, panels, etc). It makes 
  possible a friendly interaction with an informatic system through
  the visual lenguage. This concept is divided in backend (it does all the hard work behind-the-scenes to make the
figures) and frontend ( i.e. the plotting code used by the user).}. This allows a high-quality and versatile graphical user
interface that can work interactively (without menus) using events handling. To be able to update the
plots quicker while working with a large
amount of data, the application have
been programmed using techniques for real-time visualization. 
\begin{figure}[h!]
                \centering
                \includegraphics [width=0.6\textwidth]{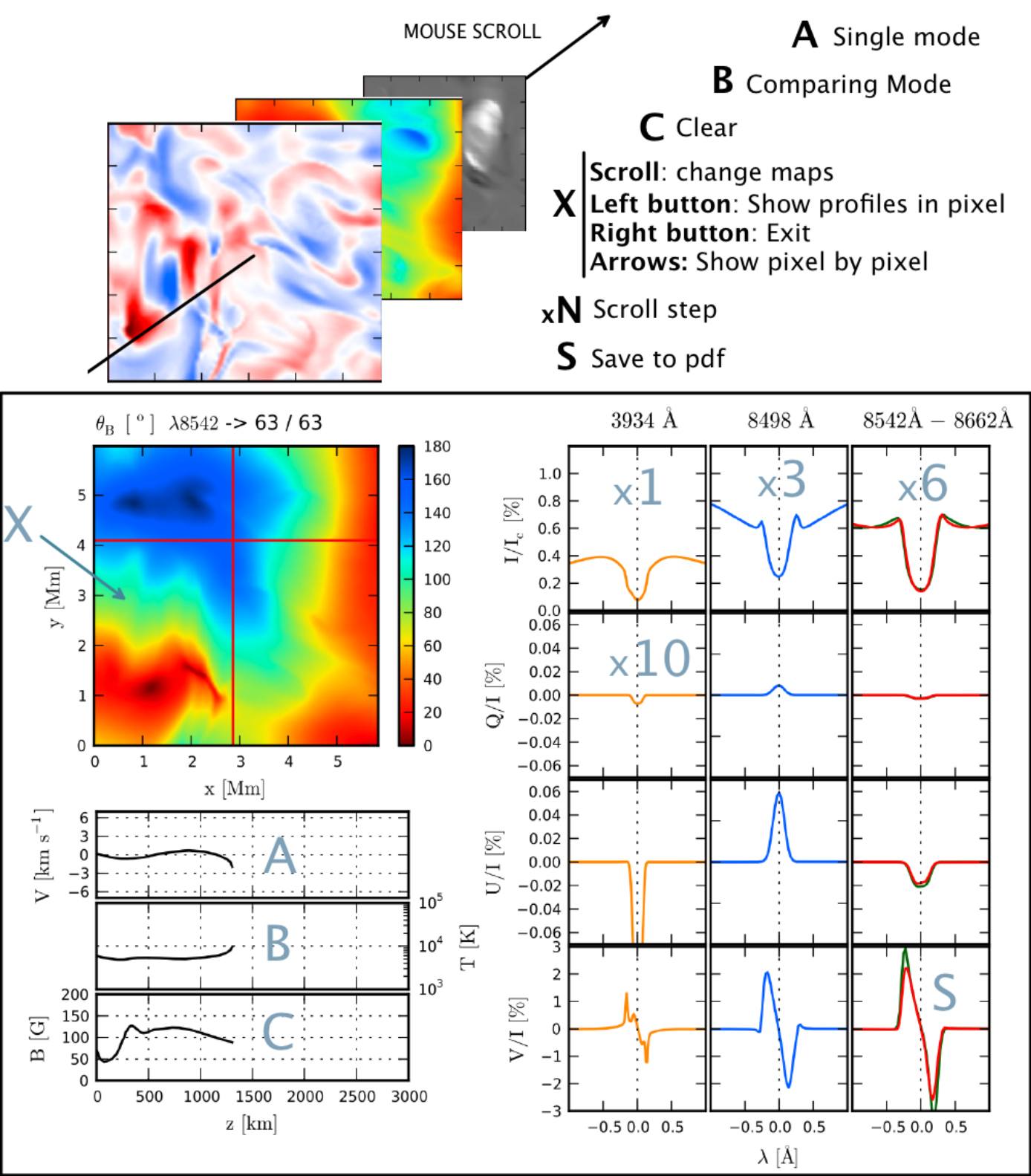}
        \caption{An example of a screen obtained with the
          Solar Inspector visualization facility. The main regions and the functionalities
          they offer are explained at the top of the figure.}
                \label{fig:chromocube}
\end{figure}

%

\section{Advanced 3D visualization with Mayavi.}\label{sec:mayavi}

Mayavi is a general-purpose scientific 3D visualization package. It uses
the Visualization Toolkit (VTK), which is by far the best visualization
library available \citep{Ramachandran:2011aa}. Unfortunately, VTK is not entirely easy to understand and many
people are not interested in learning it since it has a steep learning
curve. Mayavi strives to provide interfaces to VTK that make it easier
to use, by applying the features of Python, a dynamical programming language, to offer
simple APIs\footnote{An application programming interface (API)
  specifies how some software components should interact with each
  other and can be used to ease the work of programming graphical user
  interface components. In practice, most often an API is a library
  containing specifications for routines, data structures, object
  classes, and variables.}.

In short, some characteristics of Mayavi that have been relevant for
the development of this thesis are:

\begin{itemize}
\item Rich multidimensional representation of any object and
physical magnitudes able to be represented as points, lines, surfaces,
volumes and fields using vectors, matrices or tensors (see examples in Figure
\ref{fig:may}).
\item A standalone application for visualization through the
  so-called pipeline interface.
 \item Interactive 3D plots from the Python console IPython. The
 \textit{mlab} module contained in Mayavi provides an easy way to visualize data from a
 script or from an interactive prompt.
\item Visualization engine for embedding in user dialogs box. Mayavi allows a relatively
easy interaction with other Python applications. For instance, we could use it to
attach a suitable 3D visualization to the Solar Inspector application
described in the previous section. 
\begin{figure}[h!]
        \begin{subfigure}[a]{\textwidth}
                \begin{center}$
                \begin{array}{cc}
                  \fbox{\includegraphics[width=3.7in]{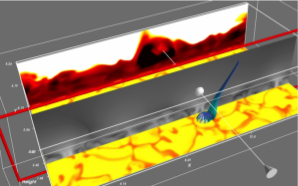} }&
                  \fbox{\includegraphics[width=2.2in]{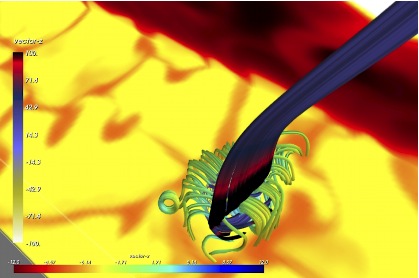} }
                \end{array}$
              \end{center}
        \end{subfigure}
        \begin{subfigure}[b]{\textwidth}
                \begin{center}$
                \begin{array}{cc}
                  \fbox{\includegraphics[width=2.0in]{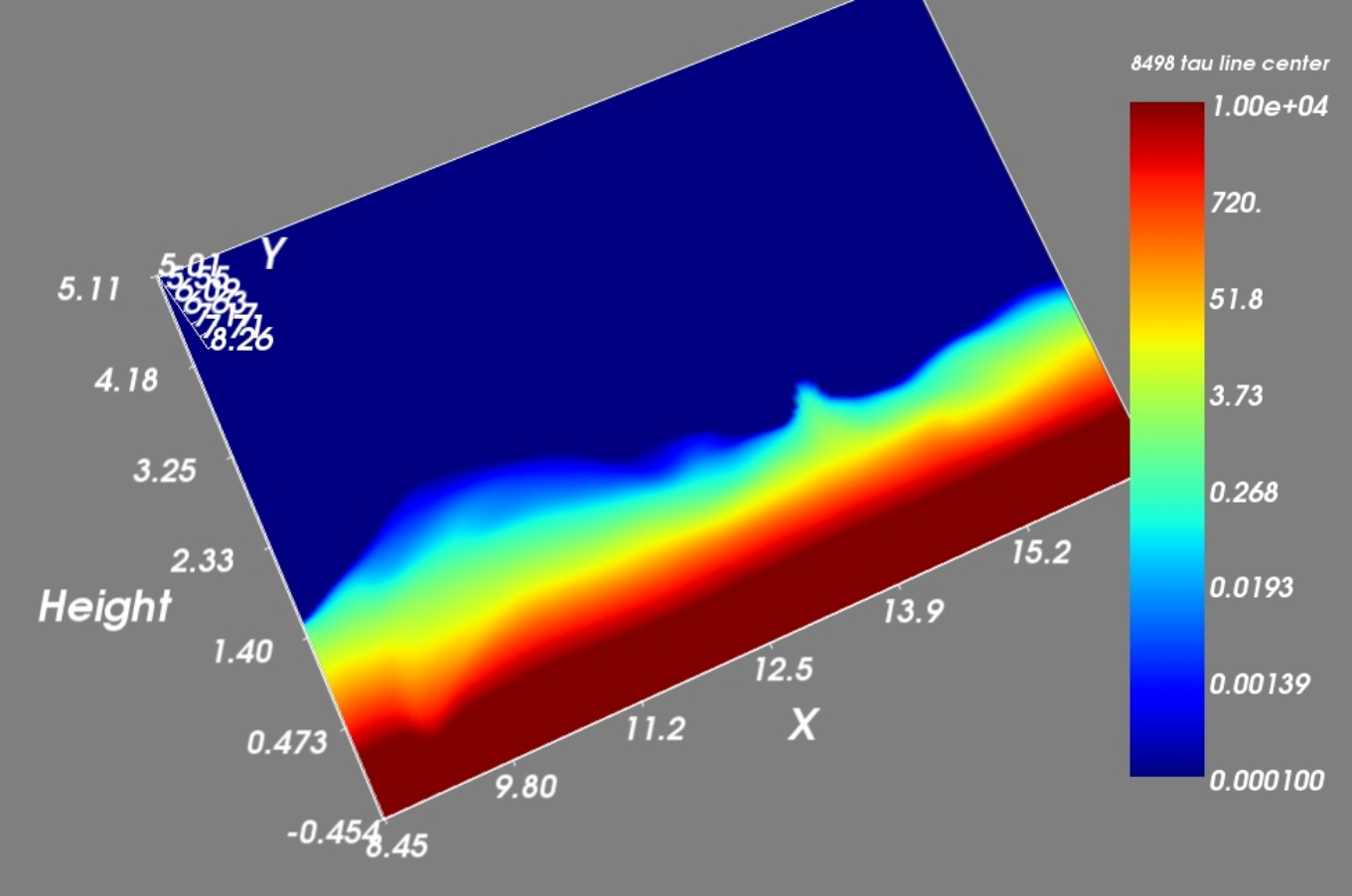}} &
                  \fbox{\includegraphics[width=3.9in]{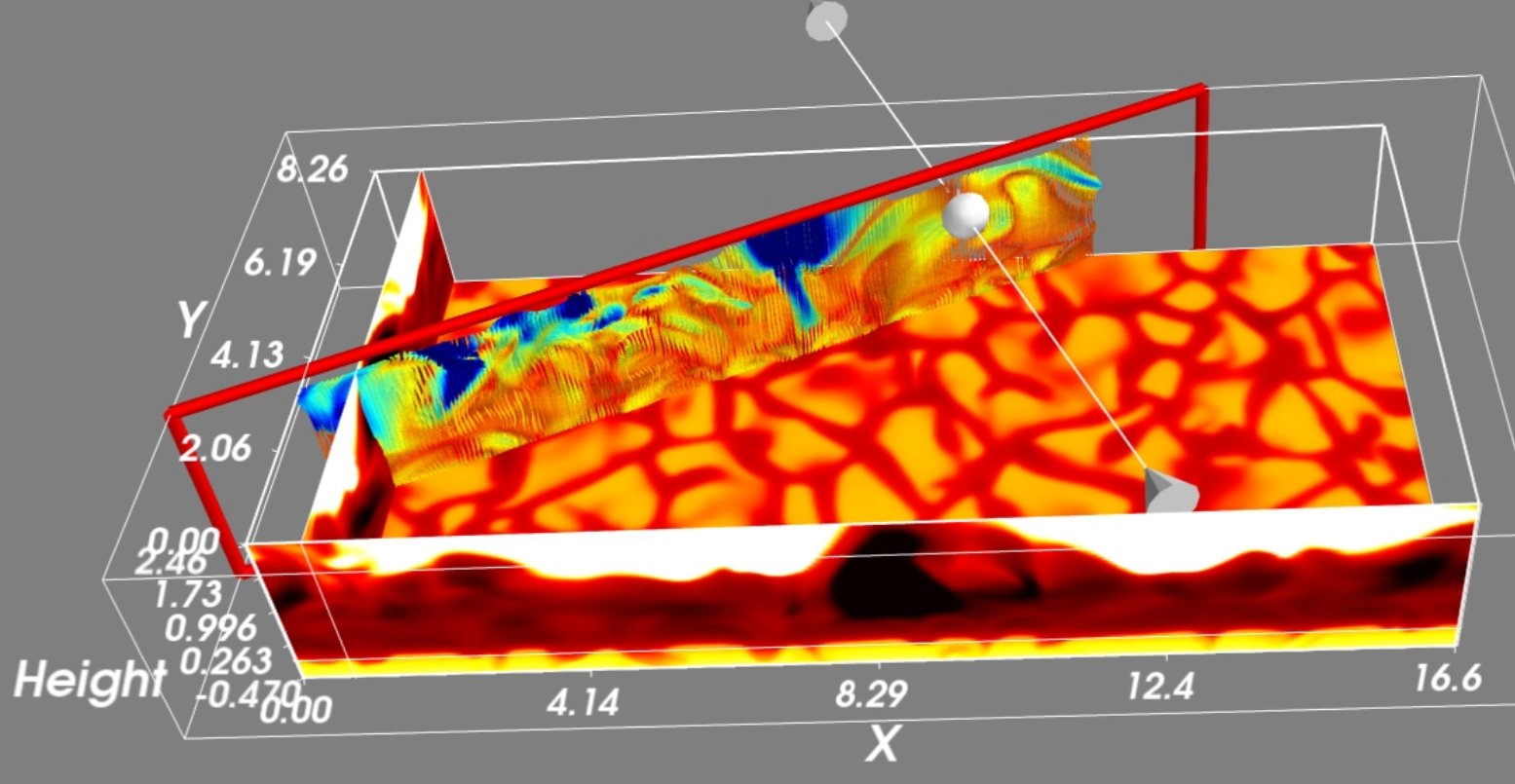} }
                \end{array}$
              \end{center}
        \end{subfigure}
    \begin{subfigure}[a]{\textwidth}
                \begin{center}$
                \begin{array}{cc}
                  \fbox{\includegraphics[width=3.0in]{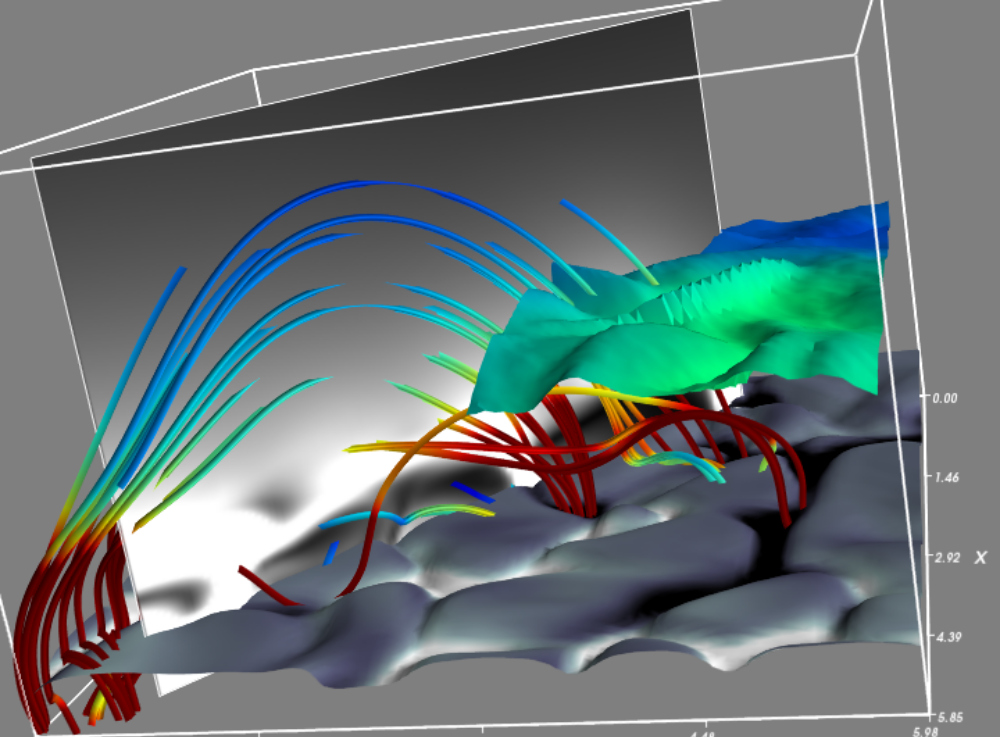} }&
                  \fbox{\includegraphics[width=2.7in]{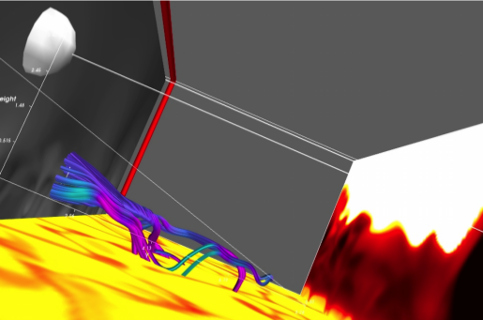} }
                \end{array}$
              \end{center}
        \end{subfigure}

        \caption{Some of the visualization works done with Mayavi to
          analyze the 3D MHD models of Chapter \ref{cap:four}. The
          model dataset has been provided by \cite{Leenaarts:2009}.}
\label{fig:may}
\end{figure}

\end{itemize}

When working with solar atmospheric models, importing the dataset to
Mayavi is crucial but non direct. It often involves transformations and interpolations
of the data to a convenient uniformly-spaced data set that is
manageable by Mayavi. Distributing or creating the atmospheric models
in VTK or other similar format suitable for visualization would avoid
such complications. On the other hand, through third-party packages Mayavi can import IDL save files (from scipy.io.idl import
readsav), which can simplify the issue.

 Our experience with Mayavi
indicates that it is very suitable for solar physics research. It seems to be able to create almost any data representation. In particular, we have used it to represent scalar and
vector fields, surfaces, flux and field lines in a dynamic
and interactive way. 

To introduce the radiative information in a 3D visualization
we proceed in the following way. We calculate the heights of the model
where the optical depth
in a certain spectral line is unity, which defines a surface in the 3D
space. We plot that surface and colour it with the value of
a given radiative quantity (an emergent Stokes parameter or similar). Other
useful representation is to calculate the line center optical depth
for a given line of sight at
each point of the atmosphere and visualize it using cutting planes,
isosurfaces or volumes having a variable opacity in the plot. This gives a visualization of the main
formation regions for the spectral lines.

 \clearemptydoublepage  

\chapter{Scattering Polarization with Velocity Gradients.}\label{cap:two}
In this chapter, we present the effects that gradients in the macroscopic vertical
velocity field have on the non-magnetic scattering polarization signals, establishing the
basis for other cases. We show that the solar plasma velocity gradients
have a significant effect on the linear polarization produced by scattering in chromospheric spectral lines.
 In particular, we calculate the impact of
velocity gradients on the anisotropy of the
radiation field and on the ensuing fractional alignment of the Ca {\sc ii} levels,
and how they can
lead to an enhancement of the zero-field linear polarization signals. This
theoretical investigation remarks the importance of knowing the
dynamical state of the solar atmosphere in order to correctly interpret
spectropolarimetric measurements, which is important, among other things, for
establishing a suitable zero field reference case to infer magnetic fields
via the Hanle effect.

\section{Introduction} 
Over the last few years it has become increasingly clear that
the determination of the magnetic field in the ``quiet" solar chromosphere
requires measuring and interpreting the linear polarization profiles
produced by scattering in strong spectral lines, such as H$_{\alpha}$ and
the 8542 \AA\ line of the infrared triplet of Ca {\sc ii} \citep[e.g., see][]{Trujillo-Bueno:2010,Uitenbroek:2011aa}. In these chromospheric lines, the maximum fractional linear polarization signal occurs
at the center of the spectral line under consideration, where the Hanle effect 
(i.e., the magnetic-field-induced modification of the scattering line polarization) operates \citep{stenflo98}. Since
the opacity at the center of such chromospheric lines is very significant, it is
natural to find that the response function
of the emergent scattering polarization to magnetic field perturbations peaks in
the upper chromosphere \citep{Stepan:2010aa}.
This contrasts with the circular polarization signal caused by the Zeeman effect
whose response function
peaks at significantly lower atmospheric heights \citep{Socas-Navarro:2004aa}. Of particular importance for
developing the Hanle effect as a
diagnostic tool of chromospheric magnetism is to understand and calculate
reliably the linear polarization profiles
corresponding to the zero-field reference case.

The physical origin of the scattering line polarization is atomic level
polarization
(that is, population imbalances and/or coherence between the magnetic sublevels
of a degenerate level with total angular momentum $J$). Atomic polarization, in turn, is induced by anisotropic radiation pumping, which can be
particularly efficient in the low-density regions of stellar atmospheres where
the depolarizing role of
elastic collisions tends to be negligible. The larger the anisotropy of the incident field,
 the larger the induced atomic level polarization, and the larger the amplitude of the 
 emergent linear polarization.

The degree of anisotropy of the spectral line radiation within the solar
atmosphere
depends on the center-to-limb variation (CLV) of the incident intensity. In a
static model atmosphere
the CLV of the incident intensity is established by the gradient of the source
function of the spectral line under consideration \citep{Trujillo-Bueno:2001aa,LL04}.
However, stellar chromospheres are highly dynamic systems, with shocks and wave
motions \citep[e.g.,][]{Carlsson:1997}. The
ensuing macroscopic velocity gradients and Doppler shifts might have a
significant
impact on the radiation field anisotropy and, consequently, on the emergent
polarization profiles. Therefore, it is important to investigate
the extent to which macroscopic velocity gradients may modify the amplitude and
shape of the emergent linear polarization profiles produced by
optically pumped atoms in the solar atmosphere. The main aim of this chapter
is to explain why atmospheric velocity gradients may
modify the anisotropy of the spectral line radiation and, therefore, the
emergent scattering line polarization. We also aim at evaluating, with the help
of
ad-hoc velocity fields introduced in a semi-empirical solar model atmosphere,
their possible impact
on the scattering polarization of the IR triplet of Ca {\sc ii}. A recent
investigation by \cite{manso10}, based on radiative
transfer calculations in static model atmospheres, shows why the differential
Hanle effect in these lines is of great potential interest for the exploration
of chromospheric magnetism.


\section{Formulation of the problem and relevant equations}\label{sec:problem} 

\subsection{The atomic model and the statistical equilibrium equations (SEE)}

We assume an atomic model consisting of the five lowest energetic, fine
structure levels of Ca {\sc ii} (see Figure 1). 
The excitation state of the atomic system is given by the populations of its 18
magnetic sublevels and the coherences among them.
We neglect coherences between different energy levels of the same term
(multilevel approximation). 
Moreover, since the problem we consider here (plane-parallel, non-magnetic
atmosphere with vertical velocity fields) is axially symmetric around the
vertical direction, no coherences between different magnetic sublevels exist when
the symmetry axis is taken for quantizing the angular momentum. 
We use the multipolar components of each $J$-level, 
\begin{equation}\label{eq01}
\rho^K_0=\sum_{M=-J}^{+J}(-1)^{J-M}\sqrt{2K+1}\left(
\begin{array}{ccc}
J & J & K \\
M & -M & 0
\end{array}
\right)N_{M},
\end{equation}
where $K=0, ..., 2J$, $N_M$ is the population of the sublevel with magnetic quantum number $M$, and 
the symbol between brackets is the Wigner $3j$-symbol (e.g., Brink \& Satchler 1968). 
Due to the symmetry of the scattering process (no magnetic field, no polarized
incident radiation in the atmosphere's boundaries), in a given level
$N_{+M}=N_{-M}$, 
and the excitation state of the system is described by just 9 independent sublevel
populations.
Consequently, odd-$K$ elements (orientation components) in Eq. (\ref{eq01}) vanish for
all five levels, and the only independent variables of the problem in the
spherical components formalism are the total populations of the five levels
($\sqrt{2J+1}\rho^0_0$); the alignment components ($\rho^2_0$) of levels 2, 3,
and 5; and $\rho^4_0$ of level 3, whose role is negligible for
our problem.

The statistical equilibrium equations accounting for the radiative and
collisional excitations and deexcitations in the 5-level system of Fig. \ref{fig:elevels} are
given explicitly in \cite{manso10}. We have particularized them to the no-coherence case (only $\rho^K_0$
elements) in Appendix \ref{app:B}.
The statistical equilibrium equations for the $\rho^0_0$ components contain terms 
that are equal to those appearing in the statistical equilibrium equations for the populations in a
standard (no polarization) NLTE problem (e.g., Mihalas 1978), plus higher order
terms $\sim J^2_0\rho^2_0$ (see Eqs.~\ref{see01}-\ref{see05}).
The statistical equilibrium equations for the alignment ($\rho^2_0$ components)
are formally similar to the ones for the populations with additional terms $\sim
J^2_0\rho^0_0$ accounting for the generation of alignment from the anisotropy of
the radiation field, and (negligible) higher order terms $\sim J^2_0\rho^2_0$
and $J^2_0\rho^4_0$ (see Equations~(\ref{see06}), (\ref{see07}) and
(\ref{see08})). These equations are expressed in the atom reference frame
(comoving system).
\begin{figure}[t!]
\centering
 \includegraphics[scale=1]{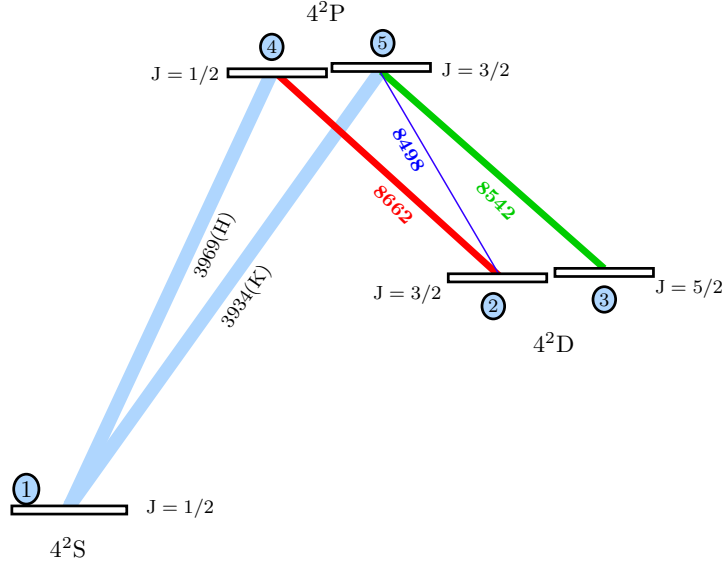}
\caption{Atomic model with energy levels for Ca{\sc ii}. The labels indicate the spectroscopic terms. 
Numbers inside blue filled circles identificate each level. Solid lines connecting 
levels show the allowed radiative transitions and the numbers at the middle of each 
segment are their wavelength in \AA. The wider the width of each connecting line the larger the spontaneous emission rate $A_{ul}$ of the transition. Atomic data for each spectral 
line are shown in Table \ref{tab:atomic}.\label{fig:elevels}}
\end{figure}
 \begin{table}[h!]
\centering
\begin{tabular}{cccccc}
\hline
 $\lambda$ (\AA\ ) & $u$ &
${\ell}$ & $A_{u{\ell}}$
(s$^{-1}$) & $ w_{J_{\ell}J_u}^{(2)}$ & $
w_{J_uJ_{\ell}}^{(2)}$ \\
\hline
\multicolumn{6}{c}{Allowed transitions} \\
\hline
3934 (K)    & 5 & 1 & $1.4\times 10^8$  & $0 $ & $\sqrt{2}/2$\\
3969 (H)    & 4 & 1 & $1.4\times 10^8$  & $0$ & $0$\\
8498 & 5 & 2 & $1.11\times 10^6$ & $-2\sqrt{2}/5$ & $-2\sqrt{2}/5$\\
8542 & 5 & 3 & $9.6\times 10^6$  & $\sqrt{7}/5$ & $\sqrt{2}/10$\\
8662 & 4 & 2 & $1.06\times 10^7$ & $\sqrt{2}/2$ & $0$\\
\hline
\end{tabular}
\caption{Short list of Ca {\sc ii} atomic data parameters. From left to right: the central wavelength, the upper and the lower level of each transition, the radiative rates from NIST atomic spectra database ({\tt http://www.nist.gov/physlab/data/asd.cfm}) and the atomic polarizability coefficients introduced by \cite{Landi-DeglInnocenti:1984}.}
 \label{tab:atomic}
\end{table}

Since the radiation field is axially symmetric, just two radiation field tensor
elements ($J^0_0$ and $J^2_0$) are necessary to describe the symmetry properties of
the spectral line radiation. 

Let $I(\nu, \mu)$ and $Q(\nu, \mu)$ be the Stokes parameters expressed in the observer's frame
at a given height $z$, where $\nu$ is the frequency, $\mu=\cos\theta$ and $\theta$ is the angle that the ray forms with the
vertical direction. Then, the corresponding values seen by a comoving
frame with vertical velocity $v_z$ with respect to the observer's frame are $I'(\nu',
\mu)=I(\nu, \mu)$ and $Q'(\nu', \mu)=Q(\nu, \mu)$, where
$\nu'=\nu(1-v_z\mu/c)$ and $\nu=\nu'(1+v_z\mu/c)$ (to first order in $v_z/c$).
Therefore, the mean intensity at the considered height, can be expressed from
one or another reference frame as:
\begin{align}\label{eq02}
\bar{J}^0_0&=\frac{1}{2}\int d\nu \int_{-1}^1 d\mu \phi'_{\ell u}(\nu, \mu)
I(\nu, \mu) \notag \\
&=\frac{1}{2}\int d\nu' \int_{-1}^1 d\mu \phi_{\ell u}(\nu')
I(\nu'(1+v_z\mu/c), \mu),
\end{align}
where $\phi_{\ell u}(\nu)$ is the absorption profile (e.g., for a Gaussian profile, we would have $\phi_{\ell u}(\nu)=\pi^{-1/2}{\Delta\nu_D}^{-1}\exp(-(\nu-\nu_0)^2/{\Delta\nu_D}^2)$, with
$\nu_0$ the central line frequency and $\Delta\nu_D$ the Doppler width)  and
$\phi'_{\ell u}(\nu, \mu)=\phi_{\ell u}(\nu(1-v_z\mu/c))$, with $v_z>0$ for
upflowing material.
Analogously, the anisotropy in the observer's frame:
\begin{eqnarray}\label{eq03}
\bar{J}^2_0&=&\frac{1}{4\sqrt{2}}\int d\nu \int_{-1}^1 d\mu \phi_{\ell u}'(\mu,
\nu) \nonumber \\
&\times& [(3\mu^2-1)I(\nu, \mu)+3(1-\mu^2)Q(\nu, \mu)],
\end{eqnarray}
The important quantity that controls the ability of an anisotropic radiation
field to generate atomic polarization is the line anisotropy factor for each transition, which
can be calculated as:
\begin{equation}\label{eq:aniso}
w_{\mathrm{line}}={\sqrt{2}} \frac{\bar{J}^2_0}{\bar{J}^0_0}.
\end{equation}
Its range goes from $w_{\mathrm{line}}=-0.5$ (for an azimuthally
independent radiation field coming entirely from the horizontal plane) to $\sc w_{\mathrm{line}}=1$ (for a collimated vertical
beam). 
\\
\subsection{The radiative transfer equations (RTE)}\label{subsec:rte}

Due to symmetry, in a non-magnetized plane-parallel medium with a vertical
velocity field, light can only be linearly polarized parallel or perpendicularly
to the stellar limb. 
Therefore, chosing the reference direction for positive $Q$ parallel to the
limb, the only non-vanishing Stokes parameters are $I$ and $Q$, and they satisfy
the following radiative transfer equations:
\begin{subequations}\label{eq:rte}
\begin{align}
\frac{\rm d}{{\rm d}s}I&=\epsilon_I-\eta_II-\eta_QQ, \label{rte1}
\displaybreak[0] \\
\frac{\rm d}{{\rm d}s}Q&=\epsilon_Q-\eta_QI-\eta_IQ, \label{rte2}
\end{align}
\end{subequations}
where $s$ is the distance along the ray. The absorption and emission coefficients
are \citep{manso10}:
\begin{subequations}\label{eq:coefs1}
\begin{align}
\epsilon_I &= \,\epsilon_I^{\rm cont}+\epsilon_I^{\rm line}  \nonumber \\
&=\, {\eta_{I}}^{\rm
cont}B_{\nu}+ \epsilon_0 \left[\rho^0_0(u)+w_{J_uJ_{\ell}}^{(2)}
\frac{1}{2\sqrt{2}} (3\mu^2-1)\rho^2_0(u)\right], \label{coefs1a}
\displaybreak[0] \\
\eta_I &= \,\eta_I^{\rm cont}+\eta_I^{\rm line}  \nonumber \\
&=\,{\eta_{I}}^{\rm cont}+\eta_0
\left[\rho^0_0(\ell)+w_{J_{\ell}J_u}^{(2)} \frac{1}{2\sqrt{2}}
(3\mu^2-1)\rho^2_0(\ell)\right], \label{coefs1b}
\end{align}
\end{subequations}
\begin{subequations}\label{eq:coefs2}
\begin{align}
\epsilon_Q\,=\,\epsilon^{\rm line}_Q\,=\,\epsilon_0 w_{J_uJ_{\ell}}^{(2)}
\frac{3}{2\sqrt{2}}(1-\mu^2)\rho^2_0(u), \label{coefs3}
\displaybreak[0] \\
\eta_Q\,=\,\eta^{\rm line}_Q\,=\,\eta_0 w_{J_{\ell}J_u}^{(2)}
\frac{3}{2\sqrt{2}}(1-\mu^2)\rho^2_0(\ell), \label{coefs4}
\end{align}
\end{subequations}
where ${\eta_{I}}^{\rm cont}$ and ${\epsilon_{I}}^{\rm cont}$ are the continuum
absorption and emission coefficients for intensity, respectively. Likewise, $\eta_I^\mathrm{line}$ and $\eta_Q^\mathrm{line}$
are the line absorption coefficients for Stokes $I$ and $Q$, respectively, while $\epsilon_I^\mathrm{line}$ and 
$\epsilon_Q^\mathrm{line}$ are the line emission coefficients for Stokes $I$ and $Q$, respectively. The coefficients $w_{J_{\ell}J_u}^{(2)}$ and
$w_{J_{u}J_{\ell}}^{(2)}$ depend only on the transition and are detailed in
Table (\ref{tab:atomic}).The subscripts $u$ and $\ell$ refer to the upper and
lower level of the transition considered, respectively, and $B_{\nu}$ is the
Planck function at the central frequency  $\nu_0$ of the transition. Note also
that:
\begin{subequations}\label{eq:epseta}
\begin{align}
\epsilon_0=\frac{h\nu}{4\pi}A_{u{\ell}}{\phi^{\prime}_{\ell u}(\mu,\nu)}{\cal
N}\sqrt{2J_u+1}, \label{epseta1}
\displaybreak[0] \\
\eta_0=\frac{h\nu}{4\pi}B_{{\ell}u}{\phi^{\prime}_{\ell u}(\mu,\nu)}{\cal
N}\sqrt{2J_{\ell}+1}, \label{epseta2}
\end{align}
\end{subequations}
 where $\cal N$ is the total number of atoms per unit volume. 
 
With the total absorption coefficient for the intensity, the line of sight (los)
optical depth for each frequency is calculated by the following integral along
the ray:
\begin{equation}
\tau^{\rm los}_{\nu}=- \int \eta_I(\mu^{\rm los},\nu) \frac{dz}{\mu^{\rm los}}
\end{equation}

\subsection{Numerical method}\label{sec:numeric}
The solution to the non-LTE problem of the second kind considered here (the self-consistent solution of
the statistical equilibrium equations for the density matrix elements together with the
radiative transfer equations for the Stokes parameters) is carried out by 
generalizing the computer program developed by \cite{manso_trujillo03a}, 
to allow for radial macroscopic velocity fields (see Sec. \ref{sec:dynamic_traviata}). For integrating the RTE, a
parabolic short-characteristics scheme \citep{kunasz-auer} is used. At each
iterative step, the radiative transfer equation is solved, and $\bar{J}^0_0$ and
$\bar{J}^2_0$ are computed and used to solve the SEE following the accelerated
Lambda iteration method outlined in Sec. \ref{sec:trav1}. Once the
solution for the multipolar components of the density matrix is consistently
reached, the emergent Stokes parameters are calculated for the desired line of
sight, which in all the figures of this chapter is $\mu=0.1$.   
This final step is done increasing the frequency grid resolution to a large value
in order to correctly sample the small features and peaks of the emergent profiles.

Some technical considerations have to be kept in mind for the treatment of
velocity fields (see Sec. \ref{sec:dynamic_traviata}). Due to the presence of Doppler shifts, the wavelength axis used
to compute $\bar{J}^0_0$ and $\bar{J}^2_0$ must include the required extension
and resolution, because the spectral line radiation may be now shifted and
asymmetric. In our strategy for the wavelength grid, the resolution is larger in
the core than in the wings, keeping the same wavelength grid for all heights.

The cutoff wavelength for the core (where resolution is appreciably higher) is
dictated by the maximum expected Doppler shift. Thus, the core bandwidth is
estimated allowing for a range of $2V_{\rm{max}}$ around the zero
velocity central wavelength of the lines, with
$V_{\rm{max}}$ the maximum velocity found in the atmosphere (in Doppler units).
Apart from that, a minimum typical resolution for the core is set to 2 points
per Doppler width ($\Delta\nu_{D}$). Then, the height with the smallest Doppler
width determines the core resolution, and the height with the maximum
macro-velocity states its bandwidth. 

Furthermore, as frequencies and angles are inextricably entangled (through terms
$\nu-v_z\mu\nu/c$ appearing in the absorption/emission profiles due to the
Doppler effect, like in Eq. (\ref{eq02})), the maximum angular increment ($\Delta\mu_{\rm{max}}$) is
restricted by the maximum frequency increment ($\Delta x_{\rm{max}}\thickapprox
1/2$, in Doppler units). Thus, it must occur that $\Delta\mu_{\rm{max}} \cdot
V_{\rm{max}}\leq1/2$. In the worst case, the maximum allowed angular increment
will be smaller (more angular resolution needed) when the maximum vertical
velocity increases. Besides this consideration, the maximum angular increment
could be even more demanding because of the high sensitivity of the polarization
profiles to the angular discretization.

Finally, the depth grid must be fine enough, in such a way that the maximum
difference in velocity between consecutive points is not too large, the
typical difference being equal to half the Doppler width ($|V(z_i)-V(z_{i-1})|\leq1/2$). If the
difference is larger, the absorption/emission profiles would change abruptly with height, producing
imprecisions in the optical depth increments \citep{Mihalas:1978}. 
\section{Effect of a velocity gradient on the radiation field.}\label{sec:preview} 
As we shall see, the presence of a vertical velocity gradient in an atmosphere
enhances the anisotropy of the radiation field and, hence, of the scattering
line polarization patterns. The fundamental process underlying this mechanism can be simply understood with
the following basic examples.
\subsection{Anisotropy seen by a moving scatterer.}\label{sub:scattercloud}
Consider an absorption spectral line with a Gaussian profile emerging from a
\textit{static} atmosphere with a linear limb darkening law,
\begin{equation}
I(\nu, \mu)=I^{(0)}(1-u+u\mu)[1-a\exp(-(\frac{\nu-\nu_0}{w})^2)], \label{eq04}
\end{equation}
where $I^{(0)}$ is the continuum intensity at disk center, $u$ is the limb
darkening coefficient, $a<1$ measures the intensity depression of the line, and
$w$ its width. In this approximation we assume that all the parameters are constant. 
Now imagine that, at the top of the atmosphere, there is a thin cloud scattering
the incident light given by Eq. (\ref{eq04}) and moving radially at velocity
$v_z$ with respect to the bottom layers of the atmosphere, supposed static. 
We will assume that the absorption profile is Gaussian (dominated by Doppler
broadening), with width $\Delta\nu_D$. When $\Delta\nu_D\ll w$, the incident spectral
line radiation is much broader than the absorption profile (in fact, for
$\alpha=\Delta\nu_D/w=0$ the absorption profile is formally a Dirac-$\delta$
function).
Then, from Equations~(\ref{eq02})-(\ref{eq03}) we can derive explicit
expressions for the mean intensity and anisotropy of the radiation field as a
function of the scatterer velocity (see Appendix \ref{app:A}): $\bar{J}^0_0={\cal
I}_0(\alpha;\xi)/2$, $\bar{J}^2_0={\cal I}_2(\alpha;\xi)/4\sqrt{2}$, where the
${\cal I}_{0, 2}$ functions are defined by
Equations~(\ref{eqa04})-(\ref{eqa05}).
The behavior of $\bar{J}^0_0$ and $\bar{J}^2_0$ with the adimensional velocity
$\xi=v_z\nu_0/(cw)$ (Figure \ref{fig:basic}), is most clearly illustrated in
their asymptotic limits at low velocities:
\begin{align}
\bar{J}^0_0&=\frac{1}{4}(1-\frac{a}{\sqrt{1+\alpha^2}})(2-u) \nonumber \\
+&\frac{a(4-u)}{
12(1+\alpha^2)^{3/2}}\xi^2 + {\rm O}(\xi^3), \label{eq05} \\
\sqrt{2}\frac{\bar{J}^2_0}{\bar{J}^0_0}&=\frac{u}{4(2-u)} \nonumber \\
+&\frac{a(64-56u+7u^2)}{
120(2-u)^2(1+\alpha^2)(\sqrt{1+\alpha^2}-a)}\xi^2  + {\rm O}(\xi^3).
 \label{eq06}
\end{align}
Equation (\ref{eq05}) shows that, for an absorption line ($a>0$), $\bar{J}^0_0$
is always increasing with the velocity since the coefficient of $\xi^2$ is
positive, regardless of the sign of $v_z$ (i.e., regardless of whether the scatterers move upwards
or downwards); if the line is in emission ($a<0$), $\bar{J}^0_0$ monotonically
decreases.
These are the {\em Doppler brightening} and {\em Doppler dimming} effects
\citep[e.g.,][]{LL04}.
An analogous analysis applies to $\bar{J}^2_0$ (Eq.~\ref{eq06}).
Note that, in the absence of limb darkening ($u=0$), the anisotropy vanishes in
a static atmosphere, while the mere presence of a relative velocity between the
scatterers and the underlying static atmosphere induces anisotropy in the radiation
field ---hence, a polarization signal.  
A real atmosphere could then be understood as a superposition of scatterers 
that modify the anisotropy depending on the local
velocity gradient and the illumination received from lower shells.
An interesting discussion on the effect of velocities with directions other than
radial can be found in Section 12.4 of \cite{LL04} .

\begin{figure}
\centering
\includegraphics[scale=0.7]{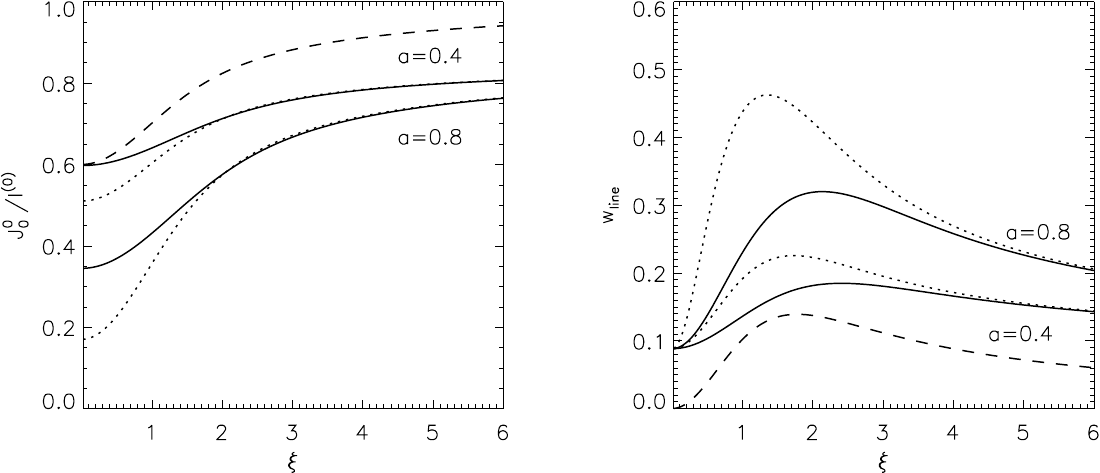}
\caption{$\bar{J}^0_0/I^{(0)}$ (left panel) and $\sqrt{2}\bar{J}^2_0/\bar{J}^0_0$ (right panel) 
as a function of the adimensional velocity $\xi$ calculated using an incident line profile 
as in Eq.~(\ref{eq04}) with $u=0.3$ and $a=0.4$ or 0.8 (see labels). 
Dotted lines have been computed for the case of an infinitely sharp absorption profile ($\alpha=0$).
Solid lines refer to the case $\alpha=0.9$ (non saturated line).
The case with no limb darkening ($u=0$) and $a=0.4$ has been plotted for comparison (dashed lines).\label{fig:basic}}
\end{figure}

\begin{figure}[!b]
\centering
\includegraphics[scale=0.6]{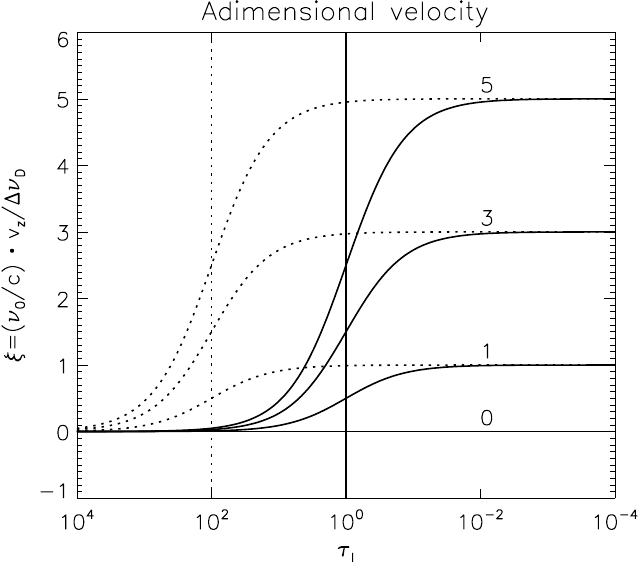}
\caption{Adimensional velocity fields considered in this section. They are
parameterized by the limiting value at small optical depths (labels) and by the location
in optical depth of the largest velocity gradient region (vertical lines marking $\tau_0$). Solid lines: $\tau_0=1$. Dotted lines: $\tau_0=100$. \label{fig:velocity}}
\end{figure}

\begin{figure}[!t]
\centering
\includegraphics[scale=0.8]{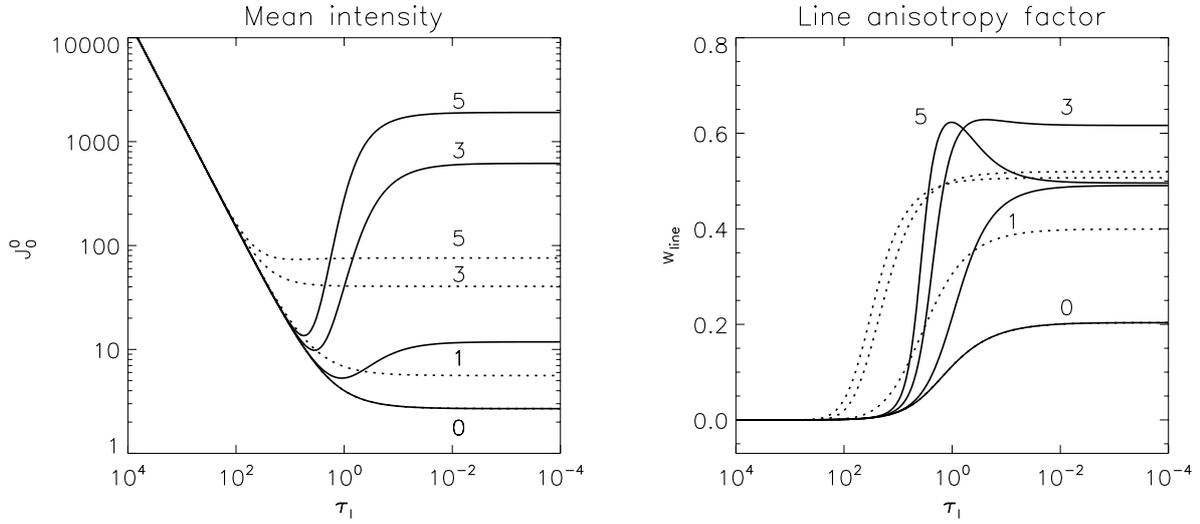}
\caption{$\bar{J}^0_0$ (left panel) and $\sqrt{2}\bar{J}^2_0/\bar{J}^0_0$ (right panel) as a 
function of the integrated static line opacity ($\tau_l$) in an expanding atmosphere with 
$S=S^{(0)}(1+\beta\tau_l)$ and different velocity stratifications $\xi=\xi_0/(1+\tau_l/\tau_0)$ (see Fig. \ref{fig:velocity}).
The parameters used in this plot are $\beta=3/2$ and $\kappa_c/\kappa_l=10^{-4}$. The labels indicate the value of $\xi_0$.
Solid lines correspond to $\tau_0=1$ while dotted lines refer to $\tau_0=100$.
\label{fig:me_fig}}
\end{figure}

Clearly, all the above discussion depends on the Doppler shift induced by the
velocity $v_z$ normalized to the width of the spectral line, i.e., on $\xi$.
A large velocity gradient on a broad line can have the same effect as that of a
smaller velocity gradient on a narrow line. 
This is important to be kept in mind since the response of different spectral
lines to the same velocity gradient will be different, what can help us to
decipher the velocity stratification. 
Even different spectral lines belonging to the same atomic species may have very
different widths, as for example, the Ca~{\sc ii} IR triplet and the UV doublet studied in the next
section.

\subsection{Calculations in a Milne-Eddington model.}\label{sec:me}
The discussion above explains the basic mechanism by means of which a velocity gradient
enhances the anisotropy of the radiation field.
Now we can get further insight on the structure of the radiation field within an
atmosphere with velocity gradients from just the formal solution of the RT
equation for the intensity \citep[e.g.,][]{Mihalas:1978}.
As before, we neglect effects due to polarization and $J^0_0$ and $J^2_0$ are
calculated from Stokes $I$ alone.
We consider a semi-infinite, plane-parallel atmosphere with a source function
$S=S_0(1+\beta\tau_l)$, where $\tau_l$ is the integrated line optical depth in
the static limit (hence, the element of optical depth
$d\tau_\nu=(r+\phi[\nu(1-v_z(\tau_l)\mu/c)])d\tau_l$, where
$r=\kappa_c/\kappa_l$ is the ratio of continuum to line opacity). 
We begin by considering a vertical velocity field
$v_z(\tau_l)=v_0/[1+(\tau_l/\tau_0)]$ (positive away from the star), shown
in Fig. \ref{fig:velocity}.
Equivalently, we may express the velocity in adimensional terms by using 
$\xi=(\nu_0/c)v_z/\Delta\nu_D$ (the width $\Delta\nu_D$ of the Gaussian
absorption profile is assumed to be constant with depth). The parameter $\tau_0$
fixes the position of the maximum velocity gradient. 
Note that the wavelength dependence of the Doppler effect ($\Delta\lambda_z =
\lambda_0 v_z / c$) is cancelled in the adimensional problem, where velocities
are measured in Doppler units.
It is easy to calculate numerically $I_z(\nu, \mu)$ at every point in the atmosphere and
thus, the mean intensity and anisotropy of the radiation field (Figure
\ref{fig:me_fig}). 

The rise in $\bar{J}^0_0$ in higher layers with respect to the static case
corresponds to the Doppler brightening discussed above. Thanks to the Doppler
shifts, the atoms \textit{see} more and more of the brighter continuum below, which
enhances $\bar{J}^0_0$. When the maximum velocity gradient takes place at 
optically thick enough layers ($\tau_0 \gtrsim 1$), $\bar{J^0_0}$ is also larger than 
for the static case, but it decreases monotonically with height in the atmosphere ($\tau_0 = 10^2$,
dotted lines in left panel of Fig. \ref{fig:me_fig}). Note that the important quantity that
modulates the increase in $\bar{J^0_0}$ is not the maximum velocity
but the velocity gradient (difference in velocity between optically thick and optically thin parts of the atmosphere). The larger the gradient, the
more pronounced the radiative decoupling is between different heights. An extreme
example of such radiative decoupling could be found in supernovae explosions, where the
vertical velocity gradients are huge.

In our case (vertical motions), the Doppler brightening implies an
enhancement of the contribution of vertical radiation to Eq. (\ref{eq03}) with
respect to the horizontal radiation, with the latter remaining almost equal to 
the static case (no horizontal motions, no horizontal Doppler brightening). This
velocity-induced limb darkening is the origin of the anisotropy enhancement. 
 
However, note that the maximum anisotropy does not rise indefinitely when
increasing the maximum velocity. If the velocity gradient in units of the 
Doppler width is larger than $\sim 3$ (see curves for $\xi_{\rm{max}}=5$ in Fig. \ref{fig:me_fig}), the
anisotropy at the surface saturates and decreases (even below the curves
corresponding to shorter velocity gradients). It forms a bump around
$\tau_l=1$ when the maximum velocity gradient is taking place at low density
layers ($\tau_0 \lesssim 1$). This behavior can be understood using Eq. (\ref{eq03}). 
When $\xi_{\rm{max}} \lesssim 3$, an increment in
$\xi_{\rm{max}}$ entails a rise in $\bar{J}^0_0$, $\bar{J}^2_0$ and
$\bar{J}^2_0/\bar{J}^0_0$ ($w_{\rm{line}}$) in the upper atmosphere, what means
that the velocity gradients enhance the imbalance between vertical and
horizontal radiation. However, if $\xi_{\rm{max}}$ is above that threshold,
$\bar{J}^0_0$ and $\bar{J}^2_0$ rise, but the ratio $\bar{J}^2_0/\bar{J}^0_0$
saturates and diminishes. The reason is that a large velocity gradient makes the
absorption profiles associated with almost horizontal outgoing rays
($0<\mu<1/\sqrt{3}$) to be so much shifted that they also capture the background
continuum radiation. Their contributions are negative to the angular integral of
$\bar{J}^2_0$ but positive for $\bar{J}^0_0$. 

Separating the contributions of rays with
angles in the range $1/\sqrt{3}<|\mu|<1$ (that we refer to with the label $+$) and angles
in the range $0<|\mu|<1/\sqrt{3}$ (that we refer to with the label $-$), the
line anisotropy can be written as $w_{\rm{line}}=w^+_{\rm{line}}+w^-_{\rm{line}}=\bar{J}^{2+}_0/\bar{J}^0_0-|\bar{
J}^{2-}_0|/\bar{J}^0_0$. Here,
$\bar{J}^0_0$ and $\bar{J}^{2+}_0$ grow always with $\xi_{\rm{max}}$, but
$|\bar{J}^{2-}_0|$ only grows appreciably when $\xi_{\rm{max}} \gtrsim 3$. Therefore, although 
$w_{\rm{line}}$ increases for all velocity gradients, its enhancement is
smaller for large velocity gradients than for smaller ones. This
effect occurs as well when motions take place deeper ($\tau_0\gtrsim 1$)
but it is less important and the anisotropy bump and saturation are reduced.

For the considered velocity fields (with a negligible gradient in the
upper atmosphere), $\bar{J}^0_0$ and $\bar{J}^2_0$ reach an asymptotic value
in optically thin regions. It does not occur if the velocity gradient is not zero at those layers.
In any case, the presence of a large anisotropy in optically thin regions barely
affects the emergent linear polarization profiles.


\subsection{Two-level model atom in dynamic atmospheres.}\label{sub:tla}
Before going to a more realistic case, a final illustrative example is
considered. In this case, we assume the same parameterization of the velocities
than in the previous example, but now we solve the complete iterative RT problem
with a two-level atom model and a specific temperature stratification. Consequently, 
the source function and the anisotropy are consistently obtained
in a moving atmosphere. The intensity source function is
$S_I=r_{\nu\mu}S^{\rm{line}}_I+(1-r_{\nu\mu})B$ \citep[e.g.,][]{rybicki-hummer}, with $r_{\nu\mu}=\phi'_{\ell
u}(\nu,\mu)/(r_c+\phi'_{\ell u}(\nu,\mu))$ and the expression for the
line source function remains formally equal to that of the static case, being
$S^{\rm{line}}_I=(1-\epsilon)\bar{J}^0_0+\epsilon B$, where $B$ is the imposed
Planck function, $\epsilon$ is the inelastic collisional parameter and
$\bar{J}^0_0$ is calculated with Eq. (\ref{eq02}).
\begin{figure}[!h]
\centering
\includegraphics[scale=0.7]{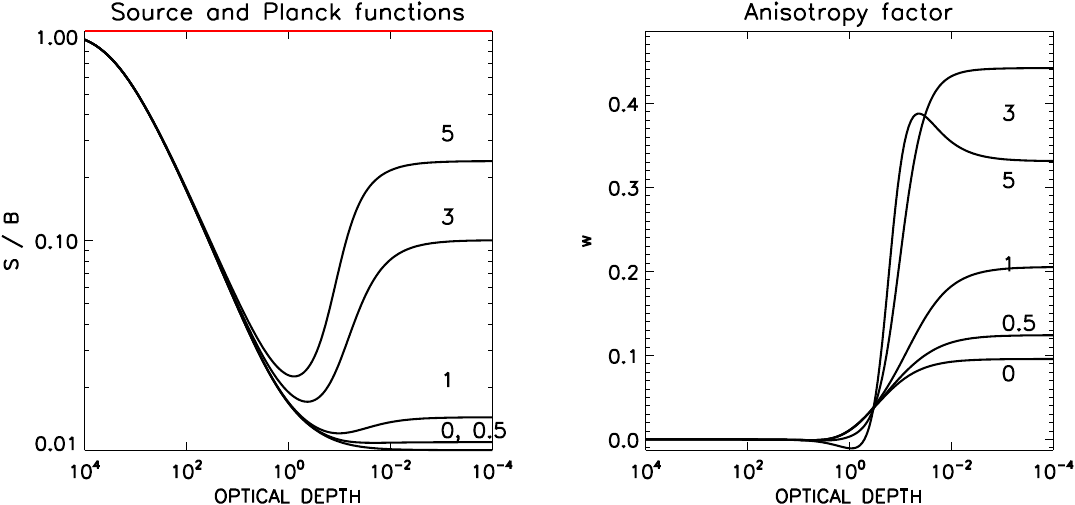}
\caption{Same as Fig. \ref{fig:me_fig} for an isothermal moving 
atmosphere with a two-level atom model using the velocity fields of Fig. \ref{fig:velocity}. We assume a very strong line ($r_c=0$)  
and $\epsilon=10^{-4}$. Left panel: the Doppler brightening effect increases
the surface source function value. In a two-level model atom in a static atmosphere (label 0) 
this value follows the well known expression $S(0)=\sqrt{\epsilon}B$. The vertical 
axis is in units of the Planck function. Right panel: we show the amplification 
of the radiation field anisotropy when the velocity gradient increases. The curve
labeled with ``5'' shows the saturation of the anisotropy and the bump
produced by the strong velocity gradient taking place in optically thin regions.\label{fig:tla1}}
\end{figure}
\begin{figure}[h!]
\centering
\includegraphics[scale=0.6]{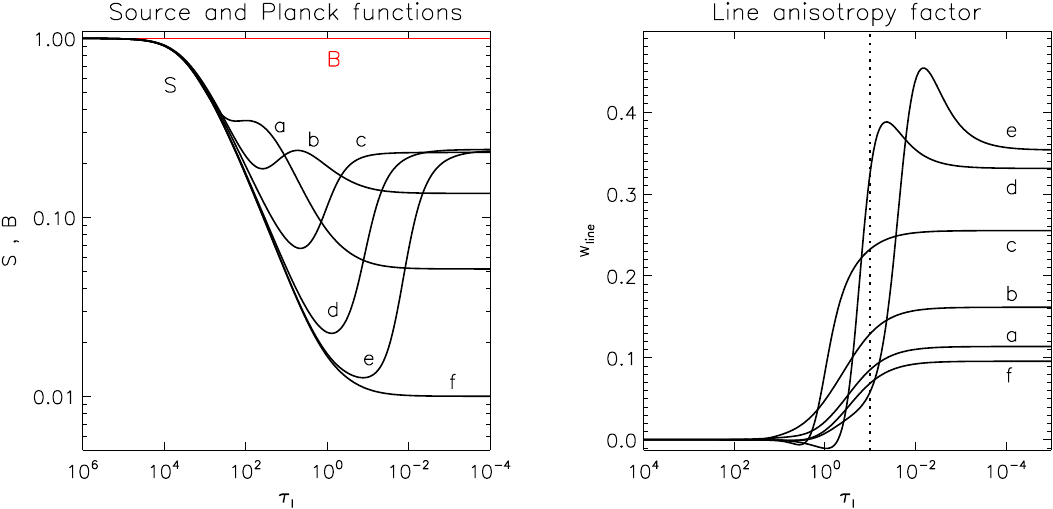}
\includegraphics[scale=0.6]{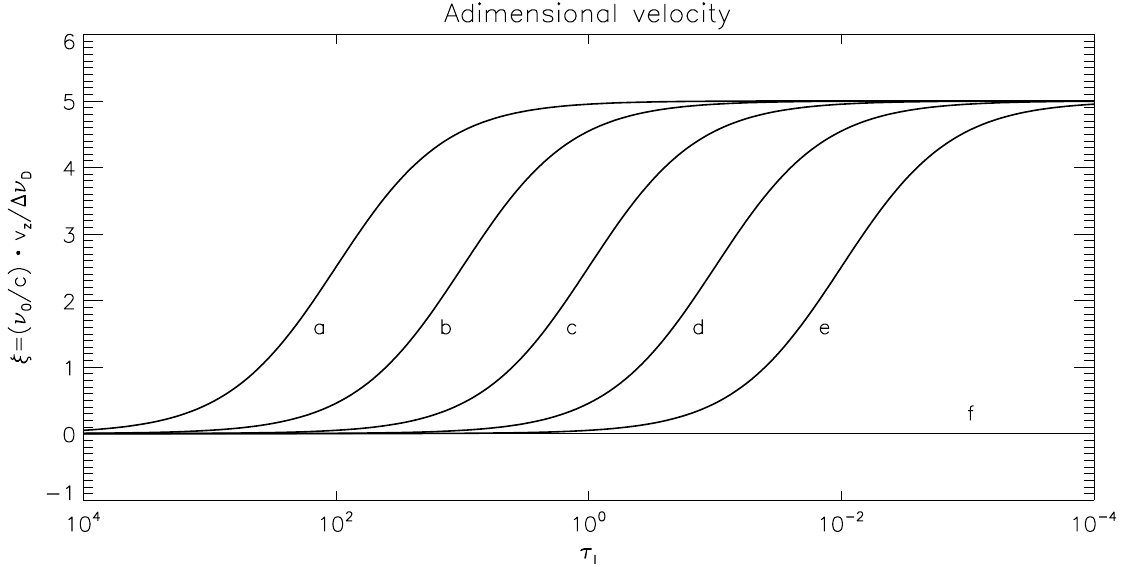}
\caption{Calculations in isothermal two-level atom moving atmospheres 
with $\xi_{max}=5$ . We assume a very strong line ($r_c=0$)  and $\epsilon=10^{-4}$. 
The highest velocity gradient occur at $\tau_l=\tau_0=100,10,1,0.1,0.01$ for a, b, c, d and e, respectively.
The case f corresponds to the solution in a static atmosphere. The vertical dotted line marks the position of  $\tau_l=0.1$.\label{fig:tla2}
}
\end{figure}
The qualitative behavior explained in the previous subsection is maintained in these
two-level atom calculations. For small $\epsilon$ values (large NLTE effects), 
the source function $S_I \thickapprox
\bar{J}^0_0$ shows Doppler brightening effects and its 
surface value depends on the maximum velocity gradient and on the
maximum background continuum set by the photospheric conditions (upper panel in Fig.
\ref{fig:tla1}). The anisotropy rises proportionally to the velocity gradient until
a saturation occurs (lower panel in Fig. \ref{fig:tla1}). A similar behavior is 
found when the maximum velocity gradient occurs higher in the atmosphere (see 
Fig. \ref{fig:tla2}).

In a static atmosphere, the radiation field anisotropy is dominated by the presence
of gradients in the intensity source function \citep{jtb-sacpeak2001,LL04}, which can be modified via 
the Planck function (equivalently, the temperature). In the dynamical case that
we are dealing with, the slope of the source function is also modified due to the
existence of velocity gradients thanks to the frequency-decoupling caused
by relative motions between absorption profiles (Doppler brightening). In general,
both mechanisms act together (velocity-induced and temperature-induced modification
of the source function gradient) and the ensuing anisotropy and the emergent linear
polarization profiles are modified accordingly.

It is important to note that the adimensional velocity $\xi$
depends both on the velocity and also on the line Doppler width because
$\xi (\tau)= \delta\nu/\Delta\nu_D= v_z/\sqrt{2k_BT/m}$, with $k_B$ the Boltzmann
constant, $T$ the temperature and $m$ the mass of the atom. In the photosphere,
where velocities are much lower than in the chomosphere, $\xi$ is expected to be
negligible. In the chromosphere, plasma motions are important and the temperature is
still comparable to that of the photosphere, inducing $\xi$ to be controlled by
the velocity field. However, for layers in the transition region and above, the high
temperatures reduce the value of $\xi$. In any case, at these heights, the density is so low that, although $\xi$ (and
consequently the anisotropy) could have a highly variable behavior, the
emergent polarization profiles of chromospheric lines will not be sensitive to
them.   

\section{Results for the Ca {\sc ii} IR Triplet}\label{sec:anisotriplet}
Now, we study the effect of the velocity field on a multilevel atomic system 
in a semiempirical atmospheric model, within the framework described in Sec. \ref{sec:problem}. 
We consider the formation of the scattering polarization pattern of the Ca {\sc
ii} infrared triplet in the FAL-C model of \cite{Fontenla:1993} in the
presence of vertical velocity fields ($\vec{v}=v_z(z)\vec{k}$, with $\vec{k}$ the unit vector along the vertical
pointing upwards). We will assume a constant
microturbulent velocity field of 3.5~km~s$^{-1}$, which is a representative value for the
region of formation that gives a correct broadening of the triplet profiles. 
\\

\subsection{Behavior of the anisotropy in the Ca {\sc ii} IR triplet}\label{subsec:aniso}
For simplicity, we set linear velocity fields (constant velocity gradient
along z) between $z=-100$ and 2150 ~km (see Fig. \ref{anisovars}). Consequently, the adimensional velocity field $\xi_z$ has a non-monotonic behavior 
due to its dependence on the temperature (upper right panel in Fig. \ref{anisovars}). In
the chromosphere, where the Ca \textsc{ii} triplet lines form, $\xi_z$ is dominated by the
macroscopic motions. Here, the velocity gradients produce variations in the anisotropy of the triplet
lines that agree with the behavior outlined in the previous sections. Namely, an
amplification and a subsequent saturation of the anisotropy factor due to the
significant velocity gradient at those heights (see the lower panels and middle
right panel in Fig. \ref{anisovars}). Above the chromosphere, on the contrary,
the temperature dominates ($\xi_z$ stabilizes and diminishes) and the anisotropy slightly
decreases with height.
If the (adimensional) velocity gradient is negligible
where the line forms (around $\tau^\mathrm{los}_{\nu_0}\thicksim 1$), the anisotropy
remains unaffected. Otherwise, if a spectral line forms at very hot layers,
where the absorption profiles are wider and their sensitivity to the velocity
gradients is lower, the Doppler brightening will not be so efficient
amplifying the anisotropy. This is the case of the anisotropy of the
Ca \textsc{ii} K line (middle left panel in Fig. \ref{anisovars}).  Compare how the slope of $\xi_z$ is smaller where the Ca \textsc{ii} K line forms (black line on Fig. \ref{anisovars}) than where the triplet lines do. Consequently, the enhancement 
of the line anisotropy through the presence of velocity gradients in this line is reduced. 

All our calculations demonstrate that the anisotropy in the Ca {\sc ii}
IR triplet can be amplified through chromospheric vertical velocity gradients.
This results suggest that the same occurs with the ensuing linear polarization
profiles.

\begin{figure*}[b!]
\centering
\includegraphics[scale=0.67]{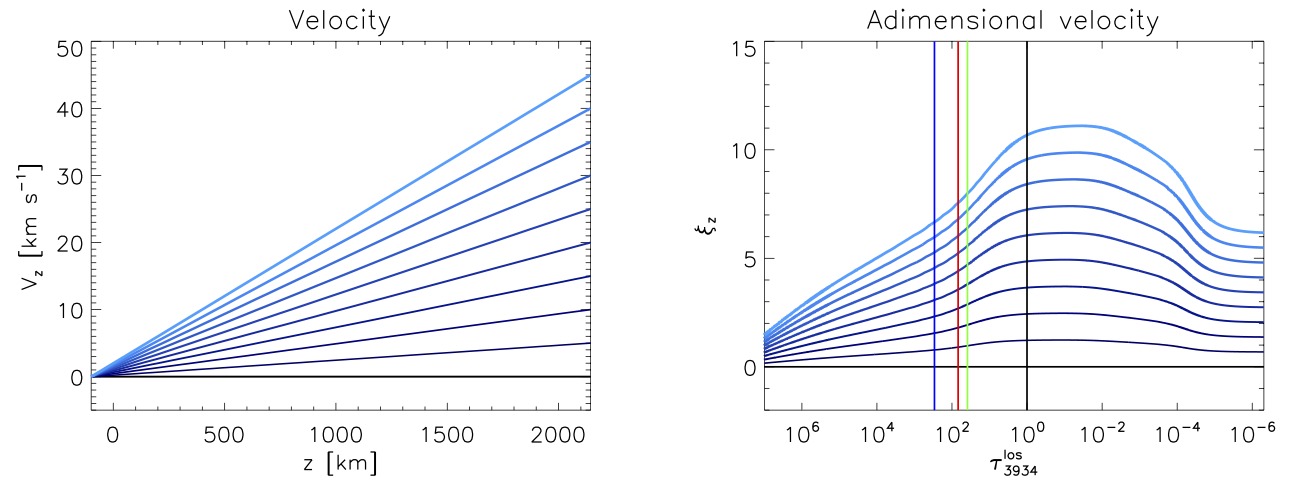}
\includegraphics[scale=0.195]{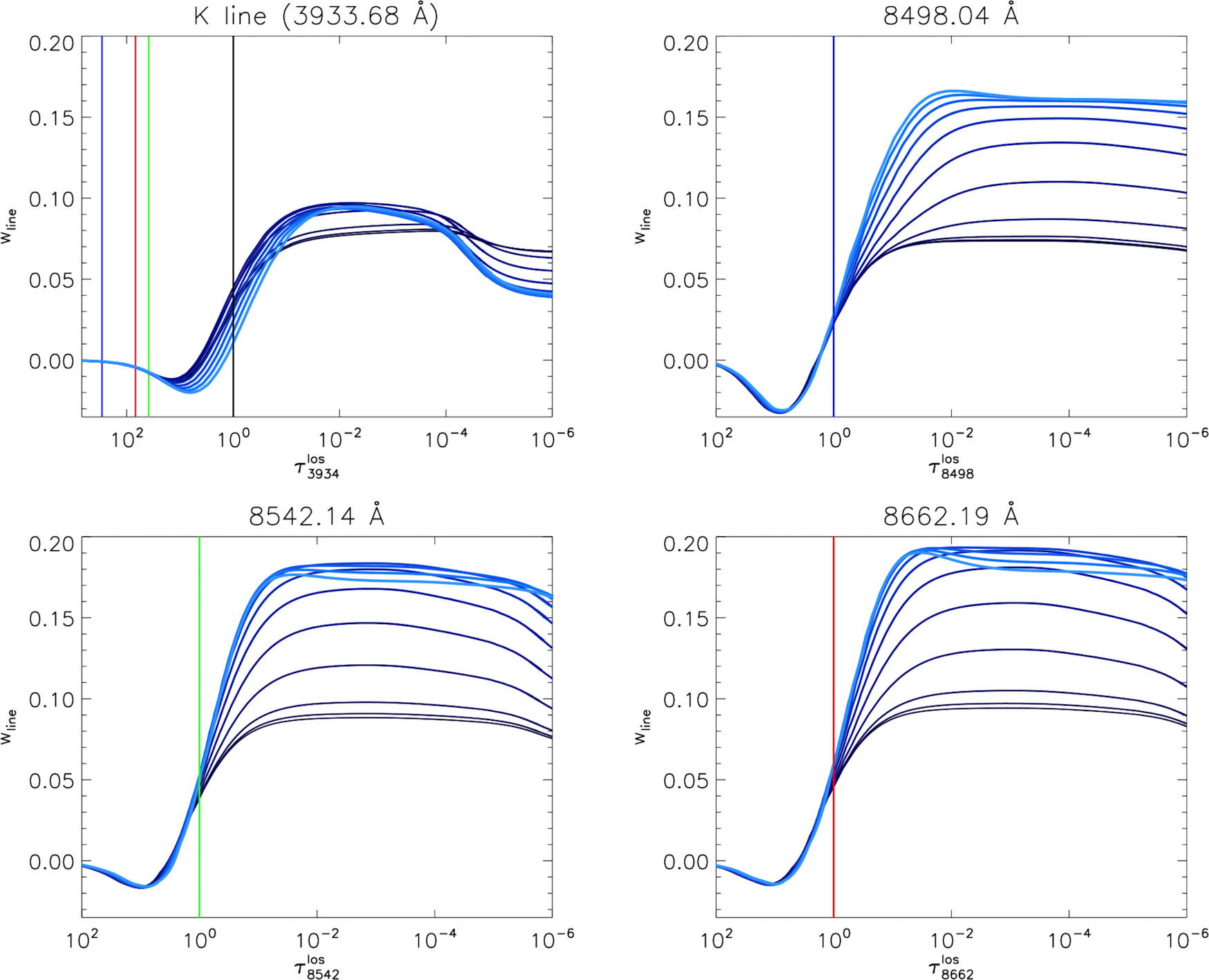}
\caption{Amplification of the line anisotropy ($\rm{w_{line}}=\sqrt{2} \bar{J}^2_0/\bar{J}^2_0$) due to vertical velocity gradients. 
Upper left panel: linear velocity fields versus height, with velocity 
gradients going from $0$ (darker lines) to $20\,\rm{m \cdot s^{-1} \, km^{-1}}$ 
(light blue lines) in steps of $2.23\,\rm{m \cdot s^{-1} \, km^{-1}}$. Upper right 
panel: corresponding adimensional velocity fields ($\xi_z$) for a FALC 
temperature stratification and a constant microturbulent velocity of 
$3.5 \,\rm{km\, s^{-1}}$. The horizontal axis is in units of the K-line 
optical depth along the line of sight (los). The vertical lines mark the position of 
$\tau^{\rm{los}}_{\nu_0}= 1$ for the transitions $8498$ {\AA} (blue),
$8542$ {\AA} (green), $8662$ {\AA} (red) and the K line (black). 
Remaining panels: corresponding line anisotropy factors plotted 
against $\tau^{\rm{los}}_{\nu_0}$ for each line. \label{anisovars}}
\end{figure*}
\clearpage

\subsection{The impact of the emergent radiation on the polarization.}\label{sec:impact}
For investigating the effect of vertical velocity fields on the emergent fractional
polarization profiles we perform the following numerical experiments. First, we
impose velocity gradients with the same absolute value but opposite signs (top
left panel of Fig. \ref{fig:gradsv_ctes}). The resulting emergent $Q/I$
profiles (remaining left panels of Fig. \ref{fig:gradsv_ctes}) are magnified by a
significant factor ($>2$ for all the transitions) with respect to the static
case (black dotted line). The linear polarization profiles have the same amplitude, independently of the
sign of the velocity gradient. Another remarkable
feature is the asymmetry of the profiles, having a higher blue wing in those
cases in which the velocity gradient is positive and a higher red wing when the
velocity gradient is negative, independently of the velocity sign. Note also
that the $Q/I$ profile is shifted in frequency due to the relative velocity between the
plasma and the observer. 

As a second experiment, we consider different velocity fields with increasing
gradients (right upper panel in Fig. \ref{fig:gradsv_ctes}). In the ensuing
$Q/I$ profiles we see that the larger the velocity gradient, the larger 
the frequency shift of the emergent profiles and the larger the amplitude. In
all transitions, one of the lateral lobes of the signal remains almost constant.
Thus, what really changes is the central part of the profiles, being a ``valley''
in the $\lambda8498$ line and a ``peak'' in the other two transitions. To quantify
these variations, we define $(Q/I)_\mathrm{pp}$ (peak-to-peak amplitude of $Q/I$) as the difference between the lowest and the highest value
of the emergent $Q/I$ signal, which is also a measure of its contrast. Note that, as
expected from the first experiment, $(Q/I)_\mathrm{pp}$ depends only on the
absolute value of the gradient. Figure \ref{fig:gradvarlaw} summarizes these
results.

The sensitivity of the linear polarization to the velocity gradient can be measured approximately as
commented in Sec. \ref{sub:scattercloud}, using a parameter
$\alpha=\Delta\nu_D/w$ that accounts for the difference in width of the absorption profile
with respect to the emergent intensity profile. If $\alpha \thicksim
1$ in the main formation region, small
adimensional velocities will produce large changes in shape; if
$\alpha \ll 1$, much larger $\xi_z$ values are needed for the same effect. In
the case of the IR triplet lines, $\alpha \thicksim 0.355$ in the formation region
of $\lambda8498$ and around $0.29$ and $0.285$ in the formation region of
$\lambda8542$ and $\lambda8662$ (having wider profiles), respectively. Then, the
former is more sensitive to velocity variations in its formation region (Fig.
\ref{fig:gradvarlaw}). Finally, the K line has $\alpha\,(\tau^{\rm{los}}_{\nu_0}=1)
\thicksim 0.015$, a low value due to its wider spectral wings.  

The enhancement of the polarization signals are a consequence of the increase in the
anisotropy. Therefore, since this increase is produced by the presence of velocity fields, the
polarization signals of the Ca {\sc ii} IR triplet are sensitive also to the dynamic state of
the chromosphere.

\begin{figure*}[!h]
\centering
\includegraphics[scale=0.65]{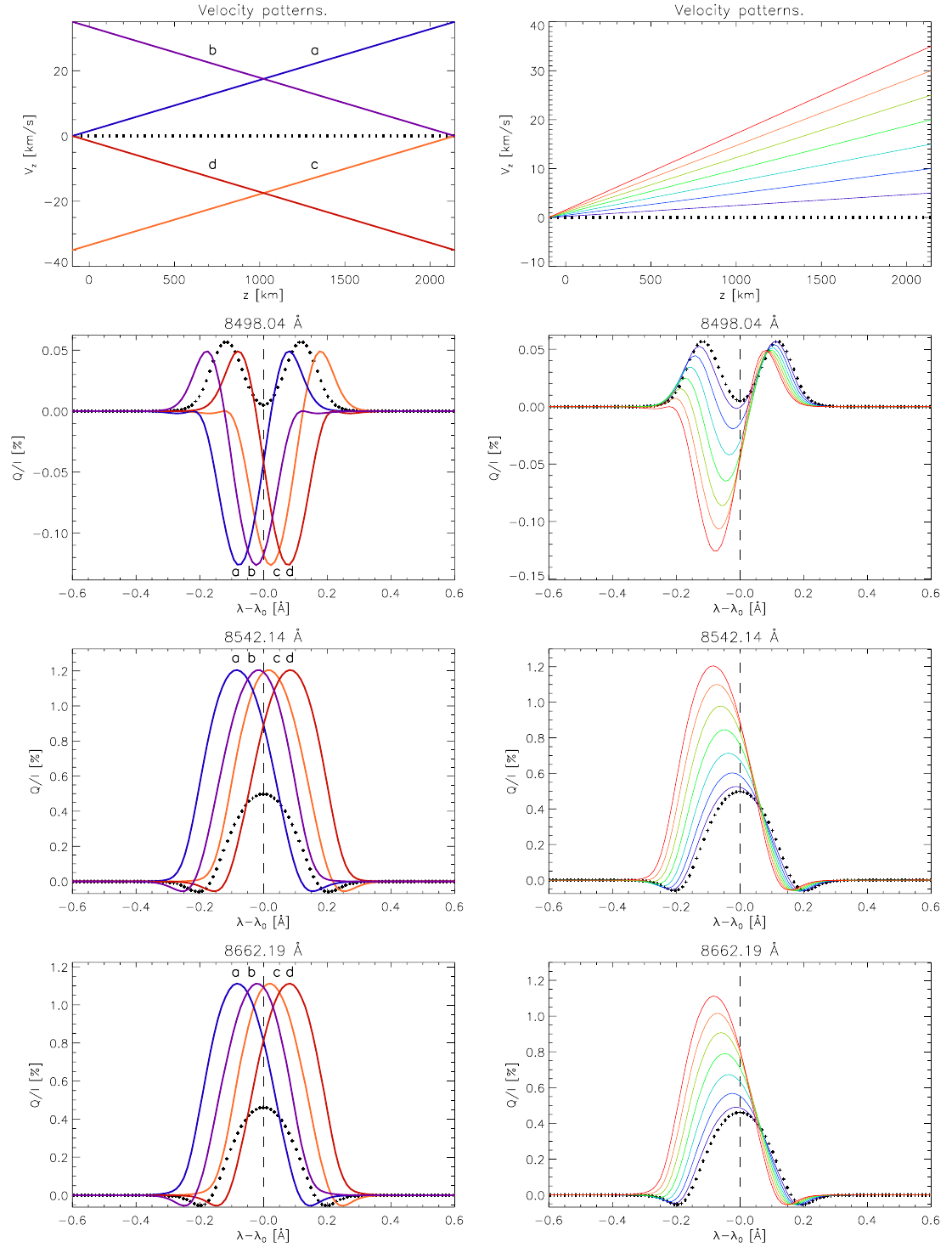}
\caption{Left panels: calculation at $\mu=0.1 $ of the 
emergent $Q/I$ polarization signals of the Ca \textsc{ii} IR triplet when four 
different choices for the vertical velocity gradients are imposed. Positive 
velocities imply upflowing plasma. Gradient ``a''/``b'' simulates an atmosphere 
where the plasma is entirely moving towards the observer increasing/decreasing 
linearly the velocity along the outgoing z axis. Gradients ``c'' and ``d'' are 
the same for plasma moving away from the observer. The black dotted line is the 
solution for the static reference case. Each curve is computed on the converged 
solution of the multilevel NLTE problem described in Sec. \ref{sec:problem}. 
Right panels: same calculations than in the left-hand panels, 
but with different velocity gradient values varying 
from $0$ to $16.3 \,\rm{m \, s^{-1}\,km^{-1}}$ in steps of $2.3 \,\rm{m \, s^{-1}\,km^{-1}}$ 
(see top right panel). These results show that the 
polarization signals are increased and shifted with respect to the static case
depending only on the absolute value of the vertical velocity field gradient and
independently of the sign of the velocity field.}
\label{fig:gradsv_ctes}
\end{figure*}
\clearpage

\begin{figure}[h!]
\centering
\includegraphics[scale=0.45]{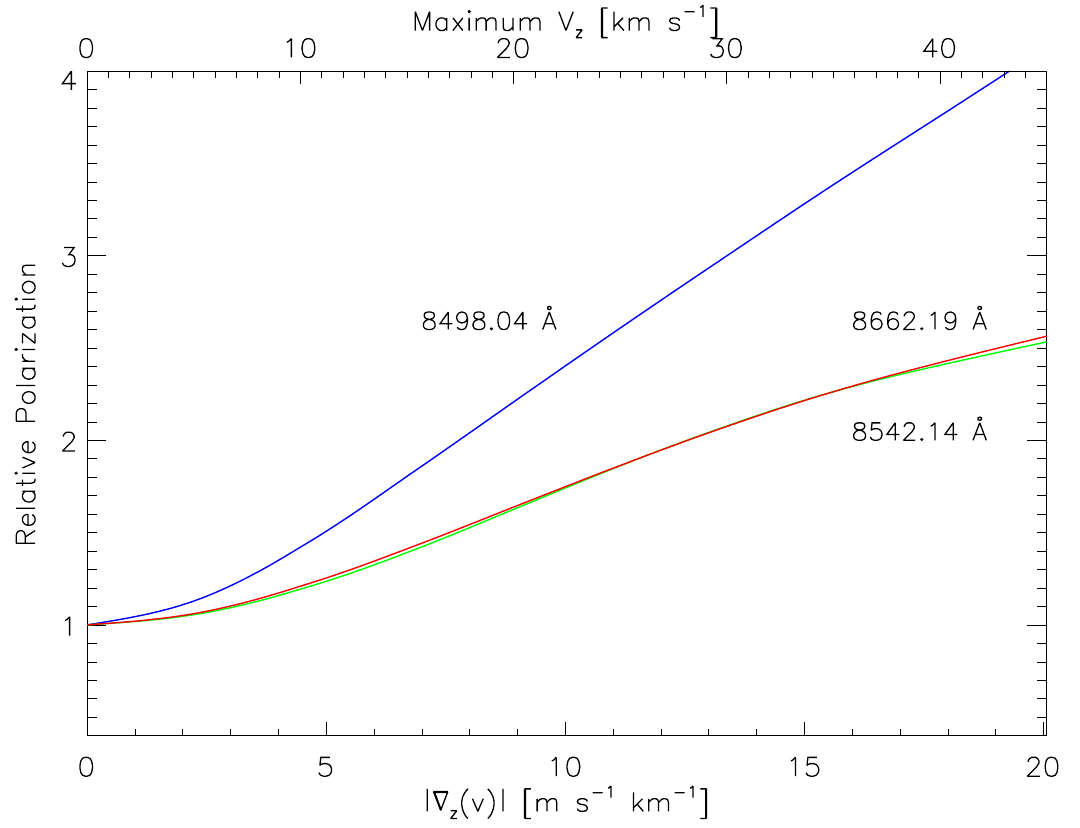}
\caption{$(Q/I)_\mathrm{pp}$ normalized to the static case solution versus the velocity 
gradient (bottom axis) and the maximum absolute velocity (top axis) for linear velocity 
fields appearing in Fig. \ref{fig:gradsv_ctes}. The results are invariant under velocity 
sign changes. The transition $\lambda8498$ is more sensitive to velocity variations due 
to its larger value of $\alpha=\Delta\nu_D/w$.\label{fig:gradvarlaw}}
\end{figure}

\subsection{Variations on the atomic alignment due to velocity
gradients.}\label{sub:EB}
In order to get physical insight on the formation of the emergent polarization
profiles, we use an analytical approximation. Following \cite{Trujillo-Bueno:2003b},
the emergent fractional linear polarization for a strong line at the central wavelength
can be approximated with (see Sec. \ref{sec:hanle_walk}):
\begin{equation}\label{eq:qestimated2}
\frac{Q}{I} \thickapprox  \frac{3}{2\sqrt{2}} (1-\mu^2) \left[ 
{\sc}w^{(2)}_{J_u J_{\ell}} \cdot \sigma^2_0(J_u)  -  {\sc}w^{(2)}_{J_{\ell} J_u}   \cdot
\sigma^2_0(J_{\ell}) \right].
\end{equation}
The symbols ${\sc}w^{(2)}_{J J'}$ are numerical coefficients already introduced
in Sec. \ref{sec:problem}. The quantities $\sigma^2_0(J_u)$ and
$\sigma^2_0(J_{\ell})$ are the fractional alignment coefficients
($\sigma^2_0=\rho^2_0/\rho^0_0$) evaluated at $\tau^{\rm{los}}_{\nu} = 1$ for the
upper and lower level of the transition, respectively. This is the
generalization of the Eddington-Barbier (EB) aproximation to the scattering
polarization and establishes that changes in linear polarization (for a static
case) are induced by changes in the atomic aligment of the energy levels.

Our calculations show that vertical velocity fields with moderate gradients
($\lesssim 10\,\rm{m\, s^{-1}  km^{-1}} $ in a linear velocity field, as the ones shown in the figures)
do indeed produce variations in the fractional alignment, which are small for 
$|\sigma^2_0(J_u)|$ and significant for $|\sigma^2_0(J_{\ell})|$ (see Fig. \ref{fig:rhosqi2}). The
lower level alignment is the main driver of the changes produced in
the analyzed $Q/I$
signals. This is strictly true for the $\lambda8662$ line, whose upper
level with $J=1/2$ cannot be aligned (zero-field dichroism polarization).
In the other transitions of the triplet, a certain influence of the upper level alignment
becomes notable only for large gradients. The reason of this behavior
is that the strong K transition
is dictating the common
upper level-5 alignment (Fig. \ref{fig:elevels}). In fact, $\sigma^2_0(J_5)$ is
driven by the K line anisotropy which, at cromospheric heights, is almost
unaffected for the considered velocity gradients, as we discussed in Sec. \ref{subsec:aniso}
(Fig. \ref{anisovars}). Thus, the strong H and K lines feed population to the
upper levels and the K line controls the alignment of the $^2P_{3/2}$ level (see Fig. 1), while the polarization signals
of the IR triplet change with velocity fields affecting $\sigma^2_0(J_{\ell})$ (through the
anisotropy enhancement). 
\begin{figure*}[!htb]
\includegraphics[scale=0.8]{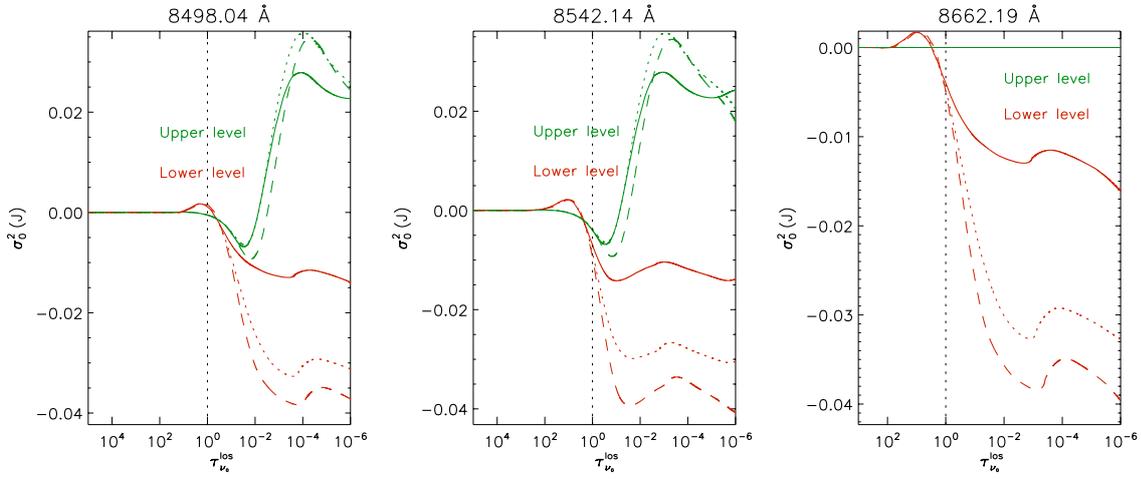}
\caption{Behaviour of the fractional alignments $\sigma^2_0(J_u)$ (green) 
and $\sigma^2_0(J_l)$ (red) of each Ca \textsc{ii} IR triplet transition
for three of the velocity fields shown in Fig. \ref{fig:gradsv_ctes}. The solid lines correspond
to the reference static case. The dotted lines correspond to the case with maximum velocity 
$15\, \rm{km \, s^{-1}}$. The dashed lines are associated with the case with maximum 
velocity $30 \, \rm{km \, s^{-1}}$. The horizontal axis is the line center 
optical depth for each value of the velocity gradient. The vertical dotted 
line marks the height where $\tau^{\rm{los}}_{\nu_0}=1.$}
\label{fig:rhosqi2}
\end{figure*}
To illustrate the well-known link between the alignment and the anisotropy we
can follow the next reasoning. For the Ca \textsc{ii} model atom
we deal with in this work, it is posible to derive a simple
analytic expression that relates the anisotropy and the alignments of the $\lambda8542$
transition. Making use of Eqs. (\ref{see04}) and (\ref{see05}) and neglecting
second order terms and collisions, we find that:
\begin{equation}\label{eq:relacionw}
{2}\sigma^2_0(J_5)-{\sqrt{7}}\sigma^2_0(J_3) \simeq  {\sc w}_{line}(3\rightarrow
5).
\end{equation}
As before, we can roughly assume that $\sigma^2_0(J_5)\thicksim \rm{constant}$
in the chromosphere because it is controlled by the K line. Then, Eq.
(\ref{eq:relacionw}) suggests that, if the radiation anisotropy increases at
those heights, an amplification of $|\sigma^2_0(J_3)|$ occurs (note that $\sigma^2_0(J_3)$
is negative for these lines). A more aligned atomic
population produces a more intense scattering polarization signal.

\section{Conclusions}

When vertical velocity gradients exist, the
polarization profiles are always shifted in wavelength, asymmetrized and enhanced in
amplitude with respect to the constant velocity case. The reason is that increments in the absolute value
of the velocity gradient increase the source function (Doppler brightening) and
enhance the anisotropy of the radiation field (Secs. \ref{sec:preview} and
\ref{sec:anisotriplet}), that in turn modify the fractional alignment (Sec.
\ref{sub:EB}) and amplify the scattering polarization profiles (Sec.
\ref{sec:impact}).
 
For this very reason, all calculations assuming static models in the
formation region might underestimate the scattering polarization amplitudes and
not capture the right shape of the profiles. In particular, it must be taken into
account that the Ca {\sc ii} IR triplet lines form under non-LTE conditions in
chromospheric regions, where velocity gradient may be significantly large due to the upward
propagation of waves in a vertically stratified atmosphere \citep[e.g.,][]{Carlsson:1997}. Probably, in
photospheric and transition region lines the effect of velocities on polarization can be
safely neglected (they will be predominantly amplified by temperature gradients
as discussed in Sec. \ref{sub:tla}), but not necessarily in the chromosphere. In our study
we see that the $\lambda8498$ line is more sensitive to macroscopic motions in the 
low-chromosphere, while the $\lambda8542$ and $\lambda8662$ lines are especially amplified when
strong velocity gradients are found at heights around $1.5 \, \rm{Mm}$ and higher in our
model. 

At the light of these results, it is obvious that the effect of the velocity might be
of relevance for measuring chromospheric magnetic fields. In
particular, the described mechanism might turn out to be important for the correct
interpretation of polarization signals in the Sun with the Hanle effect. Given
that weak chromospheric magnetic fields are inferred with the Hanle effect using the difference
between the observed linear polarization signal and the signal
that would be produced in the absence of a magnetic field, it is relevant
to compute the reference no-magnetic signal including velocities.

The polarization amplification mechanism that we have discussed in this 
chapter is not limited to plane-parallel atmospheres, although its
effect is surely more important in plane-parallel atmospheres than
in three-dimensional ones. The reason is that i) gradients in a
three-dimensional atmosphere are expected to be weaker given the
increased degrees of freedom and ii) significant variations in
velocity direction along the medium mix the spectral 
contribution of different layers to the radiative transfer and broaden
the emergent profiles, masking the Doppler features. 



 \clearemptydoublepage  
%
\chapter{Scattering Polarization of the Ca \textsc{ii} IR triplet in Hydrodynamical Models}\label{cap:three}
In Chapter \ref{cap:two} we showed that velocity gradients can significantly affect the
scattering polarization of the IR triplet of Ca {\sc ii}. Our
arguments were based on radiative transfer (RT) calculations in a
semi-empirical model of the solar atmosphere, after introducing ad hoc
velocity gradients and comparing the computed $\mathrm{Q/I}$ profiles with
those corresponding to the static case. Given the diagnostic potential
of the Ca {\sc ii} IR triplet for exploring the magnetism of the solar
chromosphere \citep[e.g.,][]{manso10,delacruz12}, and the fact that
the region where such chromospheric lines originate may be affected by
vigorous and repetitive shock waves \citep[e.g., ][]{Carlsson:1997},
it is necessary to investigate this RT problem using dynamical,
time-dependent atmospheric models of the solar chromosphere. 

In this
chapter, we show the results of such investigation. Thus, we analyze the emergent linear polarization
profiles of the Ca \textsc{ii} infrared triplet after solving the
RT problem of scattering polarization in
time-dependent hydrodynamical models of the solar chromosphere, taking
into account the effect of the plasma macroscopic velocity on the
atomic level polarization. We discuss the influence that the velocity
and temperature shocks in the considered chromospheric models have on
the temporal evolution of the scattering polarization signals of the
Ca \textsc{ii} infrared lines, as well as on the temporally averaged
profiles. We also study the effect of
the integration time, the microturbulent velocity and the
photospheric dynamical conditions, and discuss the feasibility of
observing the temporal variation of the
scattering polarization profiles with large-aperture
telescopes. Finally, we explore the possibility of using the
differential Hanle effect in the IR triplet of Ca \textsc{ii} to infer magnetic fields in dynamic
situations. This chapter is adapted from \cite{Carlin:2013aa}.

\section{Description
of the problem and the resolution procedure.}\label{sec:cal} We have
carried out RT calculations of the scattering polarization in the Ca \textsc{ii} infrared (IR)
triplet. The polarization is produced by the atomic level polarization
induced by anisotropic radiation pumping in the hydrodynamical
(HD) models of solar chromospheric dynamics obtained by
\cite{Carlsson:1997,Carlsson:2002}.

We used two time series of snapshots from the above-mentioned
radiation HD simulations, each one lasting about 3600 s and showing
the upward propagation of acoustic wave trains growing in amplitude
with height until they eventually produce shocks. The first one
corresponds to a relatively strong photospheric disturbance showing
well-developed cool phases and pronounced hot zones at chromospheric
heights \citep[see][which will be referred to as the strongly
dynamic case]{Carlsson:1997}. The second simulation corresponds to a less intense
photospheric disturbance \citep[the weakly dynamic case given by][]{Carlsson:2002}.  Thus, the thermodynamical evolution
of the atmosphere (including the chromosphere and the transition
region) is driven by the bottom boundary condition that is imposed on
the velocity. This realistic boundary condition is extracted from the
measured Doppler shifts in the Fe {\sc i} line at $3966.8$ \AA. Our
description focuses mainly on the strongly dynamic case, but in
Sec. \ref{subsec:dynamic} we compare the results with those
corresponding to the weakly dynamic case.

To characterize the simulations we can use the following
quantities. In terms of the velocity gradients, and using units
related to a representative scale height\footnote{A scale height can
be defined as the typical distance over which atmospheric magnitudes
such as the density vary an order of magnitude. Since the models contain
important temporal variations of such magnitudes, the scale height
varies. For this reason we
have defined an averaged scale height as the \textit{representative}
value used for the characterization of the velocity gradients.}
$\mathcal{ H}=275 \,\rm{km}$, the temporal average of the maximum
velocity gradient along the atmosphere is $40 \,\rm{km\, s^{-1}}$
per scale height (or $145 \,\rm{m \,s^{-1}km^{-1}}$) in the
strongly dynamic case and $13 \,\rm{km\, s^{-1}}$ per scale height
(or $47 \,\rm{m\, s^{-1}km^{-1}}$) in the weakly dynamic
case. Likewise, the temporal average of the minimum of temperature in
the atmosphere is $\rm{3976 \, K}$ in the strongly dynamic case, and
$\rm{4292 \,K}$ in the weakly dynamic case.

At each time step of the HD simulation under consideration we use the
corresponding one-dimensional stratifications of the vertical
velocity, temperature and density to compute the emergent $\rm{I(\lambda)}$
and $\rm{Q(\lambda)}$ profiles through the application of the multilevel
radiative transfer code of \cite{Manso-Sainz:2003,manso10}, after the
generalization to the non-static case described in
Sec. \ref{sec:dynamic_traviata}. Specifically, we have jointly solved the radiative
transfer (RT) equations for the Stokes I and Q parameters and the
statistical equilibrium equations (SEEs) for the atomic populations of
each energy level and the population imbalances among its magnetic
energy sublevels (equivalently, the multipolar tensor components of
the atomic density matrix, $\rm{\rho^K_0 (J_i)}$, with $\rm{J_i}$ the angular
momentum of each level i). This is the NLTE radiative transfer
problem of the second kind \citep[see Sections 7.2 and 7.13
in][]{LL04}. Once the self-consistent solution of such equations is
found at each height in the atmospheric model under consideration, we
compute the coefficients of the emission vector and of the propagation
matrix (see Sec. \ref{subsec:rte}) and solve the RT equations for a
line of sight (LOS) with $\mu=0.1$, where $\mu$ is the cosine of the
heliocentric angle. This LOS has been chosen in order to simulate a
close to the limb observation, such as that shown in Figure 13 of
\citet[][]{stenflo00b}. To account for macroscopic motions, we
have introduced the Doppler effect in the calculation of the
absorption and emission profiles for each wavelength and ray direction
(Sec. \ref {sec:dynamic_traviata}). The influence of the Doppler effect on the SEE
appears directly because the radiative rates depend on the radiation
field tensor components. Likewise, the RTE is affected because the
Doppler effect modifies the elements of the propagation matrix and of
the emission vector.

Given that the computations reported here are carried out in
plane-parallel atmospheric models, it is necessary to introduce a
microturbulent velocity, which accounts for the Doppler shifts
(inducing an effective line broadening) by moving fluid
elements below the resolution element. In order to estimate a suitable
value (assumed constant with height), we have calculated the emergent
intensities at disk center and compared them with those of the solar
Kitt Peak FTS Spectral Atlas \citep{kurucz_atlas84}. A good agreement
is obtained with $3.5$ km s$^{-1}$.

\section{Description and characterization of the
results.}\label{sec:descript_results} A standard Fourier analysis of
the atmosphere model shows that it acts as a passband filter for the
multifrequency sound waves generated in the lower boundary. The result
is that the predominating periods at chromospheric heights and higher
are around three minutes \citep{Carlsson:1997}. For practical reasons,
we divided the temporal evolution in three-minute intervals so that the
beginning of each interval coincides with the moment in which the
shock front in temperature and velocity is the sharpest in each
interval (vertical lines in figures with temporal axis, like
Figure \ref{fig:tevol_all}).  Given the power of the three-minute waves, this
division turns out to be ``natural'' and can be used to mark the most
interesting events we see in the emergent polarization.
\begin{figure*}[h!]
\centering \includegraphics[scale=0.6]{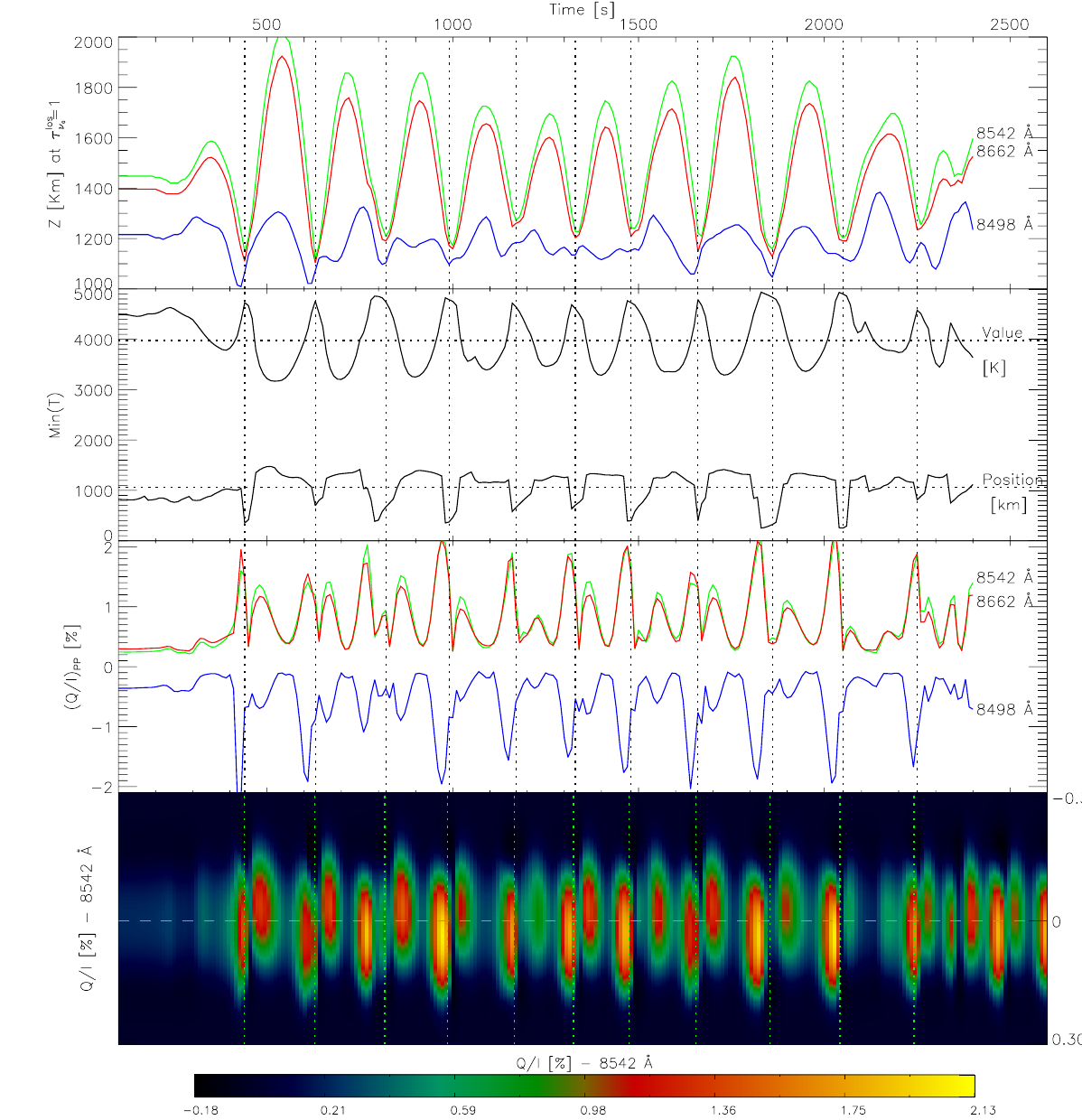} \caption{\textbf{Top row:}
time evolution of the atmospheric heights where $\tau^{\rm los}_{\nu_0}=1$
for the three Ca {\sc ii} IR transitions. The dotted vertical lines
are located at the local minimum of the $\tau^{\rm los}_{8452}=1$ curve,
and we use them to indicate the beginning and the end of
each ``three-minute" period. \textbf{Second row:} time evolution of the temperature
value and atmospheric height of the temperature minimum. \textbf{Third row:}
time evolution of the polarization contrast
($\rm{\max(Q/I)-\min(Q/I)}$) for the three lines of the Ca {\sc ii} IR
triplet. The polarization amplitude of the ${\lambda}8498$ line has
been multiplied by -5 to show the results for the three lines on the
same scale. Note that, by definition, the line contrast is always
positive. However, we added an artificial negative sign to
the 8498 \AA\ line contrast values in this figure
to illustrate that its polarization
amplitudes are predominantly negative. \textbf{Bottom row:} time evolution of the calculated
$\rm{Q(\lambda)/I(\lambda)}$ fractional linear polarization profile of the
${\lambda}8542$ line.\label{fig:tevol_all}} \end{figure*}

Inside each three-minutes cycle we distinguish between
\textit{compression} and \textit{expansion} phases. They can be easily
identified following the height at which $\tau^{\rm los}_{\nu_0}=1$, i.e.,
where the optical depth at line center ($\nu_0$) along the LOS equals
unity (upper panel of Figure \ref{fig:tevol_all}). This quantity is a
good marker of the shock fronts when they cross heights between $1$
and $2$ Mm. It is because the steep changes in opacity inside the
shocks forces the $\tau=1$ region to remain comprised within them. The
line transitions at $8542$ \AA\ and $8662$ \AA\ (green and red lines
in the upper panel of Figure \ref{fig:tevol_all}) follow a clearer
periodic pattern because they form higher, where less frequency
components of the velocity waves can arrive. Compression phases begin
when plasma falls down from upper layers ($\tau^{\rm los}_{8542}=1$ and
$\tau^{\rm los}_{8662}=1$ decrease in the top panel of
Figure \ref{fig:tevol_all}), while simultaneously a new upward
propagating wave emerges amplified into the chromosphere. At the end
of this stage a shock wave is completely developed and the
$\tau^{\rm los}_{\nu_0}=1$ position is close to $\sim$1200 km for the
three IR lines. The shock waves so created always start in such region
between\footnote{It is in this range of heights where
the Ca {\sc ii} IR triplet forms in typical semi-empirical
models} 1 and 1.5 Mm. After that, during what we term expansion phase (heights for
$\tau^{\rm los}_{8542}=1$ and $\tau^{\rm los}_{8662}=1$ arising in top panel
of Figure \ref{fig:tevol_all}), the shock fronts travel upward
increasing the plasma velocities as they encounter lower densities.

\begin{figure*}[!h]
\centering \includegraphics[scale=0.6]{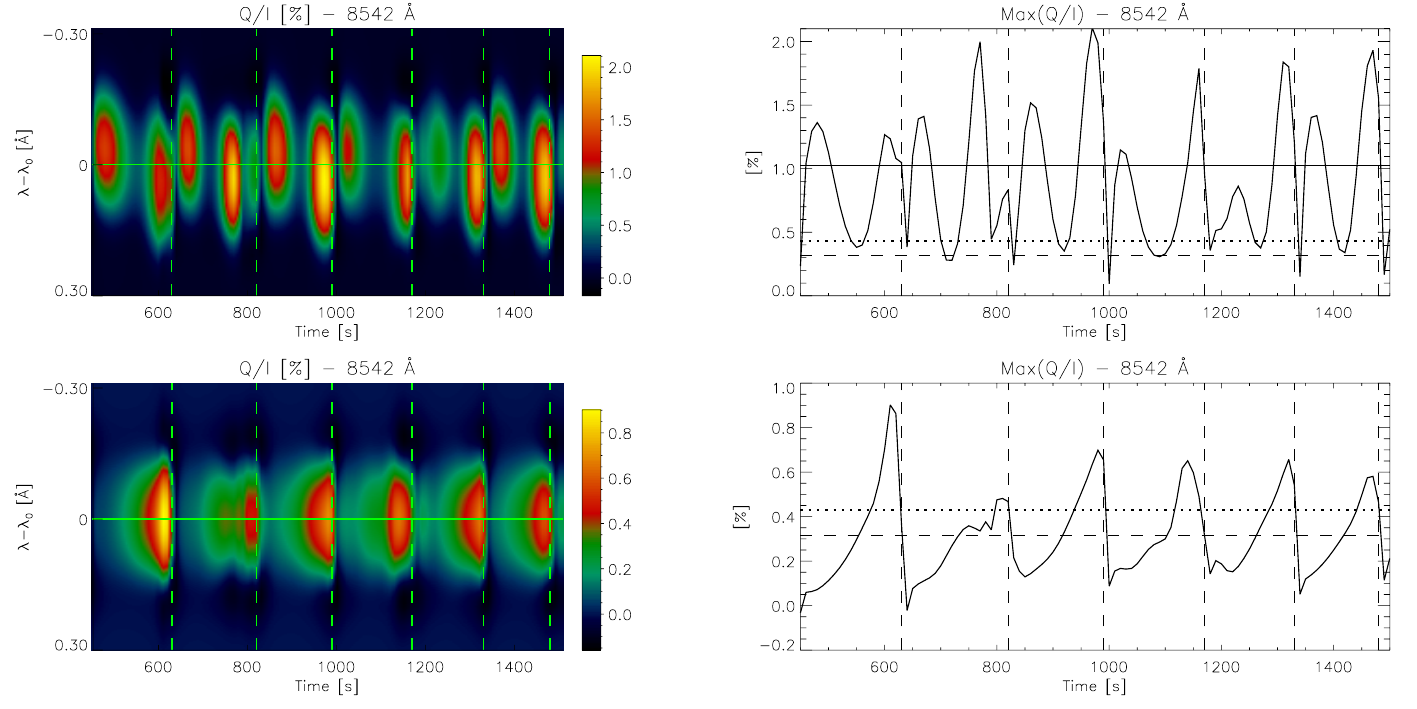} \caption{Temporal
evolution of $\rm{Q(\lambda)/I(\lambda)}$ (left column panels) and of the
${\rm{Max}}(Q/I)$ peak amplitude (right column panels) during 1040 s
(17 minutes) for the 8542 \AA\ transition.  \textbf{Upper
panels}: results taking into account the effect of the velocity gradients. \textbf{Lower panels}: results assuming no macroscopic
velocities while calculating the density-matrix elements.  The
vertical lines indicate the beginning and the end of each
three-minutes period.  The horizontal lines in the right column panels
show the temporally-averaged amplitudes ($\rm{\langle Q \rangle/\langle I
\rangle}$) for three cases: RT in the strongly
dynamic model including velocities (solid line); RT in the strongly
dynamic model but setting $v=0$ (dashed lines); RT in the (static)
semi-empirical FAL-C model (dotted lines). \label{fig:qconysinv}} \end{figure*}

Figure \ref{fig:tevol_all} also shows the time evolution of other
quantities during the first $2000$~s after the initial transient. In
the second row, the location and value of the temperature minimum are
displayed, showing a clear correspondence with expansion and
contraction phases. In the third row, we show the ensuing variation of
$\rm{(Q/I)_{pp}}$, defined as the peak-to-peak difference of the Q/I
profile for each spectral line. It is a measure of the linear
polarization signal contrast that was used in Chapter \ref{cap:two} to
characterize the polarization amplitude and discriminate their
variations with respect to static cases. In each cycle we see an
amplification of $\rm{(Q/I)_{pp}}$ occurring at expansion phases and
an usually larger amplification during contraction phases. Finally,
the time evolution of the emergent $\rm{Q(\lambda)/I(\lambda)}$ profile
for the $8542$ \AA\ line is illustrated in the lower panel (the
vertical axis shows 0.6 \AA\ around the rest wavelength of the
line). Here, we observe two distinct areas showing amplifications
inside each three-minute cycle. The first amplification is
blueshifted, because it happens in an atmospheric expansion phase
(plasma moving toward the observer). It is weaker than the second
amplification, which is redshifted and occurs during the compression
phase (plasma moving down in the atmosphere). This indicates that the
compression phase is more efficient producing a polarization
amplification than the expansion one. The reason is that during
compression we have stronger velocity and temperature gradients along
the main regions of formation. Following the results of Chapter \ref{cap:two}, the larger the gradient, the larger the enhancement of the
linear polarization signal. The behavior is similar in the other
transitions.

There is a clear correspondence between the maximum value of the
temperature minimum (hot-chromosphere time steps) and the largest
peaks of the $\rm{(Q/I)_{pp}}$ signal, taking place just before the
maximum contraction (dotted vertical lines). As the atmosphere is
compressed, the temperature increases at chromospheric heights and the
resulting gradient of the source function produces an increase of the
radiation field anisotropy in the upper layers. This directly leads to
an enhanced emergent linear polarization signal. On the contrary, in
cold-chromosphere models the expansion reaches its maximum and
$\rm{(Q/I)_{pp}}$ is near its minimum value.

Even in such complex situations, we still witness the already known
effects of amplification (with respect to the static case), frequency
shift and asymmetry in the linear polarization profiles due to
dynamics. All of them have been already explained in Chapter \ref{cap:two},
using the semi-empirical FAL-C model of Fontenla et al. (1993) with
ad-hoc velocity stratifications. The enhancement is produced as a
consequence of the velocity gradients and subsequent anisotropy
enhancements. However, some differences exist from the experiments in
semi-empirical models and the calculations presented in this
chapter. First, the velocity stratification in the HD models is, in
general, non-monotonic and with a non-constant variation with
height. Second, the maximum velocity gradients are located at shocks,
with amplitudes that reach tens or even hundreds of meters per second
per kilometer (as a comparison, in Chapter \ref{cap:two} we dealt with
velocity gradients between 0 and 20 m s$^{-1}$ km$^{-1}$).  Third, as
commented before, we have shocks in temperature that produce larger
source function gradients and additional enhancement of the radiation
anisotropy and of the linear polarization. Finally, these variations
are usually concentrated in the formation regions of the triplet
lines.  All these mechanisms act together and enhance the linear
polarization of the emergent radiation with amplification factors up
to $\sim10$ (in the $8498$ \AA line) and $\sim7$ (in the $8542$ \AA\
and $8662$ \AA\ lines), for the instantaneous values of the $\rm{Q/I}$
amplitudes with respect to the static FAL-C case. However, if we
consider temporal averages of the emergent Stokes profiles during long
periods, we get amplification factors of about a factor of two
(time-averaged Q/I amplitudes reach $\sim1 \,\%$ for 8542 \AA\ and
8662 \AA\ lines).
\clearpage
\begin{figure}[!h]
\centering 
\includegraphics[scale=0.8]{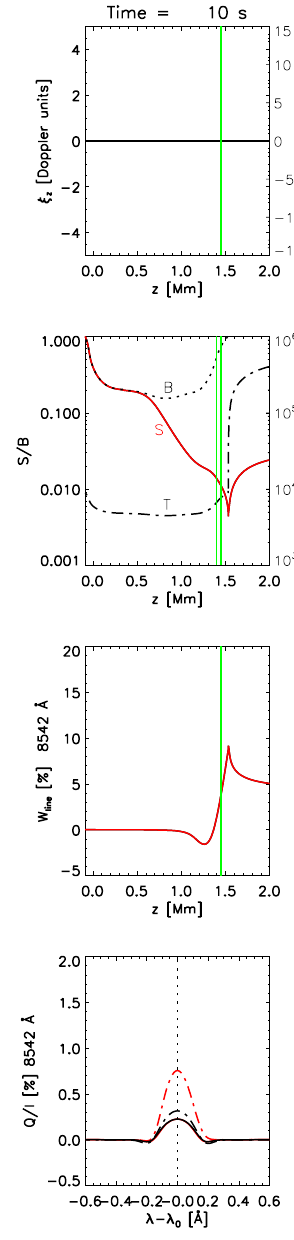} 
\includegraphics[scale=0.8]{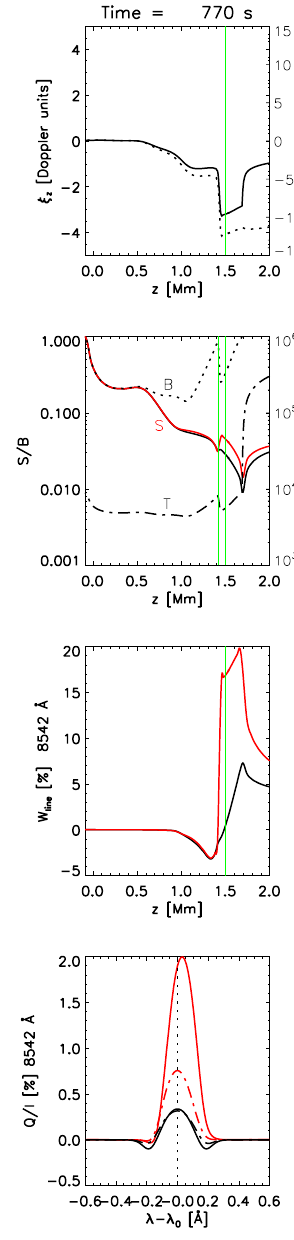} 
\includegraphics[scale=0.8]{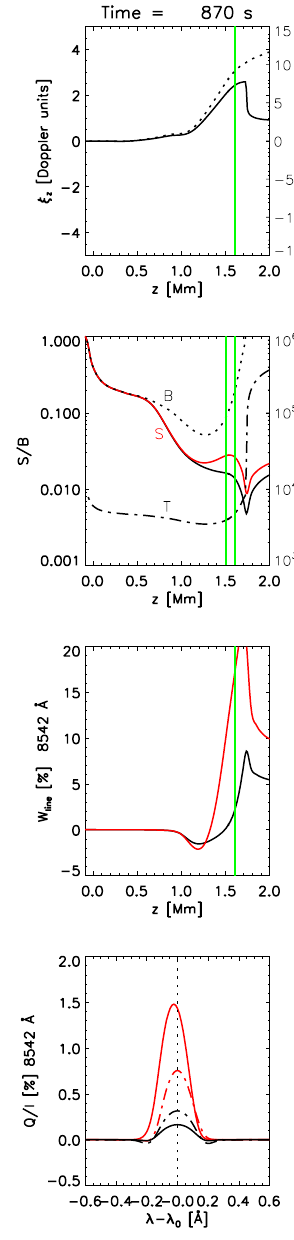} 
\caption{\footnotesize{The effect of the velocity gradient for the time steps
 10 s. (left column, a `quiet' situation), 770
s. (middle column, a compression stage) and 870 s. (right column, an
expansion phase) in the strongly dynamic hydrodynamical simulation.
\textbf{First row (from top)}: macroscopic vertical velocity (dotted line) and
adimensional vertical velocity $\xi_z$ (solid line) versus height.\textbf{Second row (from top)}: temperature (dashed
line), Planck function (dotted line), source function for the zero
velocity approximation (solid black line) and source function allowing
the influence of the model's velocity gradients (red solid line). \textbf{ Third row (from top)}: line anisotropy
factor (Equation \ref{eq:aniso})  calculated for each of
the above-mentioned cases; neglecting the effect of the velocity
gradients (black solid line) or allowing it (red solid lines). \textbf{Fourth
row (from top)}: emergent $\rm{Q/I}$ profiles versus wavelength (respect to the line center), 
with the same color code as in previous
panels. The dashed lines here are the time averages over the entire
simulation for each case (with or without
velocities). The
green lines mark the instantaneous positions of $\tau^{\rm los}_{\nu_0}=1$
and $\tau^{\rm los}_{\nu_0}=2$.}
\label{fig:termo}} 
\end{figure}
\clearpage
Summarizing, the temporal evolution of the polarization is driven by
the temperature and velocity stratifications, that in turn are a
result of the dynamical conditions set in the photosphere.


\section{Analysis and discussion of results.}\label{sec:results}

\subsection{The effect of the velocity.}\label{subsec:velocity} A way
of visualizing the effect of vertical velocity gradients on the
emergent scattering polarization is to compare the evolution of the
polarization profiles corresponding to both the static and non-static
case. In the absence of velocities (lower row of
Figure \ref{fig:qconysinv}), the maximum of the $\rm{Q/I}$ profiles is always
located at $\lambda=\lambda_0$ (i.e., line center), and its temporal
evolution presents a sawtooth shape. When the effect of velocities is
included in the calculations (upper row of Figure \ref{fig:qconysinv}),
the maximum of the $\rm{Q/I}$ signal is no longer located at the central
wavelength and its temporal evolution assumes a different shape with
two peaks every three-minute period (upper right panel). These wavelength and
amplitude modulations are produced by the Doppler effect of the
velocity gradients.

It is interesting to compare the mean $\rm{Q/I}$ amplitudes obtained
in the HD models with the one calculated in the FAL-C
model. They differ notably (see horizontal lines in right panels of
Figure \ref{fig:qconysinv}).  In the $8542$ \AA\ transition we have mean
values around $1\%$, $0.31\%$ and $0.42\%$ for the HD models with
velocities, the HD models at rest and static FALC models,
respectively. The results of these figures have been obtained using an
integration time of $1040$ s ($\backsim 17$ minutes), the duration of
the temporal interval shown in Figure \ref{fig:qconysinv}. Neglecting
the effect of the velocity gradients in the HD models, we see that the
resulting temporally averaged scattering polarization signals (which
include the impact of the temperature and density shocks) are similar
to the $\rm{Q/I}$ profiles computed in the static FAL-C semi-empirical
model.

\subsection{The combined effect of velocity and temperature on the
linear polarization.}\label{subsec:combined} In Figure \ref{fig:termo}
we display some relevant magnitudes for three different situations in
the simulation. The first column corresponds to a \textit{quiet}
time step, with no shocks, zero velocity and without any kind of
amplifications (it is the initial transient phase). The central column
shows a phase of \textit{compression}, in which shocks are
important. Finally, the last column displays an \textit{expansion}
phase, in which the atmosphere is expanded and the shocks are already
travelling over the transition region.  Furthermore, we distinguish
between the solutions when motions are taken into account (red lines)
and the solutions obtained allowing shocks in all magnitudes but
artificially setting the velocity to zero (black lines).

The normalized velocity $\rm{\xi_z=(\nu_0/c)v_z/\Delta\nu_D}$, with
$\rm{\Delta\nu_D}$ being the Doppler width of the absorption profiles (which
depends on the temperature), c the speed of light and $\rm{v_z}$ the
vertical velocity, is the quantity that controls the importance of the
atmospheric motions in relation with the radiation anisotropy and the
scattering polarization (see Sec. \ref{sub:tla}). Note that this quantity
considers the combined effect of velocity and temperature. In the HD
atmosphere models, $\rm{\xi_z}$ (solid lines in upper panels of
Figure \ref{fig:termo}) is only significant in the formation region of
the IR triplet lines ($\tau^{\rm los}_{\nu_0} \sim 1$ region with high
velocity gradients, and not very high temperatures). Although shock
waves increase the chromospheric temperature, the effect of the
velocity gradients is predominant. The contrary occurs over the
transition region, where the thermal line width is much larger than
the Doppler shifts.

The expansion and contraction can be identified also in quantities
such as the intensity source function and the Planck function (second
row in Figure \ref{fig:termo}). During contraction phases (middle column
panels), high temperatures produce a more efficient population pumping
toward upper levels, increasing the emissivity and, consequently,
the source function. Additionally, during contraction the temperature
shock occurs in optically thick and denser layers (deeper layers below
$\tau^{\rm los}_{\nu_0}=1$), forcing the source function gradient to
increase with respect to the static case at those heights. Note how in
this last case the source function rises \textit{as a whole} because
of the warming (compare the source function in the middle panel, the
black solid line that has been obtained neglecting velocities, with
the non-dynamic source function in the left column). If the
macroscopic velocity is now considered, we additionally get a jump in
the source function (red lines in middle column of
Figure \ref{fig:termo}) caused by the velocity shock that is developed
in this contraction phase. This behavior is accompanied by a
significant Doppler-induced anisotropy enhancement that amplifies the
linear polarization, as shown in the corresponding lower panels of the
same figure.

In the expansion phases, the shock waves move upward and the
chromosphere becomes cooler. This induces a lower source function and
smaller polarization amplitudes (as compared with the contraction
phase). Otherwise, as the density of scatterers is now lower around
the shock (because it moved upward to regions with
$\tau^{\rm los}_{\nu_0}<1$), the temperature gradients have smaller
effects on the polarization profiles than during the contraction
phases. In this expansion time step, the black solid line representing
the \textit{static} source function is similar to the non-dynamic
source function of the left column. However, once the motions are
introduced, and despite of the fact that the shocks have already
reached upper chromospheric layers, the remanent velocity field has
still a sizable gradient that enhances the source function (Doppler
brightening effect).

\begin{figure}[!t]
\centering 
\includegraphics[scale=0.6]{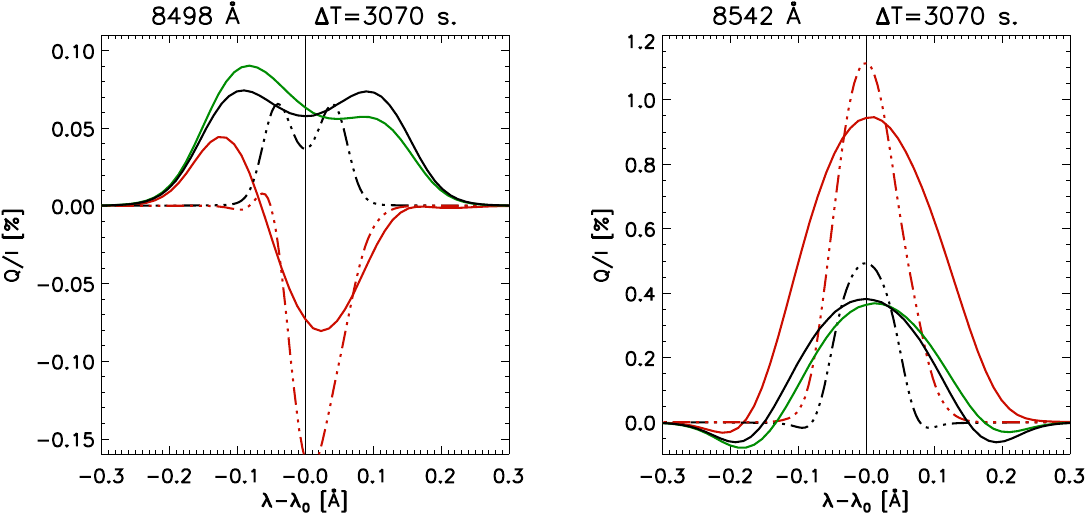} 
\caption{\footnotesize{Fractional
linear polarization profiles of the 8498 \AA\ and 8542 \AA\ lines
after temporally averaging the Stokes I and Q profiles during 3070
seconds (51 minutes). These $\rm{\langle Q \rangle/\langle I \rangle}$
profiles may be considered to emulate what can be actually observed
with today's solar telescopes. Black solid profiles: static case with
$v_{\rm{micro}}=3.5\,\rm{km\,s^{-1}}$. Red solid profiles: strongly
dynamic case taking into account the effect of the velocity gradients
and assuming $v_{\rm{micro}}=3.5\,\rm{km\,s^{-1}}$. Black dashed
profiles: strongly dynamic case neglecting the effect of the velocity
gradients and assuming $v_{\rm{micro}}=0$. Red dashed profiles:
strongly dynamic case taking into account the effect of the velocity
gradients and assuming $v_{\rm{micro}}=0$. The green solid lines show
the temporally averaged profiles obtained after applying the velocity
free approximation (VFA) with $v_{\rm{micro}}=3.5\,\rm{km\,s^{-1}}$
(i.e., neglecting the Doppler shifts of the macroscopic velocities
when computing the density matrix elements, but taking them into
account when calculating the emergent Stokes
profiles).} 
}\label{fig:qprom} 
\end{figure}

\subsection{Averaged values of the polarization
profiles.}\label{subsec:velo2} In order to compute the average linear
polarization signal that one would observe without any temporal
resolution, we average Q and I (obtaining $\rm{\langle Q \rangle /
\langle I \rangle}$) over $3070\,\rm{s}$ ($\approx 51$ minutes) for
four different cases (Figure \ref{fig:qprom}). We consider the cases
with zero microturbulent velocity (dotted lines) and a constant
microturbulent velocity of 3.5 km s$^{-1}$ (solid lines). For each
case, we distinguish between the results \textit{switching off} the
velocity (black lines) and the results allowing for macroscopic
velocity fields (red lines).

When macroscopic motions are considered, the polarization profiles
become amplified asymmetrically. Furthermore, they are
negative\footnote{This is true most part of the time. Only in some
  well-defined instants of the simulation the 8498 {\AA} profiles becomes
  positive. We study such sign variations in the next chapter.} in the case
of the 8498 \AA\ transition and positive in the other two
transitions. The asymmetry of the red profiles is a consequence of the
fact that, during the averaging period, the dynamical situations in
which the velocity gradient is negative (velocity field mostly
decreasing with height) dominate over the situations with velocity
gradients that are mostly positive. This predominance is not because
the situations with negative velocity gradients are more frequent but
because such situations are more efficient on amplifying the linear
polarization and thus, their effect prevails. This happens during the compression phase because i) the
velocity gradients are larger, ii) there is also a shock in
temperature affecting the formation region and iii) the shock fronts
are located just below the $\tau^{\rm los}_{\nu_0}=1$ height. The results
are qualitatively the same independently of the microturbulent
velocity value, but, when it is not considered, the amplification of
$\rm{(Q/I)_{pp}}$ is larger and the profiles are narrower.

If we decrease the averaging interval to $9$ minutes, we obtain
profiles that are essentially similar to the ones obtained by
averaging during $51$ minutes (showed in Figure \ref{fig:qprom}). If we
integrate less than that, significant variations appear in the shape
and amplitude of the emergent profiles. This indicates that,
concerning the linear polarization, there is still reliable dynamic
information contained in a time interval corresponding to a few three-minute
cycles.


 \subsection{The velocity free approximation.}\label{subsec:velo3} An
approximation that is sometimes applied to solve radiative transfer
problems in dynamical atmospheres (either taking into account the
presence of atomic polarization or not) is the velocity free
approximation (VFA). It is based on solving the SEE and RTE
simultaneously but neglecting the effect of plasma motions.  However,
once they are consistently solved, such plasma motions are included in
the synthesis of the emergent Stokes profiles (along $\mu=0.1$ in our
case). Consequently, the density matrix elements are calculated as if
plasma motions did not affect them, reducing the complexity and
computational effort of the problem since a reduced frequency grid is
used to compute the mean intensity and the anisotropy. The results of
applying it to each time step of our HD evolution is the temporal
average illustrated as the green line in Figure \ref{fig:qprom}. This
approximation is clearly not appropriate in our case, given that the
profiles just become asymmetric (with respect to the static profiles)
but without the amplification. The reason for this lack of
amplification is that the anisotropy controlling the linear
polarization is not correctly enhanced (see Chapter \ref{cap:two}). On the
other hand, the asymmetry is purely due to the asymmetric absorption
with respect to the line center that motions produce along the ray
under consideration. Hence, in order to obtain reliable results it is
mandatory to include the effect of Doppler shifts in the whole set of
equations, and we conclude that the VFA should not be applied.

\subsection{The effect of photospheric
dynamics.}\label{subsec:dynamic} Given that the small velocity fields
appearing in the photosphere are amplified because of the exponential
decrease in the density while the perturbations travel outward, the
properties of the bottom boundary condition are determinant in the
behavior of the emergent Stokes parameter of chromospheric lines. We
compare the strongly dynamic case that forms the core of this chapter
with the weakly dynamic case that has been already introduced in
Sec. \ref{sec:descript_results}. Although the mean maximum velocity
gradient is three times smaller in the weakly dynamic case and the
averaged polarization amplitudes are also smaller than in the strongly
dynamic one, we still find comparable or even slightly larger
instantaneous $\rm{(Q/I)_{pp}}$ amplitudes (see
Figure \ref{fig:tevol_all_weak}). The resulting averaged polarization
profiles are qualitatively the same but they differ in amplitude
(Figure \ref{fig:weak}). This is a reasonable result because in the
weakly dynamic scenario the instantaneous velocity gradients are
smaller in general. Differences are especially critical for the 8498
\AA\ line, whose linear polarization profiles can be positive, but
also adopt significant negative values at redder wavelengths (when
velocity gradients are mainly positive with height) or at bluer
wavelengths (when velocity gradients are mainly negative with
height). This behavior produces cancellation effects with integration
times larger than a three-minute period. Furthermore, the central
depression produced in the 8498 \AA\ average profile when the velocity
is neglected in the strongly dynamic simulation (solid black line in
left panel of Figure \ref{fig:weak}) do not appear in the average
profile corresponding to the weakly dynamic case (dashed black line in
the same panel) because of the differences in the instantaneous
temperature stratifications. The sensitivity of this spectral line to
the instantaneous photospheric perturbations and to the developed
chromospheric shocks is larger than in the other two lines.

\begin{figure*}[!h]
\centering \includegraphics[scale=0.6]{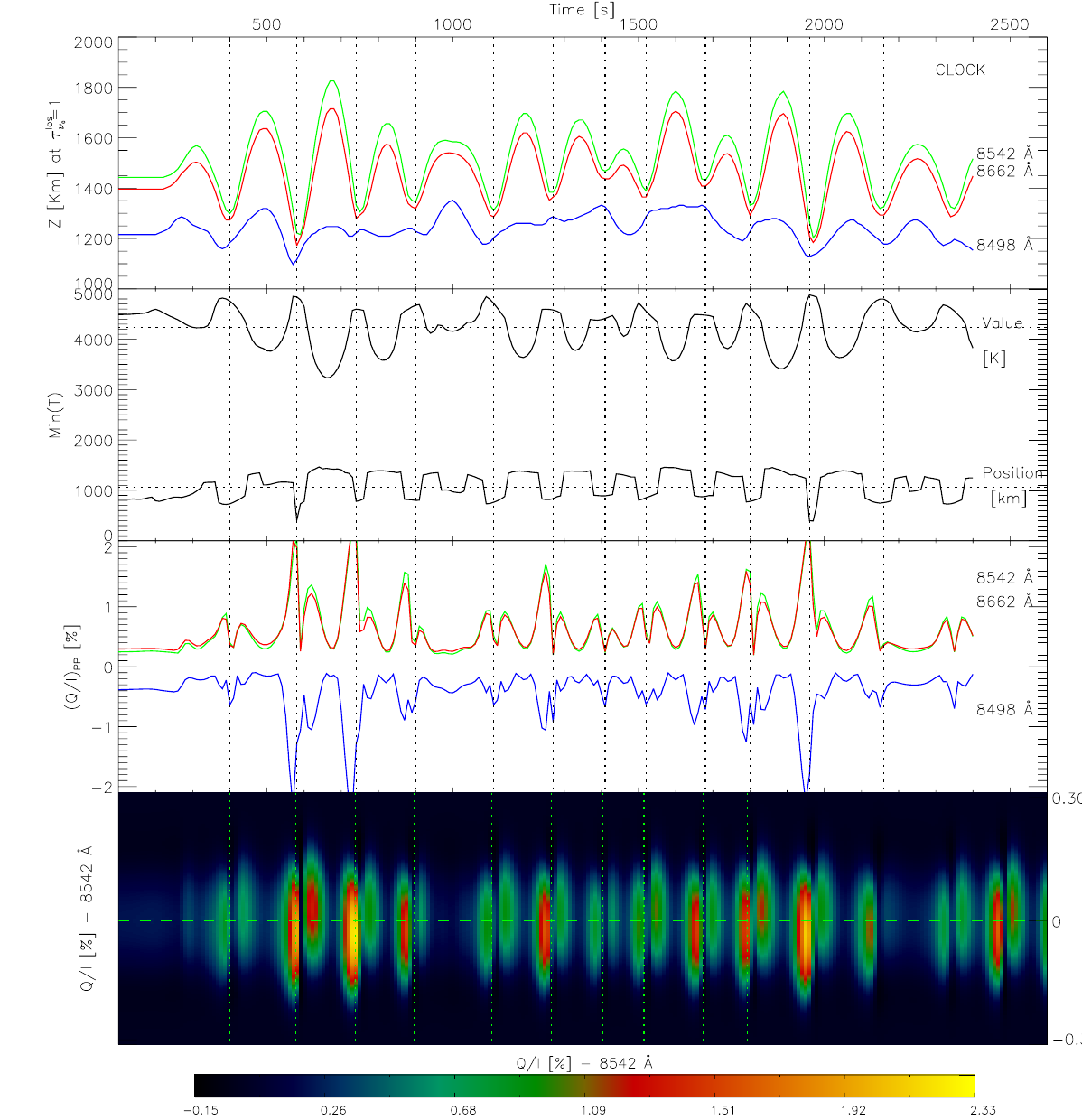} \caption{ Same as
Figure \ref{fig:tevol_all}, but for the weakly dynamic case. Remember
that the blue line amplitude has been multiplied by -5 for scale
reasons. \label{fig:tevol_all_weak}} \end{figure*}

\begin{figure*}[!t]
\centering \includegraphics[scale=0.55]{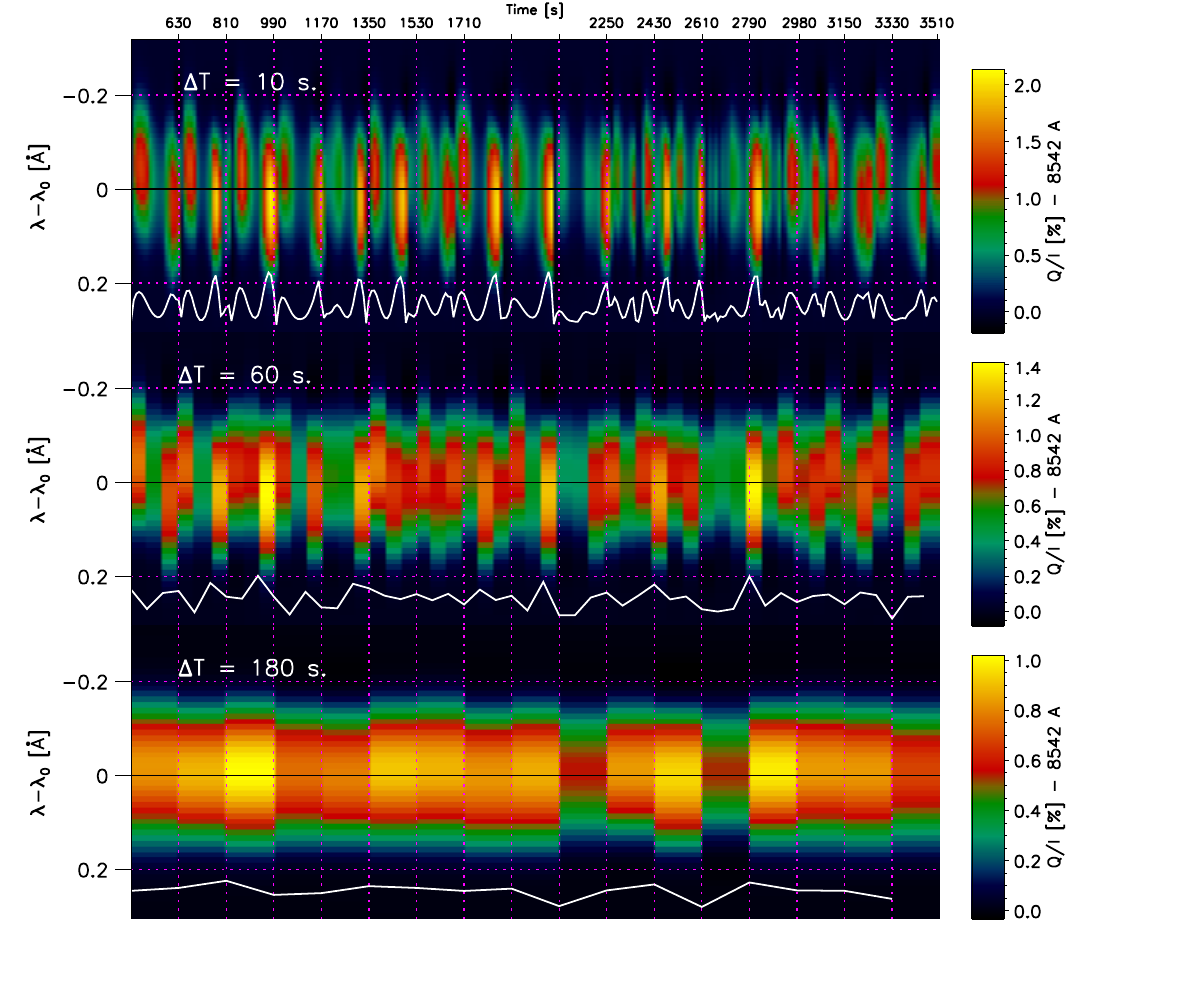} \caption{Temporal
evolution of the emergent $8542\lambda$ $\rm{Q/I}$ profiles for different
integration times. Wavelength is in the vertical axis. From top to
bottom we have a 10, 60 and 180 s. temporal resolution,
respectively. The white solid lines show the temporal evolution of
$\rm{(Q/I)_{pp}}$ (i.e., the amplitude contrast at each
time-step). Vertical dotted lines mark each three-minute
period. \label{fig:int_effect}} \end{figure*}

\begin{figure*}[!h]
\centering \includegraphics[scale=0.37]{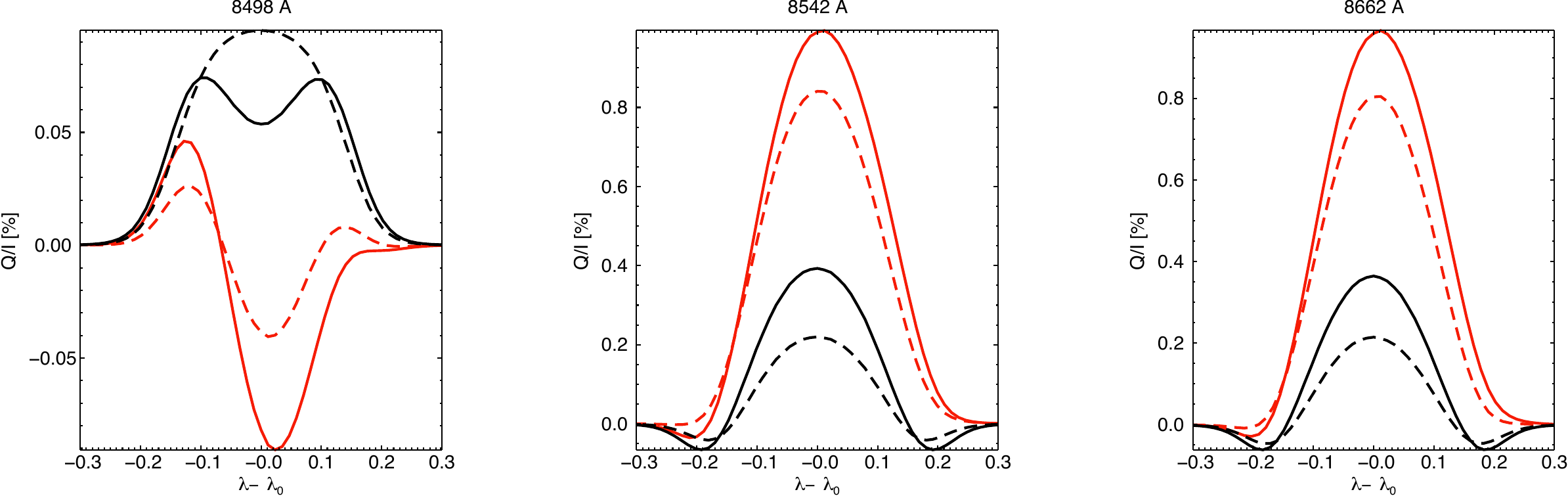} \caption{Comparison
of the results in the strongly and weakly dynamic cases (see
Sec. \ref{subsec:dynamic}). All profiles are the resulting $\rm{\langle Q
\rangle/ \langle I \rangle}$ profiles obtained averaging Q and I
during 15 minutes of the considered simulation. Solid lines: strongly dynamic case. Dashed lines: weakly dynamic case. Black lines: results allowing
variations in all magnitudes but neglecting the velocity. Red lines:
results including the velocity
gradients. \label{fig:weak}} \end{figure*}

\subsection{The effect of the integration time.}\label{subsec:tint} 
In
order to detect in the Sun the time evolution of the linear
polarization signals, the observations must have enough time
resolution, signal-to-noise ratio and spatial resolution. A sufficient
spatial coherence is important to avoid cancellations of the
contribution from different regions in the chromosphere evolving with
different phases. If we consider the expected capabilities of the next
generation of solar telescopes (like the European Solar Telescope,
EST, or the Advanced Technology Solar Telescope, ATST), we can aim at
observing the emergent Stokes profiles of Figure \ref{fig:int_effect}
with a 10 s cadence (upper panel)\footnote{Using EST (telescope
diameter of 4 m, instrumental efficiency around $10\%$) and considering
a spectral resolution of 30 m\AA, a spatial resolution of $0.1$
arcsec and an integration time of 1 s (ten times better than needed),
it would be possible to observe the linear polarization of the 8542
\AA\ line (line to continuum ratio of $\sim0.2$) at the level of $\rm{Q/I}
\sim10^{-3}$ with a confidence of $3 \sigma$ over the
noise.}. However, with the present telescopes and instrumentation, we
are forced to integrate in time and/or space to detect the scattering
polarization signals. If we degrade the temporal resolution of our
results to an integration time of 1 minute (middle panel of
Figure \ref{fig:int_effect}) and 3 minutes (lower panel of
Figure \ref{fig:int_effect}) we clearly see that the time evolution
becomes more difficult to detect. In the last case, the profiles are
already so smoothed that the original features are completely lost,
both in the spectral and temporal domains. The amplitude of the
integrated signals are lower than in the original 10 s sequence by a
factor of $2$ (see the color scales). However, integration during time
intervals ${\sim}1$ minute could reveal the amplification/modulation
effect if we capture spectro-polarimetric signals similar to the ones
showed in the middle row panel.
\begin{figure}[!h]
\centering \includegraphics[scale=0.6]{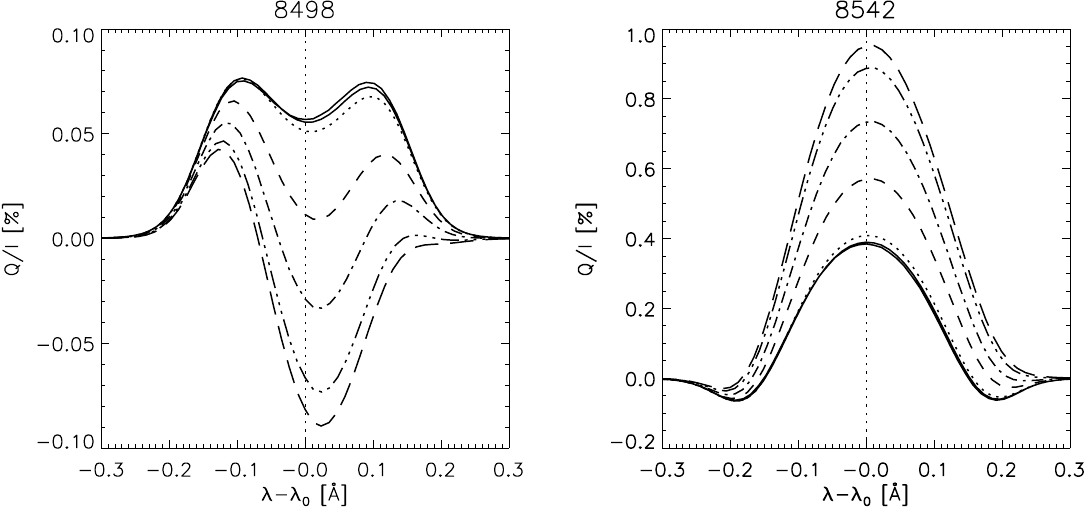} \caption{Resulting
$\rm{Q/I}$ profiles after averaging Q and I during $15$ minutes in the
strongly dynamic simulation, using different values of $F$. $F$ is the
scaling factor by which we have multiplied the modulus of the
macroscopic velocity at each atmospheric height. The curves correspond
to $F=0,\,0.1,\,0.2,\,0.5,\,0.7,\,0.9,\, 1$, going from solid lines
(static case and also $F=0.1$) to long dashed lines
(fully dynamic case). The results
for the 8662 \AA\ line are very similar to the ones obtained for the
8542 \AA\ line. \label{fig:consistency_test}} \end{figure}

\subsection{The effect of a decreasing velocity on the averaged
profiles.}\label{subsec:veloeffect} We also calculated what happens to
the emergent averaged profiles in the strongly dynamic case (with
$15$ minutes of integration, emulating an observation) when we
gradually reduce the velocity field by a constant scaling factor $F$,
keeping the rest of atmospheric magnitudes unperturbed (see
Figure \ref{fig:consistency_test}). As expected, we find that the
polarization amplitudes decrease in proportion to $F$, from the
original case, with $F=1$, towards the static case, with $F=0$. Note
that the core of the 8498 \AA\ line goes through zero for a certain
$F$ value (near $F = 0.6$). Thus, depending on the magnitude of the
velocity gradients, its linear polarization amplitude will be positive
or negative. This fact suggests an additional way to diagnose velocity
gradients along the line-of-sight. However, it is important to keep in
mind that this sensitivity also depends on the variations in density
and temperature, as shown in Sec. \ref{subsec:dynamic}.

Furthermore, the variation of the $\rm{Q/I}$ amplitudes is not linear with
$F$.  The change is small for small $F$, is larger for intermediate
values of $F$, and again becomes smaller for the largest $F$, tending
to saturation.



\section{Considerations on the Hanle effect}\label{sec:hanle} For
magnetic field diagnostics with the Hanle effect it is often necessary
to know the zero-field polarization reference \citep[e.g.,
][]{Stenflo:1994,Trujillo-Bueno:2004}. That is what we have tried to
do in previous sections, calculating and explaining the temporal
evolution of the linear polarization profiles in chromospheric dynamic
simulations. Ideally, this reference has to be computed under the same
thermodynamical and dynamical conditions than in the real Sun but
without magnetic field.  As the Hanle effect often depolarizes the
linear polarization signals, the difference between the observation
and the zero-field calculation can be associated with a magnetic field
by adjusting the magnetic field topology and intensity. The key point
is that the reference amplitude must be as precise as possible. If it
is imprecise, variations in the Stokes profiles can be associated with a
magnetic field when they were really due to uncertainties in another
magnitudes, like the temperature or the velocity field. Due to this
reason, the fact that the solar chromosphere is a highly dynamic
medium brings some complications for the use of the Hanle effect as a
diagnostic tool.

\begin{figure*}[!t]
\centering \includegraphics[scale=0.7]{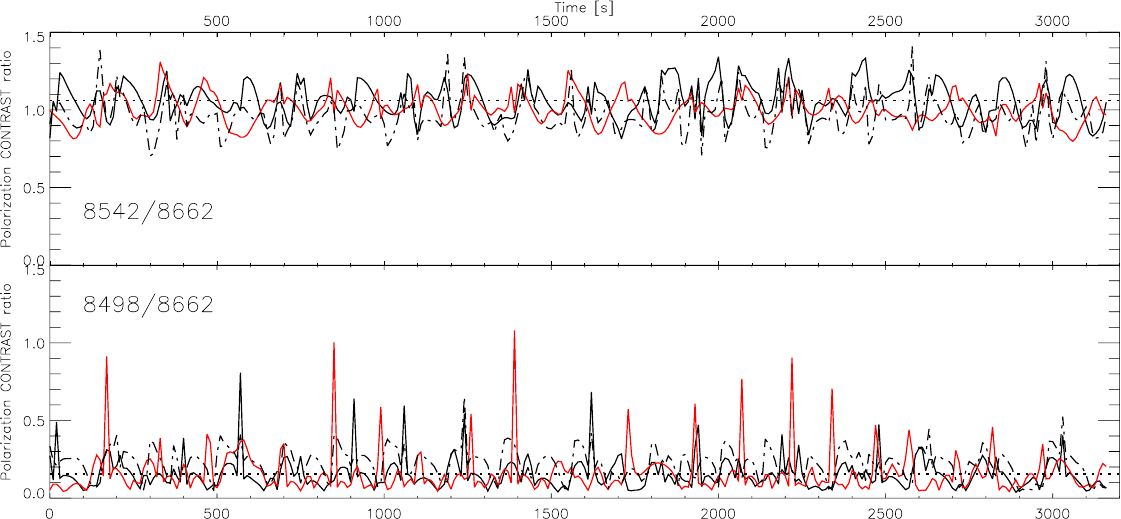} \caption{ Black
solid lines correspond to the strongly dynamic case. Red lines
correspond to weakly dynamic case. Black dashed lines correspond to
the strongly dynamic case but doing the calculations without
microturbulent velocity. The line ratio $\varrho_3 =
(\rm{Q/I})^{8498}_{pp}/(\rm{Q/I})^{8542}_{pp}$ is not shown because it
is very similar to $\varrho_1$ and because can be obtained from the
other two ratios. \label{fig:lineratios}} \end{figure*}
A strategy to avoid the above-mentioned problem is known as the line
ratio technique. It consists in finding a pair of spectral lines whose
thermodynamical behavior is identical but whose sensitivity to the
magnetic field is different in some range of magnetic field intensity
or inclinations \citep[e.g.,][]{stenflo98,manso04}. In that case, the
ratio between the polarization amplitudes should only change due to
variations in the magnetic field, thus allowing us to measure it after
a suitable calibration. As shown by \cite{manso10}, the main magnetic
sensitivity difference among the lines of the Ca II IR triplet is
between the ${\lambda}8498$ line (which is sensitive to field
strengths between 0.001 G and 10 G) and any of the ${\lambda}8662$ and
${\lambda}8542$ lines (which react mainly to sub-gauss magnetic fields
and up to 10 G in the latter spectral line). Unfortunately, while the
line-cores of the ${\lambda}8662$ and ${\lambda}8542$ lines originate
in similar atmospheric layers, the ${\lambda}8498$ line-core
originates at significantly deeper chromospheric layers (see e.g. Fig. \ref{fig:tevol_all}}). Nevertheless, we have found useful to plot in
Figure \ref{fig:lineratios} the time evolution of the following
polarization line
ratios: \begin{subequations}\label{eq:hanle1} \begin{align} \varrho_1
=
\frac{(\rm{Q/I})^{8498}_{pp}}{(\rm{Q/I})^{8662}_{pp}}, \label{hanle1a}
\displaybreak[0] \\ \varrho_2 =
\frac{(\rm{Q/I})^{8542}_{pp}}{(\rm{Q/I})^{8662}_{pp}},\label{hanle1b} \end{align} \end{subequations}
where the super-index indicates the central wavelength of the
transition in \AA.  These quantities were calculated for each
simulation considered before (weakly and strongly dynamic cases). The
more stable they are, the more useful they will be for inferring the
magnetic field.

We obtain that , on average, $\bar{\varrho_1} =0.15\,\pm\, 0.10$ and
$\bar{\varrho_1} =0.16\,\pm\, 0.14$ for the strongly and weakly
dynamic cases, respectively (lower panel in
Figure \ref{fig:lineratios}). The sudden shape variations (including
maximum amplitudes passing through zero) of the 8498 \AA\ line induce large
instantaneous excursions on $\varrho_1$ . As expected, a more stable
line ratio is obtained for the second pair of transitions, which are
precisely the ones that originate at similar chromospheric heights. We
find $\bar{\varrho_2} =1.06\,\pm\, 0.11$ and $\bar{\varrho_2}
=1.00\,\pm\, 0.09$ for the strongly and weakly dynamic cases,
respectively (upper panel in Figure \ref{fig:lineratios}). If we repeat
the calculations setting to zero the microturbulent velocity in the
strongly dynamic case we obtain $\bar{\varrho_1} =0.22\,\pm\, 0.09$
and $\bar{\varrho_2} =0.97\,\pm\, 0.12$ (dashed black lines in
Figure \ref{fig:lineratios}). These results indicate that the
$\bar{\varrho_2} $ line ratio shows a relatively stable behavior
against variations of the velocity and temperature in the solar
atmosphere. Consequently, in principle, $\varrho_2$ could be used as a
suitable line ratio to estimate the magnetic field from
spectropolarimetric observations of the ${\lambda}8662$ and
${\lambda}8542$ lines.

Regarding the sensitivity of these lines to the magnetic field and
their applicability for the diagnostic of magnetic fields through the
Hanle effect, several considerations have to be taken into account.
First, the microturbulent velocity has a small influence on the
averaged amplitudes and line ratios.  Second, once the magnetic field
is included in the calculations, the Hanle effect typically operates
at the line center for static cases. However, in a dynamic situation
there is not a preferred line center wavelength. As the maximum of the
absorption and dispersion profiles occurs at different Doppler shifted
wavelengths, the Hanle effect will operate in a small spectral region around
the line core.  Third, according to the static calculations by
\cite{manso10}, for chromospheric magnetic fields stronger than $0.1$
G in the ``quiet'' Sun, the $\rm{Q/I}$ signal of the 8662 \AA\ line is
expected to be Hanle saturated. Thus, variations between $0.1$ and
$10$ G could be measured with $\varrho_2$, being produced by changes
in the linear polarization of the 8542 \AA\ line.  Unfortunately, the
fluctuations we see in Figure \ref{fig:lineratios} (exclusively due to
the dynamics) have amplitudes of the same order of magnitude than
those expected from the investigations of the Hanle effect in static
model atmospheres (exclusively due to the magnetic field). More
realistic results will be obtained when carrying out calculations of
the Hanle effect of the Ca {\sc ii} IR triplet in dynamical model
atmospheres. In any case, it is clear that for exploiting the
polarization of these lines, we need instruments of high polarimetric
sensitivity. 

\section{Conclusions}
The results presented in this chapter indicate that the vertical
velocity gradients caused by the shock waves that take place at
chromospheric heights in the HD models of Carlsson \& Stein (1997;
2002) have a significant influence on the computed scattering
polarization profiles of the Ca {\sc ii} IR triplet. They show changes
in the shape of the $\rm{Q/I}$ profiles of the three IR lines and clear
enhancements in their amplitudes, as well as changes in the sign of
the $\rm{Q/I}$ signal of the ${\lambda}8498$ line. Such modifications with respect to the static case are evident, not
only in the temporally resolved $\rm{Q/I}$ profiles (e.g., see Figure \ref{fig:qconysinv}), but
also in the temporally averaged $\rm{\langle Q \rangle / \langle I
\rangle}$ profiles (e.g., see Figure \ref{fig:qprom}). This is true even with moderate
macroscopic plasma velocities, simply due to the presence of strong
vertical velocity gradients like the ones produced by shock
waves. This explains why the above-mentioned modifications of the
scattering polarization profiles of the Ca {\sc ii} IR triplet are
present also in the weakly dynamic simulation of \cite{Carlsson:2002}.

Our investigation points out that the development of diagnostic
methods based on the Hanle effect in the Ca {\sc ii} IR triplet should
take into account that the dynamical conditions of the solar
chromosphere can have a significant impact on the emergent scattering
polarization signals. This complication could be alleviated with
the application of line ratio techniques. In Sec. \ref{sec:hanle} we
have concluded that the ratio between the polarization amplitudes of
the ${\lambda}8542$ and ${\lambda}8662$ transitions would be the best
line-ratio choice. However, even in the absence of magnetic fields,
the small fluctuations we see in the calculated values of such dynamic
line ratios could be confused with the presence of
magnetic fields in the range between $0.1$ and $10$ G. Further work is
necessary to clarify this point.

In any case, the fact that realistic macroscopic velocity gradients
may have a significant impact on the scattering polarization profiles
of the Ca {\sc ii} IR triplet is interesting for the
diagnostic of the solar chromosphere\footnote{The physical mechanism
is general, but its observable effects are expected to be
significantly less important in broader spectral lines, such as the Ca
{\sc ii} K line and Ly$\alpha$. }. On the one hand, it provides a new
observable for probing the dynamical conditions of the solar
chromosphere (e.g., by confronting observed Stokes profiles with those
computed in dynamical models). On the other hand, the exploration of
the magnetism of the quiet solar chromosphere via the Hanle effect in
the Ca {\sc ii} IR triplet (either through the forward modeling
approach or via foreseeable Stokes inversion approaches) would have to
be accomplished without neglecting the possible effect of the
atmospheric velocity gradients on the atomic level polarization.

Some points are still unanswered after this work. First, we need to
investigate the sensitivity to the Hanle effect of the $\rm{Q/I}$ and
$\rm{U/I}$ profiles of the Ca {\sc ii} IR triplet using magnetized and
dynamical atmospheric models. Second, we have to investigate whether
our one-dimensional RT results remain valid after
considering realistic three-dimensional models, such as those
resulting from magneto-hydrodynamical simulations
\citep[e.g.][]{sven04,leenaarts09}.

Finally, we mention that these results could be potentially interesting
in other astrophysical contexts. For instance, the mechanism of
polarization enhancement due to the presence of shocks might well be
the explanation for the changing amplitudes of the linear polarization
signals reported in variable pulsating Mira stars \citep{fabas11}.


\clearemptydoublepage  
%
\chapter{Synthesis of Stokes profiles in a 3D MHD atmospheric model.}\label{cap:four}

In previous chapters we studied the basic effects that the temperature and the velocity structure of solar models atmospheres have on the synthetized scattering polarization profiles of the Ca{\sc ii} IR triplet. Now we shall consider the action of the magnetic field as well, trying to obtain information about the influence that the dynamical state of the atmosphere can have on the diagnosis of cromospheric magnetic fields.  To this end, we present the results of solving the NLTE RT problem of the second kind in a 3D solar atmosphere model obtained from a state-of-the-art magneto-hydrodynamical (MHD) simulation. The investigation focuses on the forward scattering geometry and the linear polarization signals produced by Hanle effect. To obtain the NLTE solutions we applied the 1.5D approach with the correct 3D populations supplied by the model at each point. 

In this chapter we will study the spatial distribution and filling factor of the linear polarization signals affected by the dynamic, temperature and magnetic fields in a simulated quiet Sun. One of the aims is to identify the circumstances leading to miss the dynamic amplifications in the observational linear polarization signals. Are we actually missing them? Such effects, explained in previous chapters, have not been clearly captured by the telescopes so far and we want to know why. Furthermore, we will investigate the spatial patterns that the magnetic field and the temperature stratification create in the scattering polarization maps as well as how we could use them for diagnosis purposes. As an exercise, we will try to infer the magnetic field topology from the synthetic maps using the Hanle and Zeeman signals. Finally, we will dedicate a section to evaluate some spectropolarimetric quantities and observational strategies applied to signal detection under the physical situation of our models. 
 
\section{Description of the MHD model and the RT calculations.}\label{sec:mhdmodels}
The atmospheric model used to perform the spectral synthesis correspond to a snapshot of a radiation MHD simulation of the solar atmosphere computed by \cite{Leenaarts:2009} with the Oslo Stagger Code \citep{Hansteen:2007}. This code solves the set of MHD equations that describe the plasma motion together with the RTE. It employs an LTE equation of state and includes non-LTE radiative losses using a multi-group opacity method and thin radiative cooling in the corona and upper chromosphere. It includes thermal conduction along magnetic field lines. The model has $256 \times 128 \times 213$ grid points and a physical size of $16.6  \times  8.3 \times 5.3$ Mm. But, to our aims, we selected a volume\footnote{The exact portion we chose from the full cube found in \cite{Leenaarts:2009} spans from $0$ to $5.85$ Mm in the $x$ direction, from $1$ to $6.98$ Mm in the $y$ direction and from $-0.5$ to $3.5$ Mm along the vertical.} with $5.85 \times 5.98 \times 4$ Mm and $91 \times 93 \times 191$ grid points. The snapshot has a mean magnetic field strength of $120$ G at $300$ Km, which is representative of the magnetization expected in the quiet regions of the solar photosphere \citep{Trujillo-Bueno:2004}.

 The (first-kind) 3D RT calculations performed by the Oslo Stagger Code to obtain the snapshot atmospheric stratifications were done under complete redistribution in frequency for all the spectral lines. We used the solutions of the 3D RT problem for the Ca {\sc ii} level populations as a guess to solve the scattering polarization problem. The electron density was computed assuming LTE ionization for all relevant species. Photoionization by hydrogen Lyman lines was not taken into account.

The hydrogen number density was not available in the supplied model and we had to compute it. To this end, we considered the stratifications of temperature, density and electron number density and we solved the chemical equilibrium and ionization equations for all the relevant atomic species, including hydrogen, as explained in \cite{Asensio-Ramos:2004}. 

We then solved the full RT problem for the Stokes vector by treating each model column as an independent planeparallel atmosphere (1.5D approach), neglecting the effect of horizontal inhomogeneities. The technical details of the core calculations and the computer program developed for performing them are described in Chapter \ref{cap:one}.

Finally, as done in previous chapters, we add a constant microturbulent velocity $\mathrm{v_{\mu}=3.5\, km\cdot s^{-1}}$ to fit the calculated average of the $\lambda8542$ line intensity of the model with the emergent intensity provided by a solar atlas \citep{kurucz_atlas84}.

\subsection{Pre-processing the models. }
\label{sec:preprocess}
 The selected data cube contains $8463$ columns (hereafter, columns or models). In order to decrease the computational load, we adaptatively truncated the colums in height. We estimated the column heights where $\tau^{8498}_{\mathrm{cont}}=10^{3}$ (setting the lower boundary position) and $\tau^{\mathrm{K\,line}}_{\nu_0}=10^{-4}$ (setting the upper boundary position). Thus, we could run the RT calculations with less grid points but without appreciable loss of accuracy in the obtention of the density matrix elements and the Stokes vector. In other words, the boundary conditions (optically thin outside and optically thick inside) are still fulfilled for all the spectral lines.

To obtain the iso-surfaces of optical depth we calculated the line optical depth along the ray with $\mu=1$, at a given frequency, and for each vertical stratification of the data cube, with
\begin{equation}
\tau_{\nu}=- \int \eta_I(z,\nu) dz,\nonumber 
\end{equation}
 approximating the total absorption coefficient as
\begin{equation}
\centering
\eta_I \approx \,{\eta_{I}}^{\rm cont}+ \frac{h\nu}{4\pi}B_{{\ell}u}{\cal
N}_{\ell}\frac{1}{\sqrt{\pi} \Delta\nu_{D}}, \label{coefsi} 
\end{equation}
where ${\eta_{I}}^{\rm cont}$ is the absorption coefficient for the continuum calculated with the same background opacity package we shall use for the RT calculation. The second term on the right hand side of Eq. (\ref{coefsi}) is the line absorption coefficient at line center, with $\cal N_{\ell}$ the overall lower level population of the considered transition and $\Delta\nu_{D}$ is the thermal witdth of the spectral line. This second term is taken equal to zero when calculating $\tau_{\mathrm{cont}}$.  Along this chapter we shall always use the line center optical depth (instead of the integrated optical depth) in order to estimate formation heights. 

We also modified the vertical resolution of the model spatial grid. The spatial resolution of the model is key for obtaining an accurate solution, especially in highly dynamical situations. In the previous chapter, the time-dependent models we used were supplied with an adaptative sampling, so the pre-treatment of the data was almost direct and we did not find any issue during the convergence of the RT calculations. However, that is not the case with the 3D snapshot under study. Here, we have to interpolate along the columns for obtaining a smaller variation of the physical quantities from point to point and thus achieve the convergence of the iterative method. 

The way we interpolate will inevitably affect the iteration process. As is well known, radiative transfer methods based on accelerated $\Lambda$-iteration suffer from a degradation in the convergence rate when the discretization of the atmosphere is very fine \citep{Trujillo-Bueno:1995}. The following dichotomy has to be considered: increasing the number of discrete points along the vertical produces a more precise solution of the RT problem but it also increases the computational time. Due to the large number of columns, a trade off between both effects must be reached. 

Initially, we chose the option of adding the minimum number of points in order to fulfill  the requeriments imposed by the velocity gradients (Section \ref{sec:dynamic_traviata}). However, in regions where velocity gradients are large, the total number of points in the vertical direction can still be too large after the interpolation. To avoid the presence of too many points, we tried to decrease their number in the atmospheric regions with smooth physical variations, typically in the \textit{plateau} around the minimum of temperature. A code (called Doppler Inspector) was written to do this before applying the RT code.  

\begin{figure}[htb!]
\centering%
\includegraphics[scale=0.7]{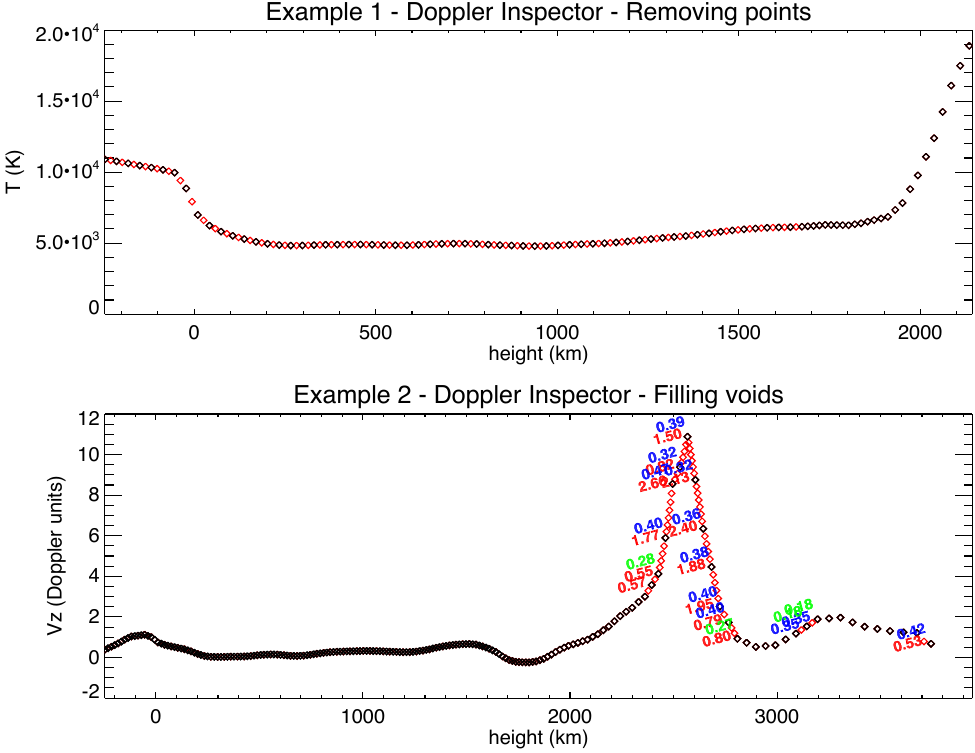}
\caption{Illustrative examples of some models that needed a pre-processing. Upper panel: stratification of temperature showing (in red color) the grid points that were removed to accelerate the performance of the RT code. Lower panel: stratification of vertical velocity $\mathrm{V_z}$(in Doppler units) showing (in red color) the grid points that were added to diminish the velocity interval between them. The numbers in red are the values of such interval (in Doppler units) before the interpolation. Numbers in blue and green give the corresponding values after interpolating (blue $\rightarrow\Delta \mathrm{V_z}\in\left[0.3,0.5\right]$; green $\rightarrow\Delta\mathrm{V_z}\in\left[0,0.3\right)$), what helps to quickly identify if the correction is valid.}
\label{fig:doppins18}
\end{figure}

With the Doppler Inspector we can identify atmospheric regions with small velocity gradients and remove there alternate points if necessary (Fig. \ref{fig:doppins18}, upper panel). We can also identify places where the velocity gradient exceeds the threshold of $0.5$ Doppler units (see Section \ref{sec:numeric} or \ref{sec:dynamic_traviata}) and add there as many points as necessary (Fig. \ref{fig:doppins18}, lower panel). We employed linear interpolation to introduce new grid points.
In this approach to a better interpolation grid we face one of the limitations of working with non-adaptative discretization methods. In general, the procedure that we have explained leads to convergence in most of the models but not in all of them. However, in some models convergence cannot be obtained, regardless of the number of interpolation points. When the grid is not locally adapted to the physical conditions, there will be\textit{ a gradient of numerical error} along each ray or characteristic. To mantain the stability of the iterative process, such error has to be similar everywhere in the atmosphere or, at least, to vary smoothly. Otherwise, the (always existing) imprecisions deriving from any integration method for the RT equation can increase without control. These imprecisions in local values of the specific intensity and source function propagate through the numerical domain and eventually develop discontinuities that do not fit into the polynomial variation assumed by the short-characteristics scheme. To alleviate this problem we can improve the interpolation approach of the formal solver (using higher-order polynomials) but, as the grid is not adaptative, the numerical error gradient still exist. In the worst case, convergence will not be attained. In the best case, the solution is achieved, but the time to reach the convergence will never be optimal because the iteration process is hampered and delayed by the errors resulting from the use of a non-adaptative grid. None of the options (including oversampling) is satisfactory, specially with the high number of RT problems that we have to solve in the 3D cube.      

These problems can be overcome in several ways. Using multigrid methods, for example, we can minimize the number of iterations and assure convergence by putting a heavier weight on the grid-related calculations at each iterative step \citep{Hackbusch:1985aa, Wesseling:1995aa}. Since a formal implementation of such methods is out of the scope of this thesis, we developed a computationally cheaper option inspired by the finite elements methods (FEM) to obtain a better performance and convergence. A detailed explanation about it was given in Sec. \ref{sec:fem}.

\subsection{Description of physical quantities in the model.}
\label{sec:physicmodel}
\begin{figure}[h!]
\centering%
\includegraphics[scale=0.25]{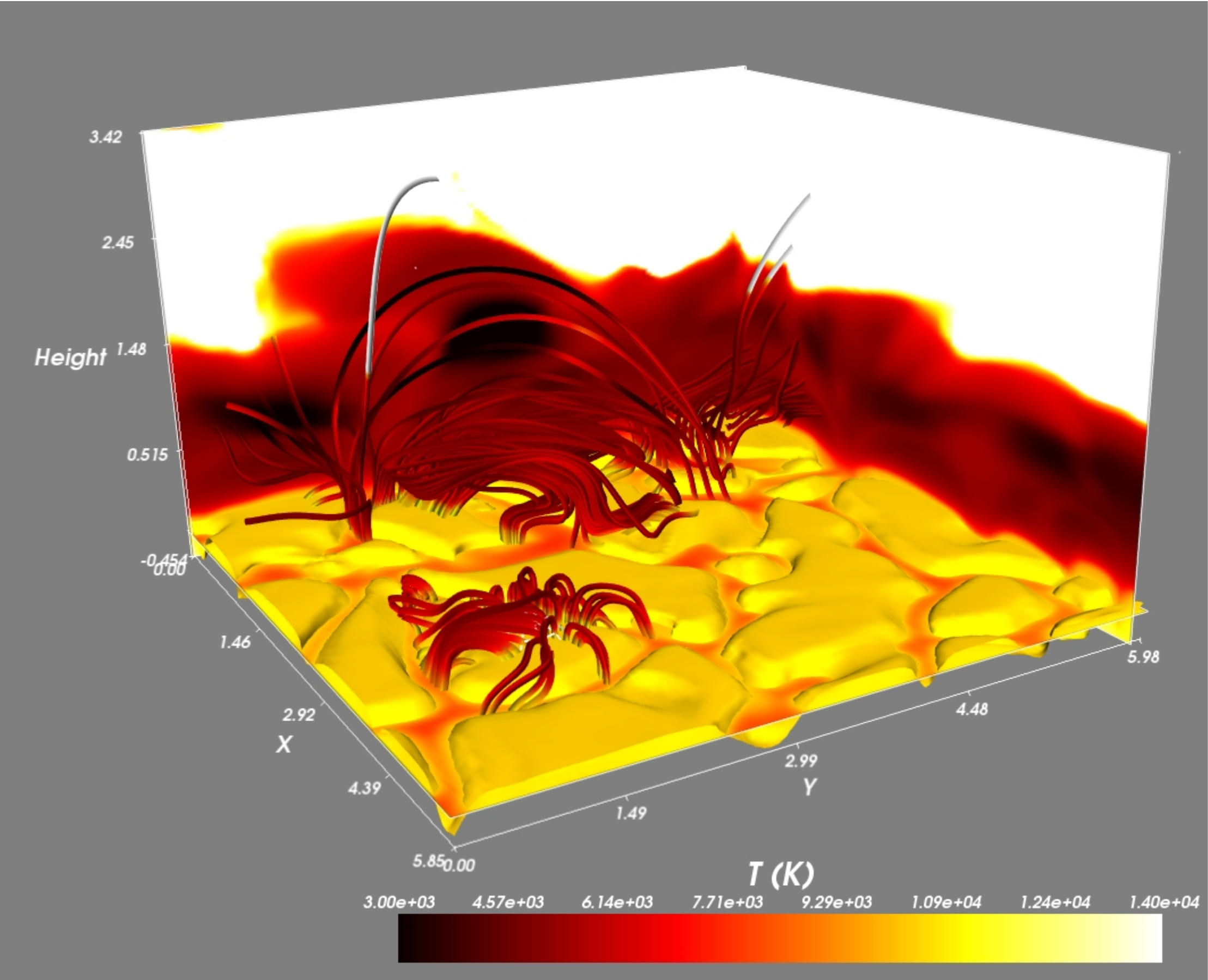}
\caption{Three-dimensional visualization of the data cube temperature. The tubes represent the magnetic field lines (background tubes) and the velocity (foreground tubes).  }
\label{fig:cuboT}
\end{figure}

\begin{figure}[ht!]
\centering%
\includegraphics[scale=0.24]{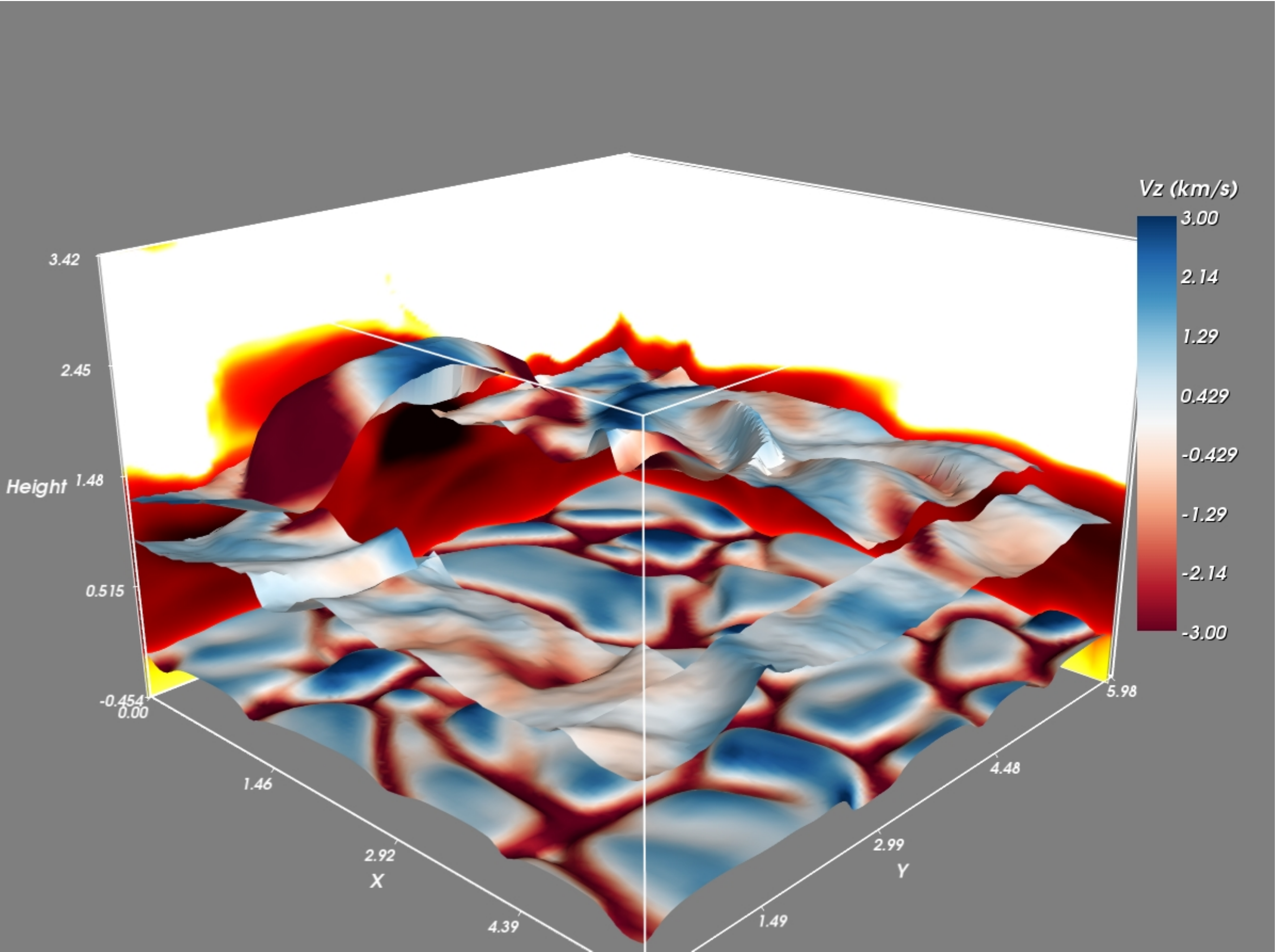}
\caption{Three-dimensional visualization of the vertical velocity in the data cube. The background planes represent the temperature variation and give a reference about where the chromosphere ends. The component $\mathrm{V_{z}}$ is represented by the color of the surfaces, showing between $-3\,\mathrm{km \,s^{-1}}$ and $3\,\mathrm{km \,s^{-1}}$. The corrugated horizontal surface with L-shape nearest to the reader is defined by the heights of $\bar{\tau}=1$ for the $8498$ {\AA} line. The other two corrugated surfaces are the same for $8542$ {\AA} (towards the right) and the K line (the big ``bubble'' shape to the left).}
\label{fig:cubovz}
\end{figure}

In the selected data volume we can observe a number of interesting features
associated to the variations in temperature, velocity and magnetic field. A good representation of the data set is provided by the temperature stratification.  In Fig. \ref{fig:cuboT} we visualize the deep layers of the model photosphere, with the isosurface of $\mathrm{T=10300}$ K enclosing the hotest part of the granules. The background planes reveal the vertical variation of the temperature until the corona (white region). In them, we identify dark regions as bubbles of minimum temperature (cool bubbles), followed by the variable chromosphere and the abrupt temperature increase of the transition region located at the beginning of the white region. The isolated tubes appearing in Figure \ref{fig:cuboT} like trapping a granule correspond to some selected velocity field lines. Their colors indicate the local temperature. They show how just above the granules the photospheric plasma gets cooler and turns downdraft through the intergranules. The magnetic field lines near the Y-Z plane also have colors indicating the temperature. 
In Fig. \ref{fig:cubovz}, we draw some portions of the isosurfaces of optical depth unity for the IR triplet lines. These corrugated surfaces are being modified at each point by the vertical velocity field (whose modulus is indicated with the color of the surface). We see that there is not a fixed height associated with the chromosphere, but a vertical range in which the plasma has the values of density, temperature and magnetic field that we identify as chromospheric layers. This region is constantly bubbling up and down as a consequence of the upward motions originated from the photosphere and amplified by the density gradient along the chromosphere. The increment of the vertical velocity is thus a result of the conservation of energy in the fluid. So, the macroscopic motion pass to be temperature-activated in the photosphere and below (granular convection) to density-enhanced from the chromosphere upward, leading to an expansion and a radiative cooling. The process is continued by gravity during the chomospheric downflows, in which the potential energy is transformed into kinetic and internal energy, arising temperature again. The model is a snapshot of that process.
\begin{figure}[h!]
        \centering
        \begin{subfigure}{0.49\textwidth}
                \centering
                \includegraphics [width=\textwidth]{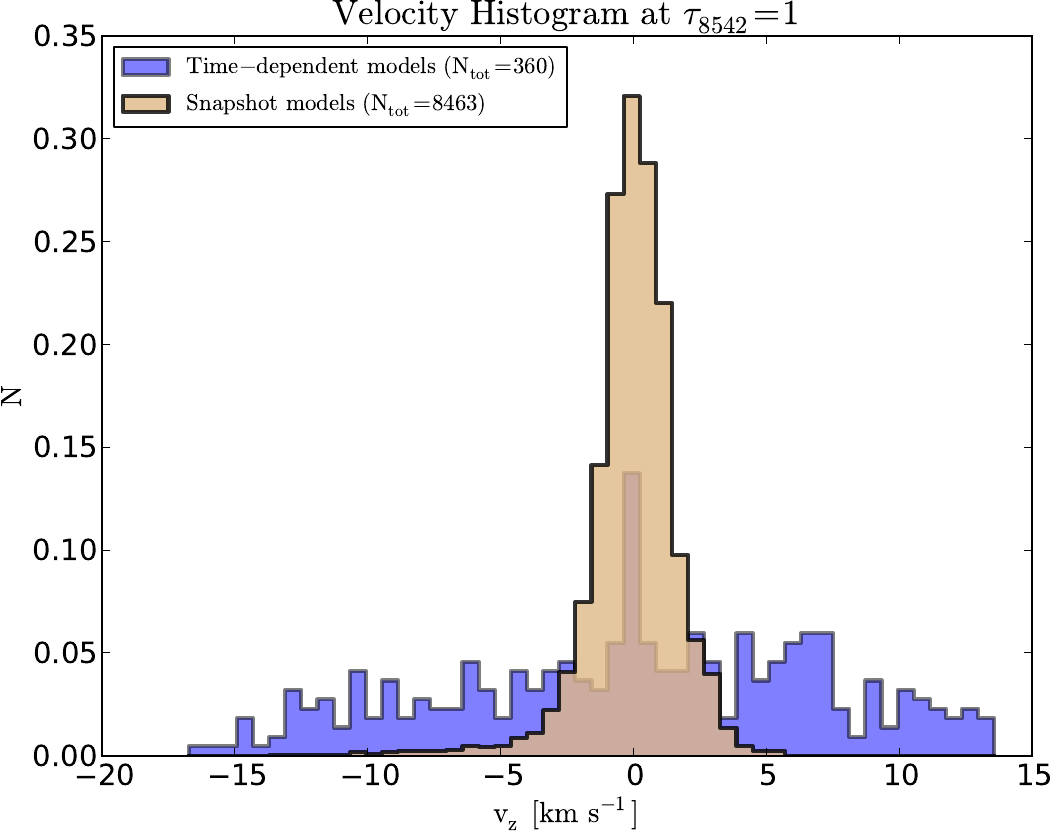}
        \end{subfigure}
        \begin{subfigure}{0.49\textwidth}
                \centering
                \includegraphics [width=\textwidth]{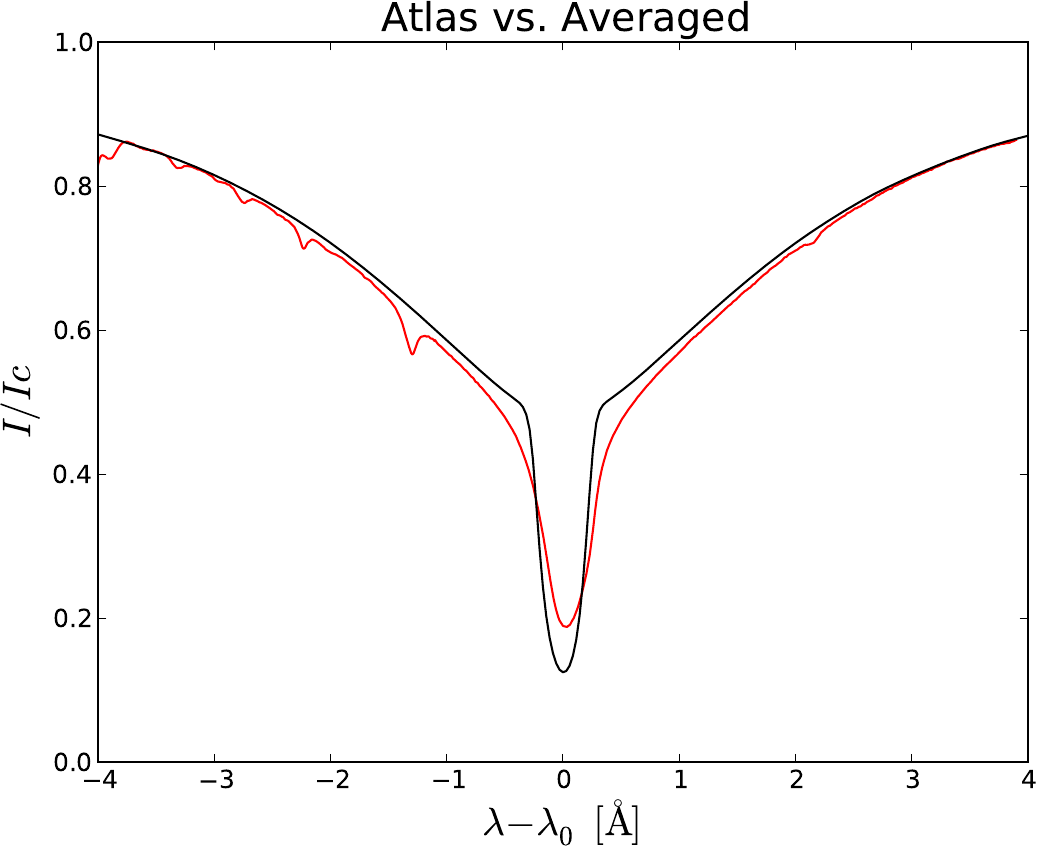}
        \end{subfigure}
        \caption{\textbf{Left:} vertical velocity distributions (normalized to unit area) of the 3D model analyzed in this chapter (orange) and of the strongly-dynamic 1D dataset analyzed in Chapter \ref{cap:three} (violet). The velocities are taken at $\tau=1$ for the $8542$ {\AA} line.  \textbf{Right:} Synthetic average intensity using a microturbulent velocity of $\mathrm{3.5 \,km\cdot s^{-1}}$ (black line) and average intensity extracted from the Kitt Peak Solar Atlas (red line). Case of the $8542$ {\AA} line at disk center. }
\label{fig:uves}
\end{figure}
\begin{figure}[h!]
\centering%
\includegraphics[scale=0.27]{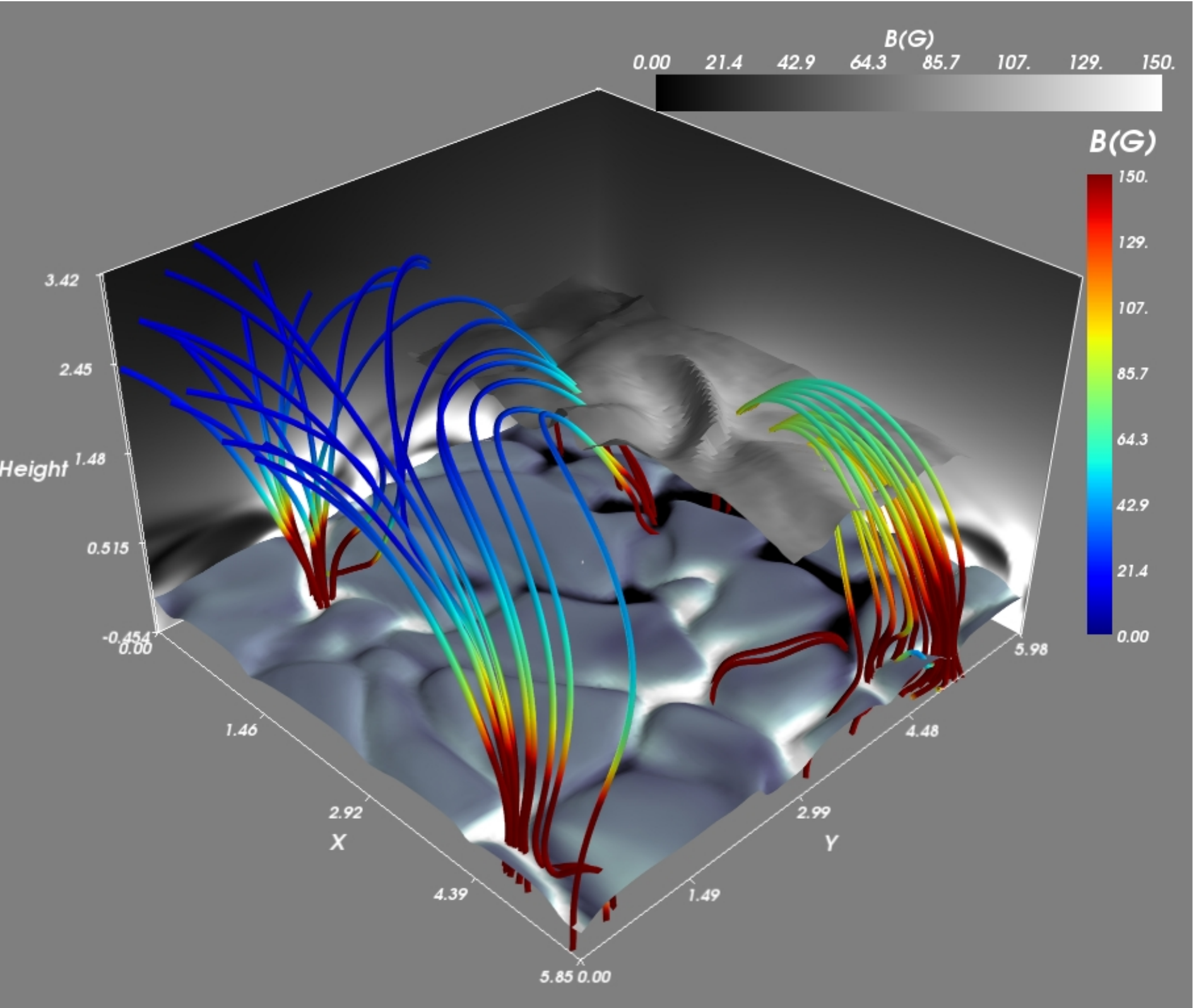}
\caption{Visualization of the magnetic field topology of the data cube. The colors in the granules surface have no colorbar, it just represent the component $\mathrm{B_{z}}$, saturated between $-500$ G (black) and $500$ G (white) to identify the photospheric flux concentrations. The rest of surfaces and lines are coloured with the magnetic field intensity, saturated to $150$ G. The tubes represent some magnetic field lines. The corrugated horizontal surface is defined by the heights of $\tau=1$ for the $8542$ {\AA} line.}
\label{fig:cubobz}
\end{figure}
The maximum velocity in the photosphere is around $4.3\,\mathrm{km \,s^{-1}}$. 
 The velocity stratification patterns are similar to the ones in the time-dependent models of Chapter \ref{cap:three}, but they differ notably in the maximum values and their statistical ocurrence. They do not have to coincide because one model is a temporal evolution and the other is a spatial mapping. Furthermore, we also have to realize that the LOS chosen in these new calculations is $\mu=1.0$, while the calculations in previous chapters were for $\mu=0.1$. Because of it, the emergent spectral lines will be now more sensitive to lower layers than they would be for a limb observation\footnote{Observations done at the limb ($\mu=0.1$) capture more strongly the behavior of higher layers because the optical path along a LOS towards the photosphere is always larger than for a disk center observation ($\mu=1.0$).}. Among other things, this means that the vertical velocities in the formation region of the spectral lines will tend to be smaller and the magnetic fields will tend to be stronger. Normalized histograms of chromospheric velocity allow a comparison between both set of models (fig. \ref{fig:uves}). The maximum velocity in the upper chromosphere reaches $12\,\mathrm{km \,s^{-1}}$, but such large values are scarce in its spatial distribution, being a value of around $5\,\mathrm{km \,s^{-1}}$ more representative. With these velocities we would not expect significant linear polarization amplifications induced by dynamics. However, we also have to take into account that the velocity component along the line of sight will be larger now because we look at disk center, directly in the direction of the motion. 

Computing the emergent intensity and comparing with observations, the authors of the atmospheric model \citep{Leenaarts:2009} extract some relevant conclusions that we reproduce here. The standard deviation of the central wavelength of the $8542$ {\AA} intensity  profiles with respect to the central wavelength of the average intensity is $\sim 50 \%$ less than in real solar observations. On one hand, this means that there is a lack of macroscopic dynamics in the model, something to remember when analyzing the polarization profiles. But they also point out that the main reason of the difference is that the individual profiles are already narrower than in the Sun, implying that the real chromosphere is also more dynamic on scales not resolved by the simulation. Hence, the added microturbulent velocity is again justified in our calculations.

Concerning the magnetic field, Fig.\ref{fig:cubobz} illustrate a detail of the general topology of the magnetic field lines and intensities. Photospheric intergranular regions with high concentrations of magnetic flux and different polarities are the source of magnetic loops crossing the cromosphere. The magnetic intensities have typical values of a quiet Sun region. 

\begin{figure}[h!]
\centering%
\includegraphics[scale=0.5]{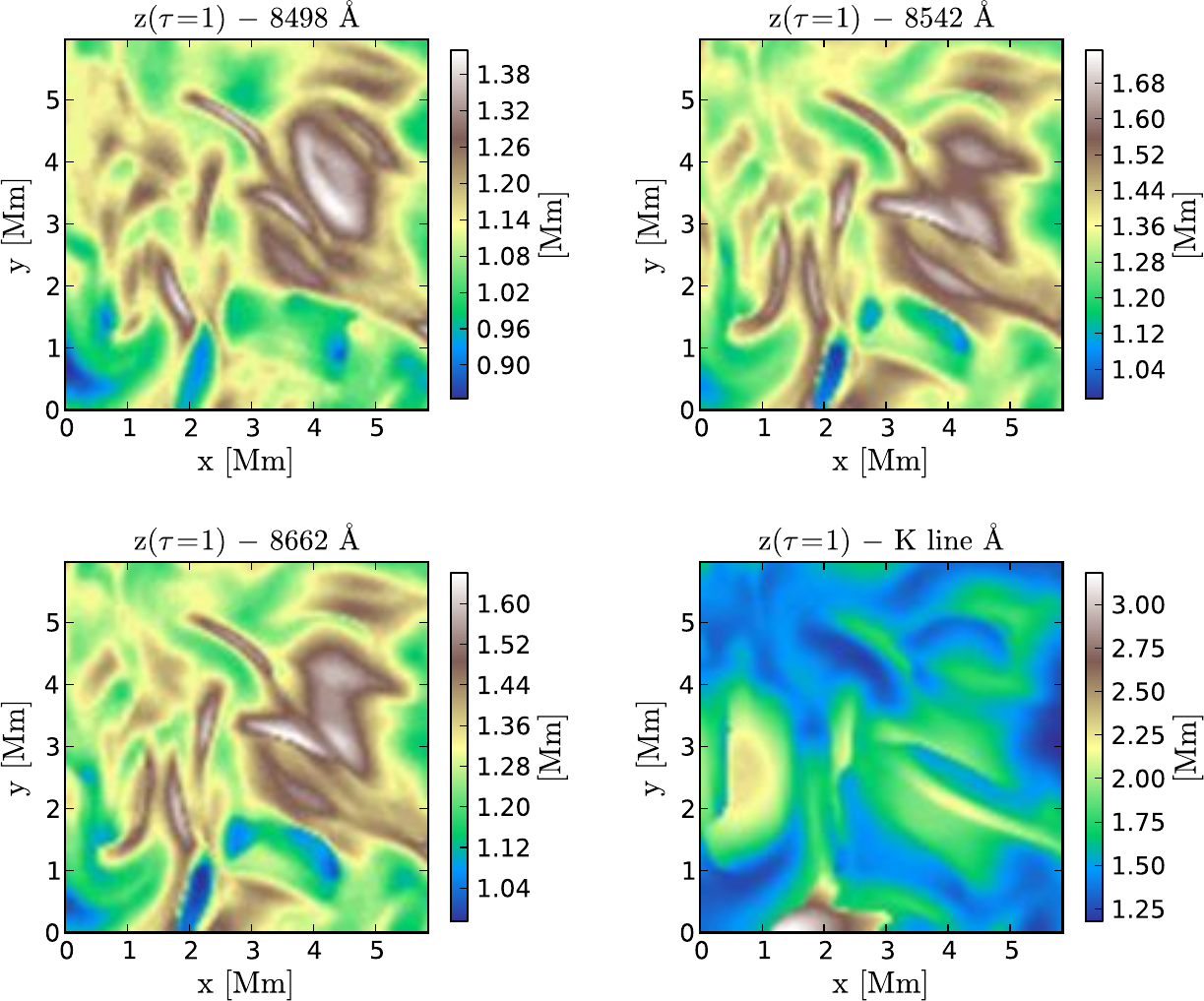}
\caption{\small{Heights of $\tau=1$ for the spectral lines under consideration. We can use the K line picture to identify expanded or compressed atmospheres because there is a correlation between this feature and the heights where the K line optical depth is unity.} }
\label{fig:toneh}
\end{figure}

A good way of grouping much of this information is drawing the main physical quantities (Fig.  \ref{fig:vtb8542} and Figs. \ref{fig:vtb8498}, \ref{fig:vtb8662} in Appendix \ref{app:D}) at the approximated formation heights for each line. The heights at optical depth unity are shown in Fig. \ref{fig:toneh}. 
\begin{figure}[t!]
\centering%
\includegraphics[scale=0.8]{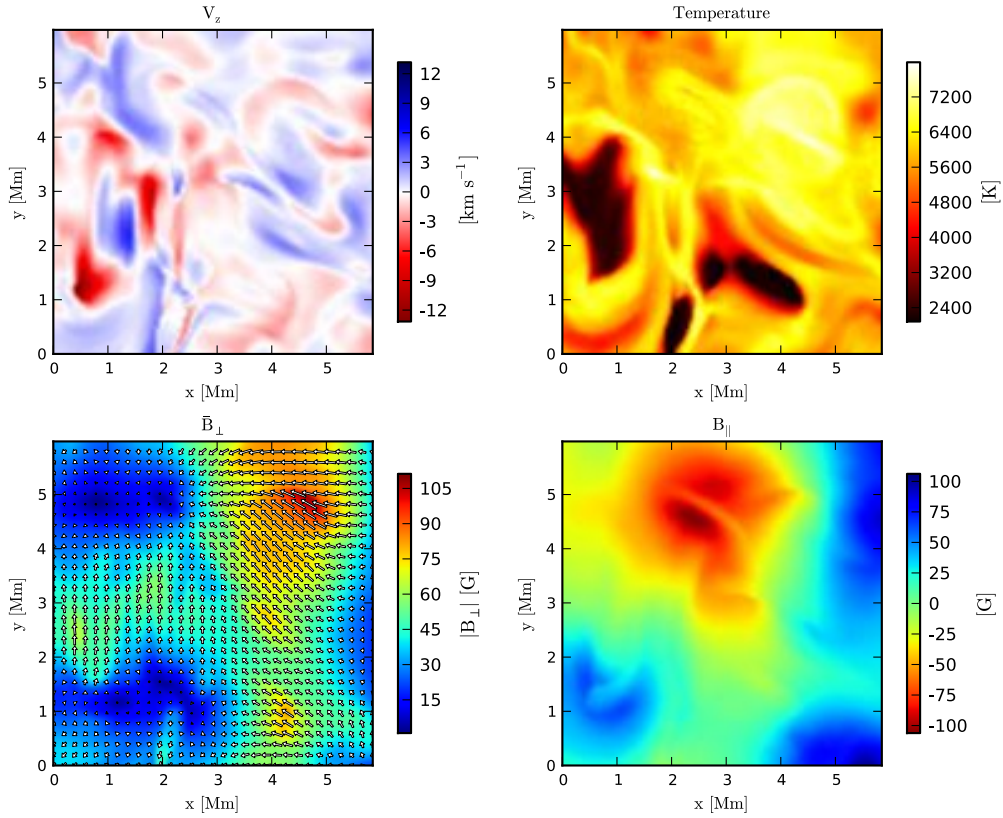}
\caption{Physical quantities at $\tau=1$ for the $8542 \,$ {\AA} line. Upper left: vertical velocity. Upper right: temperature distribution. Note the cool patches. Lower left: horizontal magnetic field intensity. Lower right: longitudinal magnetic field intensity. }
\label{fig:vtb8542}
\end{figure}

In the next section, we will need to clasify each model into three simple thermodynamic categories: static FALC-like models, which are an intermediate reference, having a hot chromosphere from the minimum of temperature to the TR, but without extra heatings like shocks and compressions and without extra cool or extended plateaus of minimum temperature; hot models, typically more compressed and hotter than FALC; and cool models, typically more expanded and cooler than the others. They are very poor representations of the large variability found in the dataset but have to be understood as extreme cases where some of the features we describe in the following sections happen very frecuently. Each of these categories can have a chromosphere significantly moving upward, significantly moving downward or without a significant movement. 

\section{Synthetic polarization profiles in forward scattering.}\label{sec:profiles}
In forward scattering geometry at disk center, the Hanle effect can create polarization in presence of inclined magnetic fields (Section \ref{sec:introhanle}). The symmetry breaking produced by such magnetic fields then gives us access to the thermodynamic and magnetic state of the plasma through the Hanle-induced Stokes Q and U signals and the effect that the anisotropic illumination has in the atomic system. In the following, we analyze the results of synthetizing the Stokes profiles in the 3D atmospheric models under such circumstances.
 
\subsection{Individual profiles.}\label{sec:profilesA}
\begin{figure}[h!]
\centering%
\includegraphics[width=1.0\textwidth]{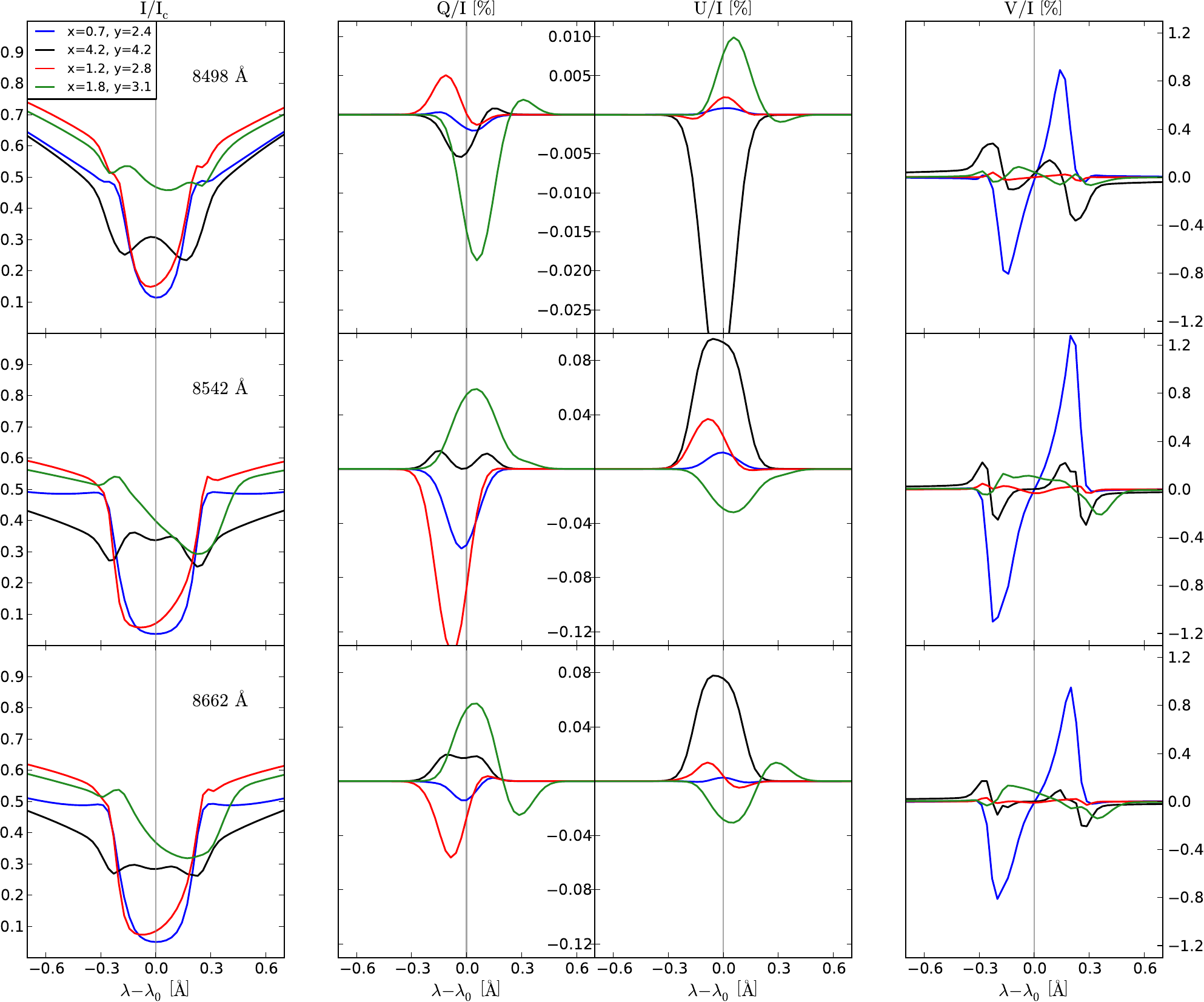}
\caption{Examples of typical Stokes profiles of the IR triplet lines at four random pixels (different color means different pixel) where the magnetic field is predominantly horizontal at the chromosphere. }
\label{fig:profiles_a}
\end{figure}

From each pixel/column of the model we obtained a Stokes vector (I,Q,U,V) as a function of wavelength for each spectral line of the atomic model. The variety of these spectral profiles is wide. A first classification of the emergent Stokes profiles can be done according to the inclinations of the magnetic fields producing them. The set of Stokes vectors with significant linear polarization is caused mainly by a predominantly horizontal magnetic field (say, with $\mathrm{|\cos{\theta_B}|< 1/\sqrt{3}}$). They define what we will call Horizontal Field (HF) regions, which have magnetic field lines with the Van-Vleck inclination at their frontiers and are located between areas whose magnetic vectors are mainly vertical. Some examples of the profiles in the HF regions are shown in Fig. \ref{fig:profiles_a}. Here, in areas with enough magnetic field inclination in the chromosphere, the pixels showing higher Doppler shifts have also larger amplitudes and qualitatively reproduce some features of the Doppler-induced modulations analyzed in previous chapters. The complementary regions will be called Vertical Field (VF) regions because the chromospheric magnetic field is there predominantly vertical. They are characterized by Stokes vectors with small Q and U but large V. 

 In general, the profile shapes in the HF regions resemble the ones analyzed in previous chapters (see Fig. \ref{fig:gradsv_ctes}). However, the shape of the linear polarization profile that was frequently associated with the $8498$ {\AA} line appears now interchanged with the shapes of the $8542$ {\AA} and $8662$ {\AA} profiles in many pixels of the snapshot models. Namely, the $\lambda8498$ Q$/$I signal resulting from FALC had always a valley\footnote{As in Chapter \ref{cap:two}, with \textit{valley} we refer to the double-peak pattern with a valley in the middle. A negative or positive single peak is simply a \textit{peak}.} at line center and the $8542$ {\AA} and $8662$ {\AA} had always a single peak. Now, the shapes are often interchanged, with the $8542$ and $8662$ {\AA} lines presenting a valley and the $8498$ line presenting a peak (black curves in Fig. \ref{fig:profiles_a}). 

Such changes are not intrinsically produced by the velocity because they are also present when the calculation is repeated in static regime. As we explain next, the variations appear when passing from limb observation to disk-center observation, but are not always visible because the Doppler shifts and the thermodynamic effects on the anisotropy also play a role. 

With Fig. \ref{fig:rhosqi2} and Eq. (\ref{eq:qestimated2}), we can understand such intrinsic shape changes related with the LOS. In that figure, the vertical variations of the upper and lower level fractional alignments are fixed by the anisotropy of the radiation field, which is intrinsic to the model and does not depend on the LOS. Then:

1) When the inclination of the LOS varies, the main height of formation also does, so it can be above or below the \textit{cross point} where the curves of the upper and lower level fractional alignments intersect (hereafter, the alignments intersection).

2) The key point is that the emerging fractional polarization at each wavelength is the result of a competition between upper and lower level alignments, whose net contributions at different layers can be positive or negative. A simplification of such competition is given by Eq. \ref{eq:qestimated2}, which only considers the layer at $\tau=1$. Thus, at layers below the alignments intersection, the net contribution of both levels alignment to the emergent polarization has a certain sign. But above the cross point, such contribution has opposite sign. 

3) If most of the layers contributing to the emergent spectral polarization in the formation region are well above (or well below) the cross point, their contributions to the fractional polarization will have the same sign everywhere. The net result will be a profile shape monotonic from the continuum to the line center. Hence, a \textit{single peak} profile.

4) However, if the height of formation is near the alignments intersection, the contributing layers situated just above this point and the ones immediatly below will have opposite signs (they somewhat cancel out). Since they form the spectral line core, the result is a depression in the emergent linear fractional polarization at line center, hence, a \textit{valley}.



In conclusion, the shape (two peaks versus one peak) of these spectral lines is modified by the relative position of the main region of formation with respect to the height of the alignments intersection. This fact could be used to evaluate the realism of the model atmospheres or to compare among them in certain observational configurations of reference, with the caution that the velocity gradients (or any other modifier to the anisotropy) are able to add their own effects. 

\begin{figure}[h!]
\centering%
\includegraphics[width=1.0\textwidth]{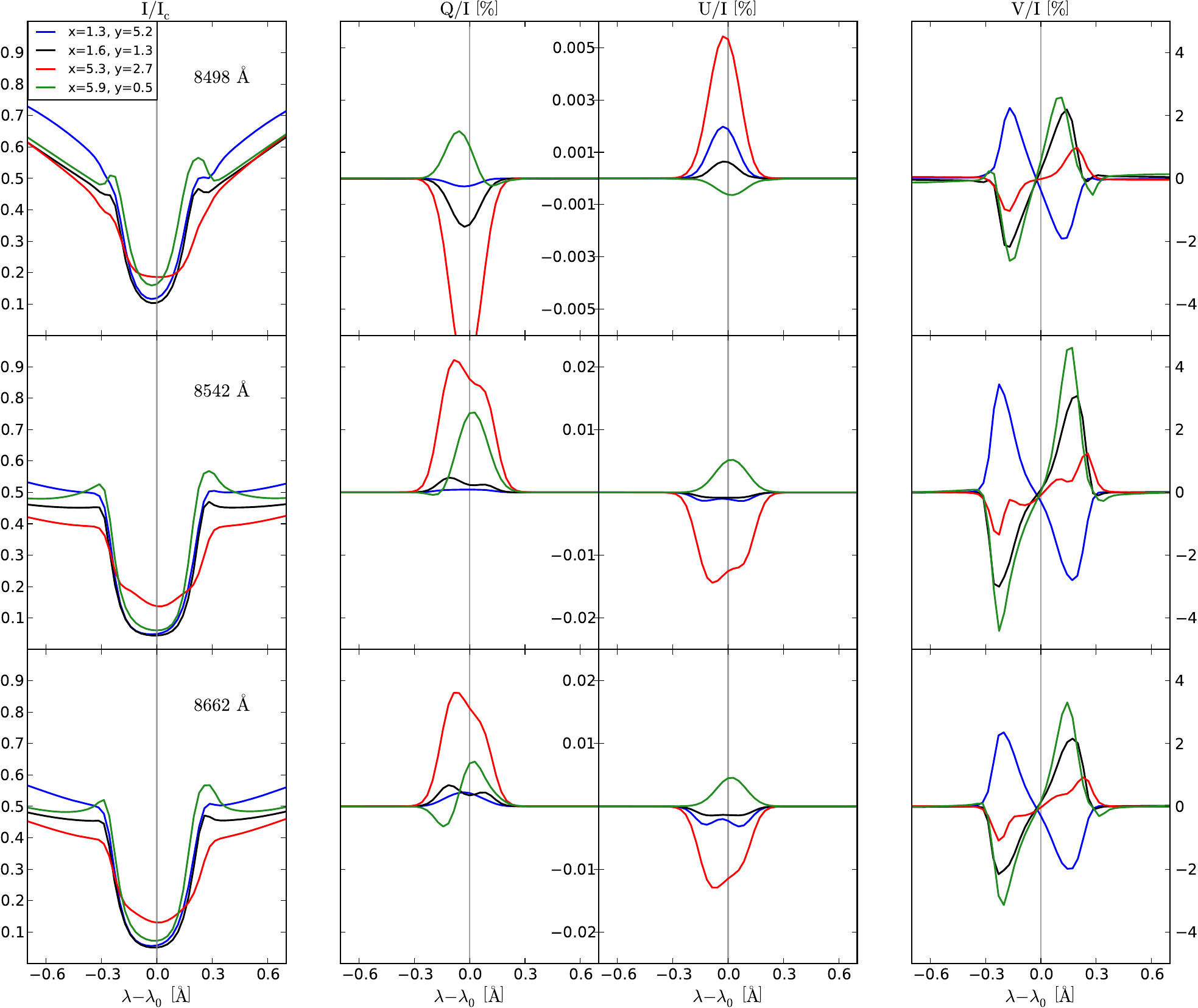}
\caption{Examples of typical Stokes profiles of the IR triplet lines at four random pixels (different color means different pixel) where the magnetic field is predominantly vertical. }
\label{fig:profiles_b}
\end{figure}

In addition to the linear polarization, we can study the Stokes V profiles generated by the Zeeman effect of the longitudinal field component (Fig. \ref{fig:profiles_b}). Their variability is also rich. It is frequent to find V profiles with two antisymmetric lobes whose signs depend on the sign of the longitudinal magnetic field component. The asymmetries they show are due to the coexistence of vertical velocity and magnetic field gradients, as is well-known from previous works \citep{Sanchez-Almeida:1989,LL04}. 
More complicated patterns arise in the shape of Stokes V (red profiles in Fig. \ref{fig:profiles_b}) in the presence of velocity gradients (intensity shape very assymetric) or heatings in the plasma (intensity in emission). 

\clearpage

\subsection{Slit profiles.}\label{sec:profilesD}
\begin{figure}[h!]
  \centering%
\includegraphics[scale=0.5]{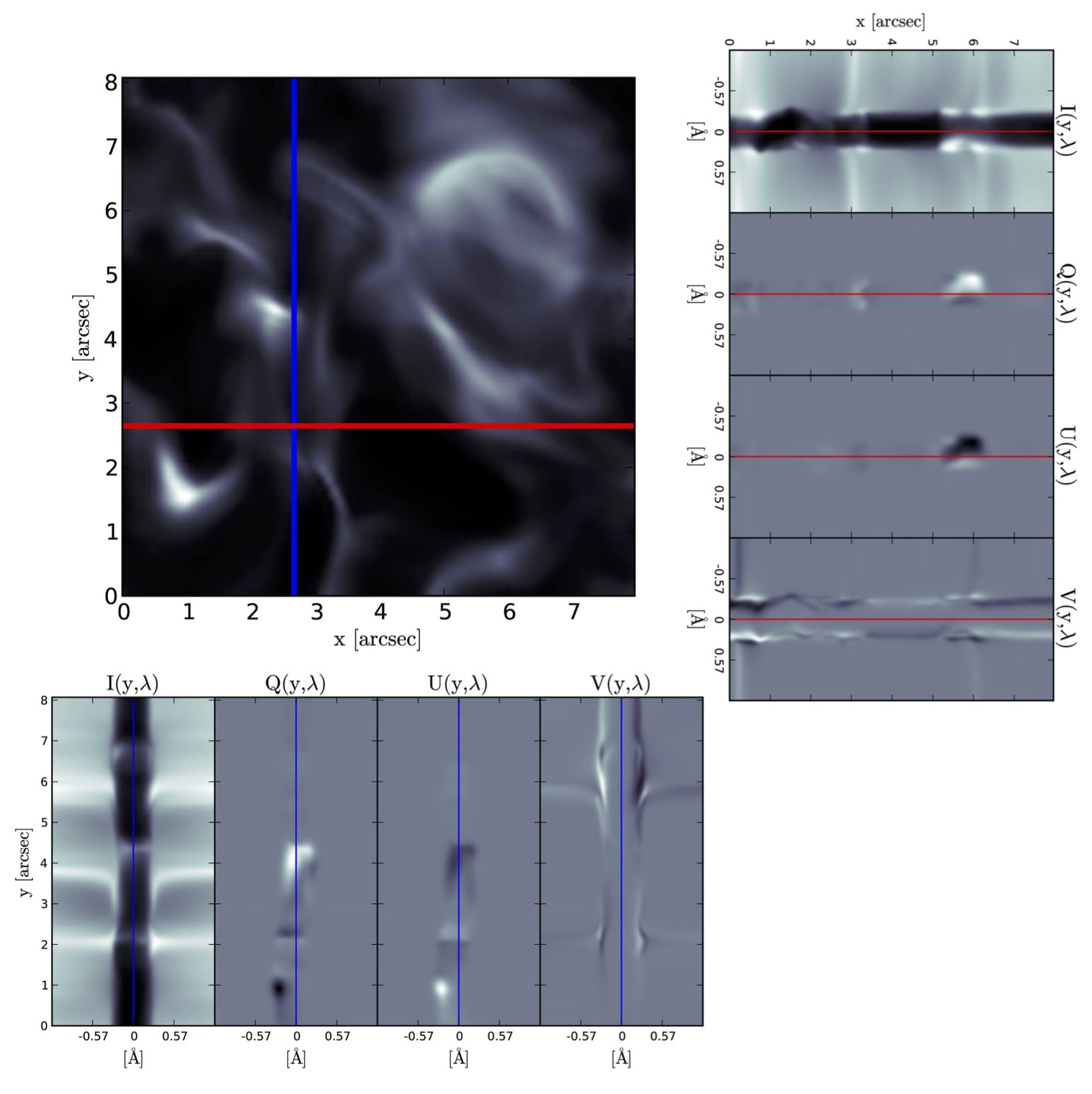}
\caption{Emergent intensity map at $8542$ {\AA} and two sets of synthetic slit-like Stokes profiles along the near-core wavelengths. }
\label{fig:slitA}
\end{figure}

 Simulating the profiles along a spectrograph's slit, we can take a quick look to many spectral profiles at the same time. In Fig. \ref{fig:slitA}, we show the synthetic map of the Ca {\sc ii} $\lambda8542$ intensity at line center, with coloured lines representing two slits. The corresponding slit profiles in the adjacent panels are identified with the coloured line at central wavelength. We have analyzed their main features using the \textit{Solar Inspector} tool (see Sec. \ref{sec:chromocube}).

 The intensity spectra of Fig. \ref{fig:slitA} show regions with emission in the wings and in the core, not always coinciding both features in the same profile. These intensity increments have different origins. The extra emission in the wings are due to heating by photospheric compression (sometimes associated with a photospheric bright point). In such a case, the downward plasma is compressing the photospheric layers while most of the low chromosphere and the temperature minimum region are almost at rest. When a pixel is associated with a bright point there is an additional increment of the magnetic field strength in lower atmospheric regions that produce large Stokes V signals. That is the case at $\mathrm{y\sim2\arcsec}$  and  $\mathrm{y\sim 6\arcsec}$ in the blue slit.

The extra emission in the line core (e.g., at $\mathrm{y\sim 4.5\arcsec}$) is linked to Doppler shifts produced by motions in the chromosphere. In expanding atmospheres with motions located at the lower chromosphere, the blue shifts produced in the spectral radiation make those layers more transparent at line center wavelengths. Consequently, the excess of  photons in upper layers produce medium and faint extra emissions in the intensity core. If medium/large velocities are compressing and shifting the chromosphere downward the effect is similar, but additionally the compression heats the plasma and can make the intensity to be completely in emission.    

In pixels where the temperature gradient along the chromosphere is high enough, we also can appreciate reversal peaks at both sides of the line core intensities in the strongest IR lines ($\mathrm{y\sim 3\arcsec}$). These features are similar to the K2 and H2 reversals appearing in the solar H and K lines \citep{Linsky:1970aa} forming at top chromospheric layers in the Sun, but do not appear in solar observations of the $8542$ and $8662$ {\AA} lines. Their presence in our results can indicate an unrealistic temperature stratification in the models (or a lack of temporal and spatial resolution in the observations). Other option is that they appear in our calculations because the microturbulent velocity is constant with height. Effectively, a variable microturbulent velocity at heights where the source function decouples from the Planck function can mixes the contributions from different layers and suppresses the reversals.  

Some profiles with many of the mentioned features are shown in the blue slit panel of Fig.  \ref{fig:slitA}, at $\mathrm{y\sim 2\arcsec}$. It corresponds to a region over a bright point. We see intensity emission in the core as well as in the wings. Accordingly, the atmospheric stratification shows a photospheric heating by compression (downward velocities below the temperature minimum) with upward velocities in the lower chromosphere that produce blue shifted radiation. A relatively intense \textit{vertical} magnetic field at the upper \textit{photosphere} is producing a notable Stokes V signal. Let us remind that the response function to the magnetic field in Stokes V is larger in lower layers for the $8542$ {\AA} line. On the other hand, the abrupt change of sign in Stokes Q and U at $\mathrm{y\sim 2\arcsec}$ has nothing to do with the previous reasons. It is indeed due to a weak magnetic field inclined almost horizontally in the chomosphere, whose azimuth changes from one pixel (showing $Q>0$ for example) to the next one (showing opposite sign). Thus, one Stokes vector can give clear information about the main physical magnitudes both at the chromosphere and the photosphere.

The largest linear polarization signals in such ``slits'' of Fig. \ref{fig:slitA} are mainly due to  velocity gradients ocurring at places with a chromospheric magnetic field that is significantly horizontal. For instance, in the blue-slit panel, this happens at $\mathrm{y\sim 1\arcsec}$ and $\mathrm{y\sim 4\arcsec}$. The difference between the LP features at those points is that the latter corresponds to a compressive atmosphere and the former to an expansive one, which is less efficient increasing the anisotropy of the radiation field, but has in this case a strong velocity. In these pixels, the magnetic field strength is weak ($\mathrm{\sim 20 G}$) at the chromosphere but also at the upper photosphere, which explains the absence of circular polarization.
\clearpage
\section{Maps of radiative quantities.}\label{sec:maps}
In this section we describe in more detail the spatial variations of the scattering polarization amplitudes taking into account different maps. Let us remind that in the forward scattering geometry we are considering in this chapter, the Stokes Q and U signals produced by an inclined deterministic magnetic field allow us to estimate the azimuth $\chi_B$ of the magnetic field. Setting the orientation of the LOS along the y axis ($\chi=-\pi/2$) in Eq. (\ref{eq:step8}) we have:
\begin{equation}\label{eq:chiuq}
 \centering
    2\,\chi_B = \, \arctan{\left(\frac{U}{Q}\right)} .
\end{equation}
\begin{figure}[h!]
\centering%
\includegraphics[scale=0.50]{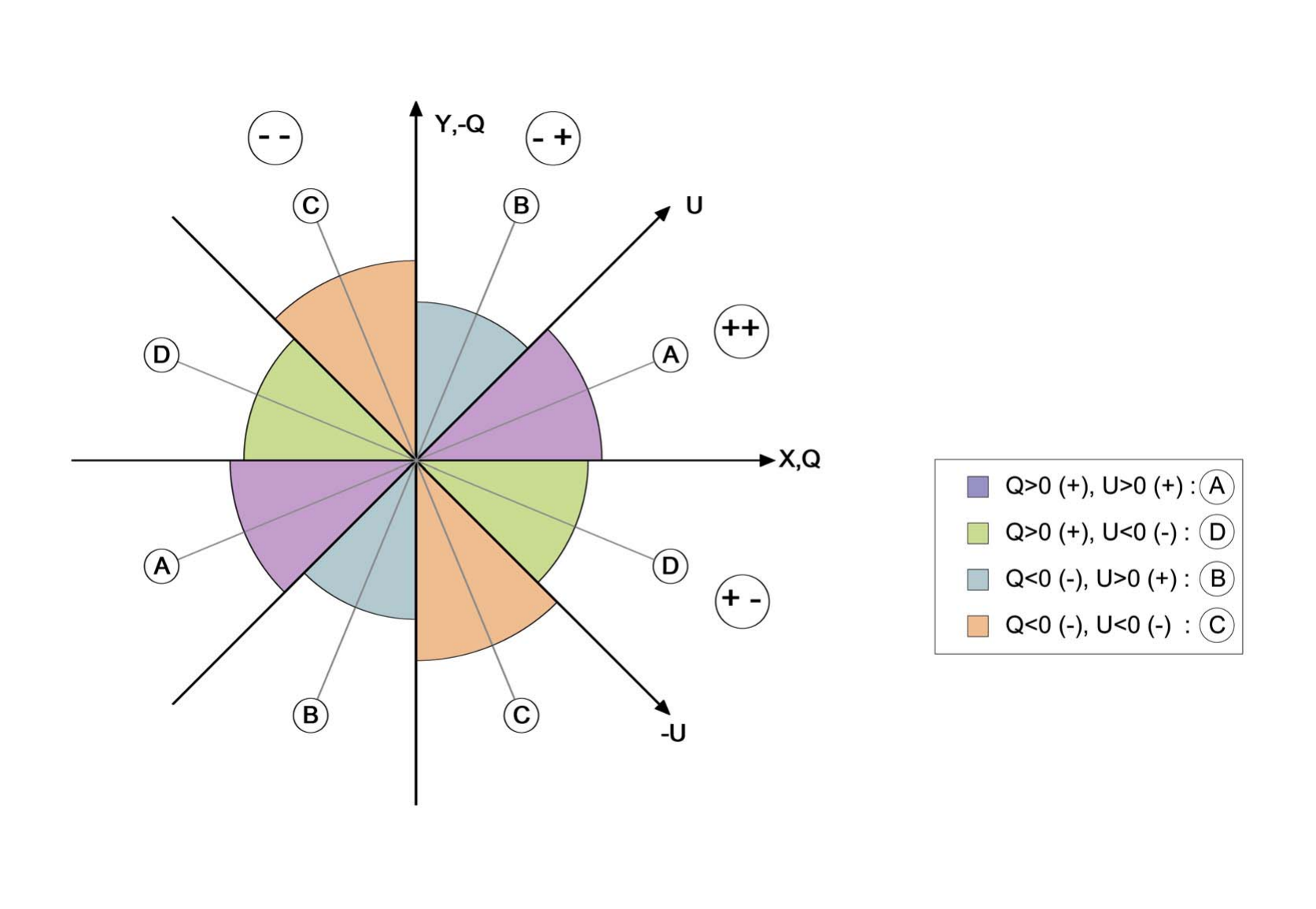}
\caption{Observer reference frame for the linear polarization in forward scattering. The axes for Stokes Q and U specify the directions along which they are positive and negative in the reference system. The polarization plane is divided in coloured regions where the signs of the coordinates (Q,U) do not change. The directions A, B, C, D defined in those regions represent the $180\deg$ ambiguity.     }
\label{fig:qusigns}
\end{figure}

 The Hanle effect in forward scattering produces linear polarization along the projection of the magnetic field vector on the solar surface or along the perpendicular direction. In our calculations, we have chosen the positive reference for Q and U as indicated in Fig. \ref{fig:qusigns}, positive Q lying along the x axis (horizontal) and positive U inclined at $45$ degrees from the x axis. The directions of negative reference are perpendicular to the positive ones. As polarimetry is differential photometry, Stokes Q and U will be the difference between the number of photons oscillating in the corresponding positive axis and the number of them oscillating perpendicularly. 


\subsection{Maps of polarization amplitudes.}\label{sec:ampmaps}
\begin{figure}[h!]
\centering%
\includegraphics[width=\textwidth]{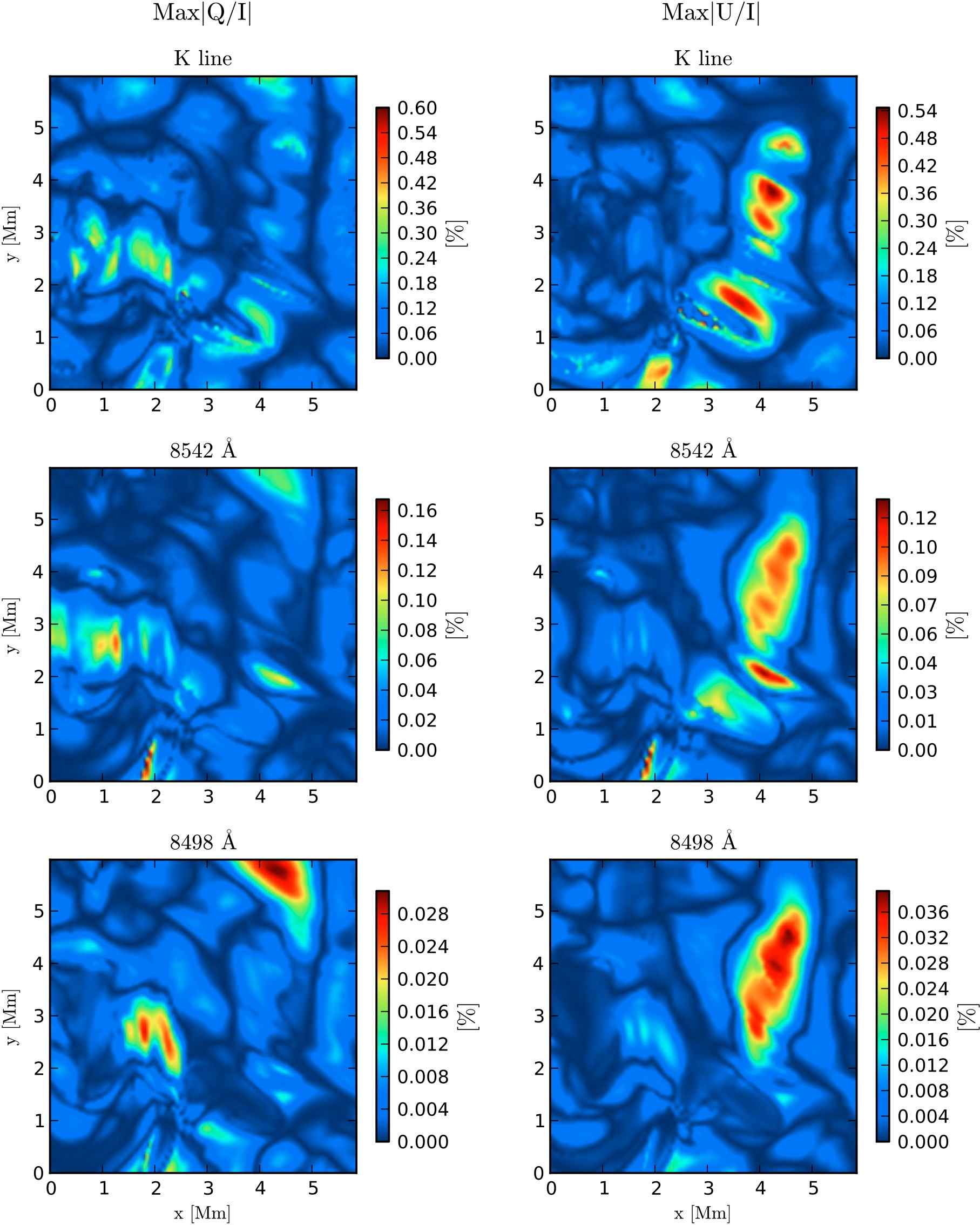}
\caption{Maximum fractional linear polarization amplitudes across the solar model at $8498$ {\AA} (lower row), $8542$ {\AA} (middle row) and $3934$ {\AA} (K line, in the upper row).  }
\label{fig:mapqu}
\end{figure}

Figure \ref{fig:mapqu} illustrates the maps of maximum Stokes Q and U separately for the $3934$ {\AA}, $8542$ {\AA} and $8498$ {\AA}
 lines, following Eqs. (\ref{eq:pamp_l0}). Line $8662$ {\AA}  is not shown here because it is very similar to $8542$ {\AA}. For making these maps, the choice of other wavelengths (typically the line center) different from the ones corresponding to the maximum amplitudes results in notable signal losses in areas with substantial velocities. 
\begin{subequations}\label{eq:pamp_l0}
  \begin{align}
    \mathrm{\left(\frac{Q}{I}\right)_{max}} &= \, 100 \cdot  \mathrm{Max \left|\frac{Q}{I}\right|}\quad [\%]\\
    \mathrm{\left(\frac{U}{I}\right)_{max}} &= \, 100 \cdot  \mathrm{Max \left|\frac{U}{I}\right|} \quad [\%]
  \end{align}
\end{subequations}
In general, areas with significant linear polarization (LP) have always a magnetic field notably inclined (HF regions). Out of those areas, the LP amplitudes are always below $1/5$ of their maximum in the map.

In many pixels, the amplitudes have the same order of magnitude as those calculated by \cite{manso10} in semiempirical models but, where we find velocities above $\mathrm{\sim 5 \,km \cdot s^{-1}}$ in the chromosphere, we see relative enhancements that reach one order of magnitude in the amplitudes of the $8542$ {\AA} and $8662$ {\AA} lines\footnote{Here, we find an association between larger velocities and larger velocity gradients.}. This is also true for the $8498$ {\AA} line only if the model is not too cool. In sufficiently cool models, the $8498$ {\AA} line always presents tiny linear polarization amplitudes (low-temperature patches in Fig. \ref{fig:vtb8498} coincide with places of almost-zero polarization in both Q and U). The largest LP amplitudes in this line appear in Stokes Q and U in the absence of large velocities (Fig. \ref{fig:mapqu}, patches over $\mathrm{x \sim 4.5}$ Mm) and are twice larger than in the FALC model. Thus, the $8498$ {\AA} anisotropy is affected by the velocity but seems to be dominated by temperature in the current models. 
 
The differences between the Q and U maps are understood with Fig. \ref{fig:qusigns}. When the magnetic field existing in a region experiences a change in its azimuth, the polarization patches stand out in Q and attenuate in U or viceversa, so giving us an idea of the likely approximate directions of the magnetic field just by comparing regions. Patches with large Stokes Q and low Stokes U indicate that the field is chiefly oriented along some of the reference axes for Q (vertically or horizontally in the maps). If the opposite holds,  the field is then mainly oriented in directions lying at $\pm45$ degrees. If Stokes Q and U have similar amplitudes, it means that the field is in between the previous four directions. 

\begin{figure}[h!]
\centering%
\includegraphics[width=\textwidth]{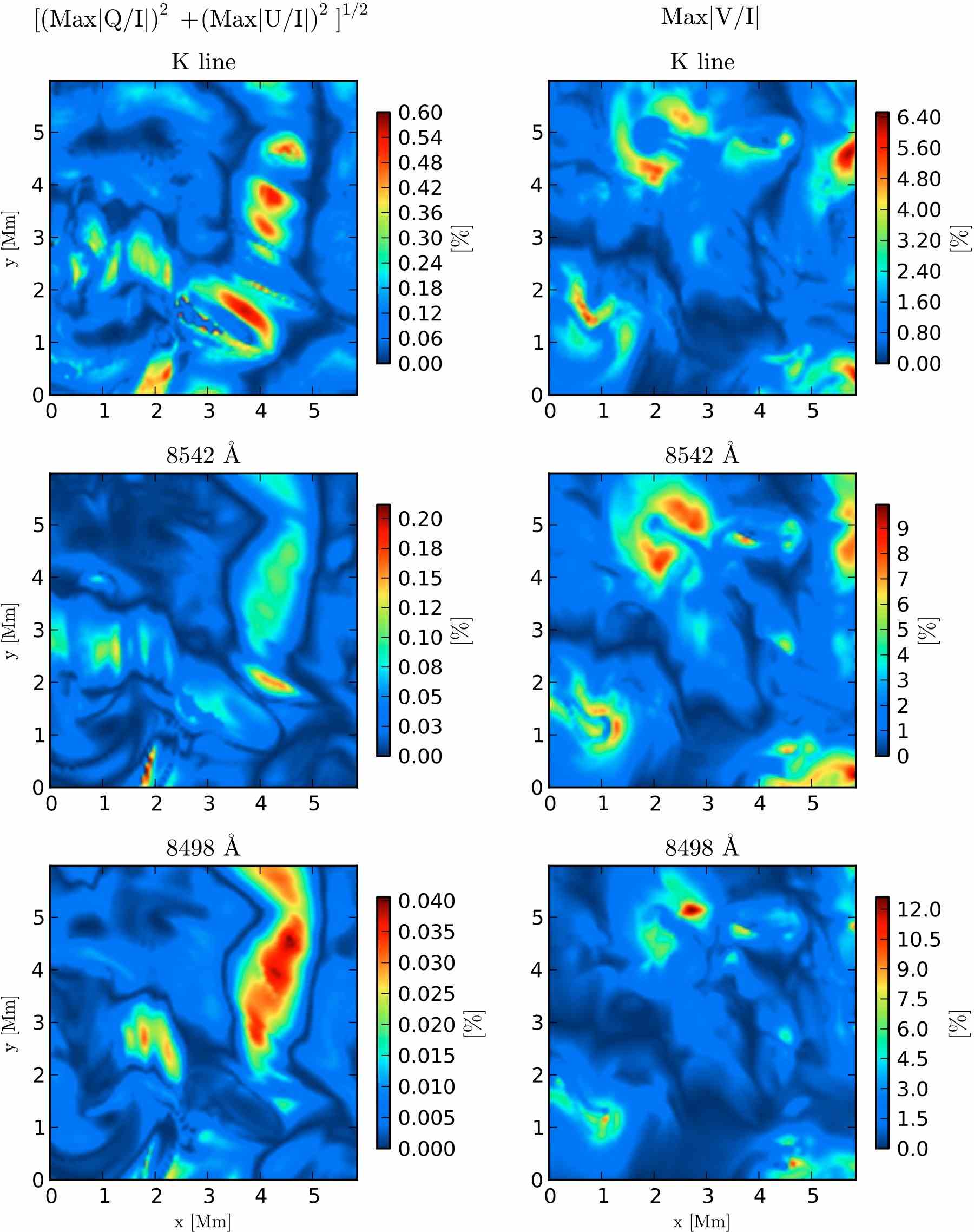}
\caption{Calculated polarization amplitudes across the solar surface at $8498$ {\AA} (lower row), $8542$ {\AA} (middle row) and $3934$ {\AA} (K line, in the upper row). Left: total linear polarization amplitudes calculated as $\mathrm{\sqrt{Max|Q/I|^2+Max|U/I|^2}}$. Right: The maximum of the absolute value of the fractional circular polarization.  }
\label{fig:mapmax}
\end{figure}
\clearpage

The total linear and circular polarization are shown in Fig. \ref{fig:mapmax}. In that figure we see the spatial distribution of the maximum values (that is, at the wavelength of the maximum in each pixel) of the total fractional linear polarization (left panels) and fractional circular polarization (right panels) for the same three spectral lines. The places of negligible circular polarization are the ones where the magnetic field is almost horizontal at the heights of formation of the corresponding spectral line. The same region has larger linear polarization in the left panels (HF region).
The strongest linear polarization patches in the $8498$ {\AA} line are where the temperature is larger. However, the strongest linear polarization patches in the $8542$ {\AA} line are where the vertical velocity is larger. The circular polarization in the $8542$ {\AA} line is specially sensitive and clear marking the vertical magnetic field concentrations. Its value at each pixel is roughly twice the corresponding ones in the K line. 
 
\subsection{The Hanle Polarity Inversion Lines.}\label{sec:nullpaths}	

In the maps of Stokes Q and U (Fig. \ref{fig:mapqu}) we observe something interesting: the lines of zero linear polarization in Hanle forward scattering. They are groove-like regions whose emergent Stokes Q and/or Stokes U are virtually zero. They encode the topology of the magnetic field pervading the solar model.

We identify three kinds of Hanle Polarity Inversion Lines (Hanle PILs) produced by three different sources that can act together. Eqs. (\ref{eq:step6}) evaluated in the frame of reference of Fig. \ref{fig:qusigns} (and using $\chi=-\pi/2$ for the LOS\footnote{This is chosen for maintaining the consistency with the close-to-limb observational configuration that will be mentioned later on.}) give a mathematical support for the following analysis in forward scattering:
\begin{subequations}\label{eq:hanlesat}
\begin{align}
\frac{Q}{I} &\simeq \,- \frac{3}{4\sqrt{2}}\cos{2 \chi_B}\cdot\sin^2{(\theta_B)}\cdot(3\cos^2{\theta_B}-1)\cdot\mathcal{F}, 
\label{eq:hanlesata}
\displaybreak[0] \\
\frac{U}{I} &\simeq \, -\frac{3 }{4\sqrt{2}}\sin{2 \chi_B}\cdot\sin^2{(\theta_B)}\cdot(3\cos^2{\theta_B}-1)\cdot\mathcal{F},
\label{eq:hanlesatb}
\end{align}
\end{subequations}
with all the r.h.s. quantities evaluated at $\tau=1$. Remind that these expressions are valid for a line in the saturation regime of the Hanle effect, which holds for the $\lambda 8662$ line in our dataset and very likely in real quiet Sun regions. If the field is weak but still strong enough, the other scattering signals of the triplet can also be saturated (especially  $\lambda 8542$). Then, their geometrical dependence will also be described by Eqs. (\ref{eq:hanlesatb}).

The first kind of Hanle PILs are due to the inclination of the magnetic field. They are the same in Stokes Q and U (hence also in the total linear polarization) because they are independent of the magnetic field azimuth. They can be found \textit{delimitating regions} in which the magnetic field is mainly horizontal (i.e., bordering areas with null longitudinal Zeeman polarization in Stokes V) or placed where the magnetic field is mostly vertical. Namely, they are always where $\theta_B= 90 \pm 35.27\deg, 90 \pm 90\deg$, as confirmed by Eqs. (\ref{eq:hanlesat}). When at $\theta_B=90 \pm 35.27\deg$ (magnetic field forming Van Vleck angles with the vertical), they are the frontiers between HF and VF regions\footnote{Remind, we already have defined the horizontal field (HF) and vertical field (VF) regions in Section \ref{sec:profilesA}}. For this reason, we call them Van Vleck HPILs. They connect pixels with the same inclination. As the magnetic field emerges in bipolar structures at all scales, the Van Vleck HPILs have to form a closed line surrounding one of the magnetic poles. 

 The second kind will be called azimuthal HPIL. Namely, a Hanle PIL appearing in a map of Stokes Q (or Stokes U) is of azimuthal type if it does not appear in the same place for Stokes Q than for Stokes U. Clearly, they depend on the azimuth of the magnetic field in the chosen reference system (terms in $\chi_B$ in Eqs. (\ref{eq:hanlesat})). In pixels defining an azimuthal HPIL in Stokes Q, the magnetic field vector is lying along the positive or negative reference directions for Stokes U. And viceversa, the pixels defining it in Stokes U have a magnetic field vector lying along the positive or negative reference directions for Stokes Q (see Fig. \ref{fig:qusigns}). Note that following an azimuthal HPIL, we connect pixels with the same magnetic field azimuth\footnote{In our definition, an azimuthal HPIL always begins and ends in an intersection of azimuthal HPILs. Thus, after such intersection, the continuation of the null line is always another azimuthal HPIL that can correspond to another azimuth.}. Note also that, when two or more of them intersect, the cross point must have a magnetic field completely vertical. Consequently, they have a radial nature, beginning in an area of concentration of photospheric magnetic flux and ending in another one.

Finally, a third possible origin of a HPIL is a particular configuration of the anisotropy of the radiation field persisting across a region in the maps. They are thermodynamically induced HPILs and appear at the same time in the Q and U maps for a spectral line, as the Van-Vleck type ones. To grasp some feeling about the conditions in which they form, note that the thermodynamical HPILs gives zero linear polarization because the non-magnetic factor $\mathcal{F}$ in Eqs. (\ref{eq:hanlesat}) is negligible. Thus, we can pose the expression 
\begin{align}
\mathcal{F}=\omega^{(2)}_{J_u J_{\ell}} \sigma^2_0(J_u)-\omega^{(2)}_{J_{\ell} J_u} \sigma^2_0(J_{\ell})=0.
\label{eq:factorf}
\end{align}
In the simplest case, given by the line $8662$ {\AA}, it yields the condition
\begin{equation}
\centering
\sigma^2_0(J_2)=\frac{\rho^2_0(J_2)}{\rho^0_0(J_2)}=0   \quad {\rm at} \quad \tau^{8662}=1
\label{eq:condicion2}
\end{equation}

How can such a condition be fulfilled?Let us suppose we have identified a thermodynamic HPIL, which appears at the same time in Q and U maps for the $8662$ {\AA} line. Thus, consider the case in which Eq. (\ref{eq:condicion2}) is satisfied in that region\footnote{In principle, the region associated to a thermodynamical HPIL might not be a line in the map, but we still term it HPIL for consistency.}. It happens when the alignment $\rho^2_0 (J_2)$ tends to zero, or when the overall population $(\varpropto \rho^0_0(J_2))$ increases or when both things happen at the same time. 
A larger level-2 population can be achieved with an increment of temperature in the upper parts of the chromosphere. Such increment strengthens the Ca {\sc ii} H  line intensity emission (forming at the top), which illuminates lower chromospheric layers from above. The extra illumination arriving at chromospheric layers immediately below (where the $8662$ {\AA} line originates) increases the population pumping from level 1 to level 4 which, in turn, produces an extra population in level 2 by spontaneous emission (see Fig. \ref{fig:elevels}). Higher temperature thus means more population in higher energy levels (levels 4 and 5) and more emission (at $8662$ {\AA}) produced by electrons falling from  level 4 to level 2. Furthermore, if at the same time the formation region of the $8662$ {\AA} line is meaningfully cool, the absorption of electrons from level 2 to level 4 will vanish, so retaining the population in level 2 (absorption to level 5 can be also neglected). If the absorption is very reduced from level 2 to level 4, the $8662$ {\AA} line cannot be polarized because its polarization can only be generated by dichroism (selective absorption).

On the other hand, to have an almost null alignment in level 2 we need a formation region illuminated with a radiation field that cancels out the component $\bar{J}^2_0$. Outside LTE, it can occur in very especific situations, when the contribution of the mainly horizontal illumination equals the contribution of the vertical one in the evaluation of the anisotropy at those layers. We can show that it happens in pixels that separate areas where the formation region is significatively cool from areas having a formation region at relatively large temperatures. If, furthermore, the HPIL is in a place without significant velocity gradients, the low existing alignment will not be enhanced by dynamics. 

The previous explanation for Eq. (\ref{eq:condicion2}) seems to be correct in the borders of a cool chromospheric plasma bubble appearing in the solar models. There, the polarization in Q and U is zero. To examinate this observation, we first identified the location of the thermodynamical HPIL in the Stokes maps. Then, with 3D visualizations (see stereographic view in Appendix \ref{app:stereo}), we verified that the cool bubble has a singular stratification at its enclosing verticals. In the interior of the bubble, the chromospheric temperature is as cool as $3000$ K and the anisotropy is dominated by vertical radiation coming from above the bubble and also from the photosphere. On the contrary, the plasma at the external surroundings of the bubble is \textit{much hotter} in the formation region of the spectral line. Consequently, the horizontal radiation dominates, changing the sign of the alignment with respect to the interior of the bubble. In the middle of both regions (the bubble's frontier), a line where the net alignment is zero must exist because it is positive at one side and negative at the other\footnote{There is also a correspondence with the velocities. The bubble interior is typically produced by an expansion cooling down the atmosphere (upward velocities) and the bubble exterior is usually a contraction (downward velocities). Thus, in the HPIL the velocity is almost zero or insignificant. }. That is a thermodynamical\footnote{As the anisotropy can be modified by the velocity, we also will talk about dynamic HPILs.} Hanle PIL and it has been induced by spatial differences of temperature.

An example of cool bubble is found around $\mathrm{(x,y)}=(3.8,1.5)$ Mm in the upper-right panel of Fig. \ref{fig:vtb8542}. The corresponding HPILs appear in Stokes Q and U maps of the $8662,8542$ and $8498$ {\AA} lines, surrounding the region. For example, a part of the null line is connecting the points $(3.5,2)$ and $(4.5,1)$ Mm. 
We conclude that a way of distinguishing a thermodynamic HPIL from a Van-Vleck HPIL is searching the former around cool patches.

The HPILs could give us extra information.
For example, the contrast and width of a Hanle PIL gives information about the variation with height of the magnetic field along the formation region of the considered spectral line. The differences in width between HPILs pertaining to spectral lines forming at different heights complement that information. Besides that, a histogram of the mean size of the regions enclosed by the HPILs could be a quick way of measuring the variability of some magnetic field parameter (helicity, inclination, azimuth) in a map. It could be used as a fine tester to a model atmosphere in order to compare the polarization fingerprints it produces with the ones in high-sensitivity solar observations. In Section \ref{sec:b_inference}, we will explain some criteria based on HPILs in order to infer the magnetic field topology. 

In principle, we see two objections to the observational potential of the HPILs. First, the practical application of these ideas is very difficult given the current polarimetric sensitivity. And second, diffuse light due to internal reflections in the optical system preceding the detector can play a role in masking the HPILs because it diminishes the contrast between regions with and without polarization. 

We conclude that the Hanle PILs in the Ca {\sc ii} IR triplet are true fingerprints encoding the orientation of the chromospheric magnetic field. Perhaps more than that, the HPILs could offer new diagnostic aids to capture very specific circumstances of the thermodynamical state in the atmosphere. Overcoming the technical impediments, we see possible to deduce the three-dimensional topology of the chromospheric magnetic field from 2D maps of the Stokes vector (Sec. \ref{sec:zh_ambig}). Some extra calculations (Sec. \ref{sec:polamps}) suggest that these structures with null polarization could likely be distinguished for the first time with the coming instrumental solar facilities (Zimpol 3, EST, Solar-C). The HPIL concept leads to an interesting question: can we get a precise magnetic field mapping using the spatial locations where the polarization amplitudes cancel? 

\subsection{ Maps of the Polarization Degree.}\label{sec:poldegree}
The degree of linear and circular polarization of the Stokes signals measure the total amount of linearly and circularly polarized light for a whole spectral line. They are calculated by integrating the contributions of photons in a certain spectral window in order to increase the signal to noise ratio. The expressions we have used to calculate them are
\begin{subequations}\label{eq:degs}
\begin{align}
    \mathrm{LPD} \,[\%] &= \, 100 \cdot  \frac{\sqrt{\left[ \overset{\lambda_2}{\underset{\lambda_1} {\sum}} Q\right]^2 +\left[ \overset{\lambda_2}{\underset{\lambda_1} {\sum}} U\right]^2}}{ \overset{\lambda_2}{\underset{\lambda_1} {\sum}} I}\label{eq:pdeglin1} ,
\displaybreak[0] \\
    \mathrm{CPD}\, \left[\%\right] &= \, 100 \cdot  \frac{\overset{\lambda_2}{\underset{\lambda_1} {\sum}} |V|}{ \overset{\lambda_2}{\underset{\lambda_1} {\sum}} I}\label{eq:pdegcir1} , 
\end{align}
\end{subequations}
where the wavelengths $\lambda_1$ and $\lambda_2$ delimit the width of the filter considered in the integration. 
\begin{figure}[b!]
\centering%
\includegraphics[scale=0.5]{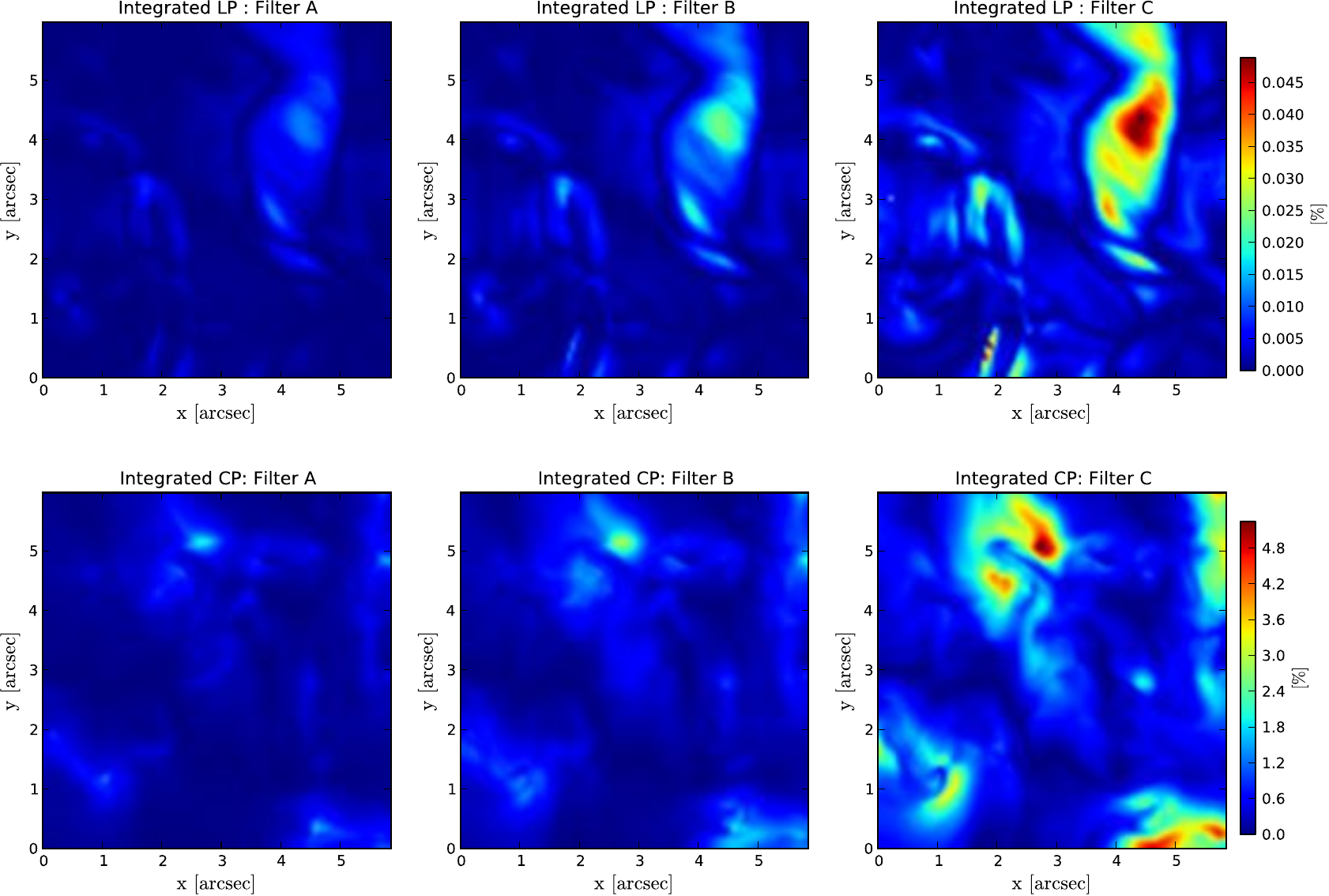}
\caption{Spatial maps of linear polarization degree (top panels) and circular polarization degree (bottom panels) calculated (Eqs. \ref{eq:degs}) by integrating around $8542$ {\AA} with three different filters. Filter A(left column): filter with a width $\mathrm{\Delta \lambda\sim2}$ {\AA}. Filter B(middle panels): $\mathrm{\Delta \lambda=1.2}$ {\AA}. Filter C(right panels): $\mathrm{\Delta \lambda=0.7}$ {\AA}. Each colorbar is common for a row of panels. }
\label{fig:8542_filters}
\end{figure}
We have obtained these quantities for our dataset in three different square filters placed at line center: the filter A ($\Delta\lambda=2$ {\AA}), which reaches wavelengths in the far wings of the line; the filter B ($\Delta\lambda=1.2$ {\AA}), which selects the core and near wing of the line; and the filter C ($\Delta\lambda=0.7$ {\AA}), which only captures the core of the line, where the linear polarization is more significant. The results for the $8542$ {\AA} line are shown in Fig. \ref{fig:8542_filters}. In Fig \ref{fig:k_filters} we show similar results for the K line.

The filter C is the most suitable because it allows a clearer contrast between regions with and without polarization. Filters A and B produce lower fractional polarization signals because the wings intensity contribution in the denominator of Eqs. (\ref{eq:degs}) is much larger than the integrated polarization at those wavelengths. We have to reach a trade off between signal gains by integration and signal losses by the excesive number of non-polarized photons at the wings. Note also that in general the resulting numbers for the polarization are not representative of the real maximum polarization amplitude. This is because the Eqs. (\ref{eq:degs}) are not linear in the Stokes parameters. Thus, the advantage of calculating the polarization degree maps is mainly to get a better signal to noise ratio, so showing structures with more contrast and definition than the fractional polarization at one wavelength.
\subsection{Maps of polarization contrasts.}\label{sec:conmaps}
\begin{figure}[t!]
\centering%
\includegraphics[scale=0.62]{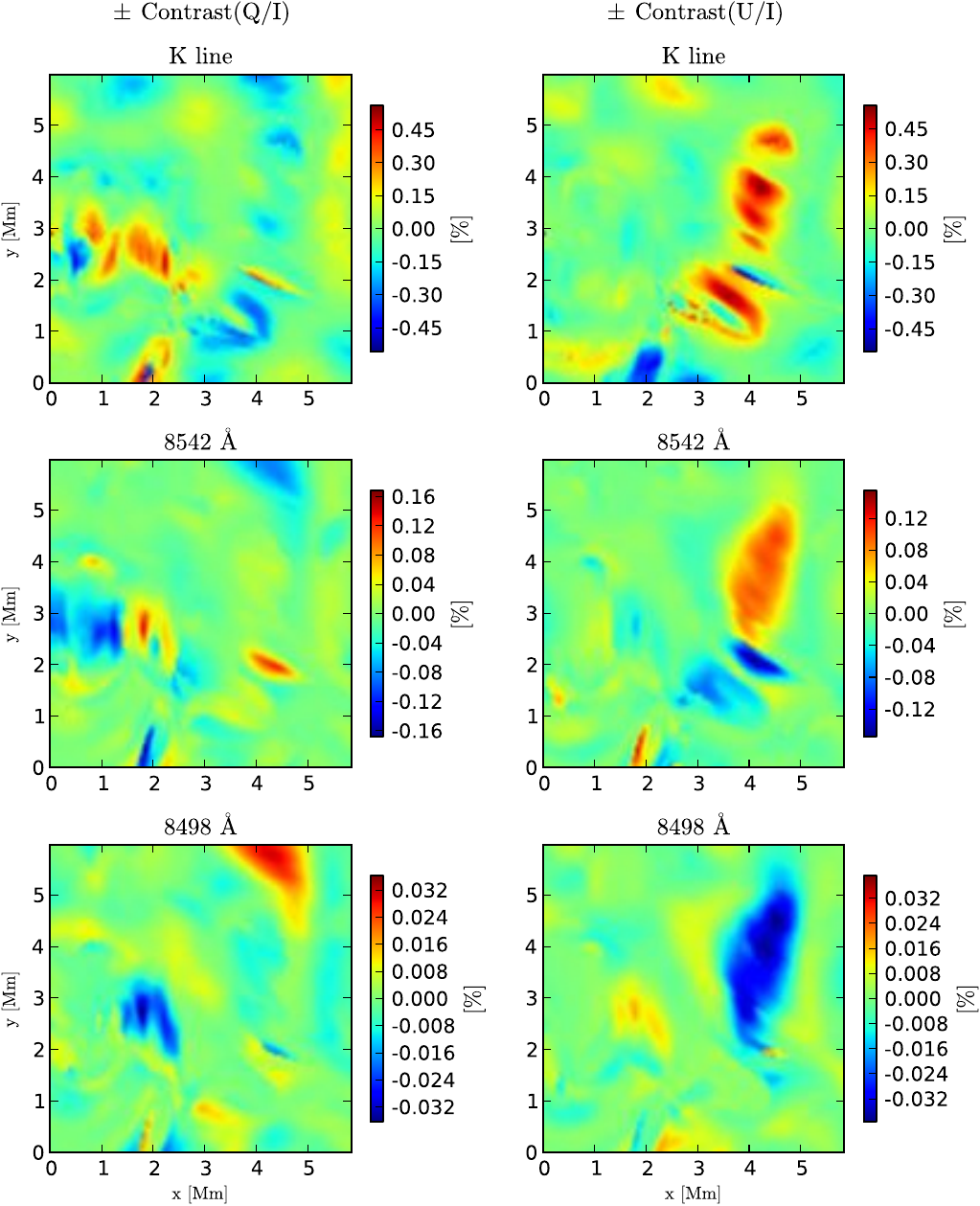}
\caption{Signed contrast amplitudes for Stokes Q and U in the same spectral lines than in Fig. \ref{fig:mapmax}. Left: the signed contrast defined as the difference between the maximum and the minimum Stokes Q, where the artificial added sign is positive/negative if the maximum Stokes Q is positive/negative. Right: the same as in left panels but for Stokes U.}
\label{fig:mapcon}
\end{figure}

In this section we show that the identification of drivers and situations changing the linear polarization signs may allow us to infer extra information about the solar atmosphere. It also helps us to deduce correctly the direction of the magnetic field, for what it is previously necessary a discrimination of all the drivers affecting the polarization sign (polarity) at each pixel. To that end, the maps of polarization contrasts  (Fig. \ref{fig:mapcon}) are useful to determine the main regions where the different drivers have to be disentangled as well as the polarization sign in those regions. We remind that the polarization contrast is basically a peak-to-peak amplitude with an artificial sign indicating whether the largest peak is positive or negative\footnote{In this chapter we apply that definition to each pixel. In Chapter \ref{cap:three} (see Figure \ref{fig:tevol_all}) we applied the same, but with a slight difference. There we multiply by $(-1)$ the $8498$ {\AA} signal because most part of the time sequence its maximum elongation was negative, but now we multiply it by the corresponding maximum sign in each pixel independently, not by the predominant sign in the map. The important point is that the polarization signal can change its sign in certain situations that we always specify in the text.}.

The first thing we note in Fig. \ref{fig:mapcon} is that the significative patches are highlighted in the HF region (to identify the HF\footnote{Reminder: the horizontal field (HF) and vertical field (VF) regions have been defined in Section \ref{sec:profilesA}} region see lower left panel in Fig. \ref{fig:bpars8662} for instance). This is because the polarization contrast, being a peak-to-peak measure, has more dynamic range than the absolute value of the polarization. Thus, for a similar color palette, the polarization contrast is more efficient (than the maximum polarization) highlighting the larger linear polarization signals, which are always in the HF region. 
\subsubsection{Reference signals in the saturation regime}
To understand and ``diagnose'' the changes of sign in the polarization patches of Fig. \ref{fig:mapcon}, we will need to define a reference model atmosphere that: (a) is static, (b) has a hot FALC-type chromosphere, (c) has no extra heatings due to shocks or compressions, and (d) has a horizontal magnetic field in the saturation regime of the Hanle effect that lies parallel to the reference direction for $Q>0$ and to the x axis ($\chi_B=0,\theta_B=\pi/2$). The LOS is set along the y axis ($\chi=-\pi/2$). This configuration of reference gives the maximum LP amplitude for a horizontal magnetic field both in forward scattering and in a close-to-limb observation. The calculations done in such model will hereafter define a reference \textit{for the signs} of the LP in the IR triplet lines.
\begin{figure}[h!]
\centering%
\includegraphics[width=0.5\textwidth]{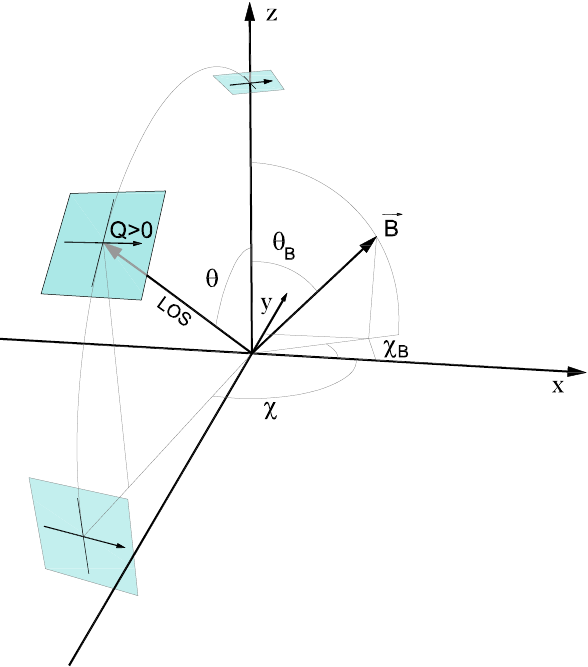}
\caption{In the reference configuration $\chi_B=0,\,\theta_B=\pi/2,\,\chi=-\pi/2$ and $Q>0$ is parallel to x. The line of sight (LOS) can be $\mu=0.1$ or $\mu=1.0$. The vertical axis
  is the solar radial direction passing through the point
  considered. }
\label{fig:refsysB}
\end{figure}
From Eqs. (\ref{eq:step5}), it follows that both the $\mu=0$ and $\mu=1$ polarization amplitudes resulting from such a reference configuration are a half of the maximum possible amplitude (given by the non-magnetic case in a limb observation):
\begin{subequations}\label{eq:relations}
\begin{align}
\bigg( \frac{Q}{I} \bigg)^{\rm ref}&\Rightarrow \bigg( \frac{Q}{I} \bigg)^{\mu=0}_{\theta_B=\pi/2}=\bigg( \frac{Q}{I} \bigg)^{\mu=1}_{\theta_B=\pi/2}
\label{eq:relationsA} ,
\displaybreak[0] \\
\bigg( \frac{Q}{I} \bigg)^{\rm ref}&=\frac{1}{2}\bigg( \frac{Q}{I} \bigg)^{\mu=0}_{\theta_B=0}=\frac{1}{2}\bigg( \frac{Q}{I} \bigg)^{\mu=0}_{B=0}.\label{eq:relationsB}  
\end{align}
\end{subequations}
In our forward-scattering case in which the polarization amplitude is roughly given by Eq. (\ref{eq:hanlesata}), note that the reference sign of a Stokes Q signal resulting from the above configuration is the sign of the thermodynamical factor $\mathcal{F}$ of Eqs. (\ref{eq:hanlesat}). That is still true in any atmospheric model with $\chi_B=0\deg$ and with a $\theta_B$ corresponding to the HF region.

\subsubsection{Polarization signs as physical references}
Now, we explain Fig. \ref{fig:mapcon} and some differences of sign found between the spectral lines analyzed.
First, the linear polarization (LP) patches in the maps of $\lambda8542$ (and $\lambda 8662$) have opposite sign to the corresponding ones for $\lambda 8498$, which is a typical behavior in the IR triplet when the calculations are done for disk center. Such signs difference holds in almost all pixels. It suggests that the \textit{difference of signs} between the $8542$ and $8498$ {\AA} lines is a essential and robust feature primarily set by the atomic coefficients contained in $\mathcal{F}$, but not significantly influenced by anisotropy variations. 

On the contrary, the difference of signs between the $8542$ {\AA} line and the K line \textit{does vary} between polarization patches. Sometimes, both lines share the same sign, sometimes not. In standard FALC-like models they should show the same sign, however.  
The reason of the change is not the magnetic field direction because it varies slowly between the medium-high ($8542$ {\AA} line main formation region) and the top chromosphere (K line main formation region). We associate such change with strong spatial differences of temperature between both heights. It happens, for instance, in the cool chromospheric bubble around $(x,y)=[3.5,1.5]$ Mm (see Sec. \ref{sec:nullpaths}), which was formed after the ascension of a shock. In that area, the LP maps show a patch with different signs in the $8542$ {\AA} line and the K line. Thus, we identify the LP sign as a simple marker of strong chromospheric temperature gradients. Concretely, patches changing their sign between the $8542$ {\AA} map and  K line map indicate a large difference in temperature between the very top and lower parts of the chromosphere. This fact allows to estimate that the cool bubble reaches a height located between the medium-height and the top chromosphere.

With respect to the reference case, there are four basic drivers that alter the polarization sign of these IR triplet lines: the inclination of the LOS; some very specific circumstances for the temperature stratification (Sec. \ref{sec:nullpaths}); the magnetic field; and the velocity gradients (Chapters \ref{cap:two} and \ref{cap:three}). 

 \textbf{Concerning the line of sight}, when passing from the solar limb to the disk center, the signs of the $8542$ {\AA} and $8662$ {\AA} lines usually change because all the main contributing formation heights shift downward, below the level alignments intersection (recall Section \ref{sec:profilesA} and Fig. \ref{fig:rhosqi2}) and it modifies the sign of $\mathcal{F}$. Thus, in our reference FALC-like model described previously, the Stokes Q sign for these lines would be $[+]$ at $\mu=0.1$ and $[-]$ at $\mu=1$. 
The $8498$ {\AA} line already had low formation heights that do not change their net contribution to the polarization because they shift yet below the alignments intersection (towards even lower layers) when passing to forward scattering. Hence, this line does not change its polarization sign, which would be $[+]$ for both LOS in the reference FALC-like model. 

\textbf{Concerning the temperature stratification}, the results of \cite{manso10} in FALC models and forward scattering suggest that the IR triplet lines change their polarization signs between hot and cool models with the same magnetic field orientation. As already commented, that is also valid in our results. For example, the pixel at $[4,3]$ Mm and the one at $[4,1]$ Mm (see Stokes U in Fig. \ref{fig:mapcon}) have opposite linear polarization signs in the triplet lines because one pixel has a FALC-like chromosphere and other has an almost completely cool chromosphere. They have the same magnetic field orientation and negligible velocity. A similar situation occurs in the surrounding pixels, resulting in two patches with opposite polarization contrasts. Another example is the patch at $[2,3]$ Mm (hot) compared with the one at $[1,3]$ Mm (cool). Such behavior is less visible for the $8498$ {\AA} line because its polarization tends to vanish in very cool  chromospheres. 

The reference model in a forward scattering geometry yields emergent Stokes-Q signals whose reference signs are $[+,-,-]$ for the $[8498,\,8542,\,8662]$ {\AA} lines. Then, under similar circumstances but in cool-chromosphere models, the result is $[-,+,+]$. An example of such reference case is the blue patch at $[4.5,5.5]$ in Stokes Q $\lambda 8542$ (Fig. \ref{fig:mapcon}). Effectively, this patch is an almost static FALC-like region with an almost homogeneous chromospheric magnetic field that is parallel to the x axis (note the absence of Stokes U).

The explanation for the sign reversal is the dependence on the radiation field anisotropy on temperature. In a cool chromospheric model, the vertical radiation is more intense than the horizontal (positive net contribution to the anisotropy). In a hot model, with or without a temperature shock or heating in the formation region, the contrary holds: the net contribution to the anisotropy is negative. The $8498$ {\AA} line is more sensitive to these changes because it forms lower (nearer the minimum of temperature and the alignments intersection), and when the plasma is cool there, its emission vanishes. The other lines of the model are mainly created nearer the transition region, for what they necessitate a cool stratification along the whole chromosphere (cool bubbles) to reverse their polarization signs. 

\textbf{Concerning the magnetic field}, the azimuthal dependence in Eq. (\ref{eq:hanlesat}) allows us to deduce the expected polarization signs for other field azimuths without calculations, just following Figure \ref{fig:qusigns}. To do it, we must note that: for a hot FALC type model in the HF region, the signs of the emerging Stokes Q and U in the $\lambda 8542$ line for any magnetic field azimuth are always \textit{opposite to the signs of the corresponding Stokes Q and U axes} that are nearest to the projection of the magnetic field vector in Figure \ref{fig:qusigns}. 

For instance, if the reference FALC-like model had the magnetic field exactly along the +U axis, the signs for Stokes U would be $[+,-,-]$ with zero Stokes Q. If it were along -Q, Stokes Q would be $[-,+,+]$ with zero Stokes U. If such hot model would have the transverse component of the magnetic field in between the $+$Q and the $-$U axes, Stokes Q would be $[+,-,-]$ and Stokes U $[-,+,+]$. Besides this, the so-given LP signs can furthermore be modified by changes in temperature and velocity.

This is helpful. For example, consider the positive U patch around $[4.5,4]$ Mm in the LP contrast map of the $\lambda 8542$ line (Fig. \ref{fig:mapcon}) and suppose we only know it is a hot (FALC-like) model in the HF region (as it is). We then see that the patch has $U > 0$ and $Q\sim 0$. Following the references, that configuration is only compatible with a magnetic field along the $-$U axis, which is correct.  If the chromosphere model were cool, $\vec{B}$ should be along $+$U. In addition to that, we recall that the LP signs in the K line core and in $\lambda 8498$ can help to establish the presence of ``cool bubbles'' and large temperature gradients after a shock wave that has crossed the chromosphere. All together gives a ``zero-order'' diagnosis based on LP references.

 \textbf{Concerning the vertical velocity}, we need to find out how it modifies the signs with respect to the static case. On one hand, we know that the velocity gradients can efficiently modulate the radiation field anisotropy (Chapter \ref{cap:three}). In the case of the Ca {\sc ii} IR lines, it directly enhances the lower level aligment \textit{at heights just above $\tau=1$} (see Fig. \ref{fig:rhosqi2}). On the other hand, we have evidences suggesting that the profiles with a valley at line center are prone to change their signs with a velocity gradient, while the single-peak ones do not (e.g., see Figure \ref{fig:gradsv_ctes}). These results suggest that a velocity gradient in the formation region of a spectral line can alter the balance between the upper and the lower level alignments that control the fractional scattering polarization. Thus, depending on the sign of the alignment at those layers (and on the intrinsic atomic polarizability coefficients, which have a certain sign), the line will have amplitudes enhanced or diminished with respect to the static results. The larger sentivity to this effect in valley-like profiles is because their formation regions are very near the alignments intersection. We have investigated this behavior in each model.

 For example, in the calculations done in dynamic non-magnetic 1D models at $\mu=0.1$ (Chapter \ref{cap:three}), the $8498$ {\AA} line was the only one that changed its linear polarization sign in expanding or compressing atmospheres\footnote{We know its sign was positive only in totally expanded models (just before starting the compression) or well in static FALC-like models}. We concluded that such changes in sign were possible due to a strong dynamics and to the proximity of the $\lambda 8498$ formation heights to the level alignments intersection, which furthermore made its LP profiles to vary their essential shape. In the current 3D models, the synthetic LP profiles of the $8498$ {\AA}  line in $\mu=1$ very rarely exhibit a valle-shape and we do not see clear indicatives of significant Doppler-induced amplifications in this line. This can be understood with the lower degree of dynamism (see Fig. \ref{fig:gradsv_ctes}) and a lower height of formation (see Sec. \ref{sec:profilesA}). On the other hand, the other two lines of the triplet can now exhibit a valley shape, and we wonder whether they can also reverse their polarization sign in presence of a velocity gradient as the $\lambda 8498$ line did at $\mu=0.1$. However, it seems that the proximity of the point of alignments intersection to the main heights of formation is not enough to produce a polarity reversal. In principle, the effect of a velocity gradient in these lines in the 3D models is only a modest variation of the amplitudes of Stokes Q and U with respect to the static case. It makes the resulting azimuth estimated from Eq. (\ref{eq:chiuq}) to vary with respect to the real azimuth.

This last conclusion is relevant because it means that Eq. (\ref{eq:chiuq}) can fail in the inference of the chromospheric magnetic field azimuth in situations where the vertical velocities are significant (dynamic fibrils, mottles and spicules). 


General rules as the ones above and the specific situations breaking them are of interest for developing chromospheric diagnosis methods based on polarization. Thus, starting from a reference situation\footnote{Reminder: hot FALC-like, with $\vec{B}_{\bot}$ along the +Q axis and $\vec{B}$ in the HF region: $90-35.27\deg<\theta_B < 90+ 35.27\deg$}, we can try to simplify the variation of the LP signals referring them to some simple qualitative changes in the reference models. Thus, in models with a weak dynamic ($v\sim 0$), the basic behavior of the LP signals in the IR triplet can be summarized with Table \ref{tab:signs}.
 
 \begin{table}[!h]
\centering
\begin{tabular}{cccc}
\hline
\multicolumn{4}{c}{$\mu=1$, $v\sim 0$ @  HF region} \\
\hline
 $\vec{B}$ & Cooler & $\leftarrow$FALC-like$\rightarrow$ &
Hotter   \\
\hline
$\boxminus$ (along $+$Q)    & $[-,+,+]$ & $[+,-,-]$ & $[+,-,-]$ \\
$\boxslash$  (along $+$U)    & $[-,+,+]$ & $[+,-,-]$ & $[+,-,-]$ \\
$\boxvert$ (along -Q)    & $[+,-,-]$ & $[-,+,+]$ & $[-,+,+]$ \\
$ \boxbackslash$ (along -U)   & $[+,-,-]$ & $[-,+,+]$ & $[-,+,+]$ \\
\hline
\end{tabular}
\caption{Signs of the LP in the Ca{\sc ii} IR triplet in forward scattering and considering static models in the HF region ($54.73\deg<\theta_B (\tau=1)<125.27\deg$). The squared symbols represent the direction of the magnetic field in the plane of the sky. The signs in brackets correspond to each spectral line ordered as [$\lambda8498$,$\lambda8662$,$\lambda8542$]. Such signs correspond to Stokes Q (when $\vec{B}$ is along $\pm$Q) or Stokes U (when $\vec{B}$ is along $\pm$U). Each column selects the stratification of temperature, which can have a cooler or hotter chromosphere than in FALC-like models. All the signs would be the opposite out of the HF region.}
 \label{tab:signs}
\end{table}

\section{Effect of the vertical velocities.}\label{sec:velocities}
We have computed the maps of polarization contrast, the maps of polarization amplitudes,  histograms of total linear polarization and the spectral profiles again, but setting the velocity to zero in all points. With this information we analyze the influence of the dynamic on the linear polarization by paying attention to the amplitudes, to the sign of the maximum amplitude in each pixel and to the shape of the profiles.

\begin{figure}[h!]
        \centering
        \begin{subfigure}[b]{0.49\textwidth}
                \centering
                \includegraphics [width=\textwidth]{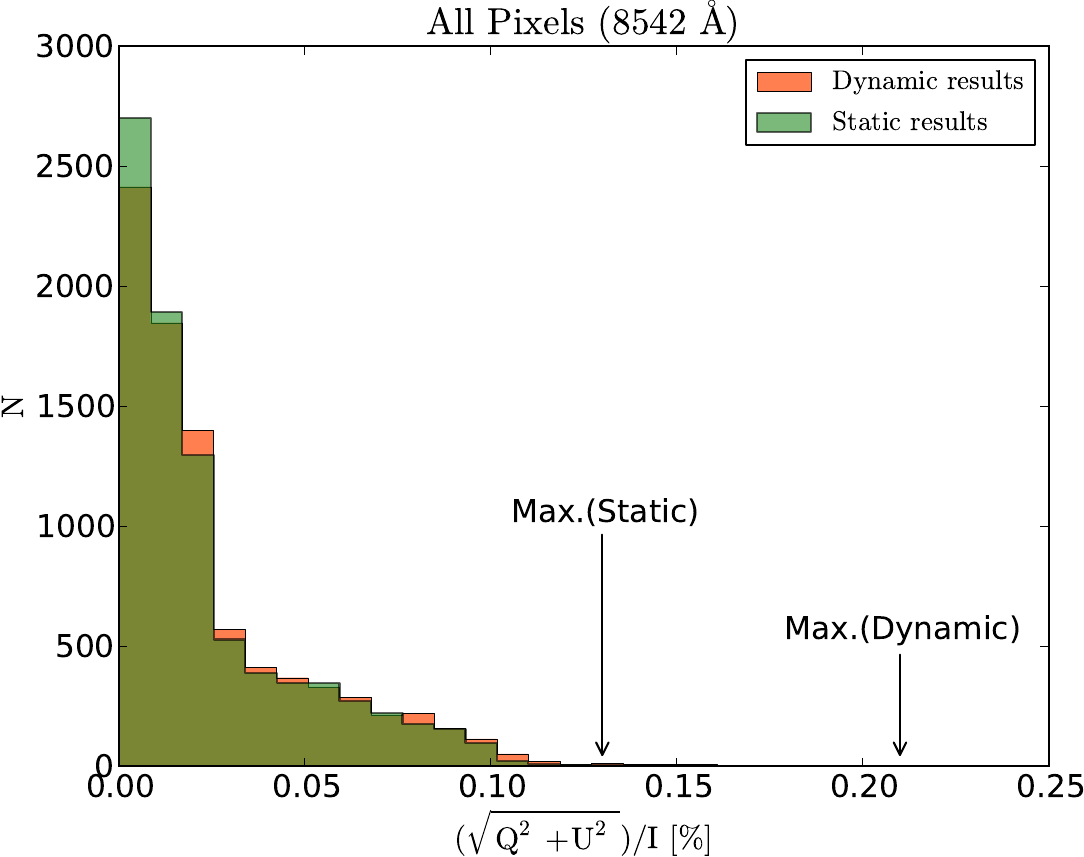}
                \label{fig:h1}
        \end{subfigure}
        \begin{subfigure}[b]{0.49\textwidth}
                \centering
                \includegraphics [width=\textwidth]{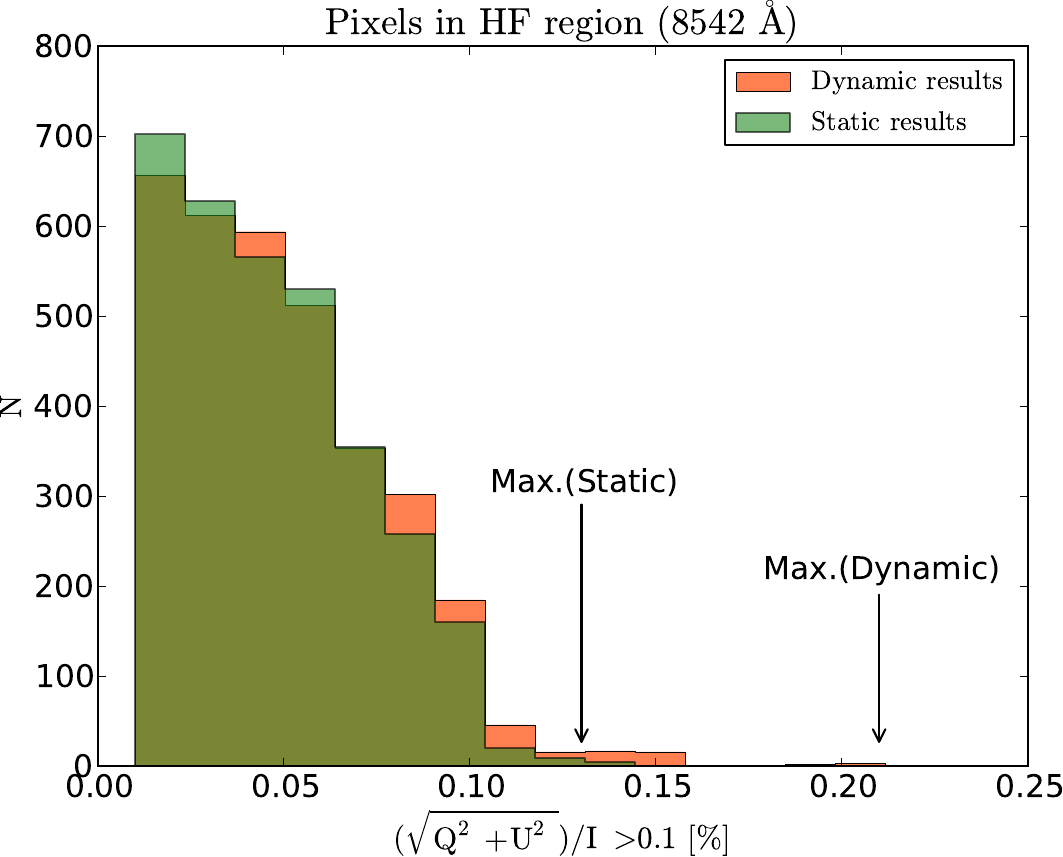}
                \label{fig:h2}
        \end{subfigure}
        \caption{Histograms of the total linear polarization for the $8542$ {\AA} line in the whole map (left panel) and only considering the HF region (right panel). The histograms are semi- transparent: a dark green color indicates that the orange distribution is behind the light green distribution.}
\label{fig:histoqu}
\end{figure}

 Statistically, the effect of the velocity in the maximum amplitudes of the LP resulting from the 3D snapshot is small in comparison with the effect of the magnetic field and the temperature gradients. The maximum LP polarization value of the distribution of velocities (histograms in Fig. \ref{fig:histoqu}) changes from $0.13\%$ to $0.21\%$ when adding the velocity, which is a typical amplification in the light of the results from previous chapters. It indicates that the modulation produced by dynamics in the profiles has a certain influence. However, the histograms in Fig. \ref{fig:histoqu} show that such effects are measurable in a low number of pixels, whose LP values ``migrate'' from lower to upper parts in the histogram (orange excesses in the distribution tails of Fig. \ref{fig:histoqu}). One reason for this relatively small number of LP signals with strong amplifications is that the velocity gradients in the formation region are not as large as in the time-dependent models. Indeed, the dynamic is comparatively much more reduced in the MHD models (Fig. \ref{fig:uves}). It has to be taken into account that we should expect stronger dynamic effects in the Sun \citep{Leenaarts:2009}. On the other hand, we have to consider that a small filling factor of enhanced signals does not necessarily mean a lack of significative effect along time.

In a non-negligible number of pixels, the ``static'' polarization amplitudes are already larger in the 3D dataset than in \textit{static semiempirical models}. The results in FALC models \citep{manso10} establish maximum LP amplitudes around $0.02\%$ in forward scattering, but we find pixels with LP values reaching $0.13\%$, an order of magnitude larger. In fact, most pixels in the HF region have amplitudes above the FALC values (Fig. \ref{fig:histoqu}, right panel). These amplitudes are a consequence of the larger chromospheric temperature gradients with respect to the FALC model. Since the linear polarization is a response to the anisotropy of the radiation field, a local variation of the limb darkening law produced by the temperature can effectively explain the new amplitudes seen in these static cases.

Irrespective of the amplification factors, the effect of the velocity gradients in forward scattering are always a modulation of the LP signals with respect to the static case. As has been anticipated, the forward scattering geometry gives decreased amplitudes in many pixels due to the proximity of the formation region to the height where the upper and lower level fractional alignments become equal (Sec. \ref{sec:profilesA}). This makes the $8542$ and $8662$ {\AA} lines alter their responses with respect to the $90$-degree scattering case, behaving similarly to the $8498$ {\AA} line at limb observations. More explicitly, when the velocity gradient starts to increase, the LP signals become systematically more asymmetric with a decreasing amplitude and tending to become \textit{antisymmetric}. If the velocity continues increasing, the signals are enhanced in amplitude but with a sign that is typically the opposite of that of the static case (reversal amplification: see Fig. \ref{fig:gradsv_ctes}, second right panel from the top). Therefore, if the velocity gradient is not very strong, the amplitude of the scattering signals can be smaller than in the static case. This behavior makes them more unpredictable because such signals are not limited to only one sign.

 We already know (see Section \ref{sec:conmaps}) that a good tracer of such process is the polarization contrast (difference between the maximum and minimum value of the signals). The reason is that the contrast is always modified (generally enhanced) because the velocity gradients generate asymmetries whose total peak-to-peak excursions are usually larger than the static amplitudes. The artificial sign we add to the contrast corresponds to the maximum peak of the signal and help us to identify if a reversal amplification is being produced. Thus, comparing the static and dynamic maps of polarization contrasts we have seen that the signs of the maximum amplitudes at each pixel are never reversed. The only change is an increment of the contrasts that strengthens the patches in Fig. \ref{fig:mapcon}. That is why we conclude that the velocity gradients in the analyzed MHD models are not strong enough for putting the LP profiles in their amplification phase\footnote{Interestingly, a bit more of dynamic will do it because the contrasts are in many cases almost twice the static amplitudes, indicating that if the velocities were a bit increased, the LP amplitudes would start to be effectively enhanced with respect to the static case.}. This calculation confirm that the \textit{polarities} of Stokes Q and U are not altered by the velocity (preliminar conclusion of Sec. \ref{sec:conmaps}). The stability of the polarities also allows to extend the validity of the reference signs in Table \ref{tab:signs} to the dynamic case, which is correct for the present models but will not be true in situations with stronger dynamics.
  
A visual summary of the situation can be given following Fig. \ref{fig:prof_velo}. The labels A, B, C, D, E and F identify the panels corresponding to some ``static'' pixels. The panel immediatly below each static case corresponds to the respective dynamic case for the same pixels. Only the F panels show the behavior of the LP in $\lambda 8498$, the other panel are for $\lambda 8542$ (the conclusions for this line are extended to the $8662$ {\AA} line). The figure illustrates the following:
\begin{description}
\item[A] In some pixels, the $8542$ {\AA} line behaves like in previous studies: it increases monotonically with the velocity (static and dynamic A panels). This would be the case of the pixels corresponding to the orange excesses in the histograms.
\item[B] But a much more frequent behavior appears for these models: double-peaked profiles become asymmetric and decrease in amplitude in presence of velocity gradients (static and dynamic B panels). 
  \begin{figure}[h!]
\begin{center}$
\begin{array}{ccc}

\includegraphics[width=2.0in]{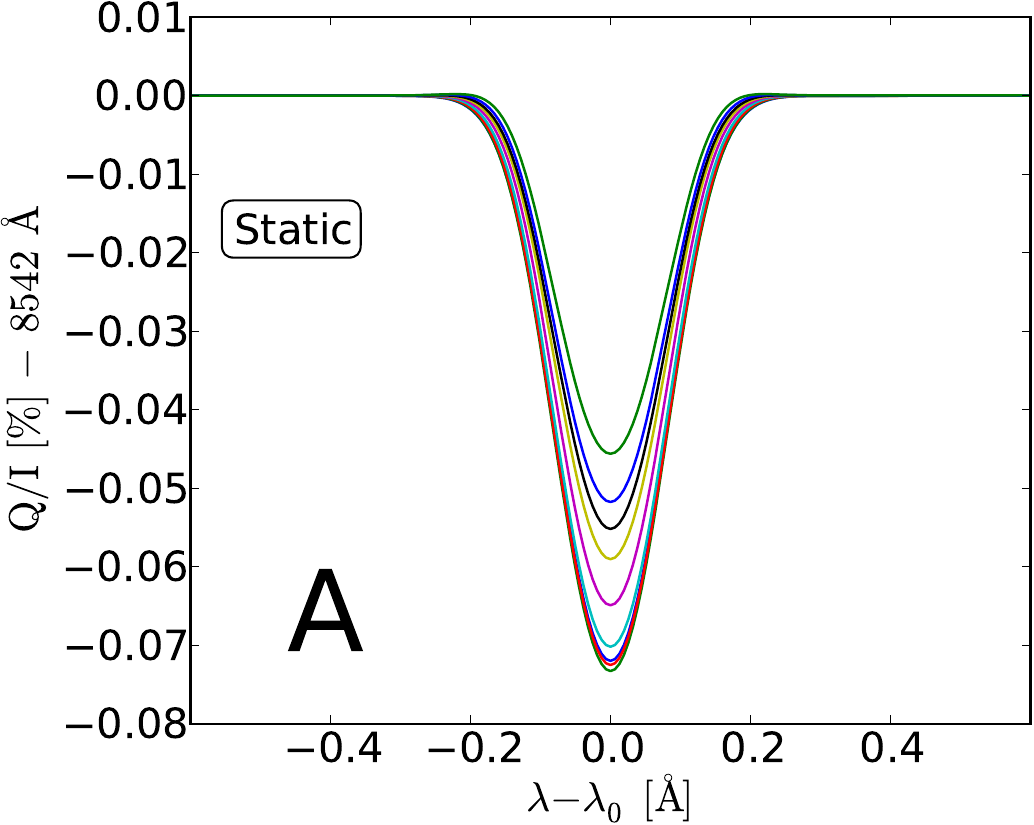} &
\includegraphics[width=2.0in]{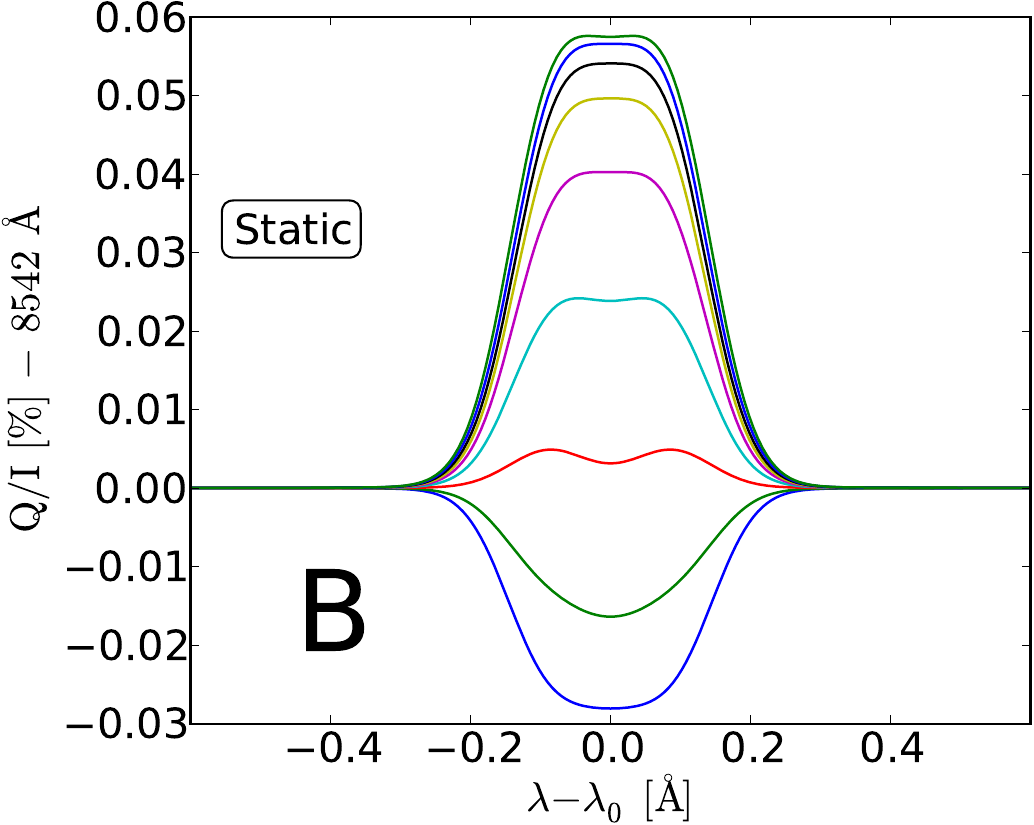} &
\includegraphics[width=2.0in]{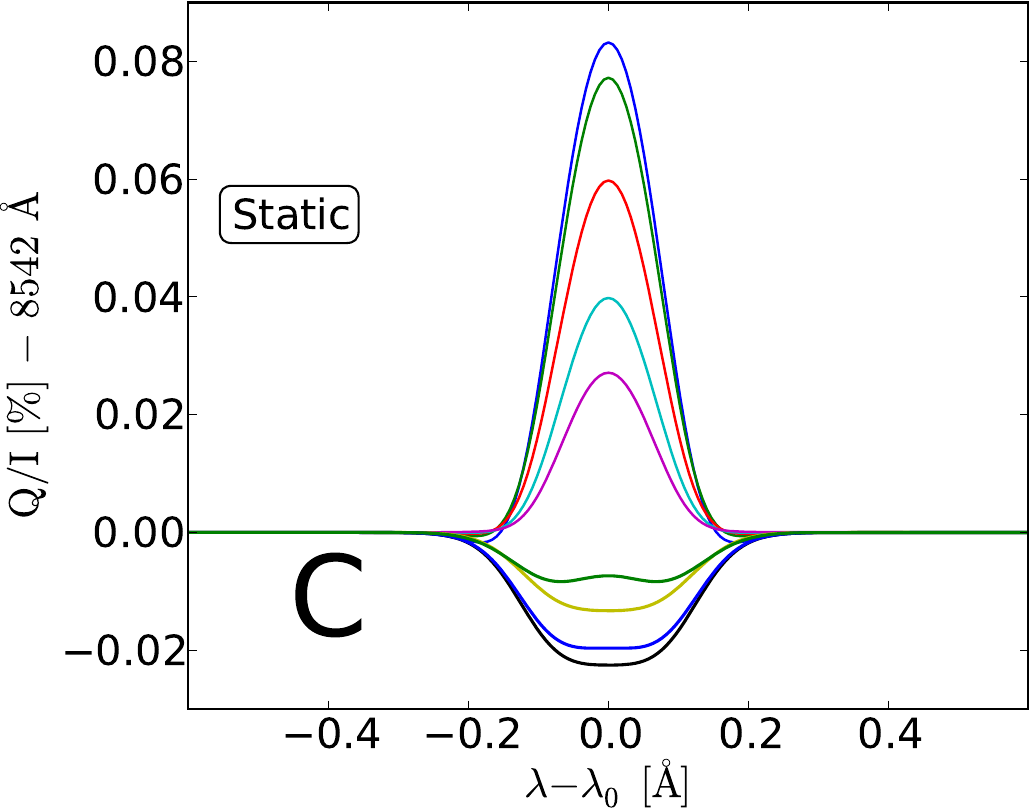} \\

\includegraphics[width=2.0in]{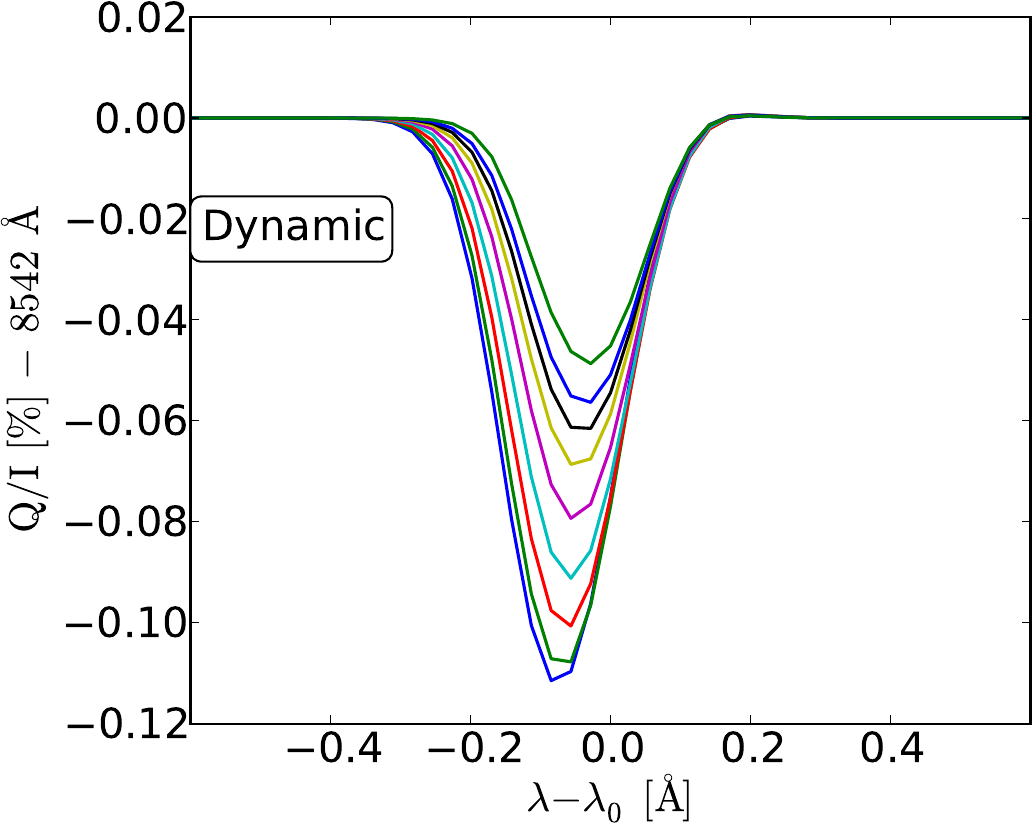} &
\includegraphics[width=2.0in]{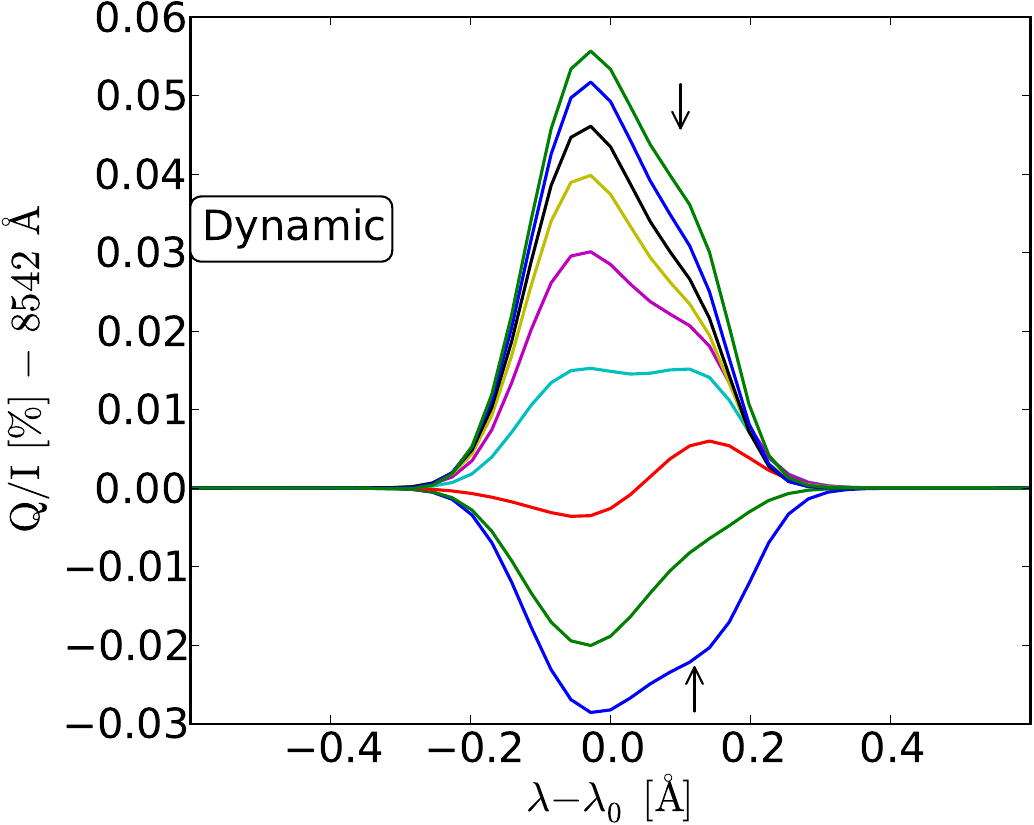} &
\includegraphics[width=2.0in]{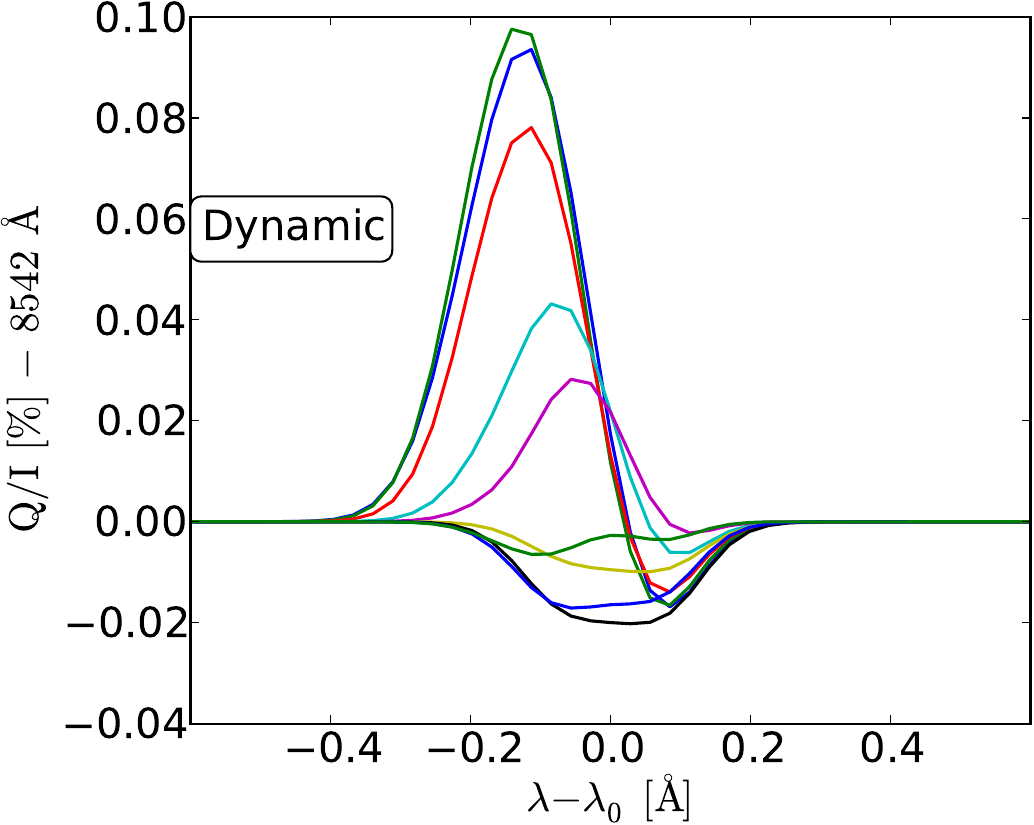} \\

\includegraphics[width=2.0in]{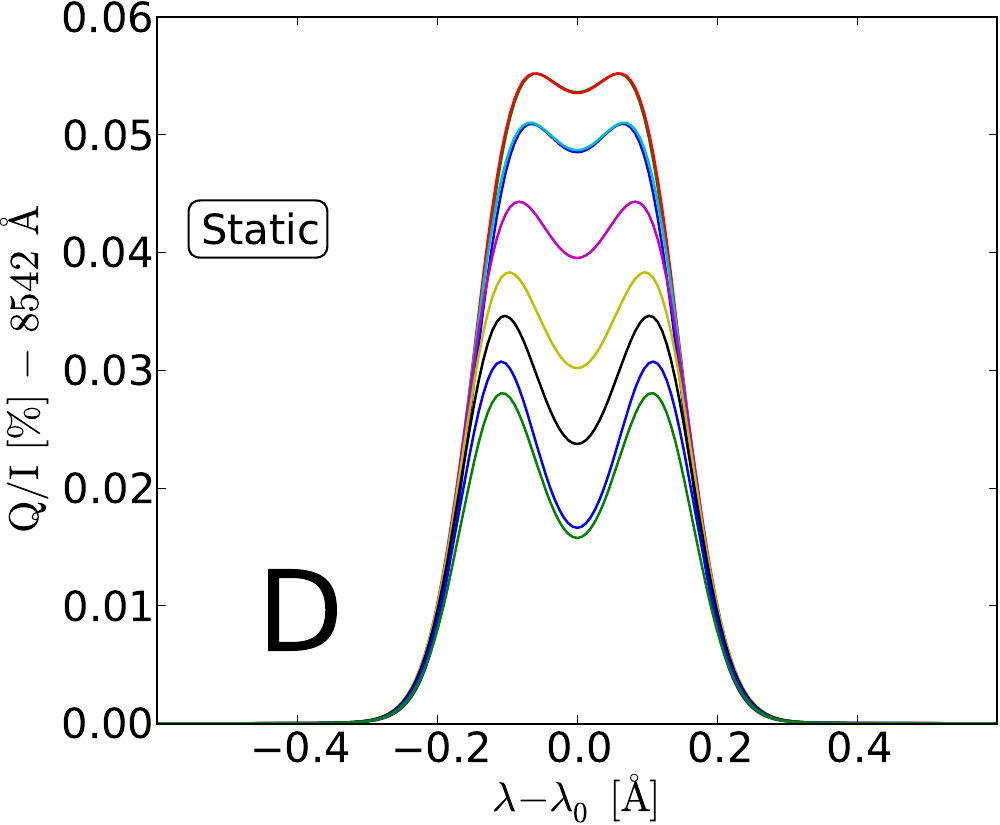} &
\includegraphics[width=2.0in]{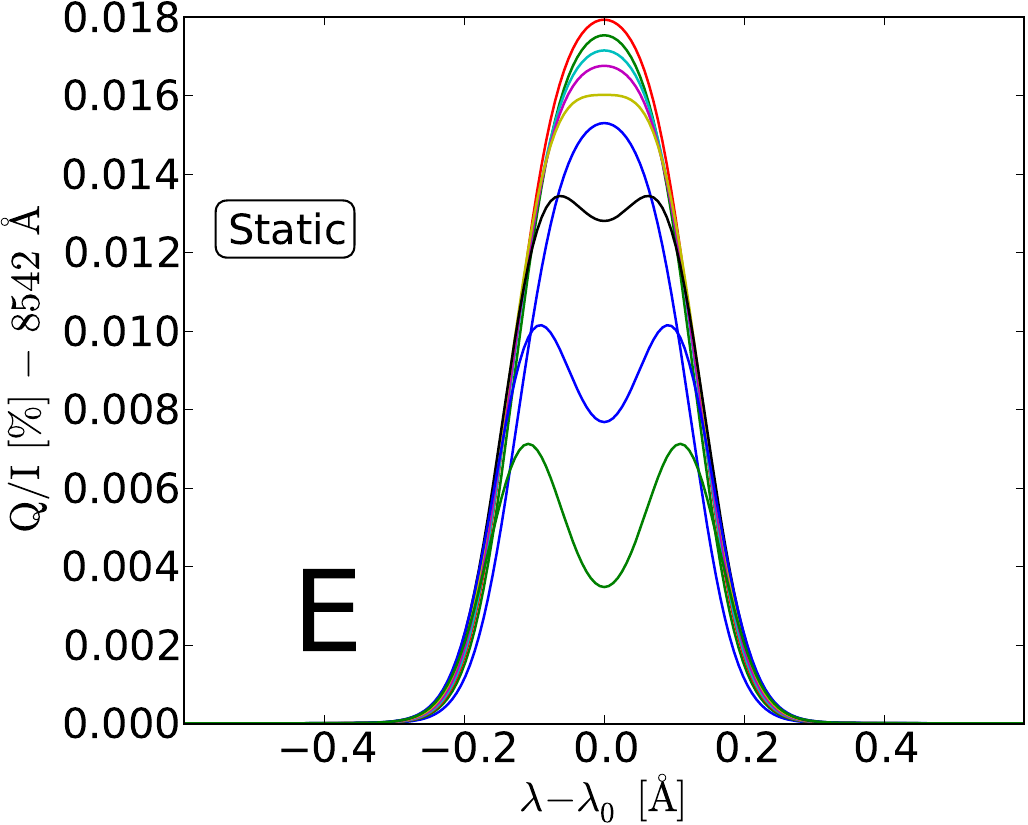} &
\includegraphics[width=2.0in]{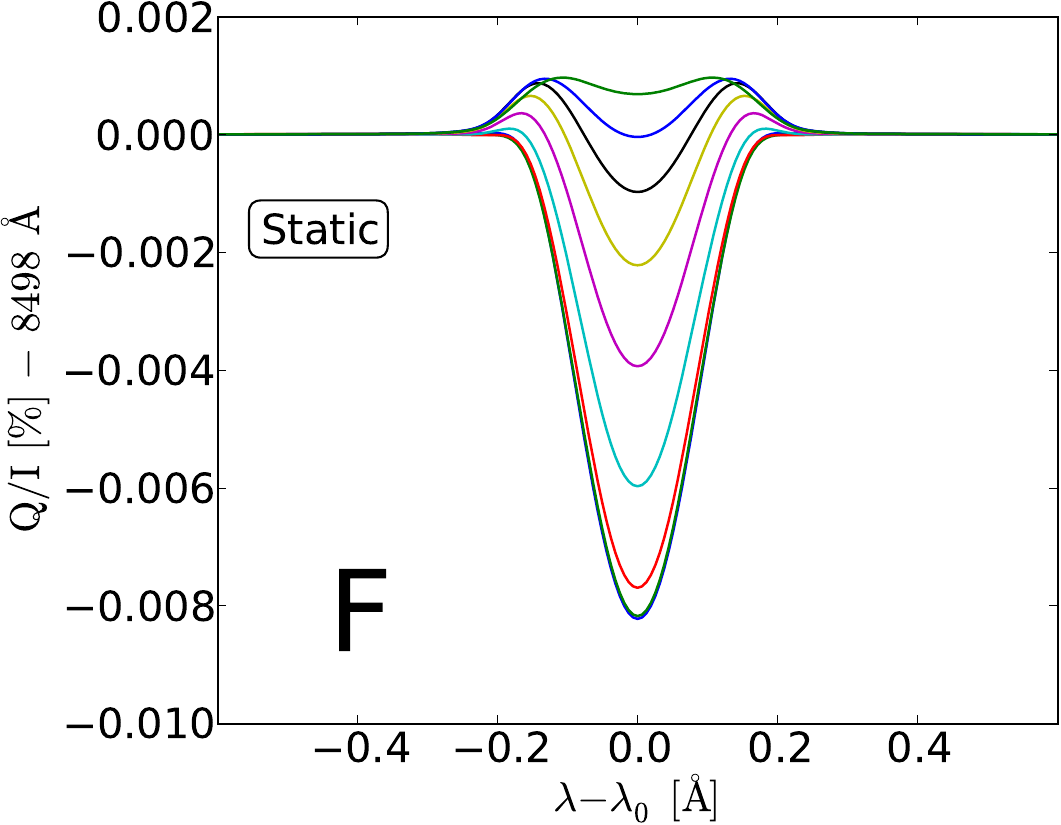} \\

\includegraphics[width=2.0in]{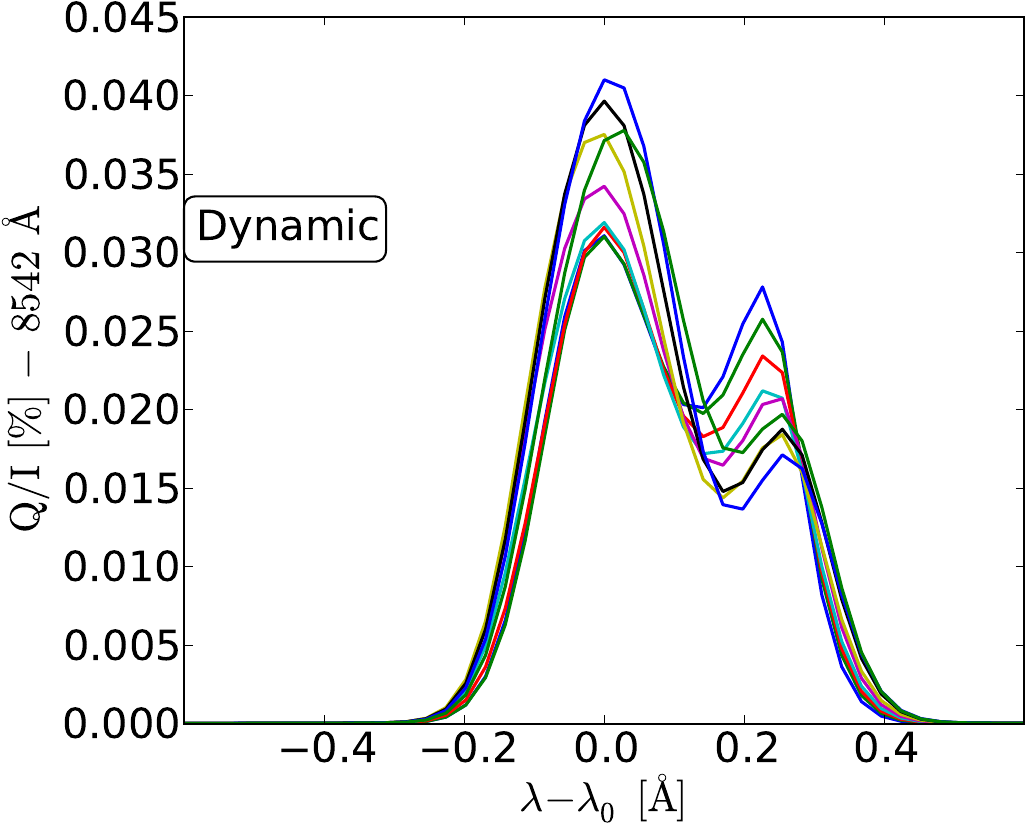} &
\includegraphics[width=2.0in]{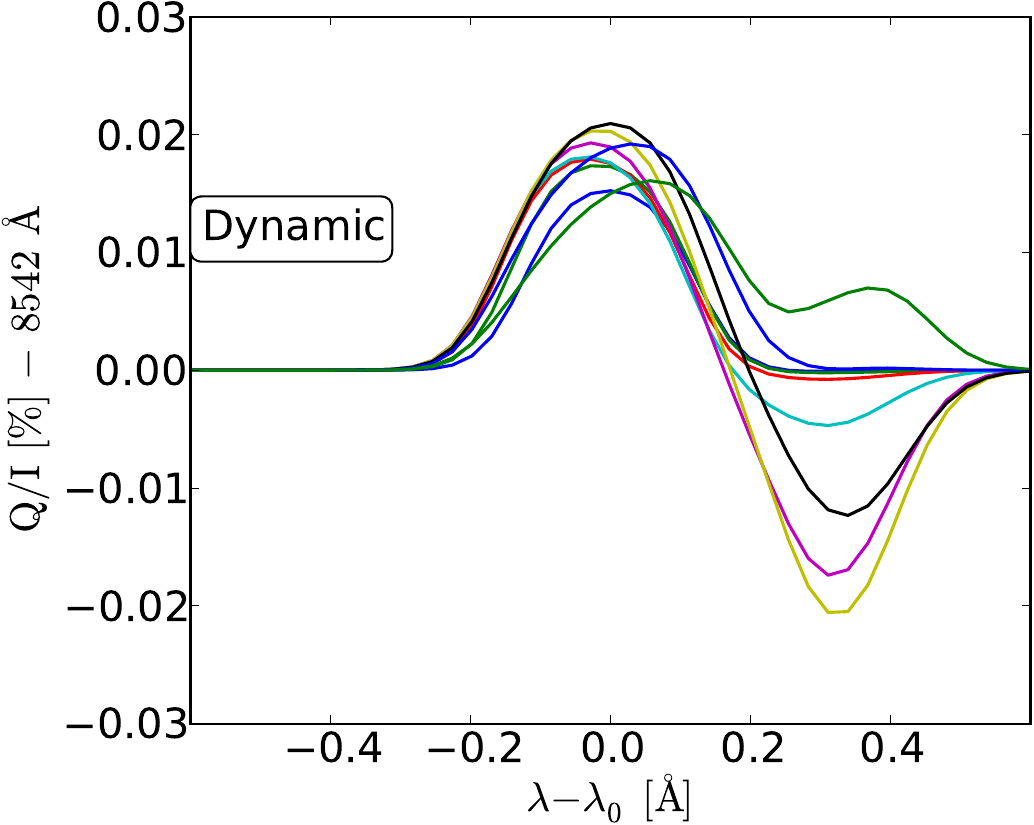} &
\includegraphics[width=2.0in]{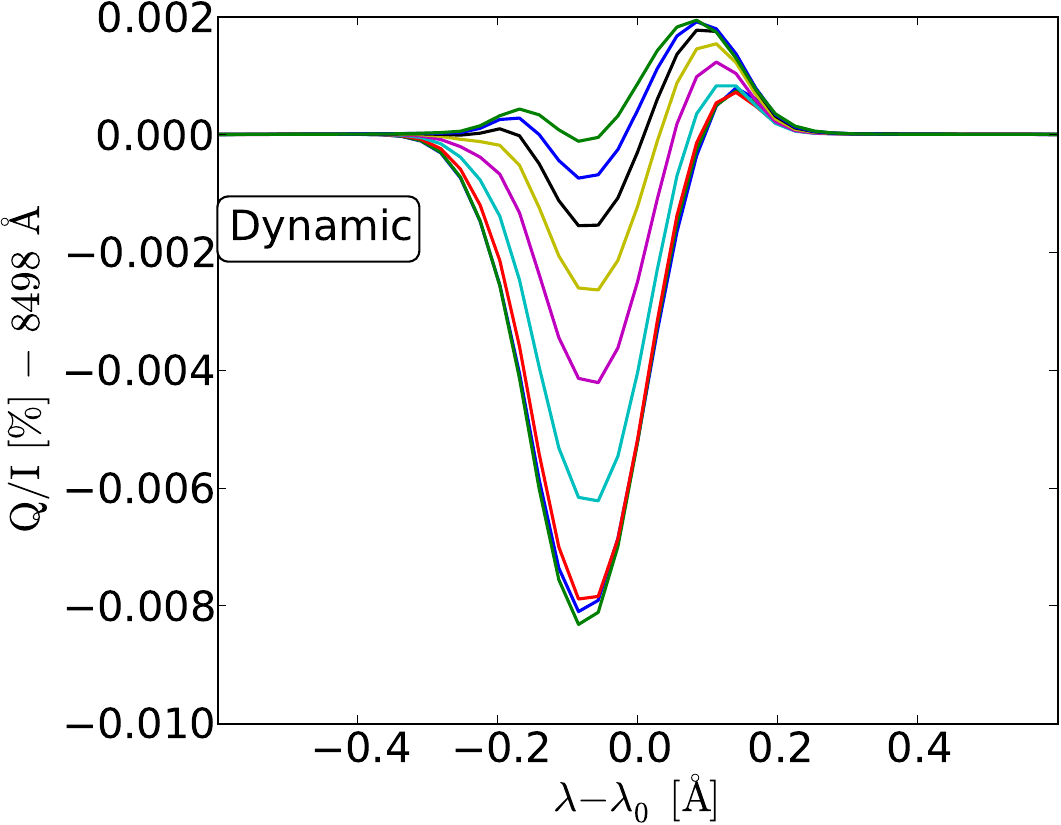} 
\end{array}$
\end{center}
\caption{Representative examples showing the effect of the velocity on the linear polarization in the considered MHD model. Grouped in pairs, the plots illustrate the $\rm{Q/I}$ signals in adjacent pixels when setting the velocity to zero (static case, labelled with a letter) and the corresponding results in the same pixels when velocity is activated (plot immediatly below the labelled one). Pairs A, B, C, D and E are for the $8542$ {\AA} line and the pair F is an example for $8498$ {\AA}. Stokes U presents similar patterns.}
\label{fig:prof_velo}
\end{figure}
\clearpage

\item[C] Both behaviors in panel pairs A and B are part of a same process that is clearly observed between adjacent pixels like being different timesteps of a similar temporal evolution (dynamic C panels). 
\item[D] The larger the amplitudes in absence of velocity, the larger the velocity gradient necessary to reverse the profile polarity. Thus, in our MHD models, the dynamic is usually insufficient to reverse the sign of the relatively large signals created by the temperature gradients. It results in the predominance of asymmetries and variable amplitudes (static and dynamic D panels). 
\item[C, D, E] In comparison with a limb observation, the larger velocity component along the LOS when looking at disk center can produce larger spectral shifts and broadened signals (dynamic panels in C, D and E). 
\item[E] Furthermore, the rich variability of the atmosphere produce extraordinary situations that can limit the amplification of the LP signals in the presence of significant velocity gradients. For instance, when there is a dense cloud or bubble with an homogeneous motion \textit{over} a region with large gradient, its opacity softens the impact of the Doppler-induced anisotropy on the emergent polarization. So, in such accumulations of material (more usual above stronger magnetic fields) we see broader spectral signals but not necessarily large amplifications (static and dynamic E panels).    
\item[F] The $8498$ {\AA} line follows a similar behavior. It also has larger amplitudes than in the FALC model due to the temperature, but it is poorly amplificated by the low- chromosphere velocities (static and dynamic F panels).
\end{description}

\begin{figure}[h!]
        \centering
        \begin{subfigure}[b]{0.49\textwidth}
                \centering
                \includegraphics [width=\textwidth]{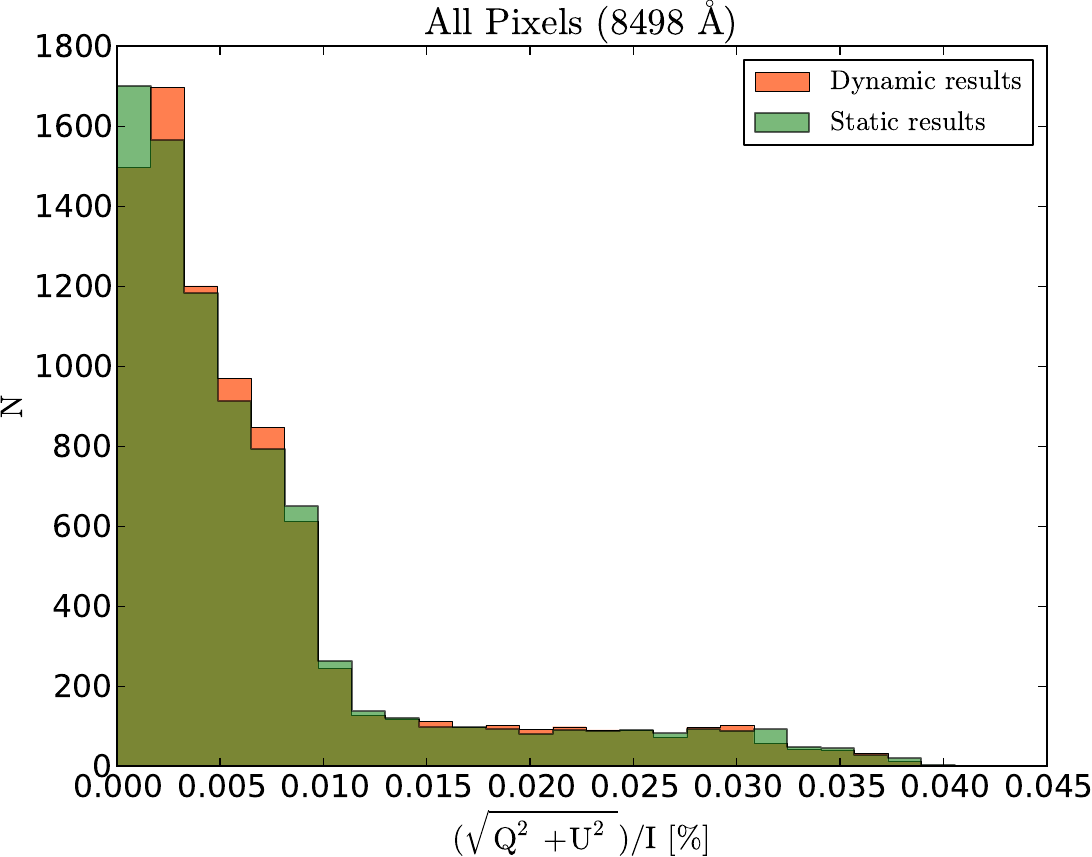}
                \label{fig:h1b}
        \end{subfigure}
        \begin{subfigure}[b]{0.49\textwidth}
                \centering
                \includegraphics [width=\textwidth]{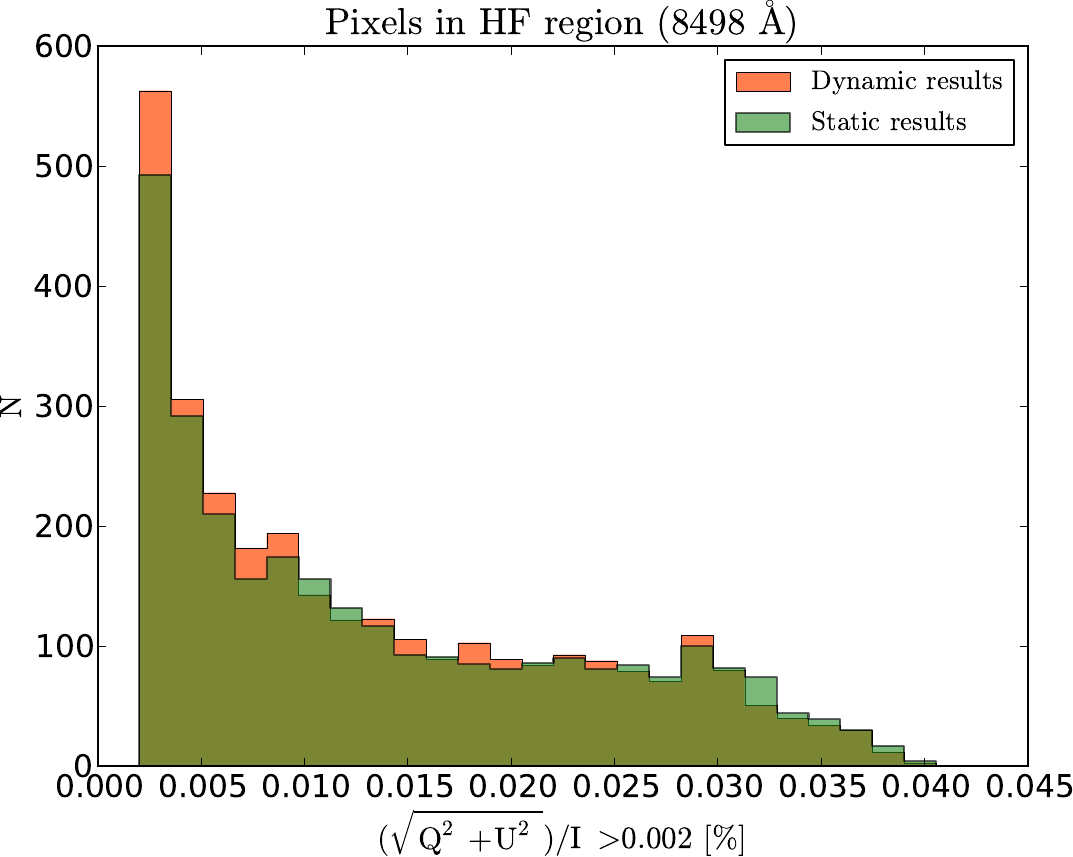}
                \label{fig:h2b}
        \end{subfigure}
        \caption{Histograms of the total linear polarization for the $8498$ {\AA} line. }
\label{fig:histoqu8498}
\end{figure}

A remarkable difference in the $8498$ {\AA} line with respect to the other spectral lines in the IR triplet is the longer distribution tail in the histograms of Fig. \ref{fig:histoqu8498}. The maximum amplitudes are produced by temperature, not velocity, and reach $0.04\%$, which is not very different to the $\sim0.02\%$ obtained in FALC and forward scattering. The statistical effect of the velocity in the LP histograms for the $8498$ {\AA} line is an increment in the number of pixels with lower amplitudes.

 We conclude that, due to the small spatial filling factors, the effects of dynamics on the linear polarization at the solar disk center can only be detected in solar observations with large  resolution and sensitivity. If the Sun were similar to our models, the temperature would be predominantly influencing the LP profiles and the action of the velocity would be practically anecdotic at disk center. The observational difficulty is to capture the temporal evolution of those small LP patches in the solar surface, specially when having place during repetitive, short and explosive propagations of shock waves along the chromosphere. Even theoretically, with only one snapshot of the atmoshere we loose the real relevance of such events. In the case the observational issues were solved, the velocity can be crucial to explain the chromospheric observations in these spectral lines. 
\section{Synthetic observations.}\label{sec:observations}

The solar chromosphere has traditionally been observed in ${\rm H_{\alpha}}$ and the Ca {\sc ii} H \& K lines. But their use for diagnostic suffers from significant drawbacks. The Ca {\sc ii} H \& K lines have wavelengths in the violet part of the spectrum, where we have a dropoff in the Planck function (decreased photon flux) and a low filter telescope transmission. Furthermore, there is an additional degradation of the chromospheric signal in the Ca {\sc ii} K filtergrams due to reduced atmospheric transparency, decreased detector efficiency and the worsening of the atmospheric seeing at shorter wavelengths. Although the chromospheric contribution of the Ca {\sc ii} H \& K is limited to a narrow core of less than $0.02$ nm wide, all the imaging has been performed with much broader filters, having FWHM passbands in the range of $0.03$ to $0.3$ nm (0.1 nm FWHM being relatively broad). Such broad filters lead to significant low-chromospheric line wing contributions, which makes the detection of small chromospheric structures difficult, and long exposure times worsing the spatial resolution \citep[e.g.][]{Reardon:2009aa,Vecchio:2007aa}.

Because of the previous issues, the on-disk images obtained in the H and K lines with relatively broad filters are always significantly different from the appearance of the chromosphere in the other prominent chromospheric line, ${\rm H_{\alpha}}$. Observations in this line, even with broad filters show a highly structured environment including fibrils, mottles, and filaments across the full solar disk. This is consistent with the growing dominance of the magnetic field and the velocity at increasing heights in the atmosphere. Most of the images taken in the Ca {\sc ii} H \& K lines, instead, do not typically show such structuring by the magnetic field \citep[see, however, ][]{Pietarila:2009aa}. In addition to the observational issues, the Ca {\sc ii} H \& K lines are subject to partial redistribution (PRD) effects, which complicates their proper modeling. The linear polarization they show can be correctly approximated only at line center but the wings polarization is purely due to PRD effects, specially at the limb. Because of that, in this section and in the Appendix we have found useful to show some calculations for the K line at line center to offer a comparisson with the linear polarization amplitudes and the spatial structure of the triplet lines.

Like ${\rm H_{\alpha}}$, the Ca {\sc ii} IR triplet lines are subordinated, but whereas the lower level of ${\rm H_{\alpha}}$ is coupled to the hydrogen ground level via the very strong ${\rm Ly_{\alpha}}$ radiative transition, the lower level of the triplet lines is metastable and only coupled to the Ca {\sc ii} ground level via electronic collisions. This makes the interpretation of the IR lines easier than that of ${\rm H_{\alpha}}$ \citep[e.g., compare the radiative transfer investigations of ][]{manso10, Stepan:2010aa}.

Observationally, the use of the IR lines provides several significant advantages with respect to H and K, including a typically better response of digital detectors in the red, a reduction in the seeing, and a higher photon flux. Indeed, in recent years, the use of the IR triplet for solar studies has increased notably \citep{Socas-Navarro:2006, Judge:2006, Uitenbroek:2006aa}, although studies combining spatial and spectral high resolution over extended fields of view (FOV) are still scarce. Observations can be done with narrow filters and short exposure times, yielding a clean and high-resolution view of the chromosphere. Furthermore, in the Ca {\sc ii} IR triplet lines, nonequilibrium and PRD effects are much less important for its formation \citep{Uitenbroek:1989aa}. This makes them an excellent choice for diagnostic, both from the observational and modeling points of view, although at a lower diffraction-limited resolution than the Ca {\sc ii} H \& K and the ${\rm H_{\alpha}}$ lines. In the triplet, the $8498$ {\AA} line is the less used in polarimetry because it exhibits a lower scattering polarization signal and that is the reason why the $8542$ and $8662$ {\AA} are usually preferred. 

The studies done by \cite{Cauzzi:2008aa} and \cite{Vecchio:2009aa} in observations of the Ca {\sc ii} 8542 {\AA} line established the suitability of imaging spectroscopy in this line for high-resolution investigations of chromospheric diagnostics. A central finding of the former work was the nearly ubiquitous occurrence of fibrilar structures. They originate from even the smallest magnetic elements, and appear to fill large portions of the chromospheric volume, even in ``quiet'' areas. Their presence indicates that even at the chromospheric heights sampled by the Ca {\sc ii} 8542 line, the atmosphere is already highly structured by the pervasive magnetic fields, entirely consistent with the picture provided by ${\rm H_{\alpha}}$ images. Thus, the picture provided by Ca {\sc ii} 8542 reflects the true and essential nature of the solar chromosphere.

Despite the goodnesses of the 8542 {\AA} line in intensity, the forward scattering observation of its predicted scattering polarization \citep{manso10} is still challenging in the quiet Sun. To obtain a good sensitivity with low integration time, instrumentation pertaining to a new generation is needed. In this section, we do the exercise of degrading the polarization maps as if we were observing with the capabilities of the satellite Solar-C for simulating a real future observation. We also explain some problems and possible solutions faced when measuring polarimetric quantities that are non-linear combinations of the Stokes vector components.

 \subsection{The space telescope Solar- C.}\label{sec:solarc}
The Solar-C mission is a project leaded by JAXA to study magnetic energy transport and dissipation governing the dynamic solar atmosphere. The mission consists of the launching of a satellite (Solar-C) with three advanced telescopes that will achieve for the first time high spatial resolution, high throughput, high cadence spectroscopic and polarimetric observations seamlessly covering the entire atmosphere (photosphere, chromosphere, transition region and corona).

Solar-C has the Solar Ultra-violet Visible and IR Telescope (SUVIT), a diffraction-limited telescope with a $1.5$ m aperture in diameter. It has potential to resolve structures with $0.1$ arcsec ($0.09$ arcsecs/pixel in the focal plane) for the first time in history of space observations. Thus, it can reveal dynamical behaviors of the solar atmosphere through elementary magnetic structures and key physics responsible for energy transfer and dissipation, with emphasis on chromospheric magnetic fields and dynamic. SUVIT covers a wide wavelength region from the UV ($\sim2800$\,{\AA}) to the near infrared ($1.1 \,\rm{\mu m}$), in which there are several of the best spectral lines suitable for diagnosing dynamics and magnetic fields in the chromosphere as well as the photosphere. He {\sc i} $10830$ {\AA} and Ca {\sc ii} $8542$ {\AA} spectral lines are prioritized as the best lines for diagnosing chromospheric magnetic fields.
\begin{figure}[h!]
\centering%
\includegraphics[scale=0.85]{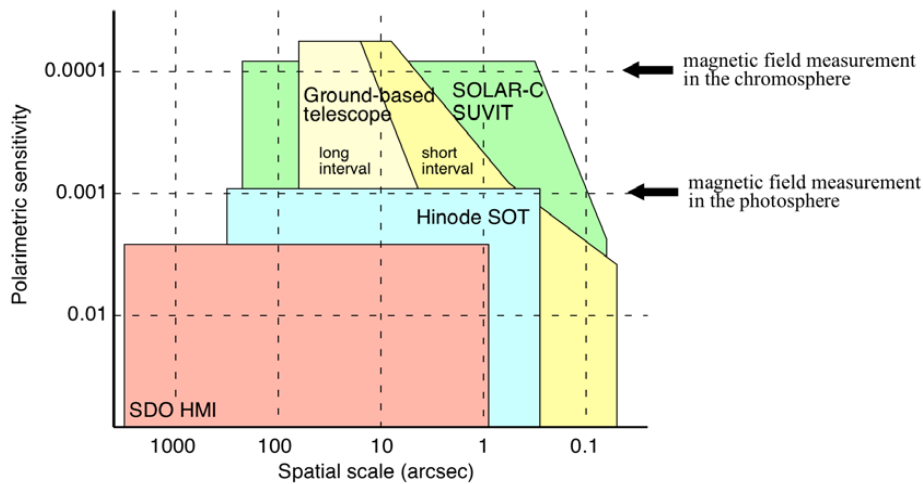}
\caption{Sensitivities of different solar facilities.}
\label{fig:sc}
\end{figure}
The instrumentation attached to SUVIT will allow to perform precise spectroscopic and polarimetric measurements (to determine temperatures, velocities, and magnetic fields) as well as imaging of intensities and magnetic fields with high temporal cadence (to capture dynamical behaviors). A narrow-band imager is planned for polarimetric and Doppler measurements at limited number of positions in spectral lines, while a broad-band imager will provide high spatial monochromatic images of the chromosphere and the photosphere.
The spectropolarimeter will reach a sensitivity of $10^{-4}$ in exposures times of $20$ s with spatial sampling of $0.2\arcsec$ per pixel, allowing the detection of the forward scattering Hanle effect in chromospheric structures (see Fig. \ref{fig:sc}). It supplies a spectral sampling of $\sim 34$ m{\AA} at $8542$ {\AA} ($\lambda/\Delta\lambda \sim 2\cdot10^{5}$). Following the calculations carried out in this thesis, Solar-C should allow the first clear detection of amplifications in the linear polarization signals produced by inclined magnetic fields and chromospheric shock waves in the Ca {\sc ii} transitions. As a consequence, we may have a suitable observable for capturing the dynamics and the energy transfer between the photosphere and the corona mapped by the variations of the Hanle signals. 

 \subsection{ Degraded and restored maps  of Polarization Amplitudes.}\label{sec:polamps}
\begin{figure}[h!]
\centering%
\includegraphics[scale=0.53]{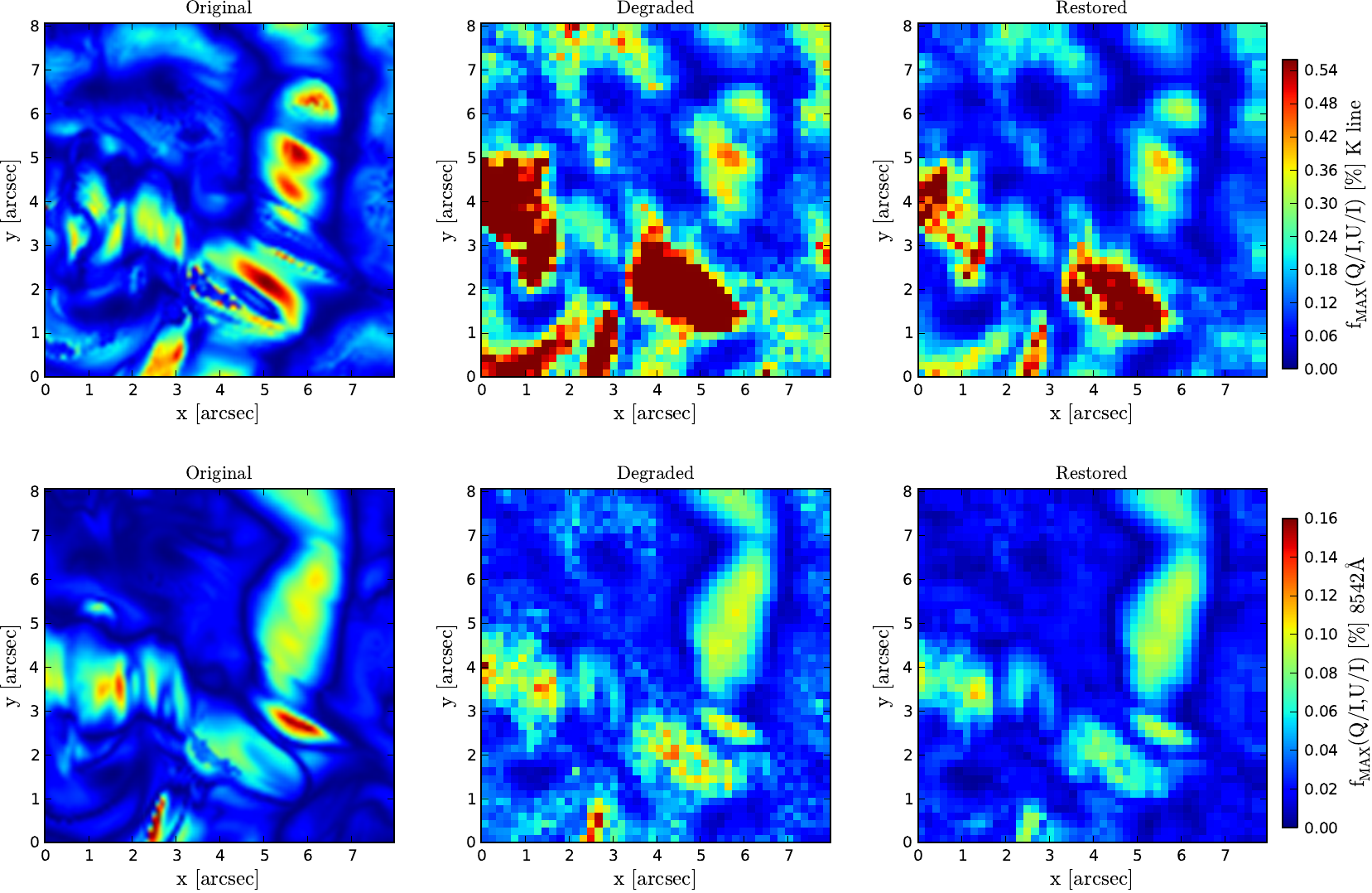}
\caption{Effect of degrading and restoring the maps of maximum total linear polarization in $8542$ {\AA} (lower panels) and K lines (upper panels). \textbf{Left column}: original synthetic maps. \textbf{Middle column}: degraded maps emulating a Solar-C observation (see text). \textbf{Right colum}: Recovered maps after integrating $0.36\arcsec$ spatially along the x axis without rebinning. Each colorbar is common for a row of maps. }
\label{fig:8542_1em4_100_3px_max}
\end{figure}

We have applied a procedure of degradation and posterior reconstruccion to the linear polarization signals for imitating a real Solar-C observation (Figure \ref{fig:8542_1em4_100_3px_max}). We followed the next steps:

\begin{itemize}

\item First, we calculated the total linear polarization applying Eq. (\ref{eq:pamp_l1}) to the Stokes vector components resulting from our RT calculations, whose original spatial resolution is $\sim0.09\arcsec/\rm{px}$. We show the results for the $8542$ {\AA} and the K lines in the left panels of Figure \ref{fig:8542_1em4_100_3px_max}.

\begin{equation}\label{eq:pamp_l1}
\centering
    \mathrm{f_{MAX}(Q/I,U/I)} = \, 100 \cdot  \mathrm{\sqrt{Max\left|\frac{Q}{I}\right|^2+Max \left|\frac{U}{I}\right|^2}}\quad [\%]
\end{equation}

\item Second, we simulate the observation with a telescope, so that we degraded the quality of the signals by convolving the original maps of I, Q and U with a spatial PSF with $\mathrm{FWHM=0.18\arcsec}$, a spectral PSF with $\mathrm{FWHM=100\,m\AA}$, making a binning of 2 square pixels to emulate a detector pixel of $0.18\arcsec$ (slightly lower than the pixel size in Solar-C in spectropolarimetric mode) and adding a gaussian noise that gives a sensitivity of $\mathrm{S/N=10^{4}}$ for an observation with 9 seconds of exposure time (worst case). Such operations were done in that order to imitate the real physical degradation at the satellite. With the resulting maps we calculated again the quantity of Eq. (\ref{eq:pamp_l1}). The results are shown in the middle column panels in Figure \ref{fig:8542_1em4_100_3px_max}.

\item Finally, we integrated photons to improve the signal to noise ratio as if we were treating the observed signals. Thus, we averaged each 2 pixels along the x axis\footnote{We did not integrate along y because the aim is to integrate the minimum possible for avoiding loss of resolution. We preferred to integrate just 2 pixels along a row instead 2 pixels square.}, doing it separately for the degraded I, Q and U maps. We preserved the pixel size avoiding binning in the operation (each pair of pixels integrated together share a common pixel with the next pair along the row). With the resulting maps we calculated again the quantity Eq. (\ref{eq:pamp_l1}). The results are in the right column panels of Figure \ref{fig:8542_1em4_100_3px_max}.

\end{itemize}

The first thing we note in the resulting maps are the dark red patches in the K line ``observations'' (Figure \ref{fig:8542_1em4_100_3px_max}). They are a product of the low number of photons that we have in the emerging core intensities of this spectral line, especially in cool chromospheric areas. Such low levels of intensity in patches of cool temperatures are able to produce spurious signals much larger than the ones observed in the true maps. The pixel integration with the aim of recovering the signals is not enough to reverse that situation to acceptable levels, even when using larger pixel sizes in the integration (pixels as large as $0.5\arcsec$ were considered) . In any case, high resolution is required to resolve the fine chromospheric structure. As a positive counterpart, we think we could use these noise patches in the K line as indicatives of a cool chromosphere. 
The $8542$ {\AA} line is also susceptible of the noise in those areas but we easily recover the Hanle signals after the simple integration. However,  note that the higher signal levels in strong polarization patches (red areas in the bottom left panel) are never recovered. We see here why the Doppler-induced amplifications studied in previous chapters are so elusive in the solar observations available in Ca {\sc ii}. Basically, the small characteristic size and the apparently elongated shape of these highly dynamic structures avoid a coherent spatial detection. The characteristics times associated with the passage of the shock waves through the region of formation are also too short to be visible in wide areas during a single snapshot. Probably, these signatures are actually everywhere, but impossible to observe with our current instrumentation given the small spatial and temporal filling factor. 
In the $8542$ {\AA} maps, we are able to restore all the LP patches in the HF region with their approximated amplitudes in most cases. We also identify the Van-Vleck HPILs in the half of the map where the horizontal field strength is larger.

\subsection{ Degraded and restored maps of the Linear Polarization Degree.}\label{sec:poldegraded}
We repeated the same steps than in Section \ref{sec:polamps}, but applying Ecs. (\ref{eq:degs}): 
\begin{subequations}
\begin{align}
    \mathrm{LPD} \,[\%] &= \, 100 \cdot  \frac{\sqrt{\left[ \overset{\lambda_2}{\underset{\lambda_1} {\sum}} Q\right]^2 +\left[ \overset{\lambda_2}{\underset{\lambda_1} {\sum}} U\right]^2}}{ \overset{\lambda_2}{\underset{\lambda_1} {\sum}} I}  , \nonumber 
\displaybreak[0] \\
    \mathrm{CPD}\, \left[\%\right] &= \, 100 \cdot  \frac{\overset{\lambda_2}{\underset{\lambda_1} {\sum}} |V|}{ \overset{\lambda_2}{\underset{\lambda_1} {\sum}} I},  \nonumber
\end{align}
\end{subequations}
with the filter C ($\Delta\lambda=0.7$ {\AA}) used in Sec. \ref{sec:poldegree} to get the maps of total linear and total circular polarization degree. We have used the same instrumental parameters than in the previous section. Since these maps are the result of integrating along wavelength, their signal to noise ratio is much better and the structures and HPILs appear clearly. In all the LP degree maps, the largest amplitudes are in places with larger field strengths and insignificant velocities, and not in areas with the largest velocities as it happens in the maps of the maximum LP of Fig. \ref{fig:8542_1em4_100_3px_max}. Concerning Stokes V, the spatial, spectral and noise degradation barely affects the results, making it unnecessary to sum pixels.
\begin{figure}[t!]
\centering%
\includegraphics[scale=0.6]{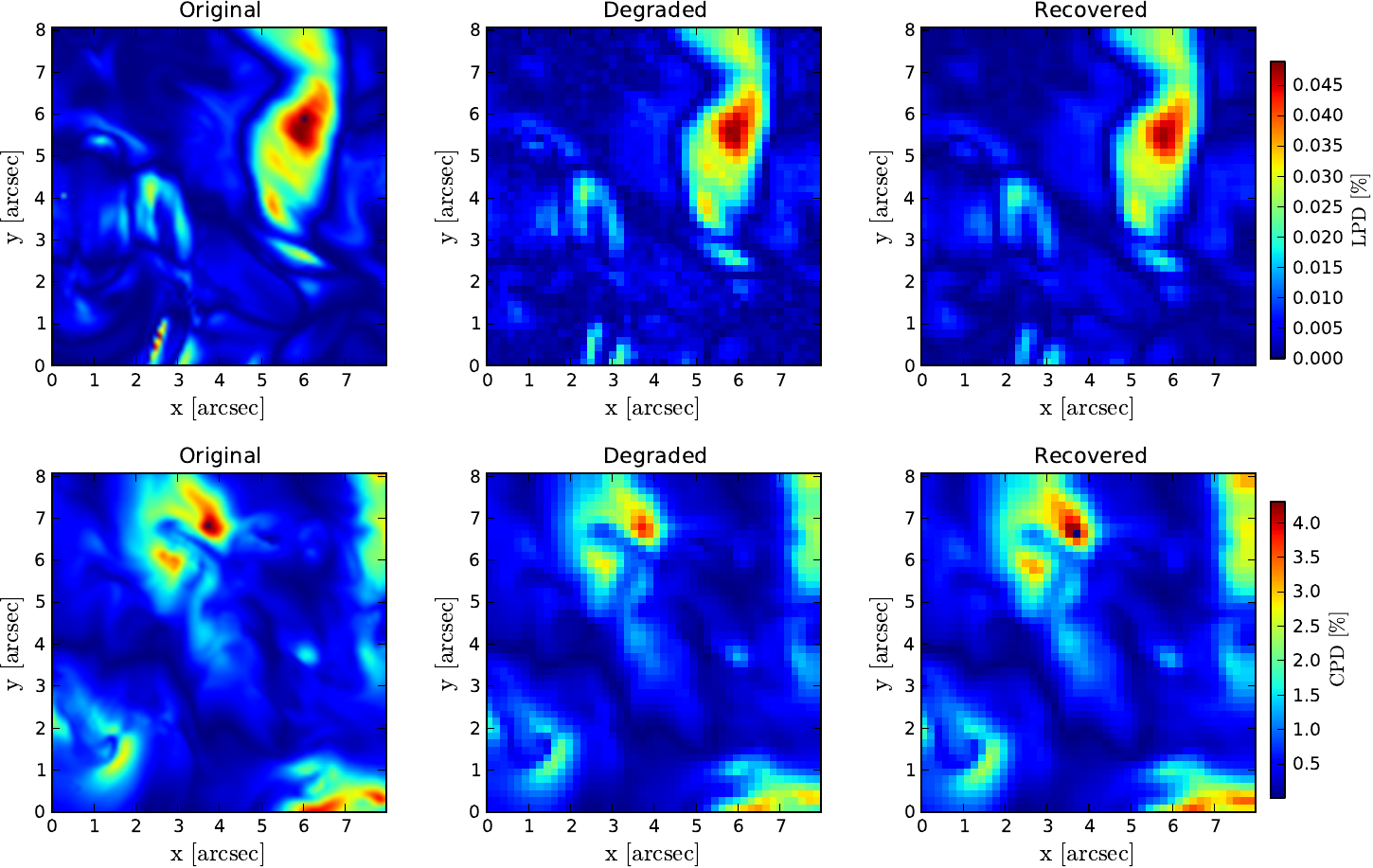}
\caption{ Effect of degrading and restoring the polarization degree maps using the filter C around $8542$ {\AA}. \textbf{Top row}: integrated linear polarization. \textbf{Bottom row}: integrated circular polarization. \textbf{Left column}: original synthetic maps.\textbf{ Center  column}: degraded maps emulating a Solar-C observation as in Fig. \ref{fig:8542_1em4_100_3px_max} (see text).\textbf{Right column}: recovered maps after integrating $0.36\arcsec$ spatially along the x axis. Each colorbar is common for a row of panels. These maps have been calculated using Eqs. (\ref{eq:pdeglin1}) and (\ref{eq:pdegcir1}).}
\label{fig:8542_integrada}
\end{figure}
\newpage
\subsection{ The practical disadvantages of the typical spectropolarimetric quantities.}\label{sec:disadvantages}
Disregarding the horizontal inhomogeneities, the only thing breaking the symmetry in a disk center observation is an inclined magnetic field. In that situation, Stokes Q and U can reach similar amplitudes irrespective of the reference for $Q>0$. Thus, to measure the total linear polarization (LP) with a single quantity, a geometrical average of Q and U is usually applied. Furthermore, it is common to use the fractional polarization, referenced to the intensity, which gives a measure that is relative to the total number of photons. Since such mathematical operations are not linear, the effect of the instrumental noise in real observations can be artificially amplified when calculating the total polarization. Then, is any of the following ways of computing the total linear polarization preferred in terms of robustness to noise? 
\begin{subequations}\label{eq:pdegrees}
\begin{align}
 \mathrm{LPD} \, [\%] &= \, 100 \cdot \overset{\lambda_2}{\underset{\lambda_1} {\sum}} \frac{\sqrt{ Q^2_{\lambda}  + U^2_{\lambda}}}{ I_{\lambda}}  \label{eq:pdegrees_lin4}
\displaybreak[0] \\
 \mathrm{LPD}\, [\%] &= \, 100 \cdot  \frac{\sqrt{\left[ \overset{\lambda_2}{\underset{\lambda_1} {\sum}} |Q|\right]^2 +\left[ \overset{\lambda_2}{\underset{\lambda_1} {\sum}} |U|\right]^2}}{ \overset{\lambda_2}{\underset{\lambda_1} {\sum}} I}  \label{eq:pdegrees_lin2}
\displaybreak[0] \\
 \mathrm{LPD} [\%] &= \, 100 \cdot  \frac{\sqrt{ \overset{\lambda_2}{\underset{\lambda_1} {\sum}} Q^2  + \overset{\lambda_2}{\underset{\lambda_1} {\sum}} U^2}}{ \overset{\lambda_2}{\underset{\lambda_1} {\sum}} I} \label{eq:pdegrees_lin3}
\end{align}
\end{subequations}

If we are interested in precisely defining the weakly polarized structures (as the Hanle polarity inversion lines defined in Sec. \ref{sec:nullpaths}), then we need a lot of photons and consequently a quantity that adds the contributions from all the wavelengths, like the linear polarization degree (LPD). But the non-linearity of the LPD makes the way we add the different wavelength contributions important. None of the Eqs. (\ref{eq:pdegrees}) are valid to do it because of two reasons. First, the physical meaning of the total linear polarization has to be preserved, in such a way that the sum operations have to be applied to photons in the same polarization state, and not to combinations of them or to derived magnitudes. For instance, the sum of fractional contibutions in Eq.(\ref{eq:pdegrees_lin4}) has not the same physical meaning that the concept of ``fractional polarization'' in the sense it is not referenced to a maximum value of $100\%$. And second, considering measures with gaussian noise (e.g., Q), their derived unsigned quantities (e.g., $Q^2$ or $|Q|$) have Rayleigh noise. When summing the Rayleigh contributions for each wavelength, the noise is added and not cancelled out (as in the gaussian case). Then, Eqs. (\ref{eq:pdegrees_lin2}) and (\ref{eq:pdegrees_lin3}) are not valid either.

The remaining basic solution would be the Eq. (\ref{eq:degs}). However, it has the disadvantage of suffering from signal cancellations in lines whose profiles have mixed polarities, as it can happen with the antisymmetric profiles produced by the velocity gradients in the LP signals of the $8498$ {\AA} line. Other disadvantage is the dependence of Eq. (\ref{eq:degs}) on the passband chosen to perform the integration (see Sec. \ref{sec:poldegree}). This is because it tends to zero as the bandwidth is enlarged, instead of tending to a fixed value proper from the LP in the line.

On the contrary, in order to capture the effects of dynamics in the polarization amplitudes,  we need quasi-monochromatic quantities. In that case, the polarization maps of Q$/$I and U$/$I should be calculated at the wavelengths of the maximum amplitudes at each pixel, which also gives a better signal-to-noise ratio. If we use the line center wavelength, the polarization amplitudes will appear decreased in map areas with chromospheric motions, avoiding the detection of dynamic amplification effects on the LP.

\begin{figure}[h!]
\centering%
\includegraphics[scale=0.52]{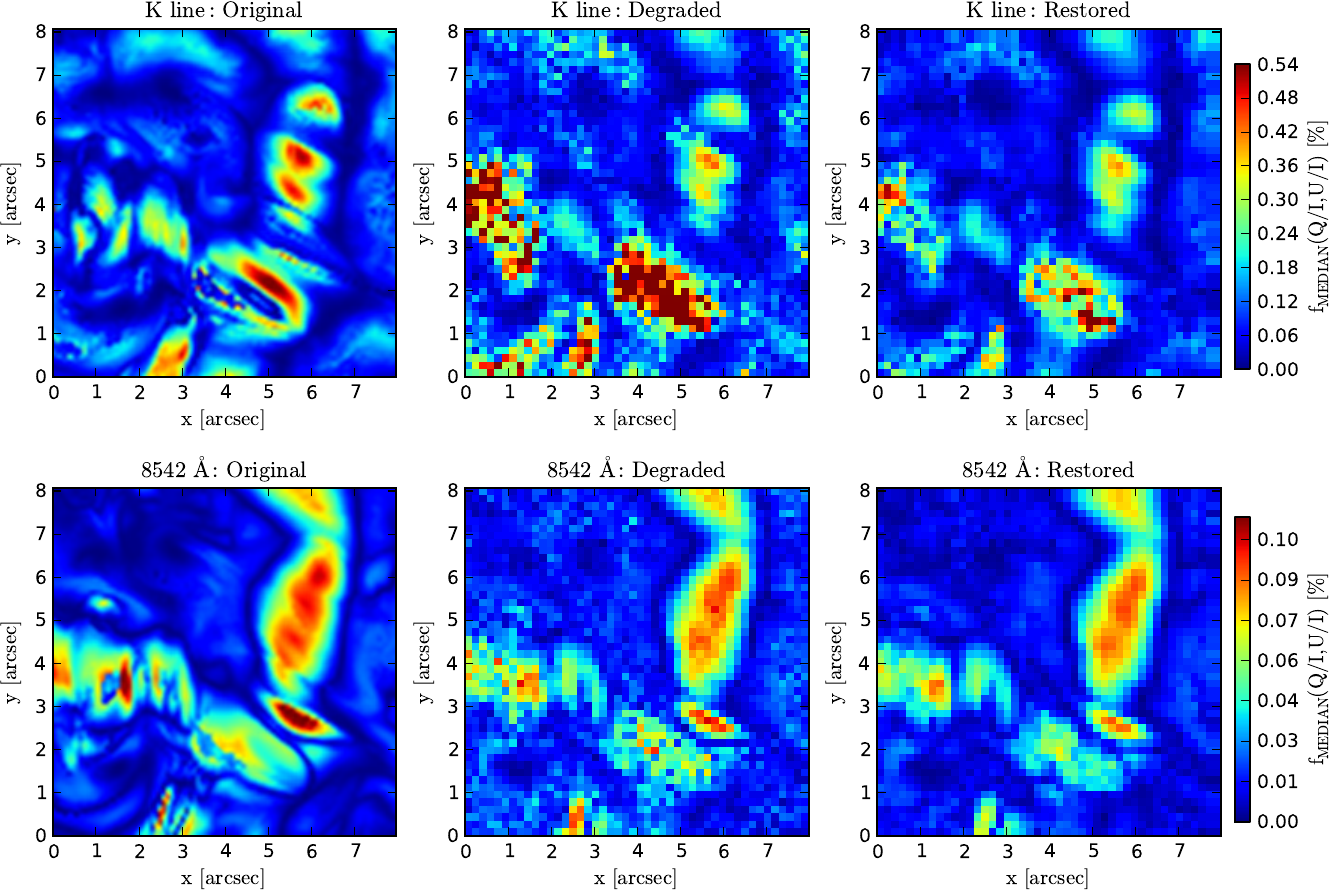}
\caption{Effect of using the median (following Eq. \ref{eq:pamp_median}) instead of the maximum value for calculating the total LP amplitudes in synthetic observations. \textbf{Upper row}: polarization amplitudes for the K line. \textbf{Lower row}: polarization amplitudes for the $8542$ {\AA} line. To be compared with Fig. \ref{fig:8542_1em4_100_3px_max}.  }
\label{fig:1em4_100_3px_median}
\end{figure}

The quantity we chose to calculate Fig. (\ref{fig:8542_1em4_100_3px_max}) is given by Eq. (\ref{eq:pamp_l1}). The polarization patches in the ``restored'' map (right panel) of the figure are a product of the noise in the faint core intensity emerging from cool- chromospheric areas. There, the LP is usually larger and the intensity is minimum, making the unstable quotients of Eq. (\ref{eq:pamp_l1}) to diverge. A possible solution is using the median (instead of the max function) to define the measure:


\begin{equation}\label{eq:pamp_median}
\centering
    \mathrm{f_{MEDIAN}(Q/I,U/I)} = \, 100 \cdot  \mathrm{\sqrt{\left[ {Median}   \left|Q/I\right|\right]^2+\left[  {Median}  \left|U/I\right|\right]^2}}\quad [\%],
\end{equation}
where the medians are calculated in a small passband ($\Delta\lambda_{\rm med}$) of a few wavelength points chosen around the common maximum for Q and U\footnote{The maximum can occur at a different wavelength for Q and U if the magnetic field azimuth varies significantly with height along the formation region. In that case, the maps do not vary too much, but at those places we would note a loss of LP signal when calculated with Eq. (\ref{eq:pamp_median}). }.
Since the median of several points around the maximum is less affected by high- frequency excursions due to the noise, it estabilizes the results without loosing the signal information at the maximum. Using three points of bandwitdh as the best choice, we got the maps of Fig. (\ref{fig:1em4_100_3px_median}). To calculate them, we repeated similar steps than in Sec. (\ref{sec:polamps}), with the same instrumental parameters, but applying Eq. (\ref{eq:pamp_median}). 

In the case of the $8542$ {\AA} map (lower panels in Fig. \ref{fig:1em4_100_3px_median}), the result of using the median is perhaps a bit better than using the maximum value (lower panels in Fig. \ref{fig:8542_1em4_100_3px_max}) because we obtain a slightly better contrast in areas with low polarization levels and recovered amplitudes that are more similarto the ones in the original map. For the K line (upper panels of the same figures), the results are also a bit better, but still unsatisfactory to distinguish between the signal and the noise in cool areas.

\begin{figure}[h!]
\centering%
\includegraphics[scale=0.45]{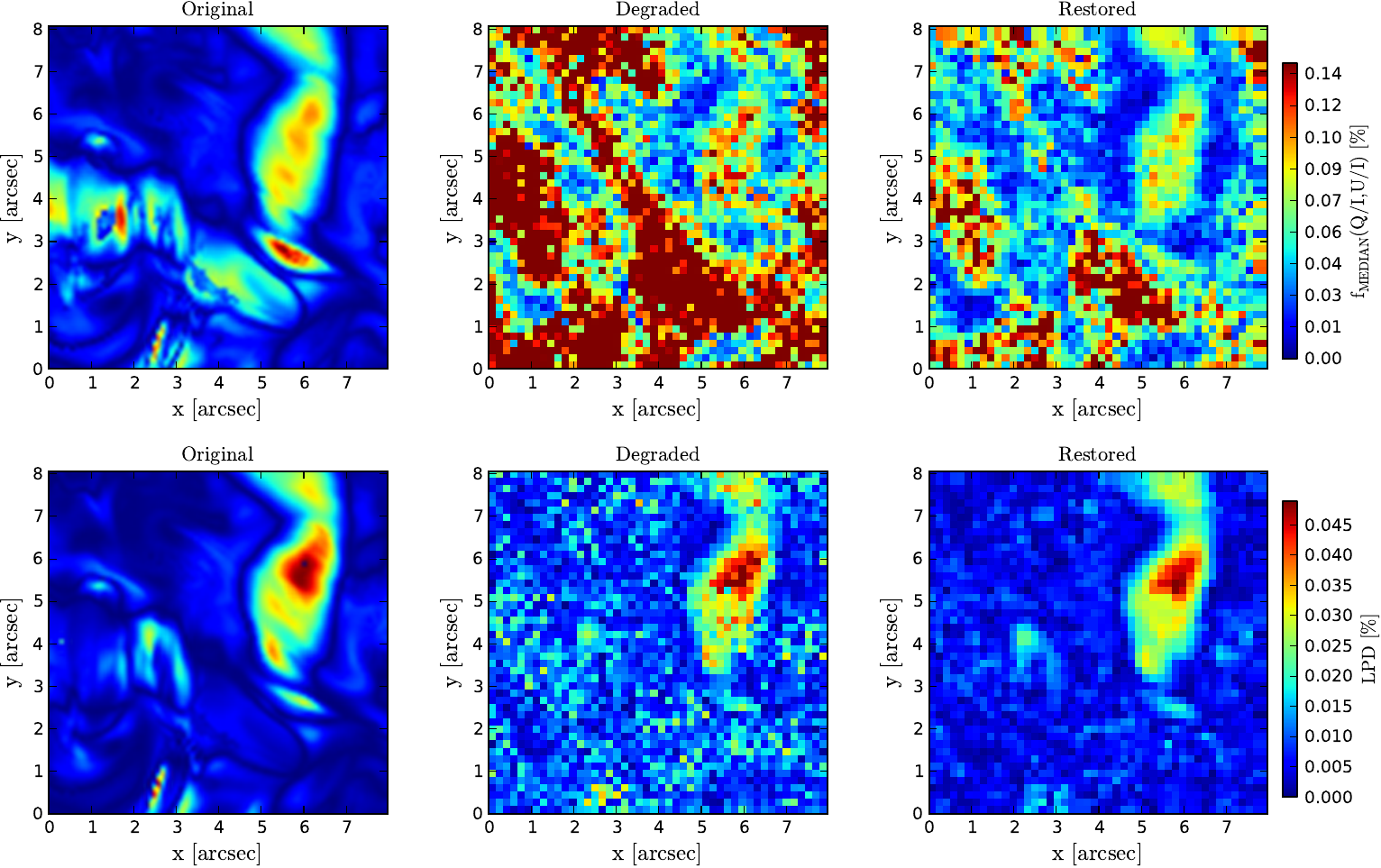}
\caption{ Synthetic observational maps of the $8542$ {\AA} line, assuming a $\mathrm{S/N=10^{3}}$ in the detector. \textbf{Upper row}: Linear polarization amplitudes calculated using the median with Eq. (\ref{eq:pamp_median}). \textbf{Lower row}: Linear polarization degree calculated following Eq. (\ref{eq:pdeglin1}). The degradation and restoration process followed from column to column is the same as in previous figures, but with $\mathrm{S/N=10^{3}}$.}
\label{fig:8542_1em3_100_3px_all}
\end{figure}

Increasing the level of noise one order of magnitude ($\rm{S/N=10^{3}}$), we do not get images with good quality for any line (see Fig. \ref{fig:8542_1em3_100_3px_all} and \ref{fig:k_integrada1em3}) or any method. In the more favorable case of the $8542$ {\AA} line, even the LPD does not allow us to distinguish the signals from the noise in a large area of the map (lower right panel in Fig. \ref{fig:8542_1em3_100_3px_all}). We would need to integrate more in time or space, so losing the required resolution and the possibility of capturing the fast and small chromospheric events.

\clearpage
\subsection{Reconstruction with PCA methods.}\label{sec:pcarec}
 We have seen that the instrumental noise can easily prevent the detection of the chromospheric linear polarization (LP) features in the Ca {\sc ii} IR triplet lines. In the past, Principal Component Analysis (PCA) has demostrated to be effective in denoising maps of Stokes profiles with much lower resolution than the ones that the Solar-C satellite and the future instrumentation will produce \citep{Rees:2000}. It can be useful when the method is applied to a large set of correlated signals (e.g., Stokes profiles in a map) that are masked by uncorrelated spurious signals (e.g., Gaussian noise). In this section, we evaluate the application of PCA to our synthetic observations of the LP in order to find out whether it is possible to effectively clean them from the noise. We experiment with two different strategies. 

In the first strategy, we obtain the PCA basis with the theoretical database of scattering polarization signals resulting from our RT calculations in the weak field regime. Our working assumption is that they can be used to explain high-sensitivity observations of the solar chromosphere as well as to validate the physical model adopted in our calculations. Later on, we use this basis to filter the noise in our synthetic observations. 

Thus, we focus on two datasets of profiles for the $8542$ {\AA} line: the dataset $1$ contains only half of the Stokes Q map calculated from the MHD models; the dataset $2$ contains all the synthetic profiles in the Stokes U and Stokes Q maps after degrading them to mimick a Solar-C observation (as we did in previous sections). We use the dataset $1$ (i.e., the well-resolved non-degraded original profiles) as the initial PCA database \footnote{Note that there are other similar posibilities. We could also create the database from U, or mixing Q and U, or even combining the Q and U of more than one spectral line, which would increment the precision and ``universality'' of the subsequent PCA basis. At disk center, Q and U should be display a very similar bahavior because their mathematical dependences only differ in a phase factor. Similarly it should be a good idea to put together the lines $8542$ and $8662$ {\AA} in order to increase the database size because their formation physics are similar. This is good because a larger database usually increases the precision of PCA (see Section \ref{sec:PCA})}. Selecting only a half of the map in Stokes Q to build the database, we can see if it is enough to explain other regions in the map and also the Stokes U signals. This database has $4185$ profiles.

\begin{figure}[h!]
                \centering
                \includegraphics [width=\textwidth]{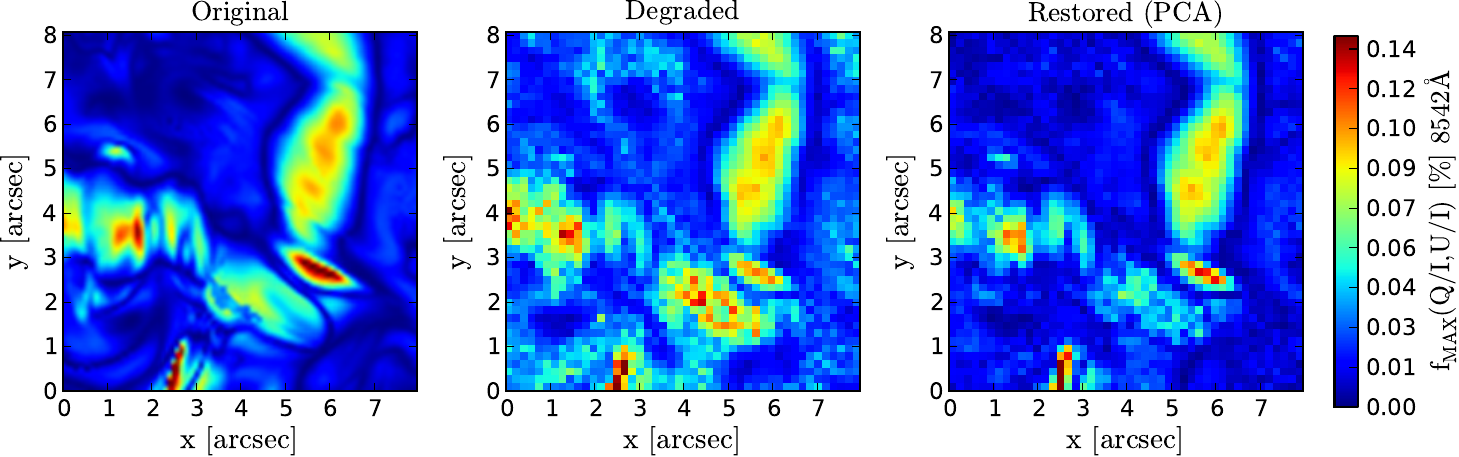}
                \caption{Simulation of the observational degradation in the maximum LP (middle panel) and its reconstruction with PCA (right panel). The detection parameters are the same as in Fig. \ref{fig:8542_1em4_100_3px_max}. The PCA reconstruction have been done independently in each Stokes parameter.}
                \label{fig:pcabest1}
         \end{figure}

\begin{figure}[h!]
          \centering%
          \includegraphics[width=0.95\textwidth]{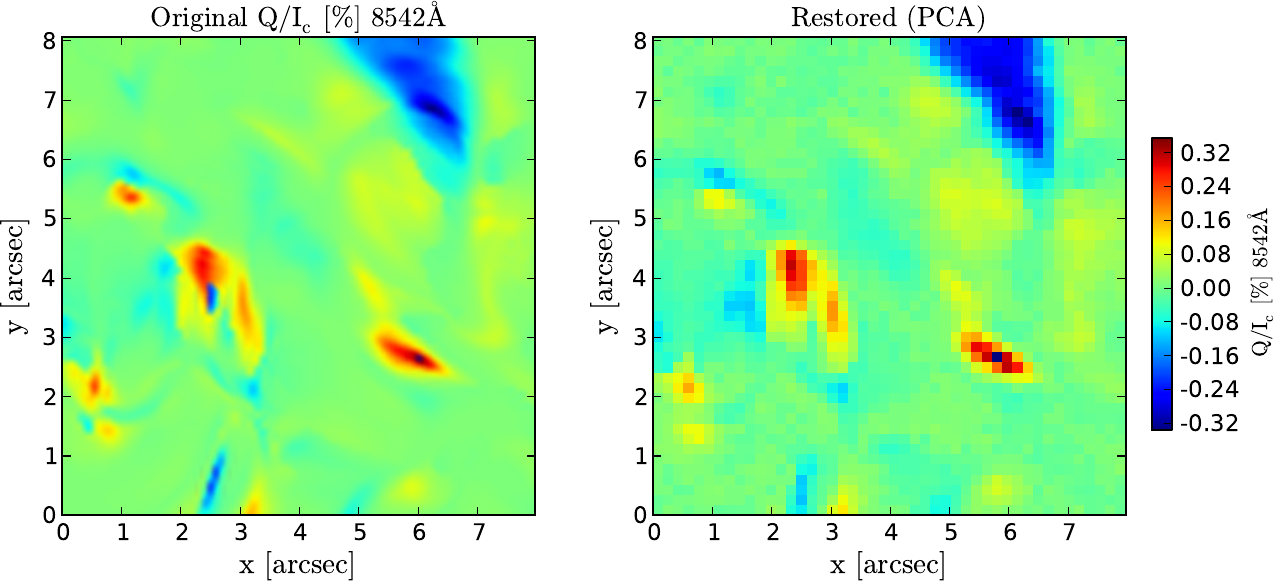}
        \caption{Maps of maximum $Q$ amplitudes at $8542$ {\AA} (including sign and normalized to the continuum intensity $I_c$) that illustrate the result of applying PCA to denoise a synthetic Solar-C observation. \textbf{Left}: original synthetic map. \textbf{Right}: Recovered map after applying PCA to the corresponding degraded map (not shown). We assumed the same instrumental parameters than in Fig. \ref{fig:8542_1em4_100_3px_max}, including a S/N ratio of $10^{4}$ in the detector.}
                \label{fig:pcaqmap}
\end{figure}

\begin{figure}[h!]
                \centering
                \includegraphics[width=0.85\textwidth]{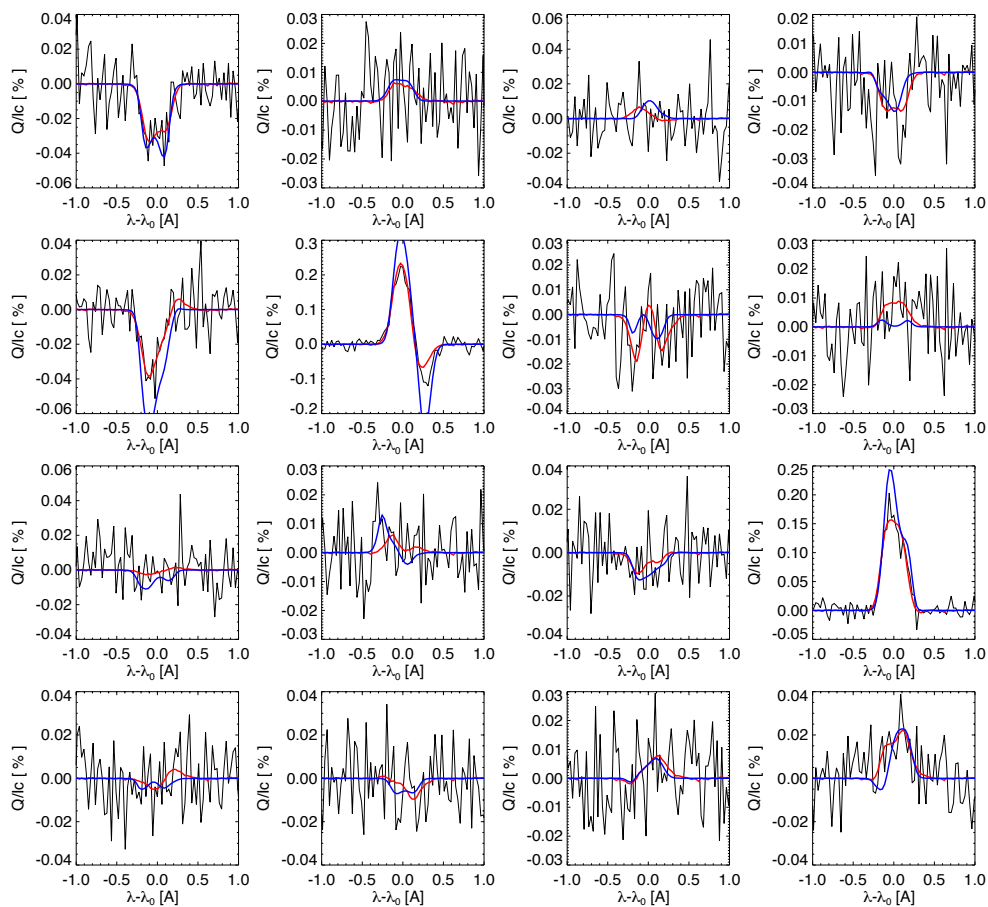}
                \caption{Examples of profiles at random pixels in the degraded observational map. In black, the Solar-C ``observation'', in red the PCA reconstruction. In blue, the real profiles that would be captured by a Solar-C pixel without introducing any instrumental degration.}        
                \label{fig:pepeb}
              \end{figure}

We use our codes (see Section \ref{sec:PCA}) to calculate the PCA basis from the profiles database. We obtain a set of $5$ eigenvectors, which explain $99,69 \%$ of the variance of the dataset $1$. Next, we project the ``measured'' profiles (dataset $2$) into the basis formed by the first $3$ eigenvectors (the more representative ones) and we reconstruct the Q and U signals with them. We repeat the procedure for the intensity and we use the outputs to calculate the maximum linear polarization with Eq. (\ref{eq:pamp_l1}). The results, comparing the original synthetic map with the PCA-reconstructed one, are shown in Figure \ref{fig:pcabest1}.
In similar circunstances, the reconstructed map is cleaner and more precise than any other one shown in the previous sections. We also obtain a clear reconstruction of the more significant amplitudes in the \textit{individual} Stokes parameters, with very small noise variability from pixel to pixel (see Figure \ref{fig:pcaqmap}). Contrarily, the maximum linear polarization, although good, still presents noisy variability. This means that, despite the effective PCA cleaning done in each Stokes parameter, the non-linear combination of them in fractional quantities still induces imprecisions in the amplitudes. The application of PCA directly to a fractional quantity, $Q/I$ for instance, does not fix the problem because the noise is still amplified before correcting it. In that case, the reconstructed maps are worse. We remark that the PCA treatment on individual Stokes parameters also shows a correct discrimination of the sign of the LP profiles in most pixels (Figure \ref{fig:pcaqmap}), which is relevant to our discussion about the orientation of the magnetic field (see Table \ref{tab:signs}).
 The signs discrimination is done from just an approximated reconstruction with regular fits to the signal shapes, as illustrated by Figure \ref{fig:pepeb}. In it, we can see the specific shape of the profiles and compare different stages during the signal processing. It is interesting to note the combined effect of the spectral, spatial and thermal degradation in the profiles (compare blue and black lines in Fig. \ref{fig:pepeb}) and the effect of applying PCA (red profiles in the same figure). Note that the blue profiles can never be obtained because they represent the ideal case before detection. 

\begin{figure}[h!]
        \begin{subfigure}[b]{\textwidth}
                \centering
                \includegraphics [width=\textwidth]{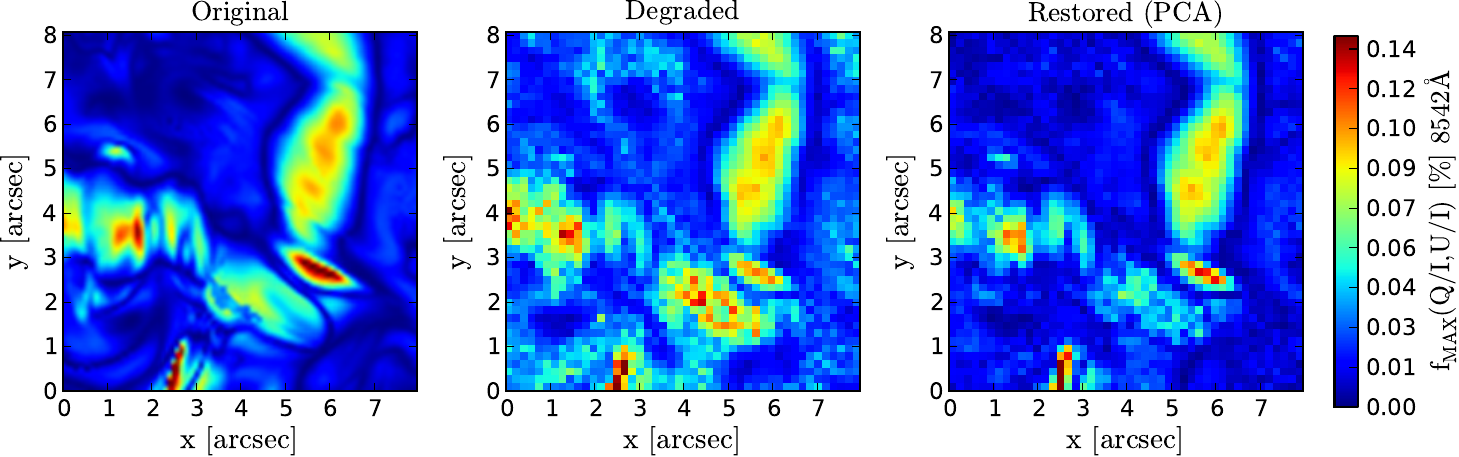}
              \end{subfigure}
        \caption{The same as Figure \ref{fig:pcabest1}, but using a PCA database formed by synthetic observations instead of by non-degraded theoretical profiles.  }
                \label{fig:pcaworse1}
\end{figure}

Our second PCA experiment consists of generating the database from the noisy synthetic observations, which are emulated from the synthetic profiles (current spectropolarimetric observations still lacks resolution and sensitivity). The procedure is similar than before. We have also used half of the map in Stokes Q (now degraded) to build the database. The result is that the so-obtained ``observational'' PCA basis also permits a similarly good signal reconstruction that matches acceptably with the original amplitudes. The PCA basis contains noise in the first eigenvectors, which is normal because our database is finite. Despite of that, from Figure \ref{fig:pcaworse1} we conclude that both strategies to create the PCA basis give a similar accuracy to reconstruct the maximum amplitudes. The use of a theoretical non-degraded database (first strategy) has been however better to fit the profile shapes and to obtain a more accurate LP sign in the reconstructions. 
PCA seems to be able to model variable and asymmetric profiles resulting from dynamic solar models. For this conclusion to be extended to real observations when using a theoretical PCA basis, it has to be proved that our physical model is a good representation of reality. It is then necessary to have a sufficiently large number of pixels covering most of the possible physical circumstances in the atmosphere. As the dynamics introduces more degrees of freedom, we expect such initial database to be larger than in the static case. On the other hand, for PCA to be useful in dynamic circumstances when using a true observational database, the number of observations has to be large enough for the noise in the first eigenvectors of the basis to tend to zero.  

Although PCA is not much better than the mere spatial integration to reconstruct the fractional polarization profiles, it gives the best results while simultaneously maintaining the spatial and temporal resolution of the dataset. Furthermore, it seems to assure a good fit to the LP signs and amplitudes. PCA provides a promising way of testing the physical models adopted in our calculations just by projecting a high-quality chromospheric observation into our theoretical PCA basis, which offers a simple feedback for the development of MHD models and the physics of polarization. Our study can be a useful anticipation for treating the analysis of well-resolved dynamic observations that the future solar facilities will provide. 
\clearpage
\section{Spectropolarimetric inference of the magnetic field.}\label{sec:b_inference}	
In this section, we use the Hanle and Zeeman effect for trying to deduce as much information as possible about the orientation, inclination and strength of the magnetic field in our synthetic quiet regions at disk center. The aim is to show some issues relevant to the inference process in the Hanle regime and how the thermodynamics and dynamism of the atmospheric plasma may affect this inference problem. 

Determinations of the vector magnetic field in the solar atmosphere are essential for understanding solar magnetic structures in general, and specifically for quantifying or even predicting solar activity. Vector magnetic fields are inferred from the Stokes parameters, I, Q, U, and V \citep[e.g., see][]{Socas-Navarro:2000, Asensio-Ramos:2008aa}. However, the component of the field that is perpendicular to the line of sight (transverse component), as inferred from observations of linear polarization produced by the Zeeman effect in magnetically sensitive spectral lines, has an inherent $180\deg$ ambiguity in its azimuth \citep{Unno:1956}. To fully determine the transverse component of solar vector magnetograms inferred by the Zeeman effect, this ambiguity must be resolved. 

Calculation of electric currents, magnetic energy and helicity budgets and most techniques of coronal magnetic field extrapolation rely on disambiguated vector magnetograms. The azimuth ambiguity \citep[e.g., ][]{Harvey:phD} continues to be an open research topic today. Nowadays, vector magnetograms are routinely provided by the ground-based Vector SpectroMagnetograph \citep[VSM; ][]{Henney:2009aa} of the Synoptic Optical Long Term Investigations of the Sun (SOLIS) facility \citep{Keller:2003aa} and by the space-based SpectroPolarimeter of the Solar Optical Telescope \citep[SOT][]{Tsuneta:2008} onboard the Hinode spacecraft. Vast amounts of seeing-free full-disk vector magnetograms are also delivered by the Helioseismic and Magnetic Imager \citep[HMI][]{Scherrer:2012aa} on board the Solar Dynamics Observatory (SDO) mission. 

Having these facilities, numerous efforts have been made toward solving the $180\deg$ ambiguity problem of the Zeeman polarization signals. There is no known method for resolving the ambiguity through direct observation using the Zeeman effect. Hence, to resolve the ambiguity, some further assumption on the nature of the solar magnetic field must be made. Typical assumptions focus on the spatial smoothness of the field or on minimizing the divergence of the field. A number of different algorithms have been developed to resolve the ambiguity, each making various assumptions on the character of the solar magnetic field.

For magnetograms obtained with photospheric or chromospheric lines \citep{Leka:2003aa} azimuth disambiguation is not a trivial problem problem because the height derivatives ($\partial/\partial z$) of most of the parameters are needed but unknown. While tackling this problem, methods have evolved gradually from a simple comparison between observed and extrapolated fields to computerized automatic procedures. These methods can be classified as ``reference field'' \citep{Allen-Gary:1990aa}, ``multistep process'' \citep{moon:etal03}, ``interactive'', such as the AZAM
utility \citep{Metcalf:2006}, ``automated minimization'' \citep{Metcalf:1994aa}, ``vertical current minimization'' \citep{Georgoulis:2005aa}, ``noniterative analytical'' \citep{Skumanich:1996aa,Cuperman:1993aa}, and ``spectroscopic'' \citep{Landi-DeglInnocenti:1993}, this last one being inapplicable for disk-center and limb observations. Except for the spectroscopic method and the non-iterative analytical methods by \cite{Cuperman:1993aa} that were tested with a MHD model, all the other methods are routinely applied to observational data by individual researchers. Some years ago, \cite{Metcalf:2006aa} provided an overview of nearly all existing algorithms for resolving the $180\deg$ ambiguity problem when applied to synthetic data at a single height level.

 The assumptions underlying each method are the ones ultimately responsible for the quality of the disambiguation results. Precisely and self-consistently disambiguating a vector magnetogram is a challenging problem, especially when considering complex magnetic structures with a multipolar, stressed, and sheared photospheric boundary. For that reason, the application of a proper authomatized disambiguation procedure and the practical issues deriving from the previously commented techniques are out of the scope of this work.
 
We restrict this section to clarify and illustrate some new results of this problem in the case the linear polarization is dominated by scattering processes and the Hanle effect, which is suitable for studying quiet regions of the Sun. The dependence of the linear polarization on the magnetic field azimuth is governed by the Hanle effect, but we show how the thermodynamic alters the inference, making the problem more complicated but more interesting. We focus on disk-center observations and highly-resolved quiet regions of the chromosphere by using the polarization of the Ca {\sc ii} IR triplet lines we have calculated in the 3D MHD model. The use of these spectral lines may give us information about the three-dimensional topology of the chromospheric magnetic field. 


\subsection{The Zeeman effect and the weak field approximation.}\label{sec:zee}	
A useful observable to infer information on the longitudinal component of the magnetic field is the Stokes V profile produced by the Zeeman effect. In the weak field regime, the Zeeman splitting ($\bar{g}\Delta \lambda_B$, with $\bar{g}$ the effective Land\'e factor \footnote{The Ca{\sc ii} IR triplet lines $[\lambda8498,\lambda8662,\lambda8542]$ have $\bar{g}=[1.06,0.83,1.1]$.}) is small in comparison with the thermal width ($\Delta\nu_D$) of the line profiles (see Sec. \ref{sec:spectral}):
\begin{equation}\label{eq:zeecond}
\centering
    \bar{g}\cdot \frac{\Delta\lambda_B}{\Delta\nu_D} \, = \, \bar{g} \cdot \frac{1.4\cdot 10^{-7} \, \lambda_0[\mathrm{\AA} ] \, B[G]}{\sqrt{1.663\cdot 10^{-2} \frac{T[K]}{m[\mathrm{u.m.a.}]} +v_{\rm micro}^2[ \mathrm{km^2s^{-2}}  ] }} \ll 1,
\end{equation}
with $m=40.06$ u.m.a for calcium. The condition (\ref{eq:zeecond}) is fulfilled at all the heights where the lines form in our models. In the weak-field regime, and assuming the absence of atomic orientation and that the longitudinal magnetic field component is constant along the formation region, we can approximate Stokes V at disk center with:
\begin{equation}\label{eq:zeeweak}
    V(\lambda) \, = \, -\bar{g} \,\Delta\lambda_B \, \cos{\theta_B} \, \frac{\partial I(\lambda)}{\partial \lambda},
\end{equation}
where $\theta_B$ is the inclination of the magnetic field\footnote{Away from disk center, the cosine of the angle between the LOS and the magnetic field vector would not be $\cos{\theta_B}$ but $\sin{\theta_B}\sin{\theta_{\Omega}}\left[ \cos{\chi_B}\cos{\chi_{\Omega}} +\sin{\chi_B}\sin{\chi_{\Omega}} \right] +\cos{\theta_B}\cos{\theta_{\Omega}}$} with respect to the LOS and $I(\lambda)$ gives the spectral profile of the emergent intensity. Note that, assuming a LOS with $\mu=1$, a line in absorption under a magnetic field with $0\deg<\theta_B<90\deg$ gives a Stokes V profile that is antisymmetric with a positive lobe at blue wavelengths and a negative one towards the red. The sign of the lobes will be the opposite when the magnetic field points away from the observer or the line is in emission. Since $\Delta\lambda_B$ is linear in B, Stokes V is proportional to the longitudinal component of the magnetic field.

Taking into account the discussion about that HPILs in Section \ref{sec:nullpaths}, we can use the regions of null linear polarization across a map to distinguish the lines where the magnetic field has an exact inclination $\theta_B=54.73\deg$ (Van-Vleck HPILs). 

\subsection{Magnetic Field Direction and Ambiguities in the Hanle regime.}\label{sec:zh_ambig}	

In order to estimate the magnetic field azimuth it is necessary to apply Eq. (\ref{eq:chiuq}). However, the inverse tangent is only defined from $0\deg$ to $90\deg$  when applied to positive angles measured from the x axis of Fig. \ref{fig:qusigns}. This is called the inverse tangent problem, because it is not directly possible to recover an angle between 0 and 360 degrees from the ratio between the two spatial vector components defining a direction in the plane $(x,y)$. That is exactly what we seek when applying Eq. (\ref{eq:chiuq}) to obtain the magnetic field azimuth. The problem results in the appearance of two ambiguities in the magnetic field azimuth: the $90\deg$ ambiguity is produced when an azimuth $\chi_B$ gives a ratio U/Q undistinguishable from the one given by $\chi^{\prime}_B\pm 90\deg$; and the $180\deg$ ambiguity is produced when an azimuth $\chi_B$ gives a ratio U/Q undistinguishable from the one given by $\chi^{\prime}_B\pm 180\deg$. To solve the $90\deg$ ambiguity we need the specific signs of each component (Q and U), and not only the ratio. Thus, the calculation of the azimuth depends on such signs as follows:

 \begin{table}[!h]
\centering
\begin{tabular}{ccc}
\hline
 $Q$ & $U$ & $2\,\chi_B$ \\
 \hline
$+$    & $+$ & $\tan^{-1}{|U/Q|}$ \\
$+$    & $-$ & $\pi-\tan^{-1}{|U/Q|}$ \\
$-$    & $+$ & $\pi+\tan^{-1}{|U/Q|}$ \\
$-$    & $-$ & $2\pi-\tan^{-1}{|U/Q|}$ \\
\hline
\end{tabular}
\caption{Signs rule to solve the Eq.(\ref{eq:chiuq}).}
 \label{table:arctan}
\end{table}
 
Whenever the signs of Q and U do not change by other factors different than the magnetic field azimuth (basically field inclination and thermodynamics) or whenever we can trace such changes back, the $90\deg$ ambiguity can be solved. However, as the argument of the tangent in Eq.(\ref{eq:chiuq}) is $2\chi_B$ (and not $\chi_B$), this procedure cannot solve the $180\deg$ ambiguity. This is because\footnote{Mathematically, it is because $2\chi^{\prime}_B=2(\chi_B+\pi)$ always in Eq.(\ref{eq:chiuq}), no matter the signs of Q and U.} the light, being an electromagnetic wave, oscillates in both senses along its direction of polarization in such a way that the information about the magnetic field orientation is missed. To deduce it, we should recall all the methods and the problematic around the resolution of this ambiguity. However, the sense of the chromospheric magnetic field can also be determined with the magnetic polarities at both ends of the magnetic field lines crossing the pixel of interest. To obtain such field lines we need to consider a map (not only one pixel) and the sign of Stokes V. We have to identify the field line crossing each pixel, just by calculating the transverse field directions given by Eq. (\ref{eq:chiuq}) and connecting them between adjacent pixels from a positive polarity region (Stokes V with positive blue lobe and negative red lobe) to a negative polarity region (Stokes V with opposite sign) or viceversa. We do not see major impediments in doing this, at least in our dataset, where we deal with smooth chromospheric structures with no significant noise added. With such \textit{topological} procedure we could solve the $180\deg$ ambiguity.
 
The application of Table \ref{table:arctan} to our LP signals has some crucial remarks that have to be clarified. In order to solve the correct azimuth, we have to know the Q and U signs. Since they do not only depend on the magnetic field azimuth, we have first to discriminate all the elements (drivers) acting on the polarities and deduce the signs that would be given only by the magnetic field orientation. Considering a Hanle-saturated spectral line and a forward scattering gemetry, the possible drivers are:

1) The term $(3\cos^2{\theta_B}-1)$ depending on the inclination of the magnetic field with respect to the vertical in Eqs. (\ref{eq:hanlesat}). It contributes with negative sign to both Q and U when the pixel is in the HF region (when the inclination is $54.73\deg<\theta_B<125.27\deg$). That makes the signs for (Q,U) in Fig. \ref{fig:qusigns} to jump together between states:  $(++) \leftrightarrows (--)$ or $(+,-) \leftrightarrows(-,+)$. The effect is always a sudden $90\deg$ shift in the inferred azimuth angle when crossing the Van Vleck HPILs (when passing from the HF to the VF region or viceversa). This is usually called Hanle ambiguity, and affects the inferred magnetic azimuth due to the magnetic inclination. We can avoid it by two means. First, with the above-mentioned topological method. This is, using the circular polarization in a map as a marker to discriminate the correct magnetic field direction between the two ambiguous perpendicular magnetic field azimuths. Second, using only one pixel, but knowing the theoretical reference sign (see Table \ref{tab:signs}) for the model that better represents that pixel and discriminating the drivers affecting it.

2) The temperature. To account for it, we classify the models in simple categories: FALC-like, cooler, hotter and so on. Thus, we associate the specific states of the linear polarization with those general macroscopical states (again Table \ref{tab:signs}). Once we know the theoretical behaviour of the polarization signs in a range of temperatures, we can use that information as a reference. In this case we need to know the sign of the thermodynamic factor $\mathcal{F}$ in static cases.

3) The velocity gradients. We concluded in Section \ref{sec:velocities} that the velocity could affect the LP \textit{signs} in the Sun or in more dynamic models, but not in our MHD models. In any case, the existence of a velocity gradient does not produce just a sign reverse and a discrete azimuth jump as it happens with the variation of the magnetic field inclination. Instead, it induces gradual alterations of the Q and U amplitudes that are translated in imprecisions when estimating the magnetic field azimuth with Eq. (\ref{eq:chiuq}). Only when the velocity gradient overcome a certain threshold, such gradual change also leads to a reversal in the linear polarization (e.g., see Fig. \ref{fig:consistency_test}). That was a common behavior for the $8498$ {\AA} line forming in the time-dependent models of Chapter \ref{cap:three}. On the contrary, the lines $8542$ and $8662$ {\AA} are unable to change their signs when they have a one-peak shape because in that case the velocity gradient always produces a positive amplification (Chapter \ref{cap:two}). Instead, if they show a valley shape at line center, a change of polarity is possible due to the contribution of the lower level alignment producing the valley. 

Our discussion on the ambiguities focuses on the saturated Hanle regime under the weak field approximation. If the field is stronger, the transverse Zeeman effect signals predominates as the sources of the linear polarization (Zeeman regime). On the contrary, if the field is on the order of a few gauss weaker, the Hanle polarization amplitudes in these lines can vary significantly with the magnetic field strength and introduce a different angular dependence in the equations.\\

\subsubsection{Ambiguities in inclination with ambiguous azimuth.} 

The previous explanations introduce the ambiguities in azimuth without specifying the corresponding values of magnetic field inclination that can produce the same Q and U. To do that, we pose the problem in a more analytical way as follows.

The lack of knowledge about the azimuth and the inclination of the magnetic field produces spectroscopic ambiguities. It means that different magnetic field orientations yield the same Q and U, making the field topology undistinguishible. To see the exact magnetic field configurations that can be mixed up, let us consider a fixed thermodynamical stratification and answer the following question: which two different magnetic field orientations, $(\chi_B,\theta_B)$ and $(\chi^{\prime}_B,\theta^{\prime}_B)$, are undistinguishable when only Stokes Q and U are used to infer them? 

\begin{figure}[th!]
          \centering%
          \includegraphics[width=0.6\textwidth]{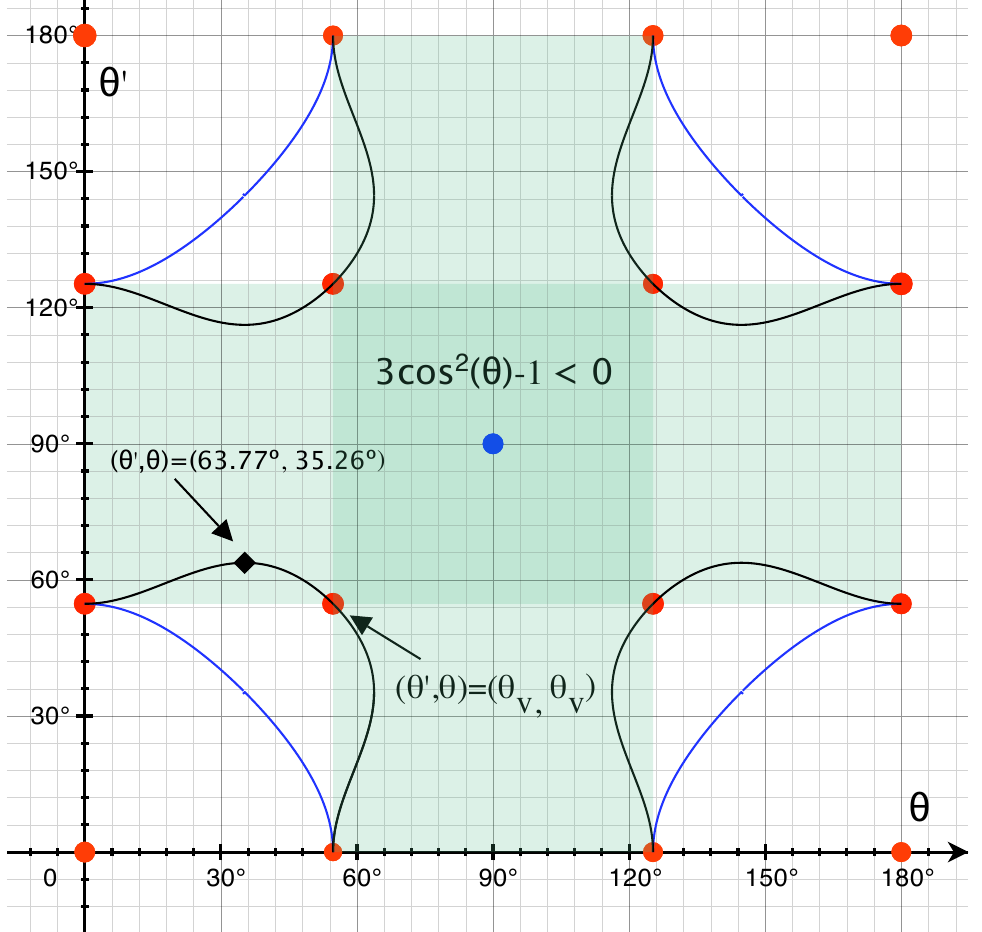}
        \caption{Theoretical curves leading to an ambiguous magnetic field inclination in the saturated Hanle effect at forward scattering. The green shadowed region is the HF region, where $3\cdot \cos^2{\theta_B}-1 < 0$. The relevant angles are $\theta=90\deg \pm 26.22 \deg ,\, 90\deg \pm 35.27 \deg $ and $90\deg \pm 90\deg)$.}\label{fig:ambig}
\end{figure}

Considering that the angular depedencies of the Stokes U profile are well described by Eq. (\ref{eq:hanlesatb}),  the answer to the question can be obtained making $U = U^{\prime}$. It yields 
\begin{equation}\label{eq:ambi1}
\centering
 \sin{2 \chi_B}\cdot\sin^2{(\theta_B)}\cdot(3\cos^2{\theta_B}-1) = \,\sin{2 \chi^{\prime}_B}\cdot\sin^2{(\theta^{\prime}_B)}\cdot(3\cos^2{\theta^{\prime}_B}-1).
\end{equation}
The ambiguities in inclination are associated with the existent ambiguities in azimuth. Thus, for the azimuthal $180\deg$ ambiguity, the corresponding ambiguous inclinations $\theta_B$ and $\theta^{\prime}_B$ are obtained when $\chi_B =\chi^{\prime}_B + 180\deg$ is substituted in Eq. (\ref{eq:ambi1}). Note it gives the same as when the azimuth is considered as solved and fixed ($\chi_B =\chi^{\prime}_B$), indicating that both situations are spectroscopically equivalent. Then, it yields:
\begin{equation}\label{eq:ambiteta1}
\centering
\sin^2{(\theta_B)}\cdot(3\cos^2{\theta_B}-1)=\,\sin^2{(\theta^{\prime}_B)}\cdot(3\cos^2{\theta^{\prime}_B}-1),
\end{equation}
whose solutions give the relations between the ambiguous inclinations $\theta_B$ and $\theta^{\prime}_B$: 
\begin{equation}\label{eq:ambiteta2}
\theta_B = \cos^{-1} \left[  \pm \sqrt{\frac{2}{3} \pm \sqrt{ \frac{1}{9}  +\sin^2{\theta^{\prime}_B}(\cos^2(\theta_B^{\prime})-1/3)   }    }  \right]   
\end{equation}
The same reasoning and results are obtained using Q. These solutions ($180\deg$ ambiguity case or solved azimuth case) are represented by the black curves in Fig. \ref{fig:ambig}. Thus, a value $\theta^{\prime}_B$ (vertical axis in Fig. \ref{fig:ambig}) can produce the same U (and Q) than other inclinations $\theta_B$ (horizontal axis), which are obtained by tracing a horizontal line that crosses in several points the black curve. We see that a value lying in the HF region (coloured region in Fig. \ref{fig:ambig}) does not have ambiguities (does not cross any curve) in the range $63.77\deg<\theta_B<116.23\deg$. That means that near-horizontal fields in that range of inclinations produce ambiguities only due to azimuth and/or thermodynamics, so highlighting the HF areas as the preferred ones to be used for estimating inclinations from Q and U. If the inclination is out that range but still in the coloured region, a horizontal line in the figure yields four ambiguities, two at a right branch and two at a left branch. Finally, if the inclination is in the VF region (white area), there are only two ambiguities, one at each branch. 

In the case of an azimuth $\chi_B$ not distinguished from $\chi_B\pm 90\deg$ (Hanle ambiguity), Eq. (\ref{eq:ambi1}) changes to:
 \begin{equation}\label{eq:ambi2}
\centering
  \sin^2{\theta_B}\cdot(3\cos^2{\theta_B}-1) = \,-\sin^2{\theta^{\prime}_B}\cdot(3\cos^2{\theta^{\prime}_B}-1),
 \end{equation}
whose solutions are the same as Eq. (\ref{eq:ambiteta2}) but with a minus sign after the number $1/9$:
\begin{equation}\label{eq:ambiteta3}
\centering
\theta_B = \cos^{-1}{\left[  \pm \sqrt{\frac{2}{3} \pm \sqrt{ \frac{1}{9}  -\sin^2{\theta^{\prime}_B}(\cos^2(\theta_B^{\prime})-1/3)   }    }  \right]               }
\end{equation}
The new solutions are represented by the blue curves in Fig \ref{fig:ambig}. Thus, under this azimuthal ambiguity, the inclination $\theta^{\prime}_B$ (vertical axis) is also ambiguous because it cannot be distinguished from $\theta_B$, but now it is only true in the VF region because the blue curves do not exist outside of it. 
\begin{figure}[t!]
          \centering%
          \includegraphics[width=1.05\textwidth]{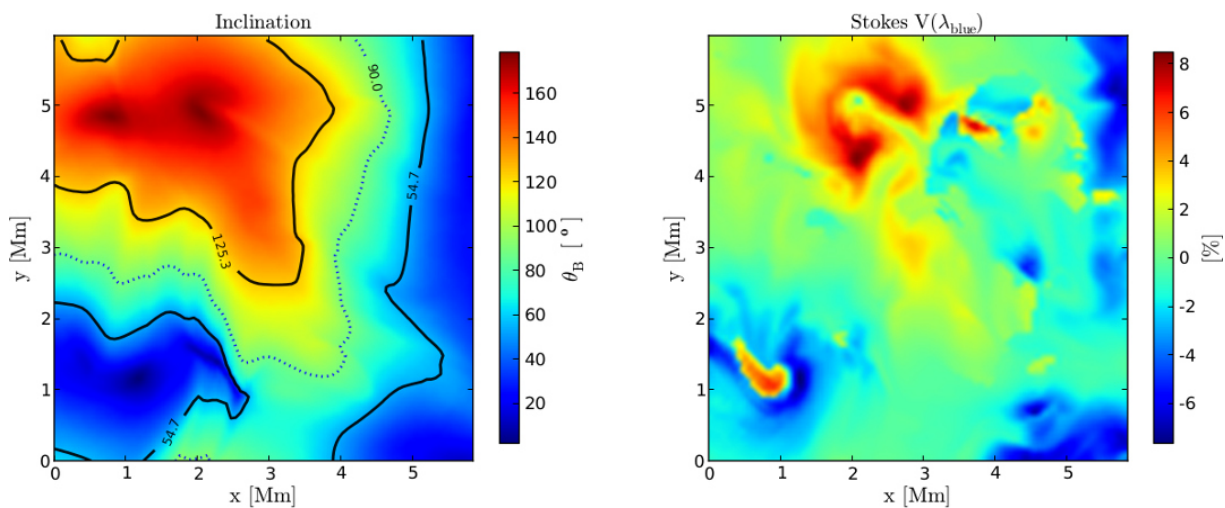}
        \caption{Left: inclination of the magnetic field at heights with $\tau^{8662}=1$. Right: Blue lobe amplitude of Stokes V, indicating the direction of the longitudinal magnetic field component (blue is magnetic field vector towards the observer).  }\label{fig:zeemincl}
\end{figure}
When the $90\deg$ ambiguity in azimuth is not solved, the black and the blue curves together give all the ambiguities in inclination. When the $90\deg$ ambiguity is solved (using Table \ref{tab:signs}), the ``blue solutions'' do not hold and only the ``black ones'' leave. In that case, the remaining ambiguities in inclination could be eliminated with just knowing if the pixel is in the HF region or not. The reason is that the black curves always connect inclinations $\theta^{\prime}_B$ in the HF region with inclinations $\theta_B$ in the VF region and viceversa.  In other words, if the pixel is clearly in one region (VF or HF), the magnetic field inclination in that pixel cannot pertain to the other region. In practice, to take advantage of that, the division between the HF and VF regions has to be located by identifying the Van-Vleck HPILs, which can be easily done calculating the degree of total polarization (see Figure \ref{fig:8542_filters}).

The above discrimination is not valid in the blue curves because they connect inclinations in the same area. We would need to divide the VF region in more subregions to avoid ambiguities. Indeed, if we were able to identify other inclination ranges spectroscopically (having acces to other LOS, for example) we could repeat this strategy of segmentation to differentiate other inclinations and then fit inclination curves between $0$ and $90\deg$ in further steps.

Note that the mere fact of interpreting the results in a map give some extra help to solve the ambiguities. We can use physical methods as the ones introduced by other authors or use topological arguments (connecting azimuth field lines, identifying regions in a map) together with spectroscopical measurements (HPILs, signs of V, Q and U, value of $\rm{U/Q}$). Without the Van-Vleck HPILs and the Stokes V profiles, the magnetic field inclination can never be univocally obtained from Q and U.

In conclusion, the calculation of the magnetic field azimuth depends on the LP signs that are modified mostly by the temperature and the field inclination. Knowing the temperature regime (cool or hot), we only need to identify the Van-Vleck HPILs separating inclination regions (VF or HF\footnote{HF and VF regions at both sides of a Van-Vleck HPIL can be identified using the strip of null Stokes V in Fig. \ref{fig:mapmax} and the areas with intense Stokes V in Fig. \ref{fig:mapmax}}) to get the correct azimuths with Table \ref{table:arctan}. At the same time, the distinction of those regions eliminates the $90\deg$ ambiguities in azimuth and the corresponding ambiguities in inclination (blue curves, Fig. \ref{fig:ambig}), leaving only the $180\deg$ ambiguity in azimuth and selecting the correct inclination as the one pertaining to the region in which the pixels is. To know if the field is pointing towards or away the observer (which is equivalent to select between right or left branches in Fig. \ref{fig:ambig}) we can use the polarity of Stokes V in the pixel. Finally, to know the sense of the magnetic field vector we estimate possible to solve the $180\deg$ ambiguity in azimuth using a topological method based in connecting all the field lines from positive to negative magnetic emergence patches (Fig. \ref{fig:zeemincl}). In any case, we remark that in a wide inner part of the HF region the azimuths are unaffected by inclinations, and inclinations itself do not show any ambiguity. All the discussion is based on that we can discriminate the non-magnetic factors altering the polarity signs, which have been presented after theoretically calculating the reference polarities in realistic maps of the solar chromopshere.
 

\subsection{The disambiguation process.}\label{sec:desambig}	
\begin{figure}[h!]
        \centering
        \begin{subfigure}[a]{\textwidth}
                \centering
                \includegraphics[width=0.58\textwidth]{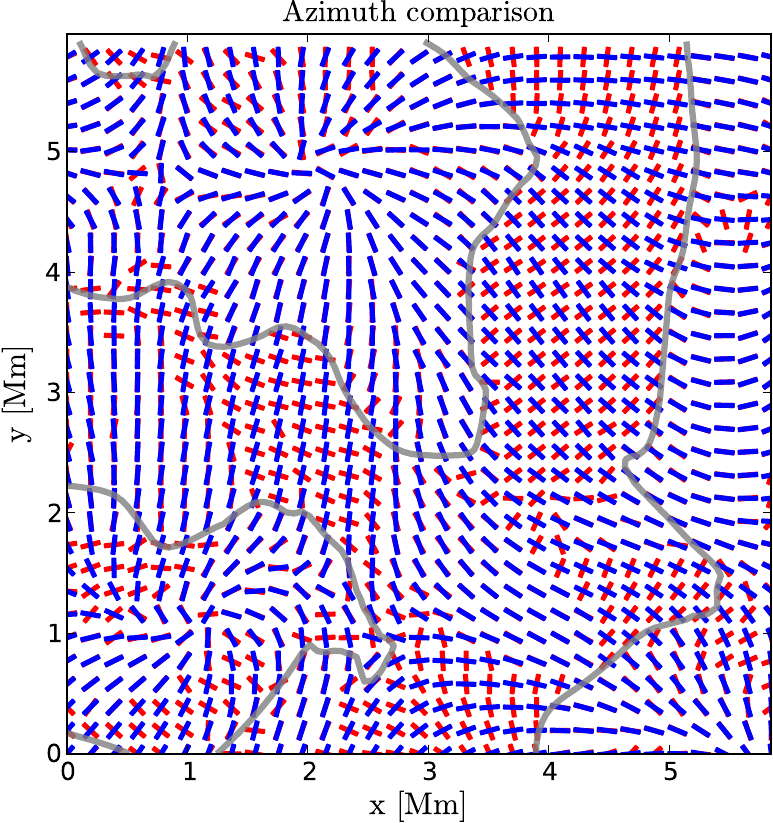}
                \label{fig:h_aci_A}
        \end{subfigure}
\\
        \begin{subfigure}[b]{\textwidth}
                \centering
                \includegraphics[width=1.0\textwidth]{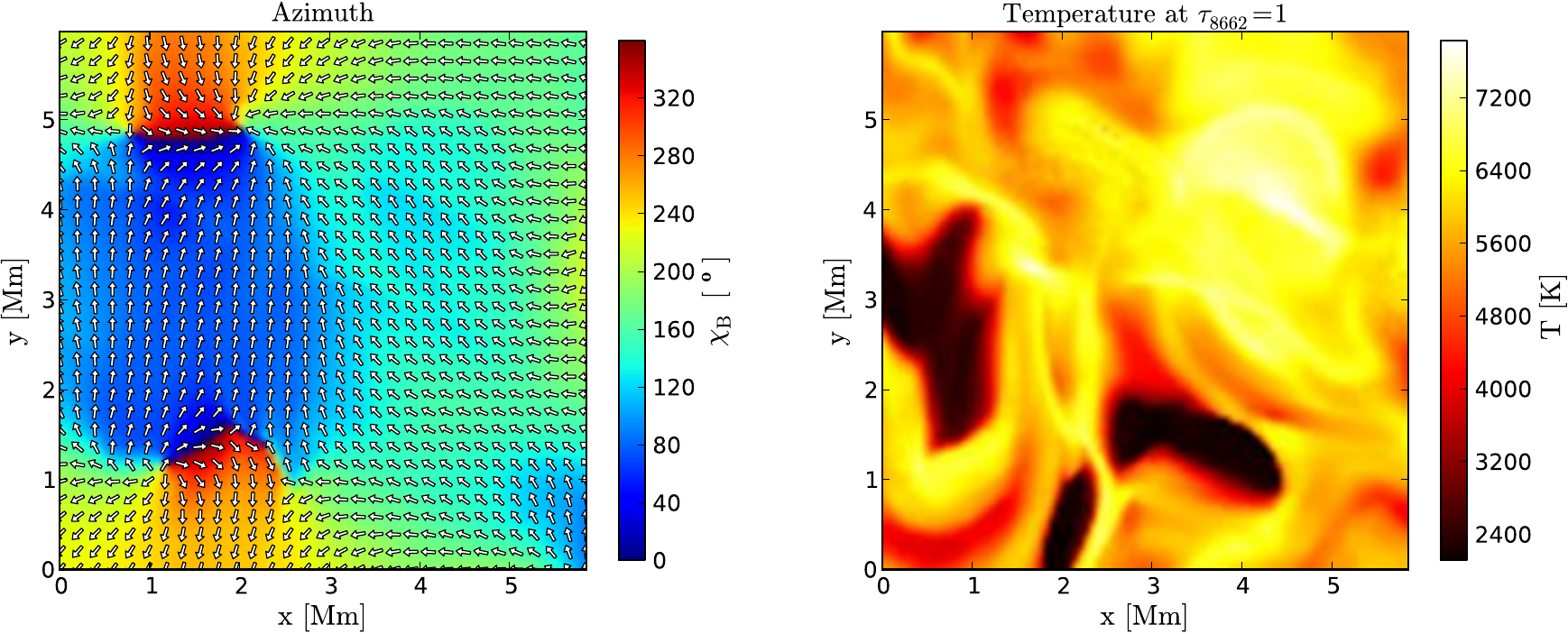}
                \label{fig:h_aci_B}
        \end{subfigure}
        \caption{Effect of the temperature, magnetic field and dynamics on a diagnostic map in saturated Hanle regime and forward scattering. \textbf{Upper panel}: Comparisson between real and calculated magnetic field azimuth. Dark blue segments indicate the real direction of the magnetic field in the model at $\tau^{8662}=1$. The red segments are the same but calculated from Q and U. Gray solid iso-contours are the Van-Vleck HPILs and connect points with $\theta_B=\theta_V$ and $\theta_B=\pi-\theta_V$. \textbf{Lower left panel}:Original magnetic field azimuth at $\tau^{8662}=1$ in the snapshot. \textbf{Lower right panel}: temperature at $\tau_{8662}=1$. }\label{fig:h_acimut}
\end{figure}

We have provided some ideas for deducing the general topology of the magnetic field. Now, we want to get the map of the magnetic field azimuth without ambiguity. To do it, we will explain the map of magnetic field azimuth in Figure \ref{fig:h_acimut} and the steps followed to create it. These steps are based only on the information provided by the ``observed'' Stokes vector.

1) The first thing we need are the maps of maximum Stokes Q and U at each pixel for a Hanle saturated spectral line (both $8662$ and $8542$ {\AA} lines are good options). In order to apply Eq. (\ref{eq:chiuq}) and estimate the field azimuth, some authors \citep[e.g.][]{Jefferies:1989aa} integrate the spectral profiles of Q and U along a certain filter width. However, the most precise way of inferring the magnetic field azimuth at optical depth unity is to consider the narrowest spectral width possible, ideally only at the wavelength where the Stokes Q and U amplitudes are maximum. If not, we will mix contributions coming from different layers or we will produce signal cancellations due to asymmetric profiles induced by velocity gradients. The central line wavelength is not a good choice because the resulting spatial maps would show some empty patches in the atmospheric regions where the velocity gradients are considerable.

2) Following table \ref{table:arctan}, we calculate the azimuth of the magnetic field. But that is not the final azimuth because we have first to consider other factors affecting the signs of Q and U. Note that in Fig. \ref{fig:h_acimut} the original magnetic field direction (blue segments) and the azimuth calculated from Q and U exclusively (red segments) are equally oriented in many patches but they differ in a non-negligible part of the map (obviously, where the red segments are visible). Most of those wrong red segments are rotated $90\deg$ from the correct directions. The reason of this almost exact and discrete shift is that either the temperature stratification or the magnetic field inclination or both at the same time are modifying the signs of both Q and U (see Sec. \ref{sec:zh_ambig}). If we perfom a $90\deg$ rotation in the red segments, the map is almost entirely correct. Thus, the major problem to calculate the precise magnetic azimuth in those pixels is the LP sign, not the amplitudes.

3) The next step is to identify the Van-Vleck HPILs in the maps of total linear polarization (Fig. \ref{fig:mapmax}) or in the maps of linear polarization degree (Fig. \ref{fig:8542_filters}). With these lines (gray lines in Fig. \ref{fig:h_acimut}) we separate with precision the HF region (strip of null Stokes V in Fig. \ref{fig:mapmax}) from the VF region (coinciding with intense Stokes V in Fig. \ref{fig:mapmax}).  

4) Since we know that the magnetic field lines go from the negative flux regions (blue areas in right panel of Fig. \ref{fig:zeemincl}) to the positive ones around $(x,y)=(2,5)$, the correct direction of the segments can be identified just connecting them from one polarity to the other. This argument of contuinity constitutes a topological inference method that can be automatized. At least, it seems feasible for our chromospheric synthetic results. 
For spectral lines forming in the chaotic photosphere it would be unsuitable because the derivatives of the magnetic field with height are larger. 

5) As we have been introducing in previous sections, other way of inferring the correct direction for the red segments is by using reference signs for Stokes Q and U. The simplest way to do it is by identifying a patch inside the HF region in Fig. \ref{fig:mapcon} whose linear polarization signal is mostly in Q or in U and whose thermodynamical stratification can be catalogued as one of the column cases in the Table \ref{tab:signs}. The example we put in the previous section was the U patch around $(x,y)=(4.5,4)$ Mm. It is a good option because the corresponding intensity map is in emission there (Fig. \ref{fig:slitA} at $(x,y)=(5.5,6.5)$ arcsec), so it is a model with a clear hot temperature stratification, and has almost null Q signal. Then, looking the corresponding contrast map (Fig. \ref{fig:mapcon}), we see that $U > 0$. As indicated by the third column and fourth row in Table \ref{tab:signs}, it is compatible with a magnetic field along the $-$U axis. So, the red segments corresponding to that patch in Fig. \ref{fig:h_acimut} are wrong due to the temperature and have to be rotated $90\deg$.  A similar inference can be done when both Q and U are not zero, just searching the projections of the magnetic field vector on the reference axes of Figure \ref{fig:qusigns} and applying Table \ref{tab:signs}. This is also a procedure that can be automatized.

6) Once we know the correct azimuth directions in only one patch (reference patch) we can deduce others from it. Different regions appear clearly differenciated around such reference patch when the inclinations jumps between HF and VF regions or when the thermodynamic structure changes significatively\footnote{Note that once we have estimated the magnetic field azimuth with the Table \ref{tab:signs}, the abrupt changes in the estimated azimuth between patches are not real. Just by continuity, they can not be produced by the magnetic field but by other factors.}. Thus, moving in the HF region, the patches adjacent to the reference patch appear necessarily only when having a very different kind of temperature stratification. If the reference patch has a FALC-like (or hotter) structure, the adjacent patch has to be cool, what means a simultaneous sign reversal in Q and U, and consequently a $90\deg$ shift in the azimuth. As a reference FALC-like patch already has a $90\deg$ rotation in the HF region (with respect to the correct azimuth), a cool patch in that region produces a double sign reversal and makes the Eq. (\ref{eq:chiuq}) to yield the correct azimuth. This can be observed in the Fig. \ref{fig:h_acimut} comparing with the temperature map below: the patches associated to cool temperatures deliver a correct inferred azimuth in the HF area. We see that the cooler, the more precise match between red and blue segments.

7) Now, we analyze the VF region. Basically, the signs inside the VF region are opposite than in the HF region for the same thermodynamic patch, as expected, just because the inclination term. Consequently, in the VF region, the hottest patches show the best alignment between the estimated and the real magnetic field vectors. For instance, note how the VF region around $(x,y)=(2, 5)$ Mm contains all the possible directions of the magnetic field and that we recover them almost exactly with the exception of some small imprecisions where the temperature is a bit cooler. It seems that a temperature between $4000$ and $5000$ K at $\tau=1$  produces these imprecisions in several parts of the map. They become directly a $90\deg$ inversion when the temperature goes down below $4000$ K at some point around $\tau=1$. An example of this is the arc- shaped region with low temperature passing by $(x,y)=(0.5,0.25)$ Mm. 

8) The effect of the vertical velocity gradients is also to diminish the precision in the estimation of the magnetic field azimuth, similarly as ocurred in the regions with medium temperatures between $4000$ and $5000$ K at $\tau=1$. In all these cases, the inferred azimuths oscillate around the correct values but without producing an abrupt $90\deg$ rotation. Thus, the thermodynamical and dynamical fluctuations are encoded in small angular variations of the magnetic field azimuth.

Despite the previous observations and procedures are still preliminar, we think that they give a satisfactory explanation to the polarization behavior in these models, being in consonance with the results and the theory exposed along this work.  

We think that the effect of the vertical velocities in the estimation of the magnetic field azimuth has to be relevant in the dynamic fibrils\footnote{There is generalized lack of a clear filamentary topology in all the emergent intensity maps resulting from the current theoretical models (e.g., Figure \ref{fig:slitA}). That is the reason preventing a proper computational study about such phenomena.} permeating the solar chromosphere becase of the same reasons by which we obtain azimuth discrepancies in Figure \ref{fig:h_acimut} where the velocities are large enough. A possible evidence of this imprecisions could have been found by \cite{de-la-Cruz-Rodriguez:2011aa} when trying to measure the magnetic field orientation in fibrils\footnote{The question is: how are the transverse Zeeman signals modified by the dynamic?}. If our result is extensible to regions with more magnetic activity, we would have a possible explanation for the discrepancies explained by those authors between the fibrils orientation and the inferred magnetic field azimuth. On top of that, \cite{Hansteen:2006aa} explains that the fibrils are basically driven by magnetoacoustic shocks. Indeed, they seems to play a crucial role in shaping the chromospheric dynamics, in particular producing a strong reduction in the oscillatory power at periods around three minutes, correlated with the absence of chromospheric acoustic shocks \citep{Vecchio:2007aa}. From those conclusions, we propose a possible estrategy to avoid the wrong azimuth estimations. First, we distinguish between the bright and dark fibrils contrasting at the line core intensity of 8542 in most solar observations. Then, interpreting the results of \cite{Vecchio:2007aa} we understand that the bright fibrils appears where the photospheric acoustic power is effectively transferred from the photosphere to the chromosphere (in presence of more vertical magnetic field lines and stronger vertical velocity fields) while the interlaced dark fibrils correspond with areas that do not transfer acoustic shocks into the chromosphere (either showing more horizontal or stronger magnetic fields or well an absence of shock waves). If this were correct, just inferring the magnetic field azimuth from the dark fibrils we would get a correct result.

\subsection{Magnetic field intensity.}\label{sec:bint}	
\begin{figure}[h!]
\centering%
\includegraphics[scale=0.52]{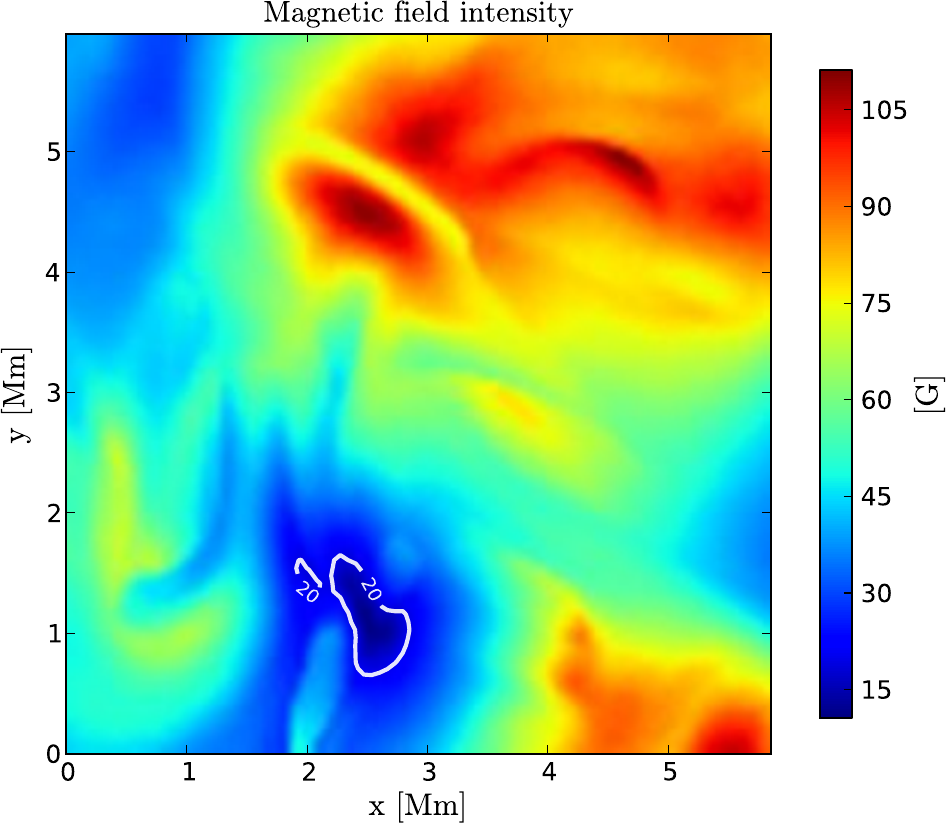}
\caption{Real magnetic field intensity in layers of the model with $\tau_{8542}=1$.  }
\label{fig:bint8542}
\end{figure}

In principle, we could not measure the magnetic field intensity in the model from our results. The reason is that we have assumed so far that we are in the saturation regime of the Hanle effect because we know, after inspecting the models, that the field strengths at $\tau=1$ are much larger than the expected critical Hanle fields for the relevant atomic levels (see \footnote{In principle, the region with lower magnetic field strength ($\sim 20$ G) in the map is the most suitable to be sensitive to the Hanle effect because its magnetic field strength is nearer the critical Hanle field. However it has an almost completely vertical magnetic vector, for what the non-saturated Hanle effect is not of practical application in that area.} Figure \ref{fig:bint8542}). But, from an observational point of view, how could we assure that? We would need to estimate the field strength with Stokes V and Eq. (\ref{eq:zeeweak}). However, with that equation we only can obtain reliable values for the longitudinal magnetic field component. To get the corresponding magnetic field intensity we have to know the magnetic field inclination. This can be done by identifying the Van-Vleck HPILs. Along them it is sure that $\cos{\theta_B}=1/\sqrt{3}$ and the magnetic field strength is then given by

\begin{equation}\label{eq:bvan}
    B \,[G]\, = \, 3.71\cdot10^{12}\,\frac{|V(\lambda_{\rm max})| \, }{ \bar{g} \,\lambda^2_0 \,\left|\frac{\partial I(\lambda)}{\partial \lambda}\right|_{\rm \lambda=\lambda_{max}}},
\end{equation}

with B in Gauss, wavelengths in Amstrongs and $\lambda_{\rm max}$ the wavelength where Stokes V is maximum. Thus, in the weak-field regime with the Hanle effect in saturation, we can measure with precision all the magnetic field parameters along the Van-Vleck HPILs (assuming we have a detector with enough polarimetric sensitivity). 

The field strengths in those points can give us an answer about whether the Hanle effect is saturated or not, but not in all the map, so the doubt can still remain in other regions. On top of that, Stokes V mainly responds to the magnetic field in significantly low chromospheric layers (usually having larger magnetic field strengths). This is known by mean of the response functions. Thus, we would tend to infer wrongly a field intensity closer to saturation. 

If the Hanle effect were not saturated, the dependences of Eqs. (\ref{eq:hanlesat}) would not hold and the amplitudes of Stokes Q and U would also depend on the magnetic field intensity. To measure the magnetic field strength in this regime we would need a reference calculation for each possible configuration of temperature, magnetic field direction and velocity. The work of \cite{manso10} is along that line, so the solution would be to repeat it including the effect of the velocity. Similarly, we can remake the calculations of this chapter but artificially decreasing the magnetic field intensity by a constant factor in the whole cube (so considering a ``more quiet'' Sun). 

Finally, the last option is to characterize the Stokes line ratios between the $8662$ and $8542$ {\AA} lines and see whether they are sensitive enough to the magnetic field intensity. To do that, note that the zero-field line ratios obtained in Chapter \ref{cap:three} can not be used because we are now in a disk center observation and the linear polarization would be zero in absence of magnetic field. The equivalent solution is to repeat those reference line ratios for a magnetic field configuration which maximizes the linear polarization at $\mu=1$: this is, for a horizontal magnetic field pointing for simplicity along the $\pm$Q or $\pm$U axes, as in the reference model defined in Sec. \ref{sec:conmaps}. In that case, the stronger the field strength, the more effective the symmetry breaking (and consequently the emergent linear polarization), reaching the limit in the saturation regime. Then, we conclude that, if the calculations presented in this chapter are really in the saturation regime, the results for the pixels where the field is horizontal directly give us useful references for the line ratios in a forward scattering geometry. In other words, these results would be the reference line ratios that can be used for measuring the magnetic field in the non-saturated case. 

\subsubsection{Does the Hanle sensitivity depend on temperature? }
For the upper level of any transition, its lifetime can be estimated with the Einstein coeffient ${\rm A_{u\ell}}$ of the transition, which is independent on the atmospheric model. In our atomic model, the only polarized upper level is the level $5$ (see Fig. \ref{fig:elevels}), which yields a Hanle critical field of $B_H=0.82$ G for the $8542$ {\AA} line. This value is not directly dependent on temperature.
However, the lower-level critical Hanle field given by Eq. (\ref{eq:hanlecri}) depends on the mean intensity, and the mean intensity could change significantly with the temperature stratification. In a model atmosphere we can directly get the number of photons per radiation field mode $\mathrm{\bar{n}=J^0_0(\nu_0)\cdot c^2/(2h\nu^3_0})$, which give us the lifetime of the level:
\begin{equation}\label{eq:lifetime}
\frac{1}{t_{\rm life}} \, = \, A_{u{\ell}}\frac{2J_u+1}{2J_{\ell}+1}\bar{n}(\tau=1) 
\end{equation}
But usually we can not access the radiation field for making estimations, during an observation for instance. For those cases, let us define a radiation temperature ${\rm T_{rad}}$ that yields a Planck function value $\mathrm{B_{\nu}(T_{rad})}$  equals to the mean intensity $\mathrm{J^0_0(\nu_0)}$. Equivalently,
\begin{equation}
\bar{n} \, = \,\frac{1}{e^{\frac{h\nu}{KT_{\rm rad}}}-1}
\end{equation}
In LTE, $\rm{T_{rad}}$ is the temperature of the plasma but in NLTE it is only a representative number that serves as an upper limit to the local plasma temperature\footnote{In general, the temperature delivered by the Planck function for a black body is lower than the temperature for a body with net radiation energy losses.}. Then, a change in temperature could produce an effective shift in the magnetic field regime making the polarization to be sensitive to the Hanle effect. For instance, to obtain a critical Hanle field of $5$ G producing a saturation around $50$ G, we would get from Eq.(\ref{eq:lifetime}) a radiation temperature $\mathrm{T_{rad}}\sim 118$ kK for the $8662$ {\AA} line and $\mathrm{T_{rad}}\sim 260$ kK for the $8662$ {\AA}.

 We see that larger temperatures are compatible with larger critical fields, what changes the range of validity of the Hanle effect. For this change to occur, the temperature increment has to be in the formation region, where the density of Ca {\sc ii} is significant. The calculated large temperature values could be possible in the the upper parts of extra hot chromospheric stratifications or inside very hot plasma bubbles (the analyzed models show such anomalous bubbles in a few regions). At least in the latter case, the population of Ca {\sc ii} is unsignificant because the ions have been abruptly ionized to Ca {\sc iii}. Other trigger suspicious of being able to bring the Hanle effect into its non-saturated regime is a shock wave. When passing through the chromosphere, it would make the emergent polarization suddenly sensitive to the magnetic field intensity. This kind of sudden Hanle effect could contribute to the small azimuth shifts appearing in the Fig. (\ref{fig:h_acimut}), in hot regions with relatively large velocities. Apart from that, a shock wave also means an increment in the density and an approach of those atmospheric layers to behave as a black body (see second row from the top in Figure \ref{fig:termo}). In such case, the radiation temperature values would be more representative of the local plasma. 

The estimation of the Hanle critical field in the above discussion should actually consider the transition rates of all the lines connected to the considered atomic level. For the upper level of the $8542$ {\AA} line, we should apply Eq. (\ref{eq:hanlecri}) summing the three Einstein emission coefficients $A_{51}$, $A_{52}$ and $A_{53}$, one per transition decaying from level $5$. In that case, we get ${\rm B_H=17}$ G (which sets the maximum  detectable field strengths around $100$ G). 
\begin{figure}[h!]
\centering%
\includegraphics[scale=0.35]{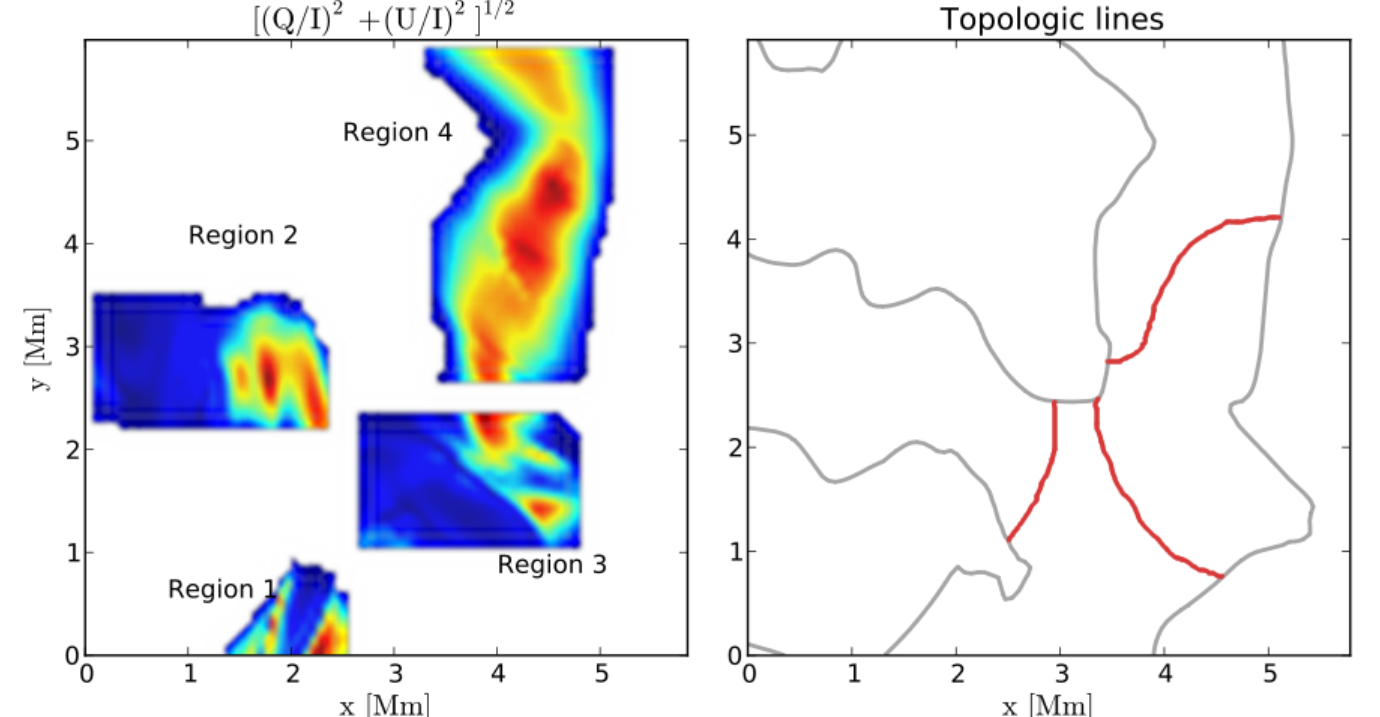}
\end{figure}
\begin{figure}[h!]
\centering%
\includegraphics[scale=0.48]{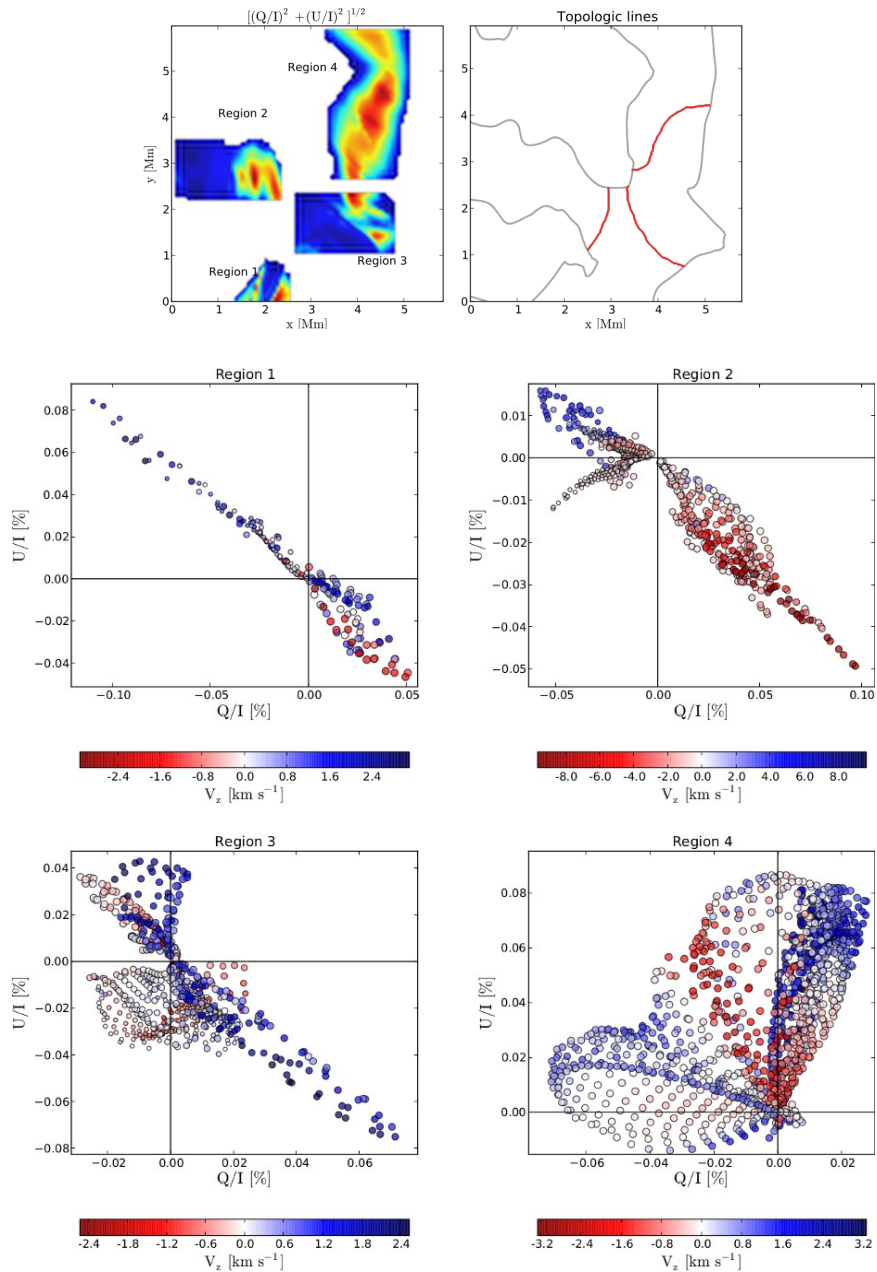}
\caption{ Forward scattering Hanle diagrams in the selected regions for the $8662$ {\AA} line. Each small circle is semitransparent and represents a pixel in the corresponding region. The color encodes the value of the vertical velocity and the size of each circle is proportional to the temperature (always at $\tau^{8662}=1$). The upper small panels show the chosen regions (left) and three arbitrary examples (in red) of topological lines (see text). }
\label{fig:hdiag1}
\end{figure}
\begin{figure}[h!]
\centering%
\includegraphics[scale=0.45]{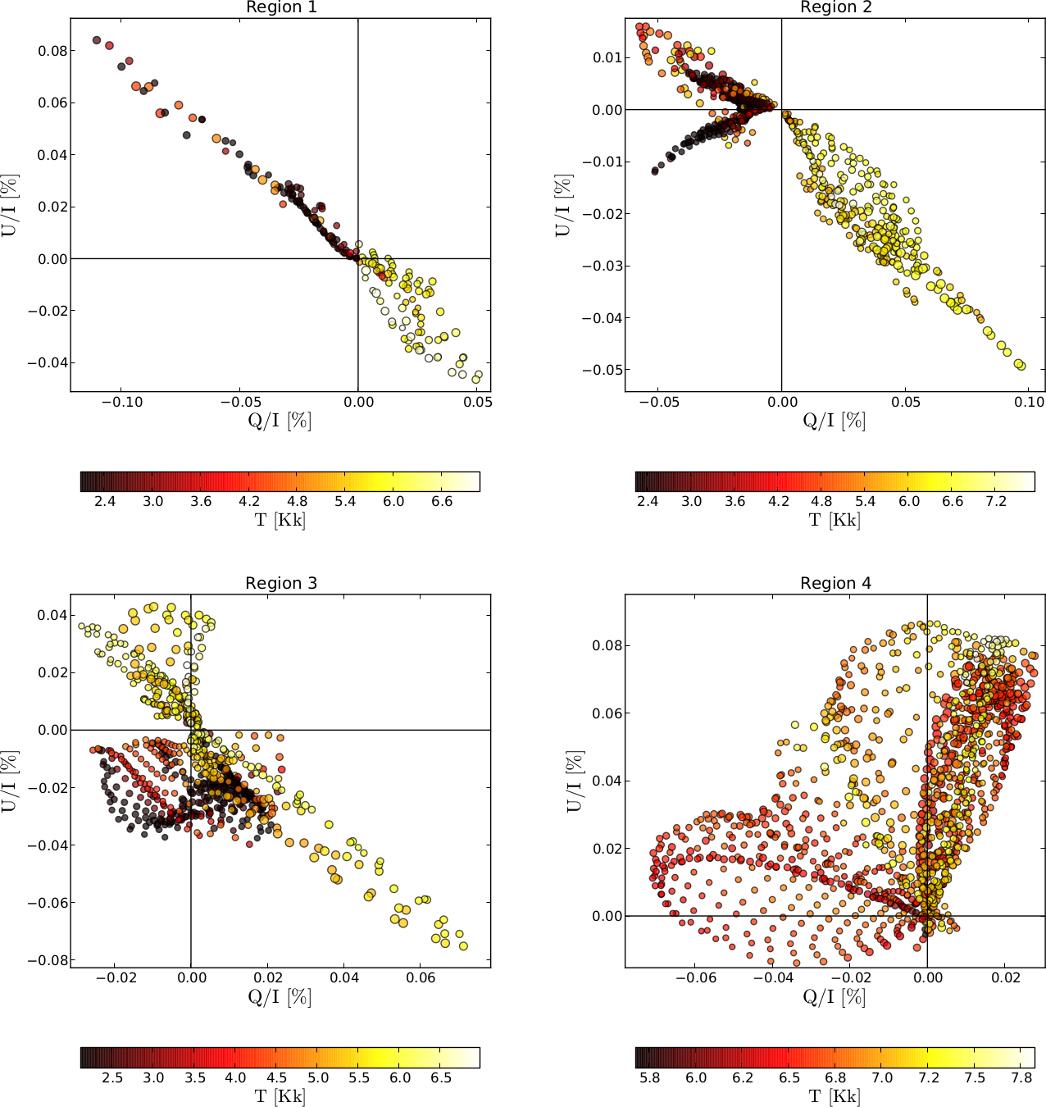}
\caption{ Similar to Fig. \ref{fig:hdiag1}, but now the colors encode temperature and the circle size is proportional to the velocity at $\tau^{8662}=1$.}
\label{fig:hdiag2}
\end{figure}
However, it seems to be more precise to estimate the critical field considering only $A_{53}$ ($A_{52}$ is negligible). This way we get $B_H=0.82$ G and a better agreement with the Hanle curves calculated by \cite{manso10} for FALC models, where the upper value for the detectable field strength is $\sim 10$ G. 


\subsubsection{Hanle diagrams and magnetic field intensity.}
More investigation is needed to contrast the previous hypothesis. Meanwhile we can try to deduce something more about the saturation of the Hanle effect through the Hanle diagrams (Figures \ref{fig:hdiag1}, \ref{fig:hdiag2}  above and \ref{fig:hdiag3}, \ref{fig:hdiag4} and \ref{fig:hdiag5} in the appendix). By comparing the Hanle diagrams resulting from our calculations in the $8662$ {\AA} line (Figures \ref{fig:hdiag1} and \ref{fig:hdiag2}) with the corresponding ones of \cite{manso10} (Figures 14 and 15), we try to know whether this line is really affected by the magnetic field intensity.

First, consider the four regions in the upper panel of Figure \ref{fig:hdiag1}, all of them with considerable magnetic field inclination (HF region); the small Region 1 has an expanding chromosphere with significant velocity; the region 2 has a cool area with patches in different states of motion: static, upward and downward (having a supersonic downflow); the region 3 is composed by a cool compressed bubble and an elongated fringe of pixels with expanding hot atmospheres; finally, the big and hot region 4, not significantly dynamic and with a strong horizontal magnetic field.  

We have included information about the vertical velocity and the temperature at $\tau=1$ in the corresponding Hanle diagrams in order to offer an aditional clarifying perspective that can not be achieved observationally (see footnotes in Fig. \ref{fig:hdiag1} and Fig. \ref{fig:hdiag2}). The colours in the diagram help to identify the different subregions in each region.

After studying these areas and their corresponding Hanle diagrams, we conclude that the Hanle effect is efectively saturated, consistently to what has been assumed in previous sections. To support it, we have shown that the variations in Q and U shown in the Hanle diagrams of Figures \ref{fig:hdiag1}, \ref{fig:hdiag2} (and similar ones in Appendix) can be understood without any variation of the magnetic field intensity\footnote{Aditionally, the comparison with the Hanle diagrams in the K line (Fig. \ref{fig:hdiag5} ) suggest important differences with the triplet ones. The reason is that the K line is not Hanle saturated, showing a totally different dependence whose origin is the variation of the magnetic field intensity.}. 

Consider that the expressions for $\rm{Q/I}$ and $\rm{U/I}$ appearing in Eqs. (\ref{eq:hanlesat}) can be seen as curves in polar coordinates (Eqs. \ref{eq:parametric_diag} a and b), where the radius depends on the inclination of the magnetic field and on the thermodynamic factor $\rm{\mathcal{F}}$:
\begin{subequations}\label{eq:polar_diag}
  \begin{align}
    \mathrm{\left(\frac{Q}{I}\right)} &=\,  r(\mathcal{F},\theta_B) \cdot  \cos{2\chi_B},\label{eq:param_a}
\displaybreak[0] \\
    \mathrm{\left(\frac{U}{I}\right)} &= \, r(\mathcal{F},\theta_B) \cdot  \sin{2\chi_B}.\label{eq:param_b}
  \end{align}
\end{subequations}

\begin{figure}[h!]
\centering%
\includegraphics[scale=0.62]{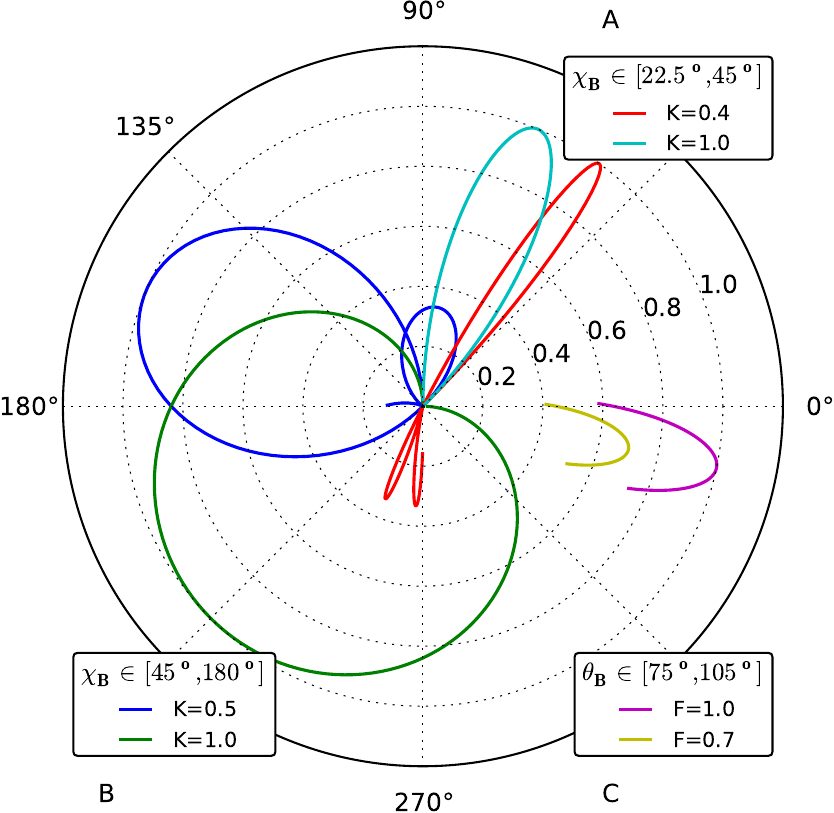}
\caption{Theoretical fit of the Hanle diagrams for three different combinations of parameters (see text and legends).  }
\label{fig:hdiagteo}
\end{figure}

 Assuming first a constant $\rm{\mathcal{F}}$, we stablished simple (linear) ligatures between the azimuth and the inclination of the magnetic field through Eqs. (\ref{eq:parametric_diag}): 

\begin{subequations}\label{eq:parametric_diag}
  \begin{align}
      2\chi_B &=\, \rm{2\chi^i_B+2 K \cdot\frac{\chi^f_B-\chi^i_B}{\theta^f_B-\theta^i_B}\cdot (\theta_B-\theta^i_B)},\label{eq:param_c}
\displaybreak[0] \\
   \rm{(\theta^i_B,\theta^f_B)} &=\, \rm{(\theta_V,\pi-\theta_V)} = \, (54.73\deg,125.26\deg),\label{eq:param_d}
  \end{align}
\end{subequations}
where K is a free parameter. The ligature tights the initial and final values of the magnetic field azimuth $\rm{(\chi^i_B,\chi^f_B)}$ with the corresponding initial and final inclination values $\rm{(\theta^i_B,\theta^f_B)}$ of the topologic lines that connect two chosen initial and final points in the spatial maps (see Figs. \ref{fig:hdiag1} and \ref{fig:cubobz}). In the following, consider that such points are at both frontiers of the HF regions (in the Van-Vleck HPILs), in such a way that the initial and final values of the magnetic field inclination are given by Eq. (\ref{eq:param_d}). Then, choosing different initial and final values for the azimuth in those two points and fixing the number K, we can fit different curves in the Hanle diagrams that correspond to different ``topologic'' lines\footnote{We term topologic lines to the pixels in the maps that follow a well- established path in the Hanle diagram. Defining and solving them we can reconstruct the magnetic field topology not only in azimuth but also in inclination.} crossing the formation region in HF areas of the spatial maps (Fig. \ref{fig:hdiagteo}). Then, the magnetic field lines can be inferred from those topologic lines using the magnetic field azimuth in the map. We leave the details for future works.
Note that for a constant azimuth, Q and U only can vary radially. In that case, the topologic lines in the spatial maps would be represented by dots radially aligned in the Hanle diagrams (see Regions $1$ and $2$ showing those radial dependencies). For the dots to describe a loop in the Hanle diagram, it is needed an azimuthal variation along the topologic lines in the spatial maps. The ligature in Eq. (\ref{eq:parametric_diag}) is thus needed to tie the variation in the azimuth relative to the variation in inclination (or in $\rm{\mathcal{F}}$), and the parametrization is used to adjust such variation to the different cases. The study of the effect of the parameter K seems to be very effective to discard magnetic lines with impossible or improbable topologies. Only a few possibilities give qualitative fits to the Hanle diagram curves. Note that, as we can match the points in the Hanle diagram with the pixels they represent in the spatial maps, we can literally trace the possible magnetic field lines in the HF region. More than that, we can infer the inclination of the field along them. This procedure is facilitated by the knowledge of the location of the main Van-Vleck HPILs because they are placed where the magnetic field topology can be determined with more precision, so acting as boundary conditions.

Some examples of the theoretical curves that explain the loops and bows of the Hanle diagrams can be compared with Fig. \ref{fig:hdiagteo}. The narrow bows (e.g., red curve) can be explained with small differences in azimuth between the beginning and the end of a field line. Allowing the action of the thermodinamical variations we can explain the modulation in amplitude of the loops (as shown by the lines in the legend C of Fig. \ref{fig:hdiagteo}). In this case, it is specially clear the signature of the strong velocity gradients. When they are present, the polarization is significatively amplified in a small region where the azimuth can not vary at the same rithm, so generating characteristic radial excursions of the points in the Hanle diagrams (Regions 1 and 2 in Fig. \ref{fig:hdiag1}). Choosing larger K values (but below K$=1$), we can explain topologic lines with larger longitudes in the spatial map, corresponded with wider loops in the Hanle diagrams (e.g., blue curves). Finally, the effect of reducing the boundary magnetic field inclinations is a truncation of the loops that result in arcs (again reproduced in the legend- C lines) as the ones appearing in the left upper corner of Region 2 panels in Fig. \ref{fig:hdiag2} for instance. This latter case is interesting because it would allow us to identify the magnetic field lines that enter in the chromosphere with lower inclinations, which is a marker of lines starting to emerge from the photosphere. There are similar combinations that seems very promising for measuring the topology of the magnetic field, at least in the synthetic maps.  

For the moment, we conclude that the Hanle diagrams are a possible solution to observationally distinguish the saturation of the Hanle effect. Furthermore, they can be also a valuable tool for fitting the magnetic field inclinations in chromospheric quiet regions and for studying the precise topology of the magnetic field in observations with good resolution. Results deriving from this ideas will be tackled in future works.

 


 \clearemptydoublepage  
 
\chapter{Conclusions and future work}\label{cap:conc}
In this thesis we have carried out a detailed radiative transfer investigation of the linear and circular polarization signals of the Ca {\sc ii} IR triplet lines produced by scattering processes and the Hanle and Zeeman effects in a variety of dynamical models of the solar chromosphere. The investigation has focussed on the effect that the macroscopic vertical velocity fields and the shock waves of the chromospheric models have on the linear polarization (LP), but taking into account also the combined influence of the thermodynamic and magnetic field topology. The thesis has been complemented with the development of radiative transfer, visualization and analysis tools.
Our results will help to confront several chromospheric models of the quiet Sun with real observations through the synthesis of Stokes profiles.

\section{Goals achieved}
The main goals of this thesis have been the following ones:

\begin{itemize}

\item We have extended the RT code developed by \cite{Manso-Sainz:2003} to treat a large number of atmosphere models, including the effect of the macroscopic vertical velocity fields on the scattering polarization signals.  

\item We have shown, quantified and explained the effects that the macroscopic vertical velocity fields can have on the scattering polarization signals, pointing out some unknown diagnostic capabilities of the Ca {\sc ii} IR triplet lines for studying the solar chromosphere.  

\item We have calculated temporal series of scattering polarization signals including the effect of the chromospheric velocity fields and temperature shocks on the atomic polarization of the Ca {\sc ii} levels.

\item We have calculated spatial maps of scattering polarization signals including the impact of  the atomic level polarization, macroscopic vertical velocities and the Hanle effect in forward scattering geometry.

\item We have obtained theoretical LP references for the amplitudes of the $8498$, $8542$ and $8662$ {\AA} lines in dynamic solar model atmospheres. First, in close-to-limb observations, the references are given by the non-magnetic situation. Second, in forward scattering, the reference polarization is given by the maximum amplitudes obtained with a horizontal field in the saturation regime of the Hanle effect. These references can be used to estimate magnetic field strengths using the line-ratio technique. 

\item We have pointed out a new strategy that can be useful for chromospheric diagnostics (Hanle Polarity Inversion Lines, references for the LP signs, magnetic field inference in forward scattering).  

\item We have studied the relevance of temperature and macroscopic vertical velocity fields on the calculation of the chromospheric magnetic field orientation, in the presence of the $90\deg$ and $180\deg$ ambiguities of the Hanle effect.  

\item We studied the feasibility of using Solar-C and EST to detect the LP features found in our calculations. 

\item We have calculated a synthetic PCA basis of scattering polarization signals produced in dynamic chomospheric models. Among other things, it can be used to explain real observations in the Ca {\sc ii } IR triplet lines.

\end{itemize}

\section{Conclusions}
The main conclusions that have been obtained in this work are: 

\begin{itemize}
\item When macroscopic vertical velocity gradients are considered in our model atmospheres, the resulting polarization profiles of the Ca {\sc ii} IR triplet lines are asymmetrized, shifted in wavelength, and modulated in amplitude with respect to the constant velocity case. We remark that these effects depend on the velocity gradient. Thus, larger velocities do not mean larger variations in the LP if no significant velocity gradient is located in the line formation region.

\item The fundamental mechanism explaining the LP signals under dynamic circumstances is that increments in the absolute value of the velocity gradient increase the source function (Doppler brightening) and enhance the anisotropy of the radiation field at some heights. The result is a subsequent enhancement of the atomic alignment of the upper and lower levels controling the linear polarization profiles of the line transition under consideration (Doppler-induced modulation).

\item A second mechanism explains why sometimes the variations in the anisotropy produce an amplification of the LP signals and why sometimes they produce a reduced amplitude or a sign reversal. The emergent LP is the result of the atomic alignment values of the lower and upper levels of the line transition at every height in the formation region.
The variation with height of such atomic aligments is such that they are equal at a given chromospheric layer in the formation region. The atomic alignment of one of the line levels dominates the LP above such intersection layer, while the one of the other level dominates below. Then, the relative location of the levels's alignment intersection with respect to the heights where the anisotropy has been enhanced makes the alignment of one of the line levels to be preferentially increased. The details of this situation alter the shapes, amplitudes and signs of the LP profiles.

\item The shapes, signs and amplitudes of the LP signals of the IR triplet lines change significantly with the line of sight, because the height of formation moves with respect to the alignment intersection height. 

\item In general, the modulation of the LP signals produced by macroscopic vertical velocity fields does not have the same importance for all atomic transitions. Strong spectral lines with wide absorption profiles (e.g., with large thermal widths) require larger vertical velocities than weaker spectral lines in order to experiment the same amplitude variation in their scattering polarization signals. Some spectral lines (e.g., the Ca {\sc ii} IR triplet) forming in chromospheric layers (where the velocity gradients are significant) and whose atomic levels can be polarized are expected to be sensitive to Doppler-induced modulations in their LP profiles. 

\item The quantitative variations of the LP signals produced by velocity gradients depend on the strength of the model dynamics. In highly dynamic chromospheric models, the variations of the LP in close-to-limb observations give clear patterns characteristic of the presence of shocks in the models.

\item The close-to-limb synthesis carried out in the considered 1D time-dependent models show that the velocity gradients produce a significant size variation of the scattering polarization signals. In this case, the maximum enhancement factors of the emergent linear polarization are $\sim10$ (in the $8498$ {\AA} line) and $\sim7$ (in the $8542$ {\AA}
and $8662$ {\AA} lines), for the instantaneous values of the $\rm{Q/I}$
amplitudes with respect to the static FAL-C case. If we
consider temporal averages of the emergent Stokes profiles during
periods of several minutes, we get amplification factors of about a factor of two
(time-averaged Q/I amplitudes reach $\sim1 \,\%$) in the $8542$ {\AA} and
$8662$ {\AA} lines, and a sign reversal with antisymmetric profiles in the LP signals of the $8498$ {\AA} line. The lack of spatial or temporal coherence and resolution in the low-chromosphere dynamics could make the $\lambda 8498$ signals to cancel out due to its variable polarity.

\item The velocity-free approximation is not valid to calculate the emergent scattering polarization signals in the presence of velocity gradients because it does not capture the effect of the dynamics on the atomic level polarization. Such approximation gives polarization amplitudes that are not enhanced with respect to the static case.

\item The forward scattering RT calculations carried out in the considered MHD models \citep{Leenaarts:2009} show a sparse spatial distribution of pixels with amplitudes significantly larger than in the static case. On the one hand, the reason is that the considered MHD model is significantly less dynamic than the 1D dynamical models we have analyzed. On the other hand, the variations in the radiation field anisotropy induced by the temperature gradients of the MHD model result in LP amplitudes that are often significantly larger than in the semi-empirical model considered by \cite{manso10}.

\item In principle, the inference of chromospheric magnetic fields with the Hanle effect can be affected by the dynamic state of the atmosphere because the LP amplitudes of the Ca {\sc ii} lines are modulated by the vertical velocities. In order to give a precise magnetic field estimation of the quiet solar chromosphere it is then necessary to consider the polarization amplitude variations sourced by the dynamics of the chromosphere. However, we expect this effect to be small in \textit{low-resolution} chromospheric observations.

\item We propose the so-called Hanle Polarity Inversion Lines (HPILs) as features that can help to deduce the magnetic field vector in quiet Sun chromospheric maps of the LP signals in IR Ca {\sc ii} lines. They give useful references to constrain the orientation of the magnetic field as well as its variation with height (when analyzed simultaneously in more than one spectral line).

\item The \textit{thermodynamic} HPILs are sensitive to very specific stratifications of the radiation field anisotropy; they allowed us to find the location of cool chromospheric bubbles of plasma using the LP signals. 

\item The\textit{ Van-Vleck} HPILs in the Ca {\sc ii} IR triplet lines give the locations where the chromospheric magnetic field has an inclination of $54.73 \deg$ or $125.27 \deg$.

\item The behavior of the scattering polarization signals in the weak field regime makes the stratification of temperature, velocity and magnetic inclination to influence the inference of the magnetic field azimuth from Stokes Q and U. To illustrate and quantify these dependencies we have simulated the process of magnetic field inference from spatial maps of the Stokes vector. This study could be considered as an extension to the work of \cite{manso10} in more realistic atmospheric models. 

\item Studying the scattering polarization signals under dynamic situations, we have given a table of signs of reference for the LP signals of the Ca {\sc ii} IR triplet lines in the most important physical circumstances describing the considered MHD models. These references can be used to break the $90\deg$ magnetic field ambiguity existing in the weak field regime. They can also be used to diagnose the basic physical state of the chromosphere in pixels showing those LP signs. The LP signs are compatible with the results of \cite{manso10}.

\item The dependence of the analyzed scattering polarization signals on temperature, velocity and magnetic field stratification has allowed the identification of some simple diagnostic techniques based on the signs of the LP profiles of the Ca {\sc ii} IR triplet under different physical situations. 

\item Our calculations show that the real relevance of the chromospheric vertical velocity fields is missed with the current instrumental sensitivities and telescope apertures. This explain that we cannot observe such effects until more powerful instrumentation is developed.

\item Our calculations point out that the Solar-C telescope and perhaps the European Solar Telescope should be able to capture the behavior of the scattering polarization signals that we have described in this thesis for the IR triplet of the Ca {\sc ii}. The best strategy to improve the signal-to-noise ratio of such measurements without losing spatial and temporal resolution is a combination of short temporal integrations and PCA postprocessing.


\end{itemize}

\section{Future work}\label{sec:pm}
The MHD models of the solar atmosphere are evolving more quickly than the radiative transfer codes with capabilities to treat the scattering polarization. 
Thus, despite the fact that the synthesis of Stokes profiles is limited by the realism of MHD models, the very first technical limitation is the intrinsic difficulty of the detailed RT problem with polarization: its highly non-linear and non-local nature, the demanding spatial and angular resolution that it requires and the involved physics to solve it. A giant leap is required in the way we solve these problems. I plan to continue the research on new methods that allow a faster resolution of three dimensional radiative transfer problems. This will allow a natural extension of this thesis to treat the more general problem in which the horizontal macroscopic flows are included. 

Using state-of-the-art MHD models, I would like to continue exploring the spectropolarimetric characterization of chromospheric solar structures. This alternative seeks to identify spatiotemporal polarization patterns (in synthetic or real profiles), such as those identified in Chapters \ref{cap:three} and \ref{cap:four} (e.g., shocks fingerprints in temporal series and Hanle PILs in spatial maps). 

An important step after this thesis will be however to carry out spectropolarimetric observations and to try to interpret them applying the results of this work. It would be of great interest to detect LP features confirming our results in the solar chromosphere as well as to apply the proposed diagnosis methods to real observations. However, the current instrumentation does not have enough sensitivity to capture the dynamic effects on the scattering polarization. In the future, with the arrival of the new solar facilities, such as SOLAR-C, ATST or EST, a new world of possibilities will be opened. 

Meanwhile, we could try to validate our results in stronger spectral lines whose scattering polarization could also be affected by shocks and motions (e.g., the Na {\sc i} D lines or the Ca {\sc i} $4227$ {\AA} line). In particular, the Ca{\sc i} $4227$ {\AA} line can be a good candidate to follow up the ideas of Chapter \ref{cap:four}, which explored the forward scattering geometry as a way of inferring information about solar magnetic fields in situations where the linear polarization is dominated by scattering and Hanle processes. 

Other issue to solve is the inference of the magnetic field azimuth in the presence of significant velocity fields or temperature gradients, which produce a lack of precision in such estimation (at least in our theoretical results). How can our results be extrapolated to chromospheres with stronger magnetic fields?Are the linear polarization signals similarly modified by thermodynamic gradients in such cases?


 \clearemptydoublepage  





\clearpage
\clearemptydoublepage

\appendix
\chapter{Appendix.}\label{cap:appendix}
\section{Special functions in section \ref{sub:scattercloud}}\label{app:A}

Introducing Equation~\ref{eq04} into Equation~\ref{eq02}
\begin{equation}\label{eqa01}
\begin{split}
\bar{J}^0_0=&\frac{1}{2}\int_0^\infty d\nu'
\frac{1}{\sqrt{\pi}\Delta\nu_D}\exp\{-(\frac{\nu'-\nu_0}{\Delta\nu_D})^2\}\\
&\int_0^1d\mu
I^{(0)}(1-u+u\mu) [1-a\exp\{-(\frac{\nu'(1+v_z\mu/c)-\nu_0}{w})^2\}].
\end{split}
\end{equation}
Introducing the variables $x=(\nu'-\nu_0)/w$, $\alpha=\Delta\nu_D/w$, and
$\xi=v_z\nu_0/(cw)$, then the mean intensity in the comoving frame
\begin{equation}\label{eqa02}
\frac{\bar{J}^0_0}{I^{(0)}}=\frac{1}{2}\int_{-\infty}^{\infty}dx
\frac{1}{\sqrt{\pi}\alpha}{e}^{-(x/\alpha)^2}
\int_0^1d\mu (1-u+u\mu)(1-a e^{-(x+\xi\mu)^2}).
\end{equation}
In passing from Equation~(\ref{eqa01}) to Equation~(\ref{eqa02}), we have
extended the integration limit on $x$ to $\infty$. 
Analogously for the anisotropy in the comoving frame
\begin{equation}\label{eqa03}
\frac{\bar{J}^2_0}{I^{(0)}}=\frac{1}{4\sqrt{2}}\int_{-\infty}^{\infty}dx
\frac{1}{\sqrt{\pi}\alpha}{\rm e}^{-(x/\alpha)^2}
\int_0^1d\mu (3\mu^2-1)(1-u+u\mu)(1-a{\rm e}^{-(x+\xi\mu)^2}).
\end{equation}

The following integrals are easily evaluated \citep[see][]{Spiegel}
\begin{align}\label{eqa04}
{\cal I}_0(\alpha;\xi)&\equiv\int_{-\infty}^{\infty}dx
\frac{1}{\sqrt{\pi}\alpha}{\rm e}^{-(x/\alpha)^2}
\int_0^1 (1-u+u\mu)[1-a {\rm e}^{-(x+\xi\mu)^2}]d\mu \notag \\
&=\frac{1}{2}[2-u+a(u-1)\sqrt{\pi}\frac{1}{\xi}{\rm
Erf}(\frac{\xi}{\sqrt{1+\alpha^2}})
+a u\sqrt{1+\alpha^2}\frac{1}{\xi^2}(\exp\{-\frac{\xi^2}{1+\alpha^2}\}-1)], 
\end{align}
\begin{align}\label{eqa05}
{\cal I}_2(\alpha;\xi)&\equiv\int_{-\infty}^{\infty}dx
\frac{1}{\sqrt{\pi}\alpha}{\rm e}^{-(x/\alpha)^2}
\int_0^1 (3\mu^2-1)(1-u-u\mu)[1-a {\rm e}^{-(x+\xi\mu)^2}]d\mu \notag \\
&=\frac{1}{4}[u-2 a (u-1) \sqrt{\pi} \frac{1}{\xi} {\rm
Erf}(\frac{\xi}{\sqrt{1+\alpha^2}}) 
+2a\sqrt{1+\alpha^2}\frac{1}{\xi^2}
\left[u+(3-u)\exp\{-\frac{\xi^2}{1+\alpha^2}\}\right] \\
&+3a(u-1)(1+\alpha^2)\sqrt{\pi}\frac{1}{\xi^3}{\rm
Erf}(\frac{\xi}{\sqrt{1+\alpha^2}})
+6au(1+\alpha^2)^{3/2}\frac{1}{\xi^4}(\exp\{-\frac{\xi^2}{1+\alpha^2}\}-1)],
\notag
\end{align}
where we have made use of 
\begin{equation}
\int_{-\infty}^{\infty}dx \frac{1}{\sqrt{\pi}\alpha}{\rm e}^{-(x/\alpha)^2}
[1-a{\rm e}^{-(x+\xi\mu)^2}]=
1-\frac{a}{\sqrt{1+\alpha^2}}\exp\{-\frac{\mu^2\xi^2}{1+\alpha^2}\}.
\end{equation}
From them, the values for $\bar{J}^0_0$ and the anisotropy
$\sqrt{2}\bar{J}^2_0/\bar{J}^0_0$ are trivially derived.

In the high velocity limit ($\xi\rightarrow\infty$), ${\cal
I}_0(\alpha;\xi)=(2-u)/2$, and ${\cal I}_2(\alpha;\xi)=u/4$ (regardless of
$\alpha$).
In the low velocity limit:
\begin{align}
{\cal
I}_0(\alpha;\xi)&=\frac{1}{2}(1-\frac{a}{\sqrt{1+\alpha^2}})(2-u)+\frac{a(4-u)}{
12(1+\alpha^2)^{3/2}}\xi^2 + {\rm O}(\xi^3), \\
{\cal I}_2(\alpha;\xi)&=\frac{u}{4}(1-\frac{a}{\sqrt{1+\alpha^2}})+
\frac{a(16-u)}{60(1+\alpha^2)^{3/2}}\xi^2 + {\rm O}(\xi^3).
\end{align}

\section{Statistical equilibrium equations in the non-magnetic case. }\label{app:B}
The rate equations for the considered problem are as follow. They have been obtained by particularizing to the model-atom of Fig.1 the equations contained in Sects. 7.2 and 7.13 of \cite{LL04}.
\begin{align}
\begin{split}\label{see01}  
\frac{\rm d}{{\rm d}t} \rho^0_0(1) =& -\biggl[\sum_{u=4}^5
B_{1u}\bar{J}^0_0(1\rightarrow u) + \sum_{i\neq 1} C_{1i} \biggr] \rho^0_0(1)
 + A_{41}\rho^0_0(4) \\
&+ \sqrt{2} A_{51}\rho^0_0(5) +\sum_{i\neq 1} C_{i1} \sqrt{\frac{2J_i+1}{2}}  \rho^0_0(i),
\end{split}
\displaybreak[0] \\
\begin{split}\label{see02}
\frac{\rm d}{{\rm d}t} \rho^0_0(2) =& -\biggl[\sum_{u=4}^5
B_{2u}\bar{J}^0_0(2\rightarrow u) + \sum_{i\neq 2} C_{2i} \biggr] \rho^0_0(2) 
- \biggl( \frac{1}{\sqrt{2}} B_{24}\bar{J}^2_{0}(2\rightarrow 4)
-\frac{2\sqrt{2}}{5} B_{25}\bar{J}^2_{0}(2\rightarrow 5) \biggr) \rho^2_{0}(2)
\\
& + \frac{1}{\sqrt{2}} A_{42}\rho^0_0(4) + A_{52}\rho^0_0(5) + \sum_{i\neq 2}
C_{i2} \frac{\sqrt{2J_i+1}}{2}  \rho^0_0(i),
\end{split}
\displaybreak[0] \\
\begin{split}\label{see03}
\frac{\rm d}{{\rm d}t} \rho^0_0(3) =& -\biggl[B_{35}\bar{J}^0_0(3\rightarrow 5)
+ \sum_{i\neq 3} C_{3i} \biggr] \rho^0_0(3) 
- \frac{\sqrt{7}}{5} B_{35}\bar{J}^2_0(3\rightarrow 5) \rho^2_0(3)\\
& +\sqrt{\frac{2}{3}} A_{53} \rho^0_0(5) + \sum_{i\neq 3} C_{i3}
\sqrt{\frac{2J_i+1}{6}}  \rho^0_0(i),
\end{split}
\displaybreak[0] \\
\begin{split}\label{see04}
\frac{\rm d}{{\rm d}t} \rho^0_0(4) =& -\biggl[ \sum_{l=1}^2 A_{4l} + \sum_{i\neq
4} C_{4i} \biggr] \rho^0_0(4) 
+ \sum_{l=1}^2 B_{l 4} \bar{J}^0_0(l \rightarrow 4) \sqrt{\frac{2J_l+1}{2}}
\rho^0_0(l) \\
&  + B_{24}  \bar{J}^2_0(l \rightarrow 4) \rho^2_0(2) + \sum_{i\neq 4} C_{i4}
\sqrt{\frac{2J_i+1}{2}}  \rho^0_0(i),
\end{split}
\displaybreak[0] \\
\linebreak
\begin{split}\label{see05}
\frac{\rm d}{{\rm d}t} \rho^0_0(5) =& -\biggl[ \sum_{l=1}^3 A_{5l} + \sum_{i\neq
5} C_{5i} \biggr] \rho^0_0(5)+ \sum_{l=1}^3 B_{l 5} \bar{J}^0_0(l \rightarrow 5) \frac{\sqrt{2J_l+1}}{2}
\rho^0_0(l) \\
&  -\frac{2\sqrt{2}}{5} B_{25} \bar{J}^2_0(2\rightarrow 5) \rho^2_0(2)
   +\frac{\sqrt{42}}{10}  B_{35} \bar{J}^2_0(3\rightarrow 5) \rho^2_0(3) 
+ \sum_{i\neq 5} C_{i5} \frac{\sqrt{2J_i+1}}{2}  \rho^0_0(i),
\end{split}
\displaybreak[0]
\displaybreak[0] \\
\begin{split}\label{see06}
\frac{\rm d}{{\rm d}t} \rho^2_0(2) =& -\biggl[ \sum_{u=4}^5
B_{2u}\bar{J}^0_0(2\rightarrow u) + \sum_{i\neq 2} C_{2i} +D_2^{(2)}\biggr]
\rho^2_0(2) -  \biggl( \frac{1}{\sqrt{2}} B_{24}\bar{J}^2_{0}(2\rightarrow 4)\\
&-\frac{2\sqrt{2}}{5} B_{25}\bar{J}^2_{0}(2\rightarrow 5) \biggr) \rho^0_0(2) + \frac{1}{5} A_{52}\rho^2_0(5) + \sum_{i=3, 5} C_{i2}^{(2)}
\frac{\sqrt{2J_i+1}}{2}  \rho^2_0(i),
\end{split}
\displaybreak[0] \\
\begin{split}\label{see07}
\frac{\rm d}{{\rm d}t} \rho^2_0(3) =& -\biggl[B_{35}\bar{J}^0_0(3\rightarrow 5)
+ \sum_{i\neq 3} C_{3i} +D_3^{(2)}\biggr] \rho^2_0(3) 
- B_{35} \bar{J}^2_{0}(3\rightarrow 5)\frac{\sqrt{7}}{5} \rho^0_0(3) \\
& + B_{35}\bar{J}^2_{0}(3\rightarrow 5) \biggl[
\sqrt{\frac{5}{7}}\rho^2_{0}(3)
- \frac{9}{2}\sqrt{\frac{3}{35}} \rho^4_{0}(3) \biggr]  +
\frac{2}{5}\sqrt{\frac{7}{3}} A_{53} \rho^2_0(5) \\
& + \sum_{i=2, 5} C_{i3}^{(2)}
\sqrt{\frac{2J_i+1}{6}}  \rho^2_0(i),
\end{split}
\displaybreak[0] \\
\begin{split}\label{see08}
\frac{\rm d}{{\rm d}t} \rho^2_0(5) =& -\biggl[\sum_{l=1}^3 A_{5l} + \sum_{i\neq
5} C_{5i} +D_5^{(2)}\biggr] \rho^2_0(5) 
 + \frac{1}{5}B_{25}\bar{J}^0_0(2\rightarrow 5)\rho^2_0(2) \\
&+\frac{\sqrt{21}}{5}B_{35}\bar{J}^0_0(3\rightarrow 5)\rho^2_0(3) 
 +\frac{1}{2}B_{15}\bar{J}^2_0(1\rightarrow 5)\rho^0_0(1) - \frac{2\sqrt{2}}{5}
B_{25} \bar{J}^2_{0}(2\rightarrow 5) \rho^0_0(2)\\
& +\frac{\sqrt{3}}{10} B_{35} \bar{J}^2_{0}(3\rightarrow
5) \rho^0_0(3) + 2\sqrt{\frac{7}{5}} B_{25} \bar{J}^2_{0}(2\rightarrow 5) \rho^2_{0}(2) -
\sqrt{\frac{3}{5}} B_{35} \bar{J}^2_{0}(3\rightarrow 5) \rho^2_{0}(3)  \\
& +\frac{9}{\sqrt{5}} B_{35} \bar{J}^2_{0}(3\rightarrow 5) \rho^4_{0}(3)+
\sum_{i=2, 3} C_{i5}^{(2)} \frac{\sqrt{2J_i+1}}{2}  \rho^2_0(i),
\end{split}
\displaybreak[0] \\
\begin{split}\label{see09}
\frac{\rm d}{{\rm d}t} \rho^4_0(3) =& -\biggl[ B_{35}\bar{J}^0_0(3\rightarrow 5)
+ \sum_{i\neq 3} C_{3i} +D_3^{(4)}\biggr] \rho^4_0(3) \\
& - B_{35}\bar{J}^2_0(3\rightarrow 5) \biggl[ \frac{9}{2}\sqrt{\frac{3}{35}}
\rho^2_{0}(3) 
+3\sqrt{\frac{11}{70}} \rho^4_{0}(3) \biggr],
\end{split}
\end{align}
where $A_{u{\ell}}$ and $B_{{\ell}u}$ are the Einstein emission and absortion
coefficients; $C_{{\ell} u}$ and $C_{u{\ell}}$ are the excitation and deexcitation
inelastic collisional rates, respectively; $C^{(2)}_{{\ell} u}$ and $C^{(2)}_{u{\ell}}$
are the collisional transfer rates for alignment between polarizable levels
(with $J>1/2$); and $D^{(K)}_{i}$ is the depolarization rate of the $K$-th
multipole of level ${i}$ due to elastic collisions with neutral hydrogen atoms.
The $\rho^K_{0}$ elements are referred to a coordinate system with the
quantization axis along the solar local vertical direction.

\section{Derivation of the fractional linear polarization under EB approximation.}\label{app:C}

We consider a non-magnetic, static, axisymmetric and semi-infinite
atmosphere. With \textit{axisymmetric} we mean an atmosphere without
inhomogeneities around its radial axis. Under this assumptions,
symmetry breaking effects are negligible, what in turn nullifies
quantum coherences ($\rho^K_{Q\neq0}=0$). Thus, only $\rho^2_0$ and
$\rho^0_0$ are non-zero, what implies as well that only I and Q are
non-zero in the Stokes vector. Then, Eqs. (6) of Sec. \ref{subsec:rte} are still valid. They are used in the next demonstration with the only difference that absortion profiles are not affected by dopplershits ($\phi^{\prime}_{lu}= \phi_{lu}$).
\\
The solar atmosphere is weakly anisotropic, so $I \gg Q$. Then, neglecting second order terms in Eqs. (5) of Paper {\sc i} , and recalling that optical depth seen by a photon along the ray path \textit{s} is
\begin{equation*}
\tau^{\rm los}_{\nu}=- \int \eta_I ds,
\end{equation*}
 we have the next RTE 's:
\begin{subequations}\label{eq:rtesimple}
\begin{align}
\frac{\rm d I}{{\rm d}\tau^{\rm los}}&=I-S_I, \label{rte1simple}
\displaybreak[0] \\
\frac{\rm d Q}{{\rm d}\tau^{\rm los}}&=Q-S_Q. \label{rte2simple}
\end{align}
\end{subequations}
Source functions for intensity and Stokes Q are
$S_I=\frac{\epsilon_I}{\eta_I}$ and
$S_Q=\frac{\epsilon_Q-\eta_QI}{\eta_I}$. Under EB approximation the
source function $S_I$ is developed in power series through optical
depth and truncated to first order as $S_I(\tau_{\nu})=a+b\tau_{\nu}$
(being a and b arbitrary constants). Thus, taking into account that
$\tau_{\nu}=\tau^{\rm los}_{\nu} \cdot \mu$, the solution to the RTE
for the emergent intensity is:
\begin{equation}\label{eq:ieb}
I_{\nu}(\tau_{\nu},\mu)=a + b (\mu \tau^{\rm los}_\nu + \mu).
\end{equation}
We see that $I_{\nu}(\tau_{\nu}=0)=S_I(\tau_{\nu}=\mu)$ or, equivalently, $I_{\nu}(\tau_{\nu}=0)=S_I(\tau^{\rm los}_{\nu}=1)$.
And from Eq. (\ref{eq:ieb}) we also note that: 
\begin{equation*}
\frac{\rm d I}{{\rm d}\tau^{\rm los}}=\frac{\rm d S_I}{{\rm d}\tau^{\rm los}}=b \mu.
\end{equation*}
Hereafter, we follow the notation $ x^{[r]}$, indicating that the
quantity $x$ have to be evaluated at $ \tau^{\rm los}_{\nu} =
r$. Thus, we can write $b \mu = S^{[1]}_I-S^{[0]}_I$.
Repeating the same considerations, we assume EB as well for Stokes Q and write:
\begin{equation*}
S_Q(\tau_{\nu})=a'+b'\tau_{\nu}
\end{equation*}
\begin{equation*}
Q_{\nu}(\tau_{\nu},\mu)=a' + b' (\tau_\nu + \mu),
\end{equation*}
\begin{equation}\label{eq:variousQ3}
Q^{[0]}_{\nu}=S^{[1]}_Q,
\end{equation}
\begin{equation*}
\frac{\rm d Q}{{\rm d}\tau^{\rm los}}=\frac{\rm d S_Q}{{\rm d}\tau^{\rm los}}=b' \mu=S^{[1]}_Q-S^{[0]}_Q. 
\end{equation*}
With this expressions we can put the emerging relative linear polarization as :
\begin{equation*}
\frac{Q^{[0]}}{I^{[0]}}=\frac{S_Q^{[1]}}{S_I^{[1]}}=\left[ \frac{(\epsilon_Q-\eta_Q I)/\eta_I}{\epsilon_I/\eta_I}\right]^{[1]}=\frac{\epsilon^{[1]}_Q}{\epsilon^{[1]}_I}-\frac{\eta^{[1]}_Q}{\epsilon^{[1]}_I}I^{[1]} .
\end{equation*}
The evaluation of $I^{[1]}$ can be done using the RTE or simply realizing that:
\begin{equation*}
I^{[1]}=I_{\nu}(\tau^{\rm los}_{\nu}=1)= S^{[2]}_I=\frac{\epsilon^{[2]}_I}{\eta^{[2]}_I},
\end{equation*}
leading to
\begin{equation}
\frac{Q}{I}=\frac{\epsilon^{[1]}_Q}{\epsilon^{[1]}_I}-\frac{\eta^{[1]}_Q \epsilon^{[2]}_I}{\eta^{[2]}_I \epsilon^{[1]}_I} .
\end{equation}
Until now, we have only imposed axisymmetry and EB. If we also particularize for strong lines ($\epsilon_I \approx \epsilon^{\rm line}_I$ and $\eta_I \approx \eta^{\rm line}_I$) and use Eqs. (\ref{eq:coefs1}) of Sec. \ref{subsec:rte} for substituting absortion and emission terms, we arrive to:

\begin{equation*}\label{eq:qestimatedEB1}
\frac{Q}{I} \approx  \frac{3}{2\sqrt{2}} (1-\mu^2) \left[  \omega^{(2)}_{J_u J_l} \cdot \sigma^2_0(J_u)^{[1]}  -  \omega^{(2)}_{J_l J_u}   \cdot \frac{\rho^2_0(J_l)^{[1]}}{\rho^0_0(J_u)^{[1]}} \cdot \frac{\rho^0_0(J_u)^{[2]}}{\rho^0_0(J_l)^{[2]}} \right] ,
\end{equation*}
Note that the superindex $[2]$ is a notation for optical depth and the superindex $(2)$ has to do with the designation of the quantum numbers decribed in Table \ref{tab:atomic}. Multiplying and dividing by $\rho^0_0(J_l)^{[1]}$ the expression read

\begin{equation}\label{eq:qestimatedEB2}
\frac{Q}{I} = \frac{3}{2\sqrt{2}} (1-\mu^2) \left[  \omega^{(2)}_{J_u J_l} \cdot \sigma^2_0(J_u)^{[1]}  -  \omega^{(2)}_{J_l J_u}   \cdot \sigma^2_0(J_l)^{[1]}  \cdot \textit{A} \right] ,
\end{equation}
where the quantity $\textit{A}$ affecting to the lower level alignment is
\begin{equation}\label{eq:qEBA}
 \textit{A}=\left( \frac{\rho^0_0(J_l)}{\rho^0_0(J_u)}\right)^{[1]} \cdot \left( \frac{\rho^0_0(J_u)}{\rho^0_0(J_l)}\right)^{[2]} .
\end{equation}
 If radiative processes dominate over inelastic collisional ones,
 $\rho^0_0(J_u)/\rho^0_0(J_l) \approx B_{lu} \cdot \bar{J}^0_0 /
 A_{ul}$ with very good approximation, and then:

\begin{equation*}    
\textit{A} \approx \frac{\bar{J}^{0 [2]}_0}{\bar{J}^{0 [1]}_0} \ \ \ \  ,
\end{equation*}

It implies that

\begin{equation}\label{eq:qestimated3}
\frac{Q}{I} \approx  \frac{3}{2\sqrt{2}} (1-\mu^2) \left(  \omega^{(2)}_{J_u J_l} \cdot \sigma^2_0(J_u)^{[1]}  -  \omega^{(2)}_{J_l J_u}  \cdot \sigma^2_0(J_l)^{[1]} \cdot \frac{\bar{J}^{0 [2]}_0}{\bar{J}^{0 [1]}_0} \right) 
\end{equation}
%
is valid at line center $\lambda_0$. The only
difference with the equation estimated by \cite{Trujillo-Bueno:1999} is the
factor A, that can be neglected in majority of cases due to
its closeness to one. The factor A can be understood as a measure of
the mean intensity gradient around $\tau^{\rm los}_{\nu_0}=1$, being
larger in presence of shocks and heatings. If the gradient of
temperature is then significant in that region we could expect this
factor to modify the lower level fractional alignment and the
estimation of the line center Q/I. Although the proximity of this idea
with the main topic of the this thesis, we have not found a clear
application to Eq. \ref{eq:qestimated3}. We always fix ${\rm A=1}$ and
use that equation for explicative purposes in the text.

\section{Other figures in Chapter 6.}\label{app:D}
\begin{figure}[h]
\centering%
\includegraphics[scale=0.78]{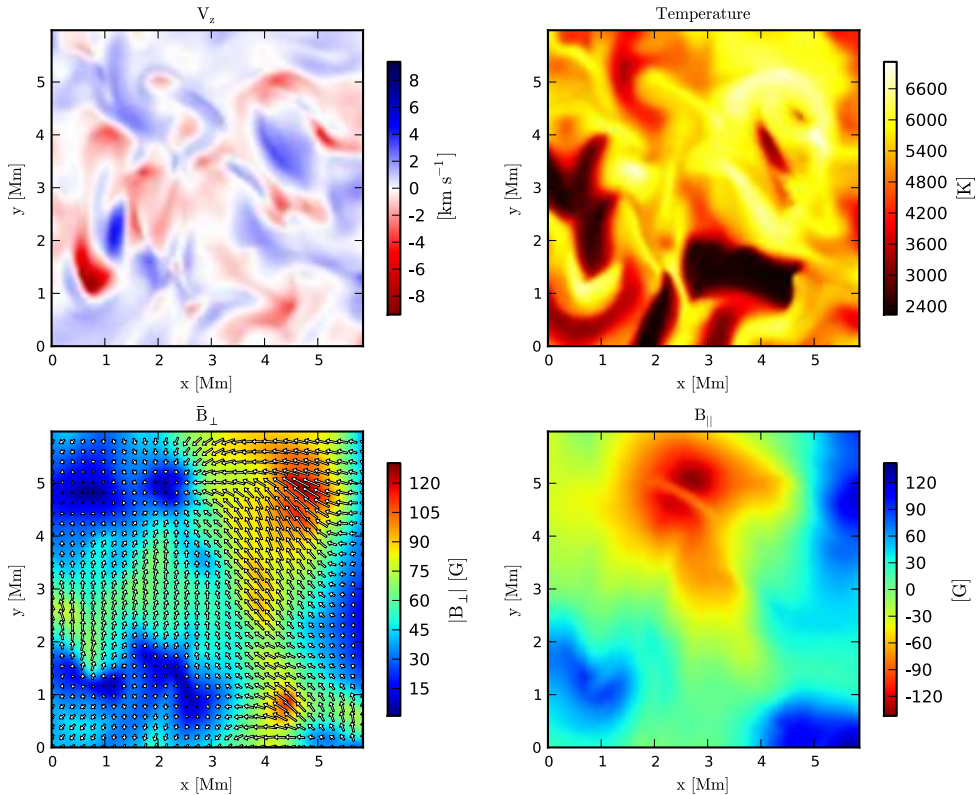}
\caption{\footnotesize {Physical quantities at $\tau=1$ for the $8498 \,$ {\AA} line. Upper left: vertical velocity. Upper right: temperature distribution. Lower left: horizontal magnetic field intensity. Lower right: longitudinal magnetic field intensity.} }
\label{fig:vtb8498}
\end{figure}
\clearpage
\begin{figure}[t!]
\centering%
\includegraphics[scale=0.73]{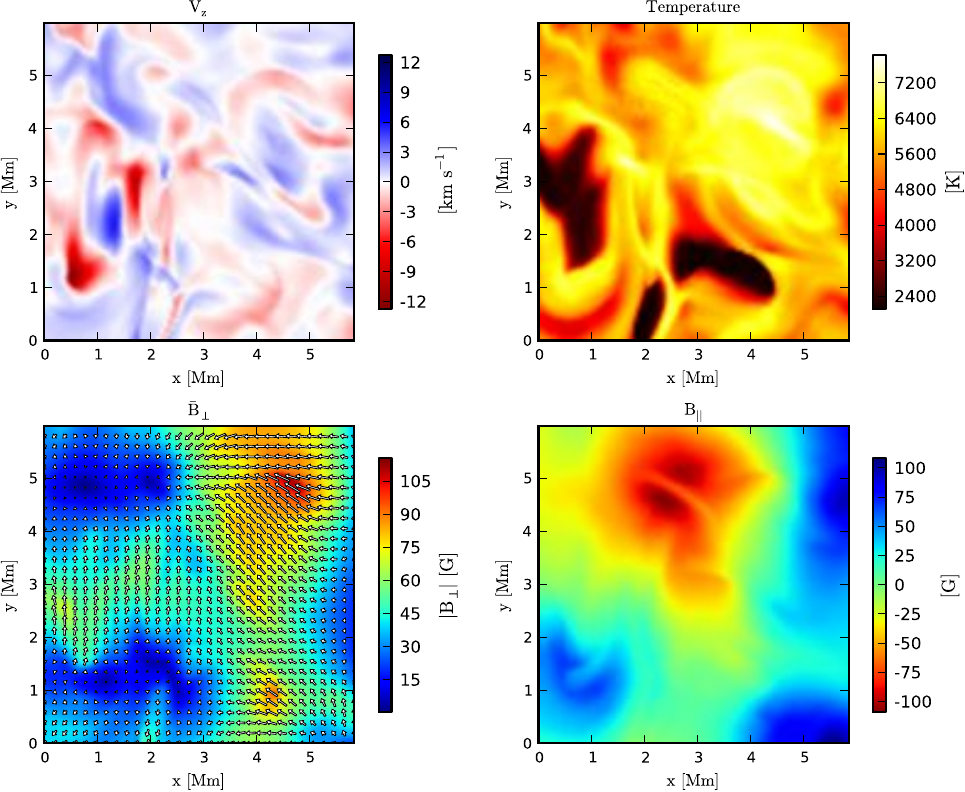}
\caption{\footnotesize {The same as in Fig. \ref{fig:vtb8498} but for the $8662$
  {\AA} line.}}
\label{fig:vtb8662}
\end{figure}
\begin{figure}[ht!]
\centering
\includegraphics[width=\textwidth]{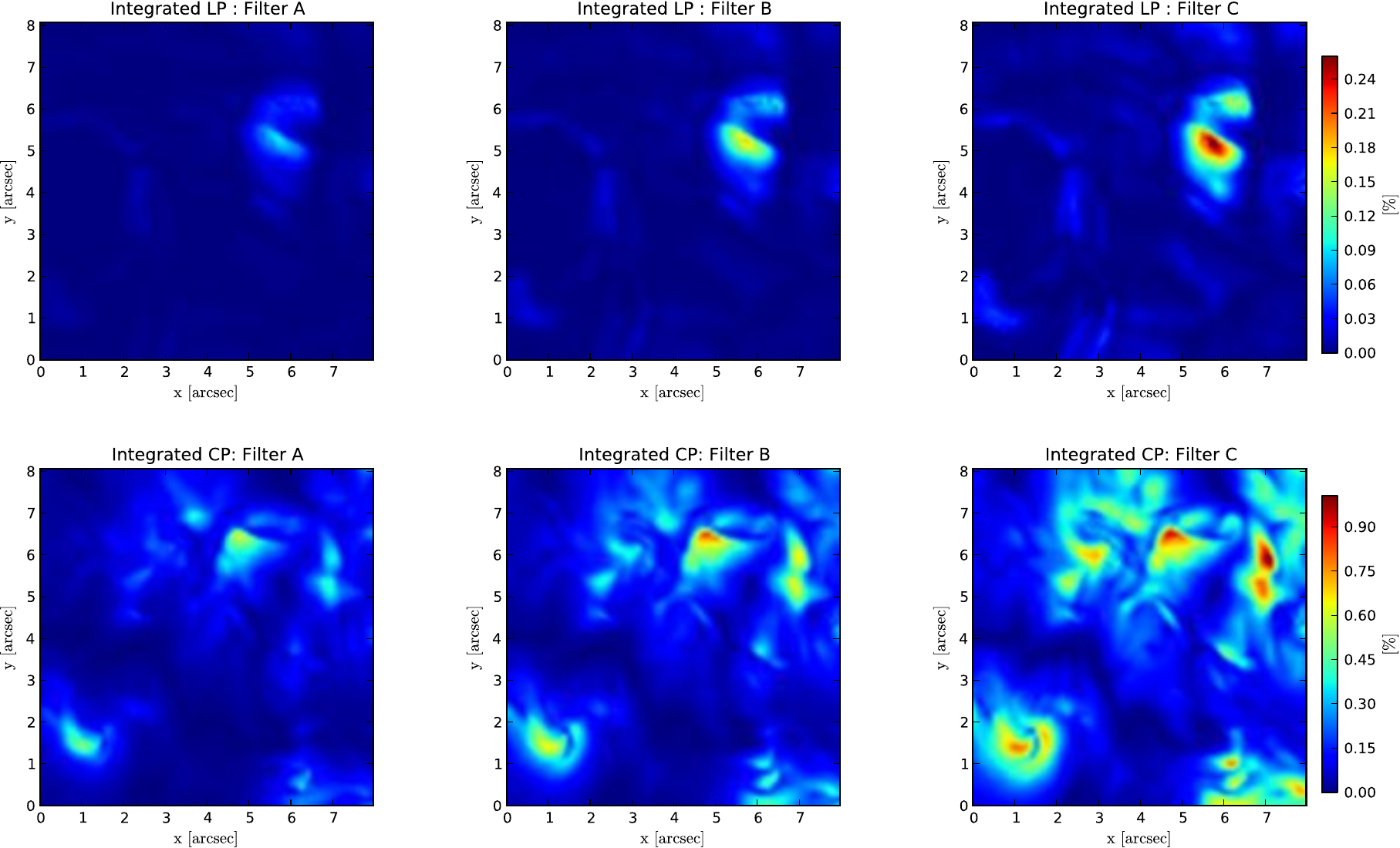}
\caption{The same as in Fig. \ref{fig:8542_filters} but for tke K line.  }
\label{fig:k_filters}
\end{figure}
\clearpage
\begin{figure}[b!]
\centering%
\includegraphics[width=0.7\textwidth]{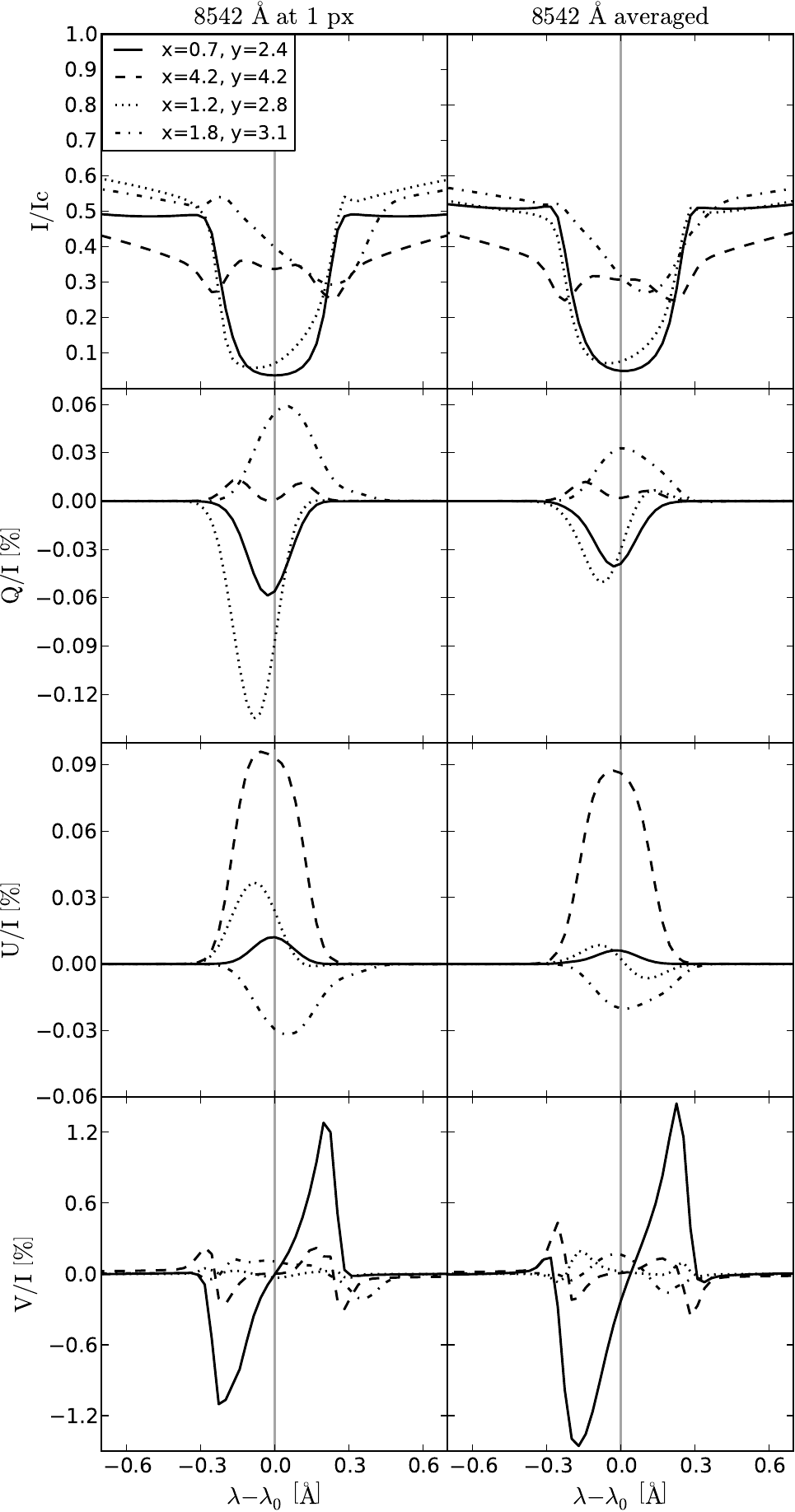}
\caption{Effect of averaging across $1$ square arcsecond centered in
  different pixels (different color identifies different center
  pixels). The average is done in each Stokes vector component before
  calculating the fractional quantities.  }
\label{fig:profiles_c}
\end{figure}
\begin{figure}[ht!]
\centering%
\includegraphics[width=\textwidth]{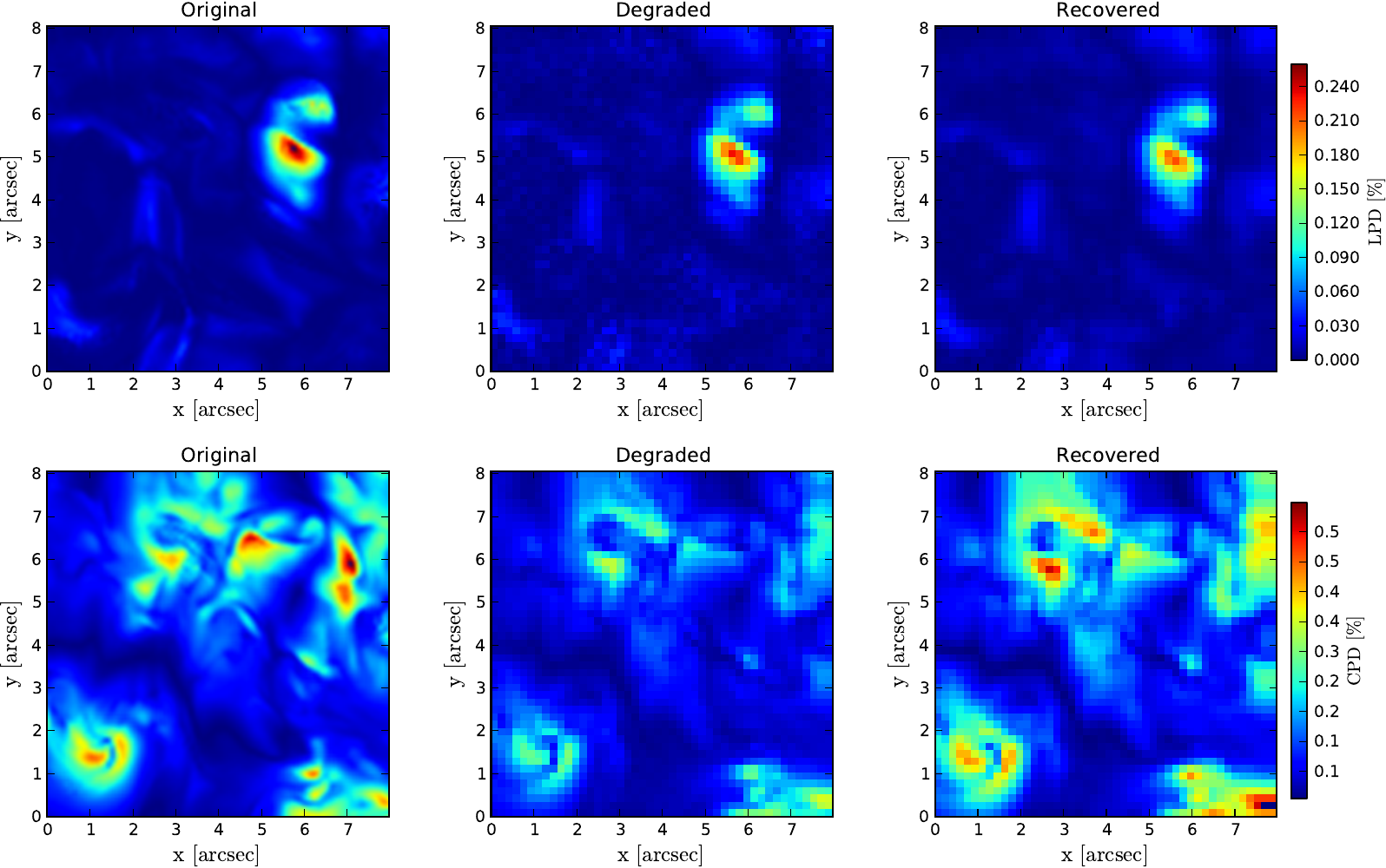}
\caption{The same as figure \ref{fig:8542_integrada} but for the K line. }
\label{fig:k_integrada}
\end{figure}

\begin{figure}[ht!]
\centering%
\includegraphics[width=\textwidth]{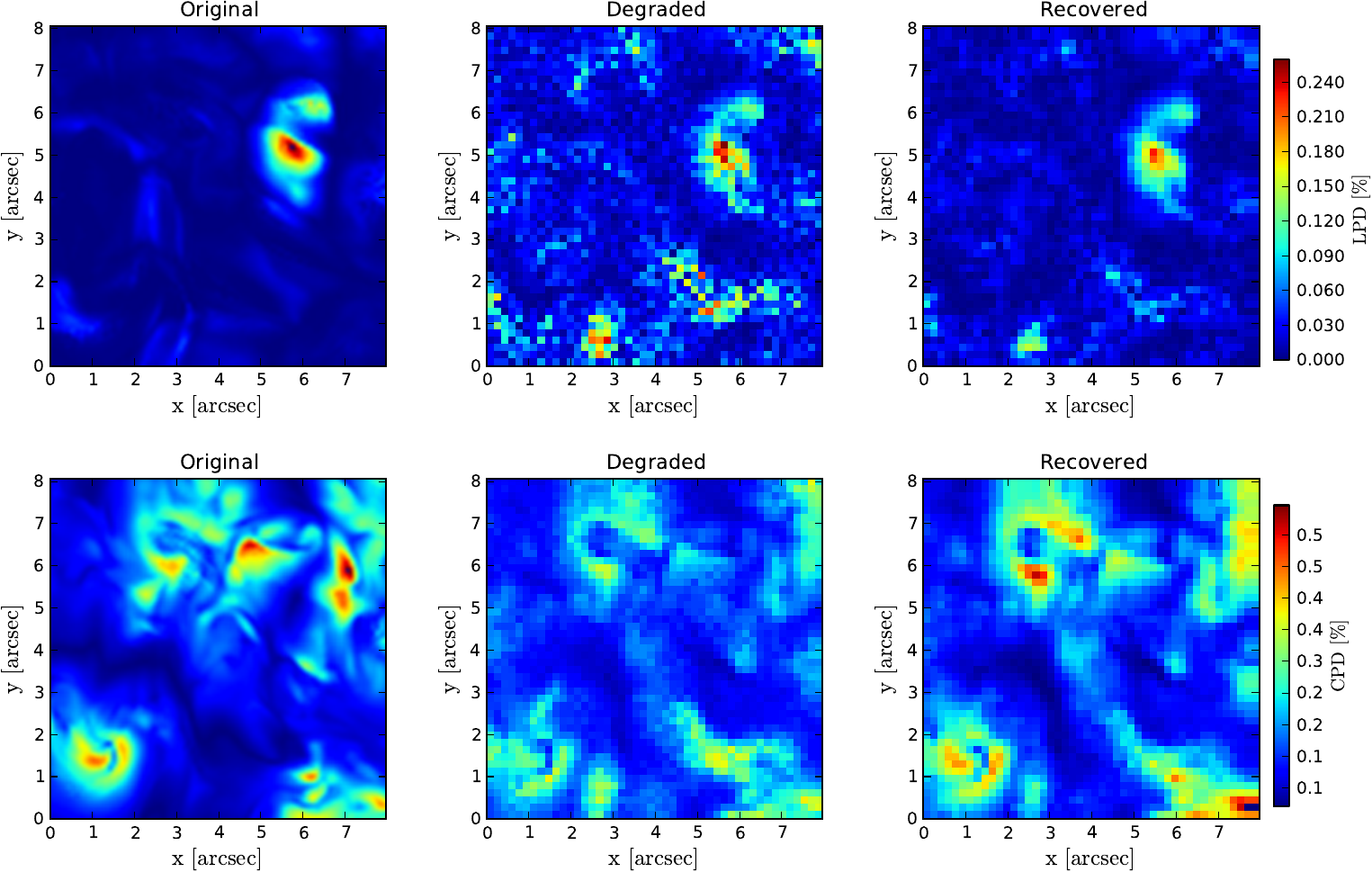}
\caption{The same as figure \ref{fig:k_integrada} but with S/N$=10^{-3}$. }
\label{fig:k_integrada1em3}
\end{figure}

\begin{figure}[h!]
\centering%
\includegraphics[scale=0.4]{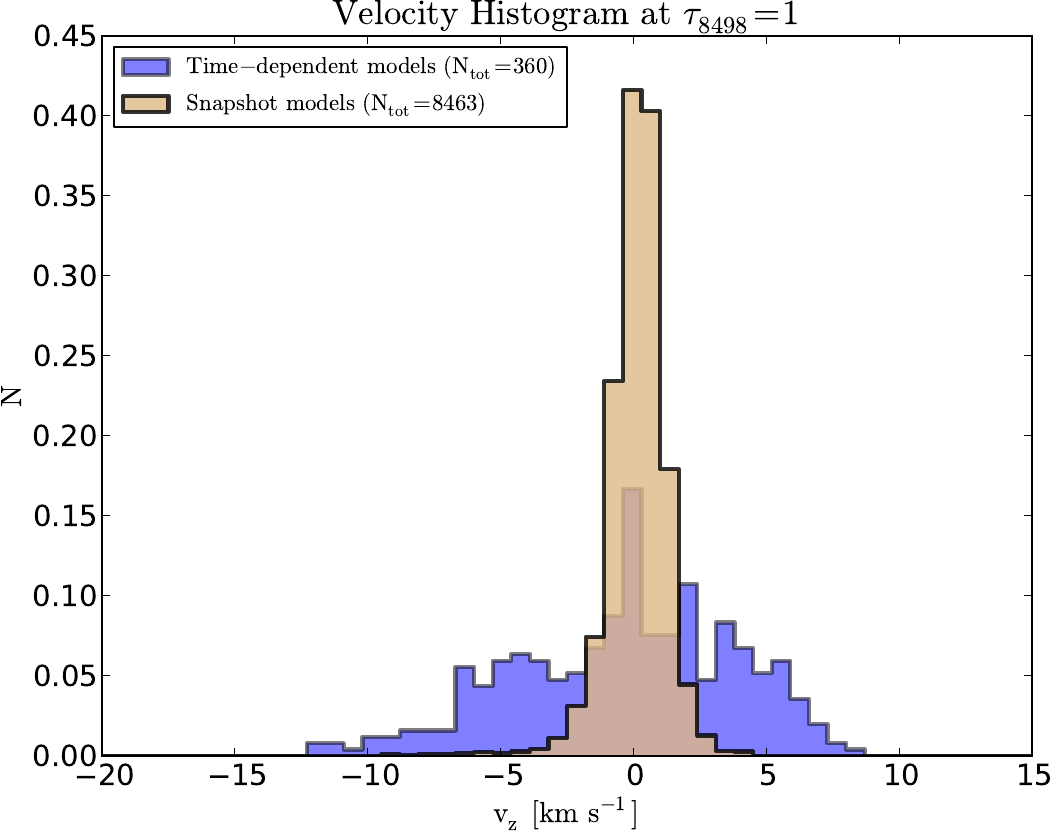}
\caption{The same as the left panel of Fig. \ref{fig:uves} but for the $8498$
  {\AA} line.}
\label{fig:hist_v8498}
\end{figure}

\begin{figure}[h!]
\centering%
\includegraphics[scale=0.5]{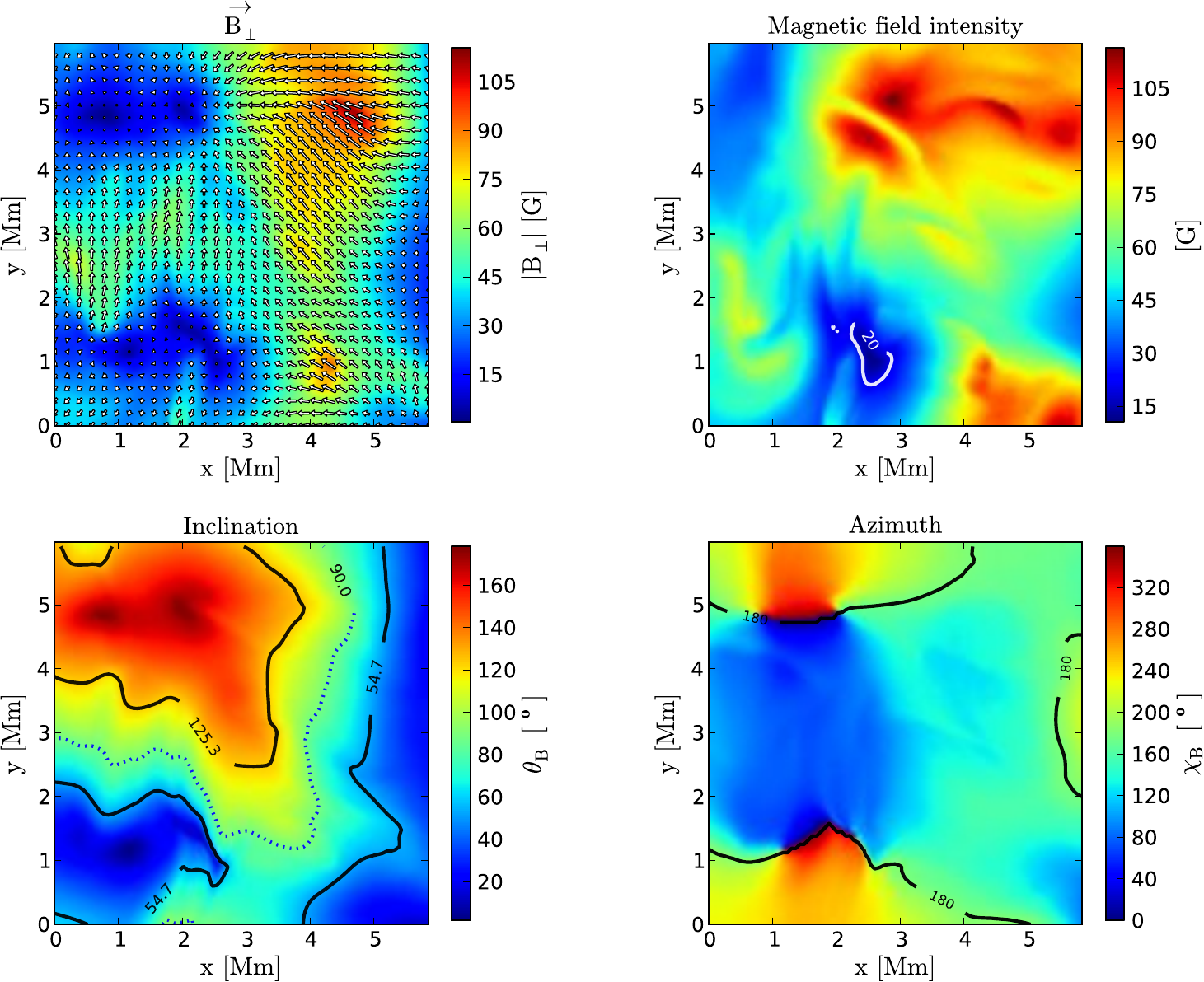}
\caption{Detailed description of the magnetic field in $\tau^{8542}=1$. \textbf{Upper left panel}: transversal component of the magnetic. \textbf{Upper right panel}: Intensity of the magnetic field. The white line encloses the area with less than $20$ G. \textbf{Lower left panel}: Inclination of the magnetic field. Solid iso-contours correspond to $\theta_B=\theta_V$ and $\theta_B=\pi-\theta_V$ (with $\theta_V$ the Van Vleck angle). \textbf{Lower right panel}: Azimut of the magnetic field and iso-contours of $\chi_B=180\deg$.    }
\label{fig:bpars8662}
\end{figure}

\begin{figure}[h!]
\centering%
\includegraphics[scale=0.34]{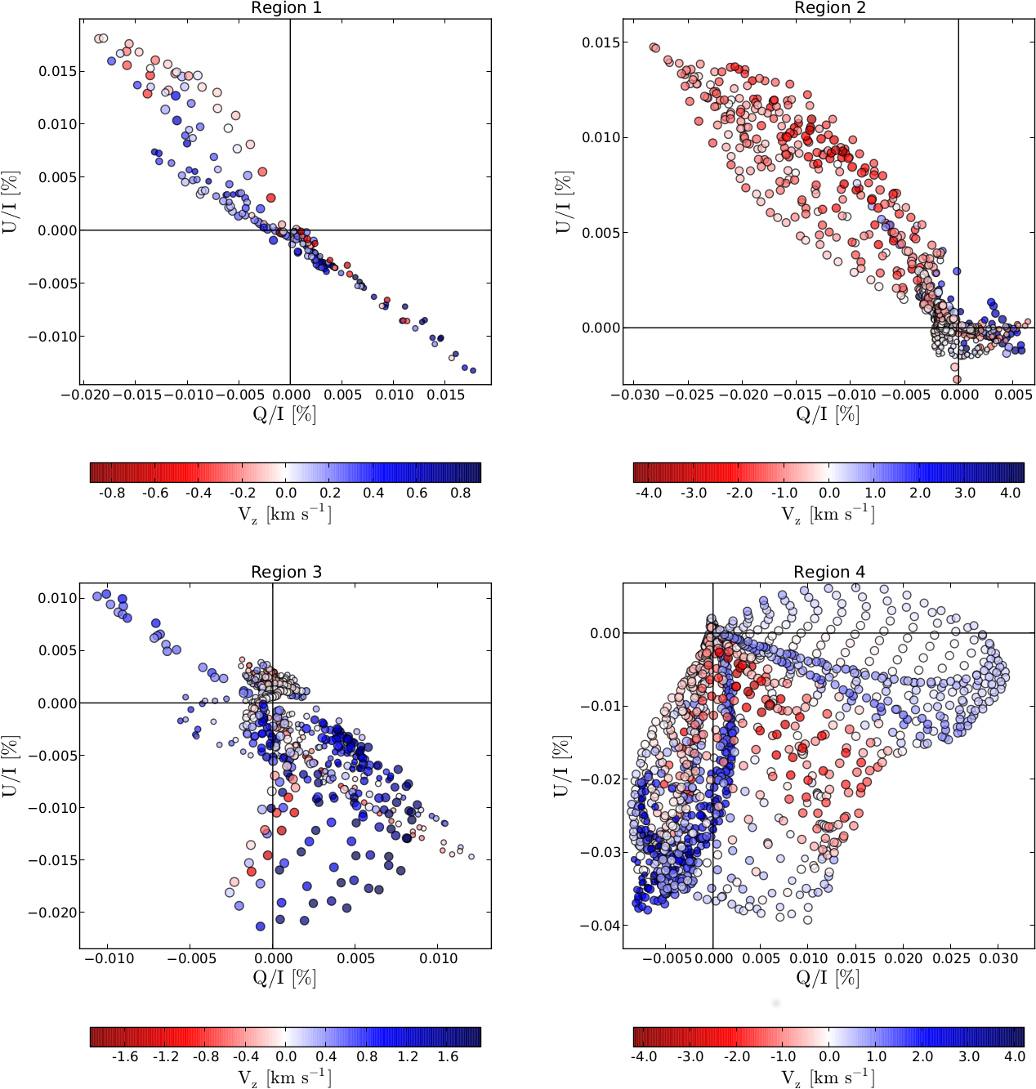}
\caption{The same as in Fig. \ref{fig:hdiag1}, but in $8498$ {\AA}. }
\label{fig:hdiag3}
\end{figure}


\begin{figure}[h!]
\centering%
\includegraphics[scale=0.36]{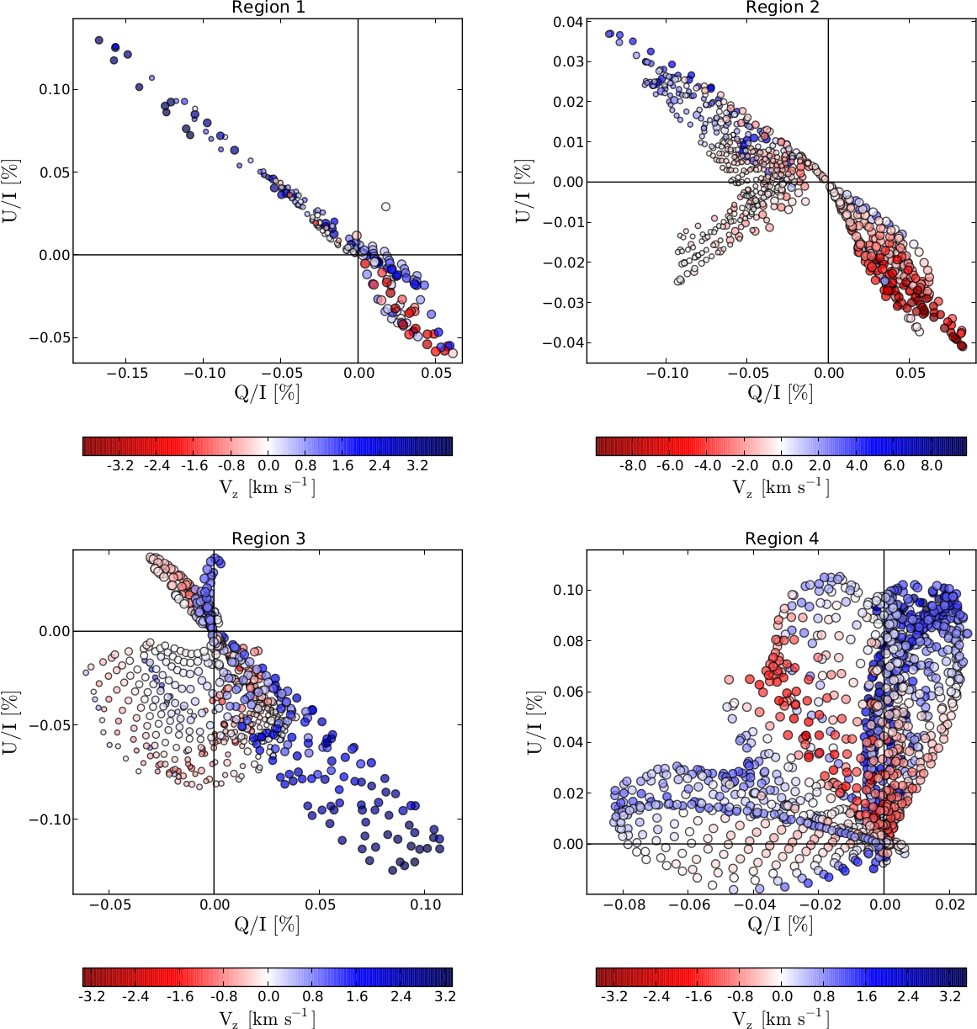}
\caption{ The same as in Fig. \ref{fig:hdiag1}, but in $8542$ {\AA}.}
\label{fig:hdiag4}
\end{figure}
\clearpage

\begin{figure}[h!]
\centering%
\includegraphics[scale=0.4]{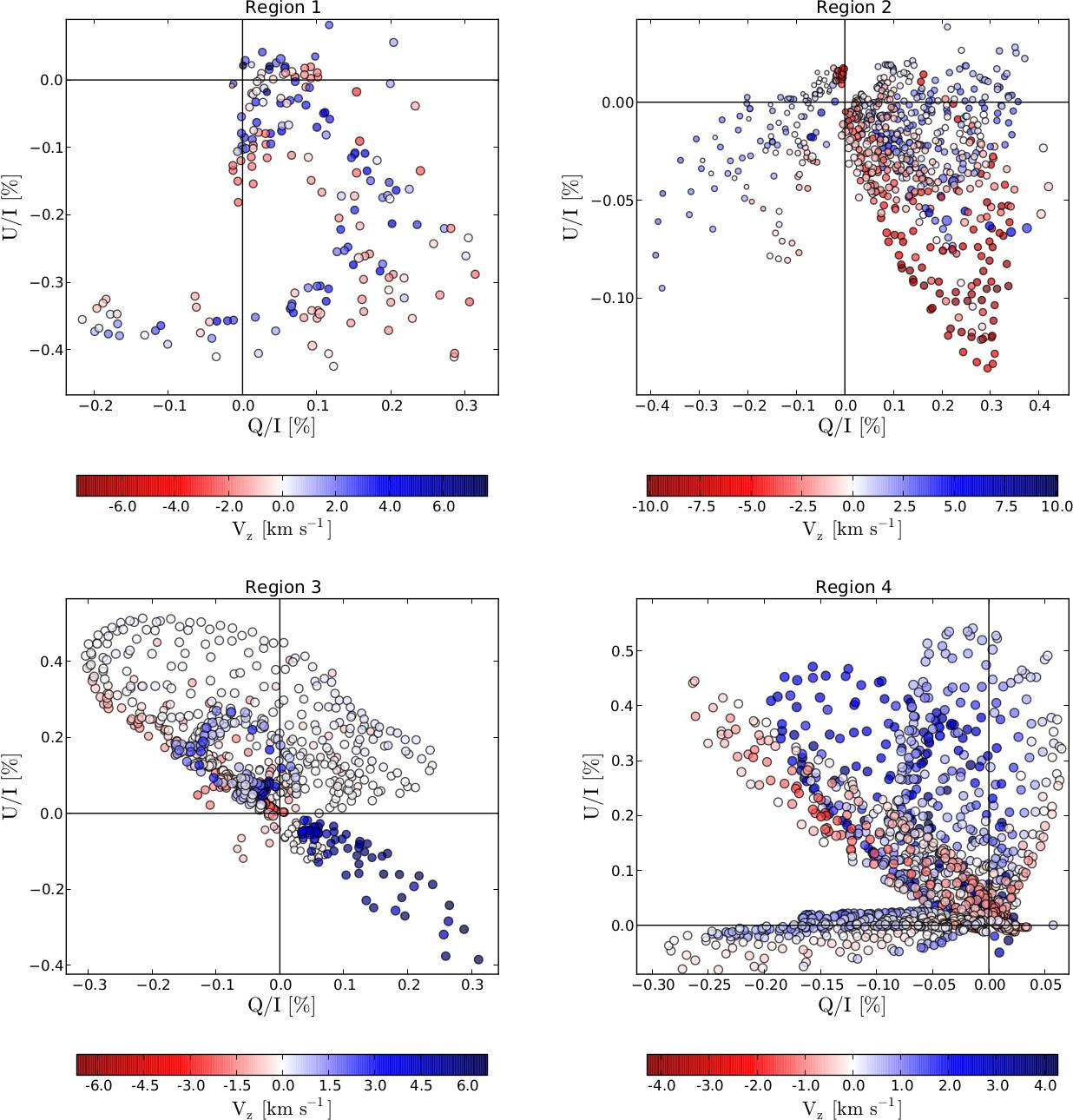}
\caption{ The same as in Fig. \ref{fig:hdiag1}, but in the K line.}
\label{fig:hdiag5}
\end{figure}
\clearpage
\section{Stereographic view of a cool chromospheric bubble.}\label{app:stereo}
\begin{figure}[h!]
        \centering
        \begin{subfigure}[a]{\textwidth}
                \centering
                \includegraphics[width=0.903\textwidth]{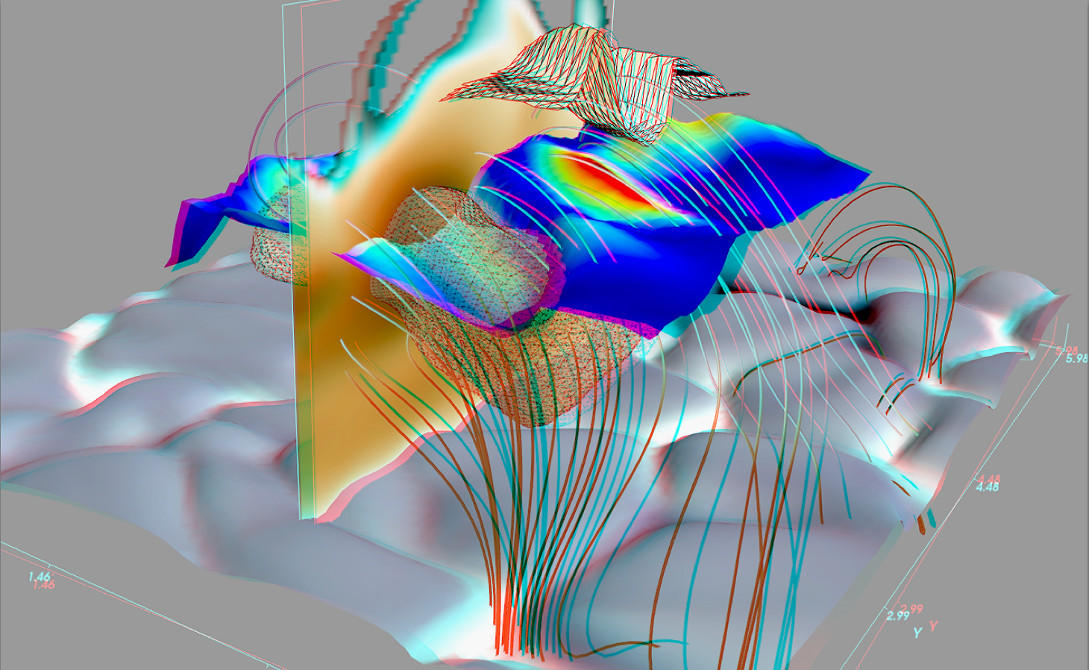}
        \end{subfigure}
\\
        \begin{subfigure}[b]{\textwidth}
                \centering
                \includegraphics[width=0.903\textwidth]{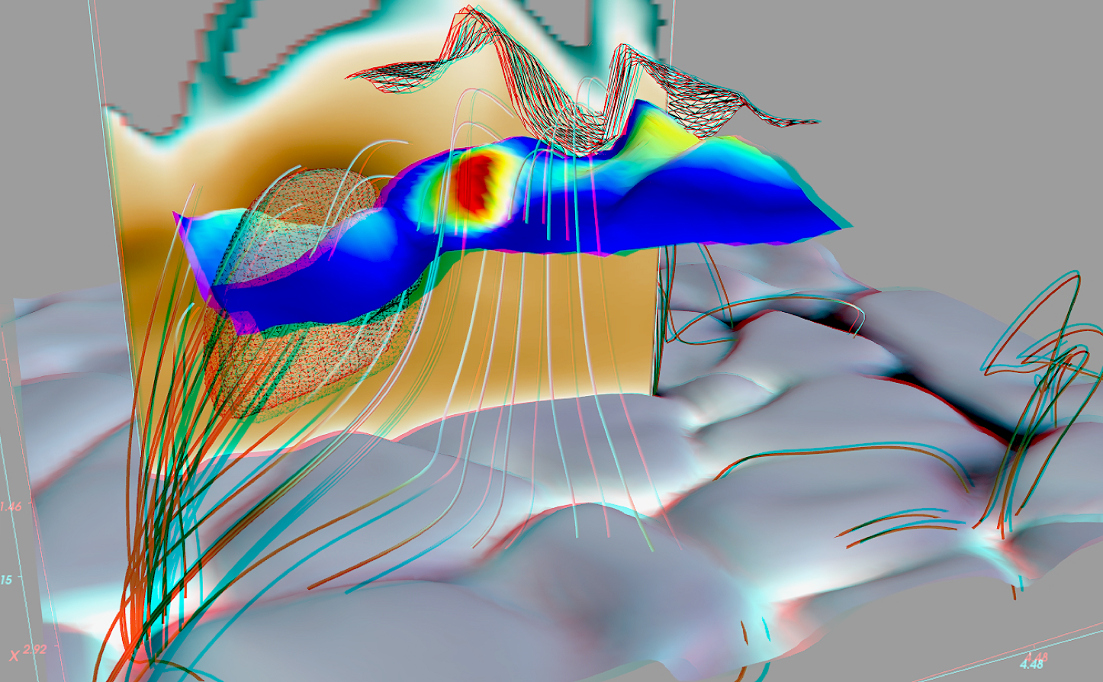}
        \end{subfigure}
        \caption{\footnotesize{Stereographic representation of a cool bubble
          (semitransparent volume). The vertical plane shows the
          temperature variation (darker brown is cooler). The 
          corrugated surface maps the heights where $\tau=1$ and its colour
          represents the emergent Q$/$I in the $8542$ {\AA} line
          (darker blue means smaller Q$/$I). The
          corrugated surface sketched with polygons represents the
          heights of $\tau=1$ for the K line.  Cyan-red glasses are
          required to see this figures correctly. Two-dimensional versions in next page.}
          }\label{fig:stereo}
\end{figure}
\begin{figure}[h!]
        \centering
        \begin{subfigure}[a]{\textwidth}
                \centering
                \includegraphics[width=0.753\textwidth]{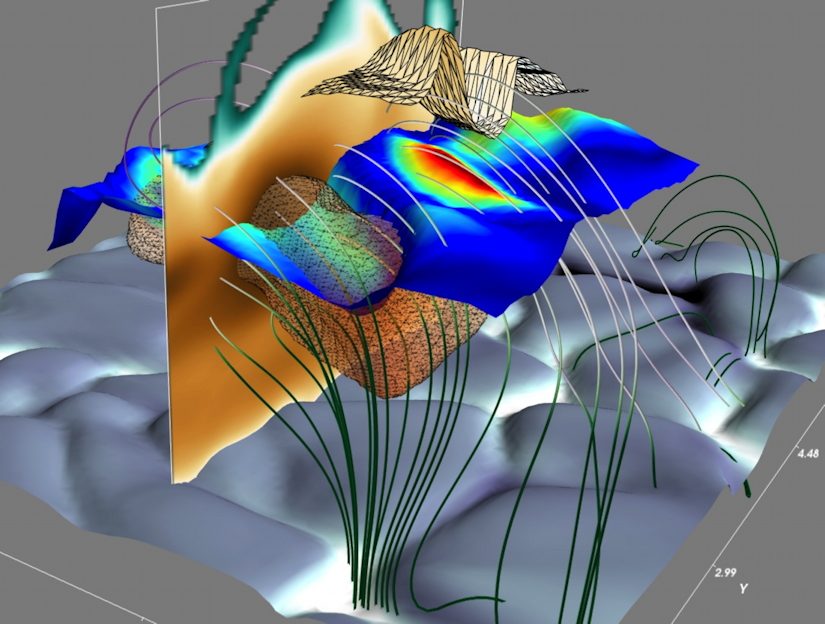}
        \end{subfigure}
\\
        \begin{subfigure}[b]{\textwidth}
                \centering
                  \includegraphics[width=0.753\textwidth]{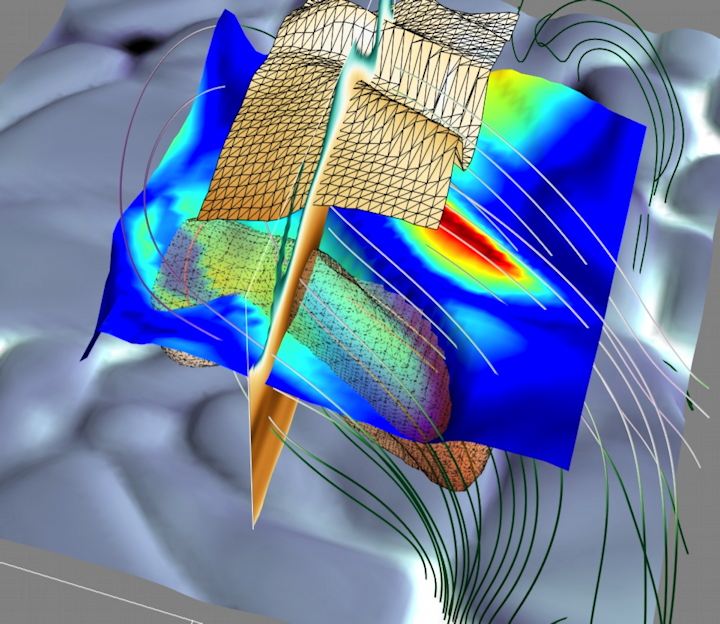} 
        \end{subfigure}
        \caption{\footnotesize{Non-stereographic visualization of a
            cool bubble (see the caption of the previous figure). The
            existence of such a cool chromospheric volume is tight to the depression in
            the formation layer of the considered spectral
            line. In the borders of the region, it can be
            appreciated that the linear polarization tends to vanish,
            so drawing a thermodynamic Hanle PIL. As explained in
            Chapter 6, this is in part favoured by the illumination of
            the K line from upper layers (polygonal surface) and by the so different
            physical conditions existing between verticals inside and outside the bubble.
            }
          }\label{fig:stereo2}
\end{figure}


\clearemptydoublepage

\def\jnl@style{\it}
\def\aaref@jnl#1{{\jnl@style#1}}

\def\aaref@jnl#1{{\jnl@style#1}}

\def\aj{\aaref@jnl{AJ}}                   
\def\araa{\aaref@jnl{ARA\&A}}             
\def\apj{\aaref@jnl{ApJ}}                 
\def\apjl{\aaref@jnl{ApJ}}                
\def\apjs{\aaref@jnl{ApJS}}               
\def\ao{\aaref@jnl{Appl.~Opt.}}           
\def\apss{\aaref@jnl{Ap\&SS}}             
\def\aap{\aaref@jnl{A\&A}}                
\def\aapr{\aaref@jnl{A\&A~Rev.}}          
\def\aaps{\aaref@jnl{A\&AS}}              
\def\azh{\aaref@jnl{AZh}}                 
\def\baas{\aaref@jnl{BAAS}}               
\def\jrasc{\aaref@jnl{JRASC}}             
\def\memras{\aaref@jnl{MmRAS}}            
\def\mnras{\aaref@jnl{MNRAS}}             
\def\pra{\aaref@jnl{Phys.~Rev.~A}}        
\def\prb{\aaref@jnl{Phys.~Rev.~B}}        
\def\prc{\aaref@jnl{Phys.~Rev.~C}}        
\def\prd{\aaref@jnl{Phys.~Rev.~D}}        
\def\pre{\aaref@jnl{Phys.~Rev.~E}}        
\def\prl{\aaref@jnl{Phys.~Rev.~Lett.}}    
\def\pasp{\aaref@jnl{PASP}}               
\def\pasj{\aaref@jnl{PASJ}}               
\def\qjras{\aaref@jnl{QJRAS}}             
\def\skytel{\aaref@jnl{S\&T}}             
\def\solphys{\aaref@jnl{Sol.~Phys.}}      
\def\sovast{\aaref@jnl{Soviet~Ast.}}      
\def\ssr{\aaref@jnl{Space~Sci.~Rev.}}     
\def\zap{\aaref@jnl{ZAp}}                 
\def\nat{\aaref@jnl{Nature}}              
\def\iaucirc{\aaref@jnl{IAU~Circ.}}       
\def\aplett{\aaref@jnl{Astrophys.~Lett.}} 
\def\apspr{\aaref@jnl{Astrophys.~Space~Phys.~Res.}}
\def\bain{\aaref@jnl{Bull.~Astron.~Inst.~Netherlands}} 
\def\fcp{\aaref@jnl{Fund.~Cosmic~Phys.}}  
\def\gca{\aaref@jnl{Geochim.~Cosmochim.~Acta}}   
\def\grl{\aaref@jnl{Geophys.~Res.~Lett.}} 
\def\jcp{\aaref@jnl{J.~Chem.~Phys.}}      
\def\jgr{\aaref@jnl{J.~Geophys.~Res.}}    
\def\jqsrt{\aaref@jnl{J.~Quant.~Spec.~Radiat.~Transf.}}
\def\memsai{\aaref@jnl{Mem.~Soc.~Astron.~Italiana}}
\def\nphysa{\aaref@jnl{Nucl.~Phys.~A}}   
\def\physrep{\aaref@jnl{Phys.~Rep.}}   
\def\physscr{\aaref@jnl{Phys.~Scr}}   
\def\planss{\aaref@jnl{Planet.~Space~Sci.}}   
\def\procspie{\aaref@jnl{Proc.~SPIE}}   

\let\astap=\aap
\let\apjlett=\apjl
\let\apjsupp=\apjs
\let\applopt=\ao


\phantomsection
\addcontentsline{toc}{chapter}{Bibliography} 
\bibliographystyle{apj}  
\footnotesize{\bibliography{main}}        
\clearemptydoublepage



\thispagestyle{plain}
\section*{\centering \LARGE{Agradecimientos}}
\addcontentsline{toc}{chapter}{Agradecimientos}
\selectlanguage{spanish}
 Tal vez el lector comparta la idea de que es importante sentirse
 inspirado para vivir con plenitud. No me
 refiero s\'olo a la inspiraci\'on cient\'ifica sino tambi\'en a esa
``inspiraci\'on vital'' que nos da impulso para evolucionar, para dar
 pasos cruciales en
 direcciones nuevas tanto en lo peque\~no como en lo profundo. Una v\'ia de inspiraci\'on insustituible es la que
 podemos obtener de otras personas que, con su manera de ser, sus virtudes y
 defectos, nos impulsan a ser la mejor versi\'on de nosotros
 mismos. Inspirar a
 otros es m\'as f\'acil cuando primero aprendemos a \textit{dejarnos
 inspirar}; por cada sutileza, por cada gesto o palabra, por personas
 cercanas y tambi\'en lejanas. Lo veo
 casi como una maestr\'ia. Esto tiene un indiscutible valor personal que
 no se suele apreciar, pero que recalco aqu\'i para
 expresar mi gratitud y hacer part\'icipes a todas las personas que de alguna
 manera me han inspirado y dado fuerzas durante estos largos a\~nos de tesis. 

En primer lugar agradezco a Andr\'es Asensio Ramos y a Javier Trujillo Bueno,
co-autores y directores de esta tesis, la
inestimable oportunidad que me han dado para formarme y aprender a su lado. Les
agradezco la ilusi\'on que pusieron en tenerme como
estudiante, la
energ\'ia y experiencia que han empleado en revisar nuestros trabajos,
el esfuerzo que
han puesto siempre en atenderme y el calvario que debe suponer tenerme
como ``pregunt\'on'' oficial a jornada completa. Me han enseñado
mucho. Adem\'as,
quiero agradecer a Andr\'es el apoyo
en mis primeros pasos como investigador con la beca de verano del
IAC. Fue precisamente con ese trabajo que tambi\'en pude colaborar e interaccionar con Arturo L\'opez Ariste y
Rafael Manso. Agradezco nuestras conversaciones sobre
f\'isica y otros ``desvar\'ios'' de la mente cient\'ifica, el apoyo recibido por Arturo y
la gran ayuda que me ha supuesto contar con la constructiva cr\'itica de Rafa
para orientar mis l\'ineas de discurso cient\'ifico a favor del campo (magn\'etico y no
magn\'etico). Gracias Rafa por ayudar a sentar las bases de trabajo
sobre las que se sustenta esta tesis y por contribuir
significativamente a su mejora. Ha sido tambi\'en una suerte y un
honor poder
contar con los conocimientos y la inspiradora
cercan\'ia de Fernando Moreno, Manolo Collados y Basilio Ruiz Cobo, ya
desde las clases en la universidad. Gracias a todos.

Damian Fabbian y Nacho Trujillo, amigos y compa\~neros de la vida,
gracias por estar ah\'i. Gracias Damian por los intercambios de ideas
sobre c\'omo lograr un mundo mejor. Gracias Nacho por tu amistad, por
compartir tu conocimiento y por mantenerme en contacto con la investigaci\'on mientras me debat\'ia
``entre la vida y la muerte'' con la gata de Schr\"odinger. Gracias por nuestros
di\'alogos sobre la \'unica revoluci\'on esencial, la de la mente y la psique.

Gracias Nayra Rodr\'iguez (siempre tan
positiva). Gracias Karla Pe\~na (vecina, por fin!, gracias por tu
apoyo). Gracias Sebas (eres un ejemplo a seguir). Gracias Carlos (siempre dispuesto a ayudar!). Gracias
Mireia (no dir\'e la palabra ``miau''). Gracias Christoph Kuckein
(hang loose!). Gracias
Manu. Gracias Judith Bakos. Gracias Anneta (Houston...hemos dejado de tener
un problema...cambio). Gracias Fernando Buitrago. Gracias Nazaret
Bello. Gracias Ariadna. Gracias Peter y
Jorge P\'erez Prieto por presentarme a Mr. Python. Gracias a todos por la
inspiraci\'on compa\~neros. 
 
Me promet\'i que ser\'ia breve pero no va a poder ser. Estoy especialmente
agradecido a Carlos Marrero, Sabrina Ruiz
Marcos, Paloma Garc\'ia Lij\'o y Mari\'an Vega, por haber sido tan
buenos amigos y mantenerme presente a\'un en la distancia. Gracias Miguel N\'u\~nez por salvarme de una lumbalgia asegurada. Gracias Mari Carmen
 Trujillo por tu positividad y por tus exquisitos platos. Thank you
 Sowmya Krishnamurthy for the scientific discussions and for your
 poppy seeds dessert recipe :D. I am also very grateful to the IRSOL staff for the support.

Muchas gracias a mi familia por hacer tambi\'en suyas mis penas y mis
alegr\'ias y por creer en m\'i. Gracias a mis padres, Melina Ram\'irez y Jose Manuel Carlin, y
a mi querida hermana Noa. Gracias Juli\'an, gracias Dafnis, gracias Ata, gracias
Cesi. Esto est\'a especialmente dedicado a todos ustedes. 

 Para finalizar quiero dar las gracias a alguien muy especial sin la que
 este trabajo se me hubiera hecho casi imposible. Gracias por inspirarme,
 por formar parte de esto y por darme razones para quedarme con la
 parte buena de las cosas. Literalmente tienes la \'ultima palabra en
 esta tesis, qu\'e mejor final que \'ese: gracias Camelia!


\clearemptydoublepage

\end{document}